\documentclass{iopjournal}

\usepackage[export]{adjustbox}
\usepackage{graphicx}
\usepackage{amssymb}
\usepackage{amsmath}

\newcommand{\br}{{\bf r}}
\newcommand{\bcr}{{\bf R}}
\newcommand{\bG}{{\bf G}}
\newcommand{\cc}{{\cal C}}
\newcommand{\cz}{{\cal Z}}
\newcommand{\cpy}{\hat{\cal P}_y}

\begin{document}

\begin{center}
  \textcolor{red}{\Large{~~~~~~~~J. Phys.: Condens. Matter {\bf 38}, 343002 (2026)}}\\
                    (Published: 1 September 2026) 
\end{center}

\articletype{Topical Review} 

\title{Artificial versus Natural Atoms:
\Large{The uncanny capability of the many-body Schr\"odinger equation to produce emergent 
behavior}}

\author{Constantine Yannouleas$^*$\orcid{0000-0002-6320-3484}}

\affil{School of Physics, Georgia Institute of Technology, Atlanta,
       Georgia 30332-0430, USA}

\affil{$^*$Retired}

\email{constantine.yannouleas@physics.gatech.edu}


\begin{abstract}
The paper reviews the theoretical and experimental progress achieved in the last 25 years
in understanding the novel physics of artificial atoms and molecules as arising from the
formation of Wigner molecules (WMs) of localized (to a stronger or lesser extent) fermionic
or bosonic particles, which are finite quantum analogs of the more familiar bulk Wigner
crystal. The term artificial atoms, as used here, encompasses a broad range of recently
fabricated quantum nanodevices and experimental apparatuses consisting of a finite number
of mutually repelling confined particles, including two-dimensional semiconductor and
moir\'e transition metal dichalcogenide quantum dots,
as well as trapped ultracold neutral atoms or ions. These nano-sized or micro-sized artificial
devices and apparatuses (in single well or multi-well of variable-shape arrangements) hold a
great promise for technological applications in the field of quantum information and quantum
computers, as well as for advances in fundamental many-body physics. Prominent quantum effects
of Wigner molecularization are the strong quenching of the spectral energy gaps, the
appearance of rovibational spectra (in analogy with natural molecules), entanglement, and
pinning due to an external perturbation. In high magnetic fields or at rapid rotation, WMs
provide an alternative theory to the fractional quantum Hall effect. The physics of Wigner
molecules is shown to derive from the solutions of the many-body Schr\"odinger equation (MBSE)
in the regime of strong interparticle correlations arising from the dominance of the potential
over the kinetic energy, or from a high magnetic field, as well as from a rapid rotation. With
the help of a hierarchical scheme of computational approaches involving group-theoretical
projection techniques beyond the mean field and exact configuration interaction, in parallel
to experimental investigations, WMs are shown to provide an ideal platform for investigating
the interplay between symmetry-preserving (stationary, referred to as rotating or sliding WMs)
and broken-symmetry (superposition-necessitating, referred to as pinned or static WMs)
solutions of the MBSE. In the process, a germane view of the phenomenon of symmetry breaking,
based exclusively on finite systems and referred to as emergent symmetry breaking, is
developed as a replacement to the formalistic spontaneous symmetry breaking that requires
invocation of a singular thermodynamic limit. Attention is drawn to the counterintuitive fact
that the very MBSE, which was inspired by de Broglie's undulatory matter waves and
successfully explained the shell structure of delocalized electrons in atomic physics, is
nevertheless capable of yielding contrasting solutions, which relate to corpuscular
geometries of localized particles. This unforeseen behavior classifies the recent
developments concerning Wigner molecularization and Wigner crystallization as an example of
weak emergence, and thus of an arrow of reductionist explanation according to Weinberg's
reasoning. Finally, this review, demonstrating the unexpected mathematical effectiveness
beyond original expectations of the Schr\"odinger equation, is dedicated as a tribute to its
100-year anniversary.
\end{abstract}

\tableofcontents

\section{Introduction: Scope of the review, controlling parameters, and overview of
essential concepts}
\label{intro}
\smallskip

\subsection{Scope of the review}
\label{intro_scop}
\smallskip

The manmade physical systems constituting the background for this topical review have been
realized recently due to phenomenal experimental advances. Amongst them the review will focus on
two-dimensional (2D) finite assemblies of $N$ strongly interacting particles (electrons, holes,
Coulombic ions, or neutral atoms) confined by an external potential under the influence or absence
of an applied magnetic field $B$ or rapid rotation $\Omega$. Such systems comprise semiconductor
quantum dots (QDs) \cite{dzur13,burk23,yann07}, periodic potential pockets in moir\'e transition
metal dichalcogenide (TMD) superlattices \cite{crom24,liu26,yann24}, and ultracold-atom and
heavy-ion traps \cite{joch12,joch24,yann25,thom15,monr21}. These nanosystems, being manmade, offer
unprecedented diversity and control of their properties due to the high degree of variability in
their physical parameters. Currently, they are attracting a lot of interest because of i) their
promise for potential applications in quantum-device fabrication, such as solid-state
qubits for quantum computers \cite{loss98,burk99,copp12,dzur13,copp14,corr21,kim21,urie21,
yann22,yann22.2,kim23,burk23,loss24,kim25}, and more generally in the field of quantum
information, and ii) they can exhibit highly correlated quantum many-body states
\cite{yann07,corr21,kim21,crom24,joch24,yann24,liu26} which are
emerging as a focus for present-day fundamental-physics explorations.
Such manmade nanostructures are often referred to as
``artificial'' atoms and molecules in order to distinguish them from the natural atoms and
molecules which mainly exhibit independent-particle physics associated with the aufbau
principle (which has the Bohr atom as the springboard).

For repulsive two-body interactions, the strong correlations in the artificial atoms and
molecules are related to particle localization and formation of crystalline-like
configurations, reminiscent of the phenomenon of Wigner crystallization (WC)\footnote{
Depending on the context, the acronym WC will also be used to refer to ``Wigner crystal''.}
in the bulk uniform electron gas \cite{wign34}. Finite systems associated with crystalline
architectures are referred to as ``Wigner molecules'' (WMs) \cite{wend96,yann07,yang07,corr21}
and their strong quantum nature contrasts with the classical Wigner crystal of point-like
electrons arranged in a bcc lattice. The corresponding process is often referred to as
``Wigner molecularization'' in lieu of ``Wigner crystallization''.

By going beyond the paradigm of intuition-based classical modeling \cite{wign34,kong02}, the
aim of this review is to reveal the quantum properties of the WMs as arising from the microscopic
solutions of the {\it many-body} Schr\"odinger equation (MBSE)
\cite{schr26,heis26,slat27,hyll29,bethe_book}.
An important corollary will be verifiable progress towards the resolution of the open questions
\cite{pwa_book,yann07,tasaki_book,mcke25} regarding the interplay of
emergent (hidden) and spontaneous (explicit) symmetry breaking,
as well as the role of
the concept of strong emergence in condensed-matter finite systems. (As is well known,
the concept of strong emergence was introduced \cite{ande72,pwa_book} in condensed-matter
physics by Philip W. Anderson to account for the phenomena of spontaneous symmetry breaking
(SSB) in the thermodynamic limit.)

\subsection{The Wigner parameter: Driving agent towards Wigner and Coulombic crystallization
at vanishing magnetic field}
\label{intro_rw}
\smallskip

In the last section of his seminal paper \cite{wign34} on interacting electrons in metals,
published in 1934, Eugene P. Wigner invoked a bold assumption which led to the prediction of
an electron crystal, a hypothetical at the time collective state corresponding to a
three-dimensional (3D) bulk lattice made out of localized electrons in a body-centered cubic
arrangement. Namely, Wigner contemplated a case when the electrons had no quantum kinetic
energy and thus they could be treated as strongly correlated classical particles localized at
the ``absolute minima of the potential energy''. Such a case could arise in the limit of a
dilute electron gas when the Wigner-Seitz radius tends to infinity ($r_s \rightarrow \infty$,
with $4\pi r_s^3/3$ being the volume occupied by a single electron).

Although a behavior not addressed in Ref.\ \cite{wign34}, it is paramount to note here that
the Wigner crystal breaks both the translational and rotational continuous symmetries of the
associated many-body quantum Hamiltonian. Indeed, building upon earlier publications (see,
e.g., Refs.\ \cite{yann07,shei21}, and references therein) the problematics and quantum
theory of symmetry breaking and symmetry restoration within the framework of condensed-matter
and atomic physics will be a major theme in this review.

Moreover, in the context of the present review,
it is important to recognize that Wigner's ``no kinetic
energy'' hypothesis for the electrons parallels the theory of the Born-Oppenheimer
approximation (BOA) \cite{born27}, widely used in chemistry. However, the justification for
the near-vanishing of the kinetic energy of atomic nuclei in chemical BOA calculations is
naturally assigned to the much larger mass of the nucleon (a proton, $m_p$, or a neutron,
$m_n \approx m_p$) compared to the free electron mass, $m_e$ ($m_p/m_e \approx 1836$).
Furthermore, due to the heavy mass, the ``no kinetic energy'' hypothesis is also immediately
relevant to the formation of ion Coulomb clusters and crystals formed in radio-frequency and
Penning traps by a finite assembly of atomic ions (e.g., $^{40}$Ca$^+$ or $^{24}$Mg$^+$) at low
temperatures; see, e.g., Refs.\ \cite{wine87,thom15,mori25}.

In the considerations above, the driving agents towards Coulombic crystallization were
identified as either the low electron density or the large mass. However, research in the
area of two-dimensional single and molecular quantum dots in the past 25 years and of
transition metal dichalcogenide (TMD) moir\'e superlattices in the last few years has revealed
that the agent that controls the propensity towards Wigner-like (or other Coulombic)
crystallization, under a vanishing magnetic field at zero absolute temperature, is a
composite parameter, referred to as the Wigner parameter in honor of Wigner and usually
denoted as $R_W$ \cite{yann07,yann99,yann23}.

The $R_W$ parameter is defined as the ratio of a characteristic value of the potential energy,
$P.E.$, associated with a pair of Coulombic particles (repelling interaction
${\cal Z}^2 e^2/\kappa r$), over the quantum kinetic energy, $K.E.$, of a single particle
participating in the system under consideration. Explicitly, one has
\begin{equation}
  R_W= \frac{P.E.}{K.E.} \propto \frac{{\cal Z}^2 e^2/(\kappa l_0)}{\hbar \omega_0}=
  \frac{{\cal Z}^2 e^2 \sqrt{m^*}}{\kappa \hbar^{3/2} \sqrt{\omega_0}}.
  \label{rw}
\end{equation}  
In the second and third step of Eq.\ (\ref{rw}), the $P.E.$ and $K.E.$ expressions employed
are those associated with the specific nanosystem of $N$ charged carriers trapped in a 2D
circular QD exhibiting (in polar coordinates) an external parabolic confinement
$m^* \omega_0^2 r^2/2$.
\begin{equation}
l_0=\sqrt{\hbar/(m^*\omega_0)}
\label{l0}
\end{equation}
is the oscillator length of the
parabolic confinement representing a characteristic length (size) of the system, $m^*$ is
the effective mass of the charge carriers (or atomic nuclei), and $\kappa$ is the
dielectric constant relevant to the nanosystem. Note that the kinetic energy
is $\hbar \omega_0/2$.

I note that, being defined as a ratio $P.E./K.E.$, the Wigner parameter is relevant to
other systems as well, including three-dimensional and one-dimensional (1D) ones. For
example in the case of the original Wigner crystal, $P.E. \propto 1/r_s$ and $K.E. \propto
1/r_s^2$, yielding $R_W \propto r_s$.
In the case of the finite system of a 2D parabolic QD [see Eq.\ (\ref{rw})], the low-density
regime is attained when $\omega_0 \rightarrow 0$, yielding $R_W \rightarrow \infty$ which
corresponds to the dominance of the potential energy. Large values of $R_W$ are also associated
with atomic nuclei and atomic ions due to their heavy mass, a fact which is consistent with
the proportionality $R_W \propto \sqrt{m^*}$ in Eq.\ (\ref{rw}).

Overall, a value $R_W > 1$
suggests a regime of strong correlations associated with charge carrier localization and
collective crystallization, whereas a value $R_W < 1$ suggests that the system under
consideration is closer to the physics associated with an assembly of independent particles
that behave as matter waves. Naturally, pure forms of classical crystallization (like the one
predicted by Wigner \cite{wign34}) or of uncorrelated matter-wave, independent-particle behavior
correspond to the limiting
values $R_W \rightarrow \infty$ and $R_W \rightarrow 0$, respectively. For values in-between,
a broad spectrum of intermediate quantum physical regimes do arise, as it will be elaborated
in this review.

Furthermore, Eq.\ (\ref{rw}) reveals that large values of $R_W$ are intimately related to
taking the classical limit by considering simply the near vanishing of the Planck constant ($\hbar
\rightarrow 0$). It is interesting that this property was recently proposed \cite{li25} to be
exploited in a potential scheme for overcoming the still open question in density functional
theory (DFT) of how to properly incorporate \cite{perd21,gori23,kotl06,jone25} symmetry
breaking and strong static correlations. 

Note that the Wigner parameter $R_W$ does not depend on the particle statistics. Indeed, it
applies equally well to both the cases of Coulombic fermionic and bosonic systems. In addition,
it can be extended to systems of trapped ultracold neutral atoms mutually repelling with a
contact interaction $g\delta(\br_1-\br_2)$ ($g>0$). In this case, the generalized Wigner
parameter is denoted as $R_\delta$ and is given by \cite{yann20}
\begin{equation}
R_\delta=\frac{g}{2\pi\Lambda^2\hbar\omega_0}=\frac{gM}{2\pi \hbar^2}.
\label{rd}
\end{equation}
$R_\delta$ expresses the strength, $g$, of the contact interaction associated with an area
$2\pi\Lambda^2$, relative to the zero-point energy, $\hbar\omega_0$, of the 2D harmonic
trap. $M$ is the mass of the ultracold atom and the oscillator length
$\Lambda=\sqrt{ \hbar/(M\omega_0) }$. Among others, relevant systems associated with
$R_\delta$ [as given in the final expression in Eq.\ (\ref{rd})] 
are the 1D assemblies of impenetrable (hard-core) bosons or fermions, designated collectively
as the Tonks-Girardeau gas and associated with the concept of fermionization; see, e.g., Refs.\
\cite{gira60,weis04,roma06,joch12}.

\subsection{The magnetic field and rapid rotating traps: Additional agents towards Wigner
crystallization}
\label{intro_mag}
\smallskip

Both the $R_W$ and $R_\delta$ parameters at vanishing applied magnetic field ($B=0$) and 
vanishing trap rotation (angular velocity $\Omega=0$), respectively, can be rewritten as
the ratio
\begin{align}
{\cal R}= \frac{\Delta E_{\rm int}}{\Delta E_{\rm sp}}, 
\label{calr} 
\end{align}
where $\Delta E_{\rm int}$ is a representative amount of repulsive energy and $\Delta E_{\rm sp}$
is an average energy spacing in the single-particle spectrum. For $B=0$, or $\Omega=0$, the
$\hbar\omega_0$ used in Eqs.\ (\ref{rw}) and (\ref{rd}) reflects indeed the average energy gap
between the single-particle states of the familiar 2D harmonic oscillator. In the case of a
Landau-level spectrum (Darwin-Fock oscillator \cite{darw31,fock28}; see Sec.\ \ref{dfll}
below), the cyclotron energy $\hbar\omega_c$ (magnetic-field case) or $2\hbar\omega_0$
(rotating-trap case) represent the energy spacing between Landau levels, and naively they
should be used for $\Delta E_{\rm sp}$, instead of $\hbar\omega_0$, in Eq.\ (\ref{calr}).
However, when the relevant many-body Hilbert space is restricted to the lowest Landau level
(LLL), the energy gap between the single-particle states vanishes due to the well-known
infinite degeneracy of the Landau levels; this is also referred to as single-particle
kinetic-energy quenching. Thus with respect to the pertinent dimensionless parameter that
controls Wigner-molecule formation in the LLL, the denominator in Eq.\ (\ref{calr}) must be
taken to be precisely zero, which results in all instances in an infinite value for
${\cal R}$. Interestingly, the single-outcome value of $R\rightarrow +\infty$ implies that
the LLL many-body case is preset for favoring the emergence of Wigner molecules (specifically
symmetry-preserving rotating Wigner molecules, RWMs; see Secs.\ \ref{lllrot} and \ref{lllfp}
below), independently of the strength or the type of the two-body interaction. In fact, in
addition to the Coulombic and contact-interaction cases, this qualitative prediction has
been confirmed by numerical calculations in the case of few fully spin-polarized LLL fermions
interacting via a dipole-dipole potential \cite{lewe07}. 

\subsection{The Schr\"odinger equation for a highly symmetric many-body Hamiltonian and its
solutions: hidden versus explicit symmetry breaking and the theory of symmetry restoration}
\label{intro_sb}
\smallskip

For a finite system of $N$ single-species $SU(2)$ Coulombic particles confined in a circular 2D
harmonic potential, the time-independent many-body Schr\"odinger equation in the absence of an
applied magnetic field can be written in dimensionless form as
\begin{equation}
  H^{\rm red}_{\rm MB} \Phi(\chi_1,\ldots,\chi_N) =E \Phi (\chi_1,\ldots,\chi_N),
\label{schreq}
\end{equation}
where the reduced many-body Hamiltonian is
\begin{equation}
   H^{\rm red}_{\rm MB} =
  \sum_{i=1}^N (-{\textstyle\frac{1}{2}} \nabla_i^2 + {\textstyle\frac{1}{2}} \br_i^2)+
  \sum_{i=1}^N \sum_{j>i}^N \frac{R_W}{|\br_i-\br_j|}, 
  \label{hmb_red}
\end{equation}
with the lengths being in units of $l_0$ and the energies in units of $\hbar \omega_0$;
$\chi_i= \br_i \sigma_i$, $i=1,\ldots,N$, with $\sigma_i=\alpha_i$ (spin up) or
$\sigma_i=\beta_i$ (spin down). 

From Eq.\ (\ref{hmb_red}), the central role played by the parameter $R_W$ is immediately
apparent. For $R_W \gg 1$, it is expected that the strongly correlated regime associated with
particle wave function localization and Wigner crystallization is reached. However, the exact
eigenstates of the Schr\"odinger equation preserve the symmetries of the many-body Hamiltonian
(and thus of the external potential). As a result, the corresponding single-particle densities
are circular and do not exhibit visible crystallization. In this case, for a finite $N$ when
exact or high-quality approximate solutions can be obtained, the Wigner molecularization in
the single-particle densities is hidden \cite{yann07}, but it remains very much present in the
intrinsic many-body correlations of the system (see below). This {\it underlying\/} ``hidden''
symmetry breaking is also referred to as ``emergent'' \cite{yann07,pape15} or ``obscure''
\cite{tasa94,tasaki_book}. As mentioned in Sec.\ \ref{intro_mag}, for a circular confinement,
the corresponding symmetry-preserving WMs are referred to as rotating WMs (RWMs).

The hidden broken symmetry underlying the WMs can become explicit in the single-particle
densities through the action of the environment, which in the technical computations is
minimally reproduced through consideration in the Hamiltonian $H^{\rm red}_{\rm MB}$ of an
additional symmetry breaking perturbing potential term $V_P$. It follows that the explicitly
broken-symmetry states are always (i.e., for both finite and bulk systems) a superposition
(wave packet) \cite{lowd62,yann07} of the unperturbed many-body symmetry-preserving
eigenstates. The corresponding symmetry-broken Wigner molecules, explicitly exhibiting $N$
localized particles (either fermions or bosons) in their single-particle densities, are
referred to as pinned (or static) WMs.

Anderson intuitively contemplated such a symmetry-breaking superposition, but only for bulk
systems in the thermodynamic limit ($N \rightarrow \infty$), when he introduced the hypothesis
that the low-energy part of the spectrum of the unperturbed many-body eigenstates resembles a
tower of quasi-degenerate states \cite{pwa_book,tasaki_book,kats23}, referred to in short as
a ``tower of states''\footnote{Up to now, the ``tower of states'' approach has been primarily
used for investigating the physics of antiferromagnets, with the Heisenberg lattice model
providing the governing Hamiltonian; see Refs.\ \cite{bern92,misg07}.}
For a finite system, the symmetry-breaking wave packet was first discussed by Per-Olov
L\"owdin in Ref.\ \cite{lowd62} in the context of his approach of restoring the broken
symmetry of unrestricted Hartree-Fock (UHF) solutions; see also Sec.\ 11.4.7 in Ref.\
\cite{rs_book} for investigations in the context of atomic nuclei, and, as a recent example,
see Ref.\ \cite{naza25} for an analysis in the context of heavy-ion collisions.

The symmetry-restoration approach, which connects symmetry-preserving wave functions to
symmetry-broken ones, and vice versa, is a major achievement of the many-body theory of finite
systems across several fields; see, e.g., Refs.\ \cite{peie57,rs_book,shei21} in the
context of nuclear physics, Refs.\ \cite{lowd55,lowd64,fuku81,maye80} in the context of
chemistry, and more recently Refs.\ \cite{yann02.2,yann07,shei21} in the context of
condensed-matter nanosystems and Refs.\ \cite{roma06,yann25} in the context of atomic and
molecular assemblies in ultracold rotating traps.

\subsection{Weak versus strong emergence}
\label{intro_emerg}
\smallskip

A crucial amplification made by Anderson to his seminal earlier paper in 1972 \cite{ande72} was
the association \cite{pwa_book,ande80} of the broken symmetry states with the concept of
emergence, leading to the reinvigorating of the philosophical theory of emergentism.   
This theory distinguishes between two varieties of emergence, namely, weak (epistemological)
and strong (ontological) emergence \cite{chal06,mcke25,elli20}. Ontology refers to the
nature of objects, whereas epistemology focusses on what is known about them. Here, for the sake
of succinctness, I use the following definitions \cite{mcke25,silb99}:
Weak emergence is associated with prediction that is ``possible in principle, but difficult in
practice'' whereas strong emergence corresponds to cases when prediction is ``impossible in
principle''.

In the context of this review, the ``principle'' is the many-body Schr\"odinger equation,
whereas the ``difficult prediction'' is the set of its exact and approximate solutions that
describe Wigner molecularization and Wigner crystallization,
including the interplay between symmetry-breaking and symmetry-restored wave functions. Put
differently, Wigner molecularization and ordered particle localization are accounted for (derived
from) within the framework of a more fundamental theoretical level, i.e., the many-body
Schr\"odinger equation. Such a state of affairs points to the following two conclusions: 1)
Wigner crystallization and its quantum analogs (including the fractional quantum Hall
effect, FQHE; see Secs.\ \ref{lllrot} and \ref{lllfp} below) belong to the domain of weak
emergence, and thus to the realm of reductionism \cite{wein87,land13}, and 2) the capability of the
mathematical form of the many-body Schr\"odinger equation to account for the physical reality
of such strongly correlated, collective behavior related to symmetry breaking is another
remarkable example of a science attribute that Wigner called ``The Unreasonable Effectiveness of
Mathematics in the Natural Sciences'' \cite{wign60}, and which Steven Weinberg described as
``... the enormous power of mathematical reasoning to explain not only idealized systems like
planets moving in their orbits, but ultimately everything'' \cite{wein87}.

It is worth noting that 1) Wigner's treatment in Ref.\ \cite{wign34}, based mainly on intuition,
does not involve solutions of the many-body Schr\"odinger equation, and thus it invites the
impression that the WC process (associated thus with spontaneous symmetry breaking \cite{ande72})
belongs to the category of strong emergence and 2) Robert Laughlin's intuitive formulation of
his famous wave function \cite{laug83,laug90,laug99} in the context of the $\nu=1/3$ FQHE
experimental discovery led him (and David Pines) to underestimate \cite{laug20} the capabilities
and the central role of the Schr\"odinger equation in condensed-matter physics and chemistry.
The present paper overcomes the methodological and numerical limitations of Ref.\ \cite{wign34}
(unavoidable in 1934) and thus it differs from the views expressed in Refs.\ \cite{ande72,laug20}
concerning the capabilities of the MBSE. At a more fundamental perspective, the present review
reveals a well-defined, but overlooked until now, ``arrow of explanation ... traced down to the
level of the quantum mechanics of electrons [and] atomic nuclei'', according to Weinberg's
reductionist reasoning \cite{wein87}. As such, the review is hoped to make a substantive
contribution to the ongoing investigations across several physics fields (as well as chemistry)
regarding the applied and fundamental-science aspects of symmetry breaking and/or restoration as
being rooted within the frameworks of emergence versus reductionism.

\subsection{Plan of the review}
\label{intro_plan}
\smallskip

Beyond the Introduction, the plan of the review is as follows:
Sec.\ \ref{metho} describes the methodologies employed, while detailed exposition of the
theoretical and experimental aspects of Wigner molecularization in artificial atoms and
related systems is contained in the three subsequent sections. Namely: 
Sec.\ \ref{resff} presents the physics of two-dimensional Wigner molecules under field-free
conditions, Sec.\ \ref{wmmagrot} elaborates on the phenomenon of 2D Wigner molecularization
under an applied magnetic field and in rotating traps (including the case of the lowest
Landau level), and Sec.\ \ref{res1d} discusses WM formation in one-dimensional finite systems. 
Secs.\ \ref{comm11} and \ref{comm12} are commentaries addressing broader fundamental-physics
aspects of Wigner molecularization and universality connections to other fields of science,
with the former discussing the analogies to chemistry and the latter the contrast between
undulatory MBSE solutions and MBSE solutions associated with corpuscular geometric architectures.
Additional commentaries, which relate the WM paradigm to some far reaching questions of quantum
physics (e.g., reductionism versus strong emergence), or simply provide a deeper explanation,
are inserted throughout the sections in
the review; they may be omitted at a first reading. The detailed organization of each section
can be seen in the listing of Contents. Finally, a summary and conclusions are given in Sec.\
\ref{conc}, whereas the three appendices provide brief guides to the literature regarding the
method of configuration interaction, the UHF approach in the form of the Pople-Nesbet equations,
and the formalism of symmetry restoration.

\section{Methods}
\label{metho}
\smallskip

\begin{figure}[t]
 \centering 
 \includegraphics[width=0.6\textwidth]{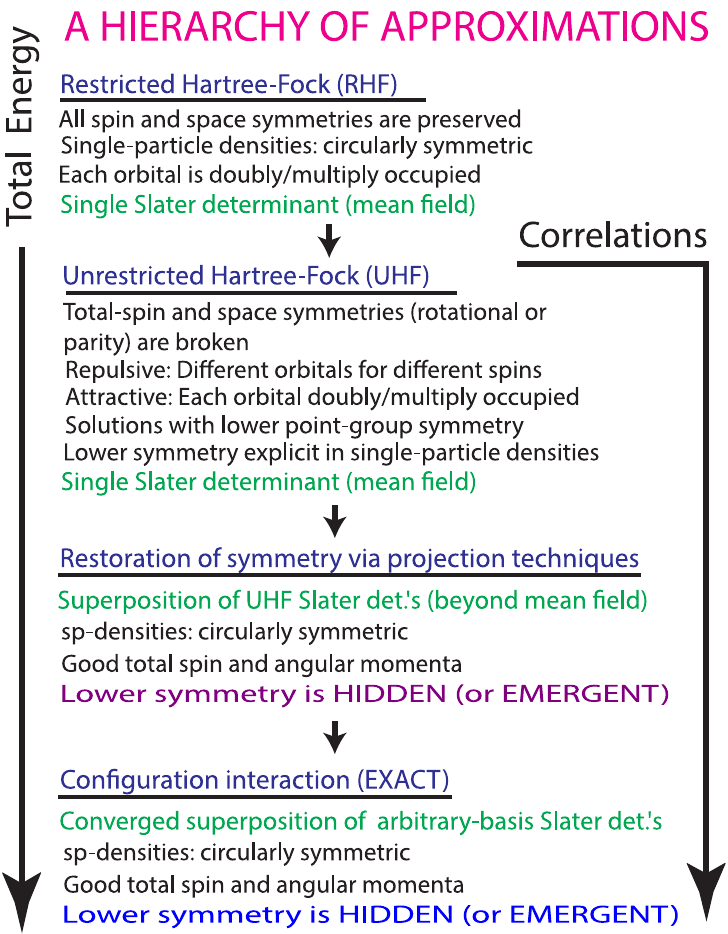}
\caption{
Synopsis of the method of hierarchical approximations, illustrating that a symmetry breaking at
the mean-field level (single UHF Slater determinant) must be accompanied by a subsequent
post-Hartree-Fock step of symmetry restoration, yielding a linear superposition of
non-orthogonal UHF Slater
determinants. The downward arrow on the left emphasizes that the total energy of the system is
lowered with each successive step (including most remarkably the symmetry restoration step
\cite{lowd62}), approaching thus from above the exact full-configuration-interaction total
energy; see Fig.~\protect\ref{He-qd-ener} below for a simple example. The arrow on the right
emphasizes that the steps beyond the restricted Hartree-Fock introduce correlations.
Reprinted figure with permission from Ref.\ \cite{yann07},
Copyright (2007) by the Institute of Physics and IOP Publishing.
}
\label{hiera}
\end{figure}

\subsection{A hierarchy of approximations establishing a bridge between broken-symmetry and exact
solutions}
\label{hieras}
\smallskip

Fig.\ \ref{hiera} presents a synoptical chart of the hierarchy of approximations that establish
the bridge between broken-symmetry and exact (symmetry preserving) solutions of the many-body
Schr\"odinger equation, in particular for the case of charge carriers trapped in 2D semiconductor
quantum dots (electrons) or in QDs formed in TMD moir\'e superlattices (both electrons and holes).
(A similar synopsis can also be envisaged for the case of other fermionic or bosonic systems,
e.g., neutral atoms and ions in ultracold traps.)
This hierarchical methodology yields lower-energy wave functions at each level of approximation
(a fact portrayed on the left side of the figure by a vertical downward arrow).
It is also referred to often as the ``two-step'' method of symmetry breaking and subsequent
symmetry restoration.

The starting approximation level (being associated with a higher energy because of the absence
of correlations) is occupied by the restricted Hartree-Fock (RHF), which is characterized
by the restriction of double occupancy (with antiparallel spins) for each space orbital. The
RHF wave function is a single Slater determinant that embodies the concept of a ``central mean
field.'' All the spin and space symmetries of the governing many-body Hamiltonian are preserved
by the RHF. Thus, for 2D quantum dots, the RHF single-particle densities [known also as charge
densities (CDs) or electron densities (EDs)] do exhibit the symmetries of the potential
confinement.

The next level of approximation consists of the unrestricted Hartree-Fock, which uses
{\it different space orbitals for the two different spin directions\/} to build a
single Slater determinant associated with a ``non-central mean field.'' The UHF preserves
the total-spin projection, but permits the breaking of the total-spin and continuous space
symmetries (i.e., rotational symmetries or parity). The broken-symmetry UHF single-determinantal
solutions, however, may still exhibit characteristic lower symmetries. For example, for 2D QDs
with a circular confinement, the UHF solutions exhibit point-group symmetries, which can be
explicitly identified in the charge densities.

Succeeding approximations restore the broken symmetries via projection techniques. This
post-HF second overall step of symmetry restoration goes beyond the mean field approximation
by producing a many-body wave function $|\Phi^{\rm PRJ}\rangle$ [often referred to as projected
(PRJ) wave function] which is a {\it linear superposition\/} of non-orthogonal Slater
determinants [see detailed account in Sec.\ \ref{2eqd} and Commentary 2 (Sec.\ \ref{comm2})
below].
The wave function $|\Phi^{\rm PRJ}\rangle$ preserves all the symmetries of the governing many-body
Hamiltonian; it has good total spin and angular momentum quantum numbers, and consequently the
continuous rotational symmetries (including the circular symmetry of the EDs in the case of
parabolic 2D QDs) are restored.

However, keeping with circular 2D confinements, the lower (i.e., point-group) spatial symmetry
arising at the UHF broken-symmetry level (associated with the first step of the two-step
approach) does not vanish into thin air. Instead, it becomes {\it intrinsic\/} or {\it hidden\/},
and it can be uncovered with the help of conditional probability distributions (CPDs), 
which are second-order correlations and are defined as (within a proportionality constant)
\begin{equation}
P({\bf r},{\bf r}_0) =
\langle \Phi^{\rm PRJ} |
\sum_{i \neq j}  \delta({\bf r}_i -{\bf r})
\delta({\bf r}_j-{\bf r}_0)
| \Phi^{\rm PRJ} \rangle,
\label{cpds}
\end{equation}
where $\Phi^{\rm PRJ} ({\bf r}_1, {\bf r}_2, \ldots, {\bf r}_N)$
specifies the many-body projected wave function under consideration.

In case that the intrinsic spin distribution of the localized fermions needs to be delved
into, spin-resolved two-point correlation functions [spin-resolved CPDs, (SR-CPDs)] must be
considered. The SR-CPDs are defined as follows:
\begin{equation}
P_{\sigma\sigma_0}({\bf r}, {\bf r}_0)=  \langle \Phi^{\rm{PRJ}} |
\sum_{i \neq j} \delta({\bf r} - {\bf r}_i) \delta({\bf r}_0 - {\bf r}_j)
\delta_{\sigma \sigma_i} \delta_{\sigma_0 \sigma_j}
|\Phi^{\rm{PRJ}}\rangle.
\label{sponcpd}
\end{equation}

The spin-resolved CPD returns the spatial probability distribution of finding a
second electron at position $\br$ with spin projection $\sigma$ given that 
a first electron is located (fixed) at ${\bf r}_0$ with spin projection
$\sigma_0$; the individual spin projections $\sigma$ and $\sigma_0$ can be either
up $(\uparrow$) or down ($\downarrow$). The space-only (spin-unresolved) CPD in Eq.\
(\ref{cpds}) has
an analogous meaning, but without consideration of the spins.

Most remarkably, signatures of the intrinsic lower symmetry occur in the excitation spectra
which are constrained by the symmetry breaking. For example, in circular quantum dots under a
vanishing magnetic field and for large values of $R_W$, the excitation spectra (determined
both microscopically and through symmetry restoration) exhibit a ro-vibrational character
conditioned by the intrinsic Wigner-molecular structure (quasi-rigid rotor behavior)
\cite{yann00,yann04.2}.

\begin{figure}[t]
\centering\includegraphics[width=13.0cm]{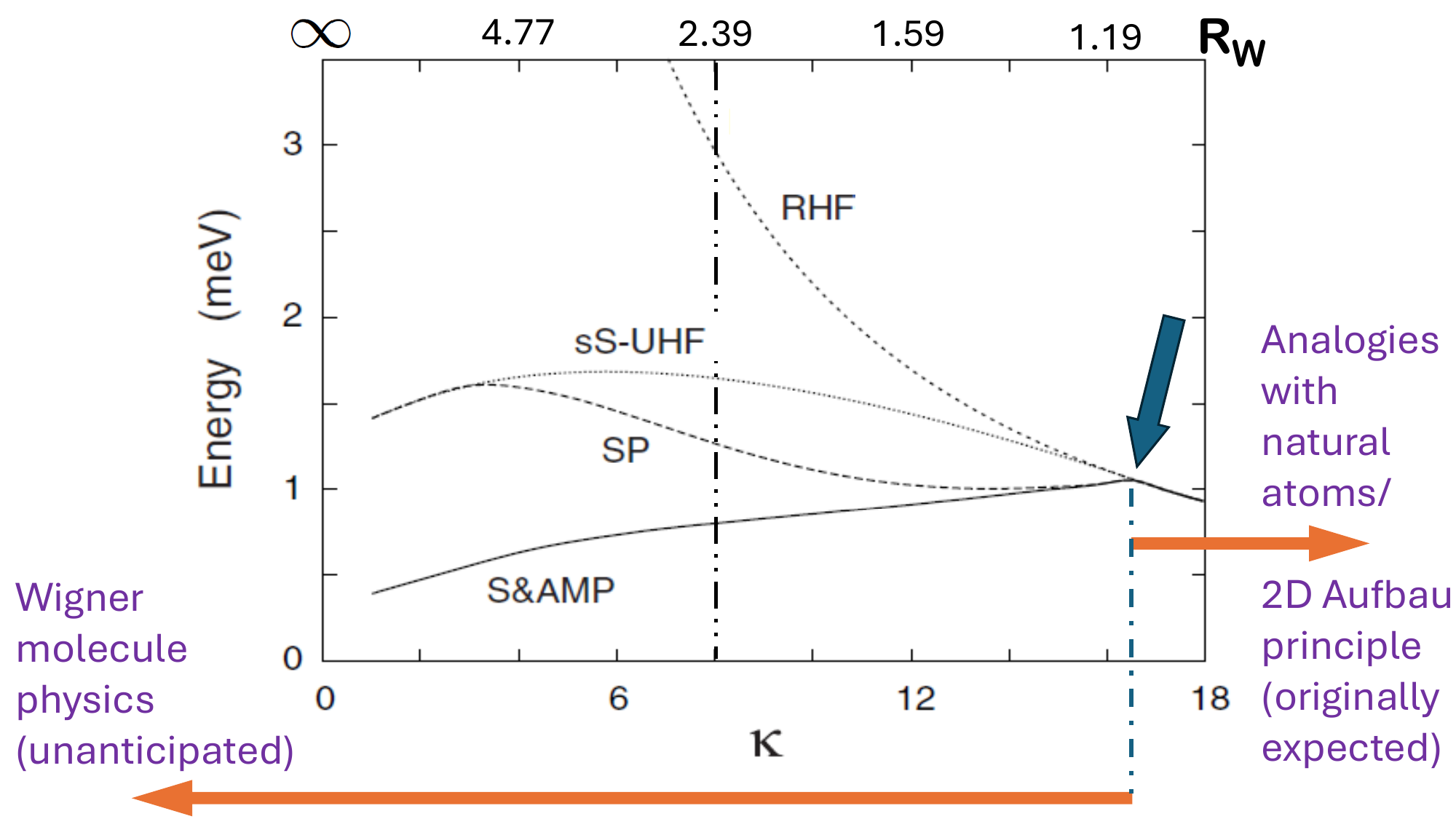}
~~~~~~~~~~~\\
\caption{
He-QD at vanishing magnetic field (two electrons confined in a 2D parabolic QD):
Energy differences between the ground states of successive approximations and the exact one,
plotted vs. $\kappa$ (the dielectric constant, see the bottom horizontal axis) or $R_W$
(the Wigner parameter, see the top horizontal axis).
From top to bottom, the curves correspond to the RHF, the spin and space
unrestrited Hartree-Fock (sS-UHF), and two subsequent steps of symmetry restoration,
i.e., the total-spin projected (SP) alone, and the combined total-spin and
angular-momentum projected (S\&AMP) approximation. The bifurcation point between
the RHF and sS-UHF curves is denoted by an arrow; it happens at $R_W=1.16$.
To the right of this point, one expects the validity of the Aufbau principle and
analogies with the natural atoms \cite{kouw98,kast93}. To the left of the same point,
new physics emerges associated with formation of quantum Wigner molecules \cite{yann07}. 
The curve labeled RHF represents also the correlation energy [see Eq.\ (\ref{ecorr})]. The
parameters were chosen as follows: parabolic confinement $\hbar \omega_0=5$ meV and
effective mass $m^*=0.067 m_e$ (GaAs).
The dashed vertical line indicates the value $\kappa=8$ that was used in the calculation
of the energies and charge densities displayed in Fig.\ \ref{succapp}.
Reprinted figure with permission from Ref.\ \cite{yann02.2},
Copyright (2002) by the Institute of Physics and IOP Publishing.
}
\label{He-qd-ener} 
\end{figure}

As pointed out in the chart displayed in Fig.\ \ref{hiera}, the mean-field HF equations are
non-linear and the symmetry breaking is associated with bifurcations that appear
in the HF total energies (for an example, see Fig.\ \ref{He-qd-ener}). Initiated
by the work of David J. Thouless \cite{thoubook}, such bifurcations have been
extensively investigated in chemistry under the general theme of HF instabilities;
see, e.g., Refs.\ \cite{pald67,fuku81,pald85}. A consequence of the bifurcations
is the fact that the HF symmetry-broken wave functions have a lower energy than
the symmetry-preserving ones; this behavior is known as L\"owdin's symmetry
dilemma \cite{lowd63}.

The happenstance of a HF bifurcation cannot be foreseen {\it a priori\/} from a mere
inspection of the many-body Hamiltonian itself; it is a genuine many-body effect that
may be revealed only {\it a posteriori\/} through the UHF solutions themselves (if
obtainable) or through appropriate experimental signatures.
I note that the wave functions $|\Phi^{\rm{PRJ}}\rangle$,
generated in the symmetry-restoration second step, recover the
linear behavior\footnote{
The non-linear character of the integro-differential HF equations contrasted to the linear
character of the full configuration interaction equations, as well as to the wave functions
generated via projection techniques, is succinctly spelled out in the discussion section of
Ref.\ \cite{lowd55.2}.}
associated with the full configuration interaction\footnote{
In the field of artificial atoms and molecules, the FCI approach is also referred to as
``exact diagonalization'' or ``numerical diagonalization'' (of the many-body Hamiltonian).}
(FCI, see Sec.\ \ref{fci} below) approach for solving the many-body Schr\"{o}dinger equation,
e.g., they can be expressed in a Hilbert space as a superposition of Slater determinants
that involve $np-nh$ excitations with\footnote{
Such Slater determinants fall outside the scope of Thouless' theorem \cite{thou60} which
states that any Slater determinant can be expressed as a superposition of other Slater
determinants that involve only $1p-1h$ excitations and {\it vice versa.\/}}
$2 \le n \le N$, in addition to the $1p-1h$ excitations.

\begin{figure}[t]
\centering
\includegraphics[width=14.6cm]{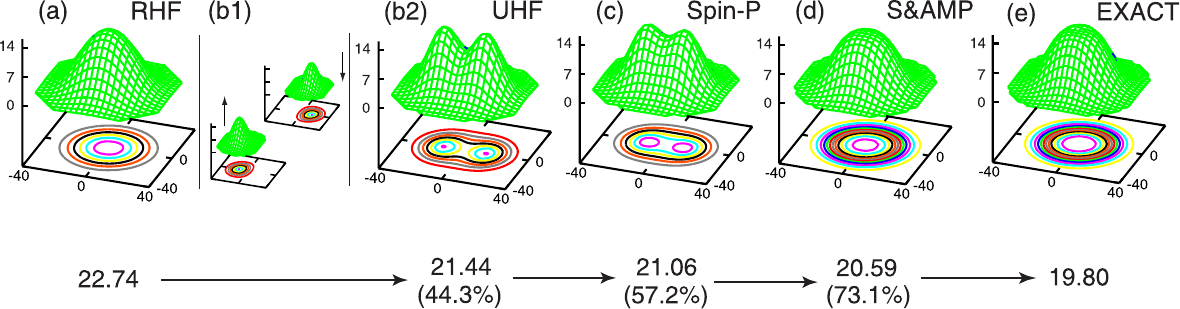}\\
~~~~~\\
\caption{
Successive approximations for the singlet ground state of a two-electron parabolic QD
with $R_W=2.39$ under a vanishing magnetic field. The associated energies (in meV) are
listed at the bottom. (a) Electron density of the RHF solution, exhibiting preservation
of the circular symmetry of the potential confinement. The correlation energy,
$E_{\rm corr} = 2.94$ meV, is given [see Eq.\ (\ref{ecorr})] by the difference between the
energy of this state and that of the exact solution [diplayed in panel (e)]. (b1) and (b2)
The two occupied spin-up and spin-down space orbitals (modulus square) of the symmetry-broken
``singlet'' UHF solution (in b1), with the associated total electron density having a
non-circular dumbbell-like shape (displayed in b2). The UHF energy captures 44.3\% of the
correlation energy.
(c) Electron density of the spin-projected (Spin-P) singlet ($S=S_z=0$), exhibiting broken
spatial symmetry, but with an additional gain of correlation energy.
(d) The ED density of spin-and-angular-momentum projected (S\&AMP) state demonstrating the
restoration of the circular symmetry. This state captures 73.1\% of the correlation energy.
The choice of parameters is: energy gap of parabolic confinement $\hbar \omega_0 = 5$ meV,
dielectric constant $\kappa = 8$, and effective mass $m^* = 0.067m_e$ (GaAs). Lengths are in
nm and the electron densities in $10^{-4}$ nm$^{-2}$.
Reprinted figure with permission from Ref.\ \cite{yann07},
Copyright (2007) by the Institute of Physics and IOP Publishing.
}
\label{succapp}
\end{figure}

The increasing weight of quantum correlations accounted by the successive steps of the
method of hierarchical approximations is illustrated further by the downward vertical
arrow on the right of Fig.\ \ref{hiera}. Indeed, as both of the two downward-pointing
arrows in Fig.\ \ref{hiera} suggest, starting with the broken-symmetry UHF solution,
each further approximation captures successively a larger fraction of the correlation
energy, which is expressed \cite{lowd58} as the difference between the
restricted Hartree-Fock and exact ground-state energies, i.e.,
\begin{equation}
E_{\rm corr}=E_{\rm RHF} - E_{\rm exact}.
\label{ecorr}
\end{equation}
A specific example of this process for a circular QD is given in Figs.\ \ref{He-qd-ener} and
\ref{succapp}.

In addition, Fig.\ \ref{succapp} illustrates how the approximate charge density (and thus
the corresponding wave function) for the same QD approaches in successive steps the
symmetry-preserving exact one, in spite of the intermediate step (UHF step) of breaking of the
circular symmetry which exhibits a qualitative, as well as quantitative, difference between
the broken-symmetry\footnote{
The earliest publication that reported broken-symmetry UHF solutions for 2D QDs at zero
magnetic field appears to be Ref.\ \cite{yann99}. For broken-symmetry UHF solutions in the
lowest Landau level, see also Ref.\ \cite{koon96}.}
and exact charge densities; contrast Fig.\ \ref{succapp}(b2) and Fig.\ \ref{succapp}(e).

\begin{figure}[t]
\centering
\includegraphics[width=14.8cm]{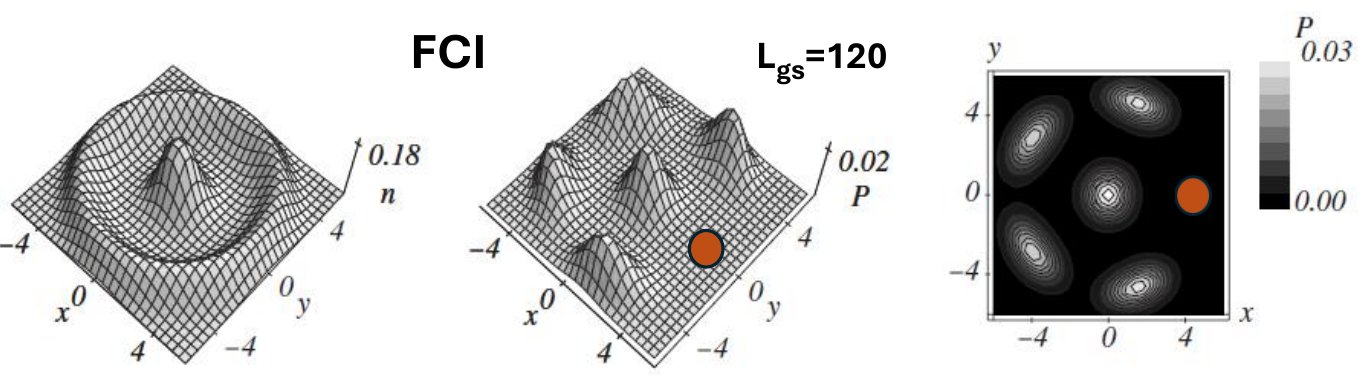}\\
~~~~~\\
\caption{
An example of the hidden (intrinsic) lower symmetry that can be extracted from an FCI
calculation. The case portrayed is that of the ground state (with total angular momentum
$L_{\rm gs}=120$) of $N=6$ scalar bosons interacting via a repulsive Coulomb potential
in a rapidly rotating trap. The many-body Hilbert space corresponds to the lowest Landau
level, namely, the angular velocity $\Omega$ of the trap equals the confinement
frequency $\omega$ of the trap, $\Omega=\omega$ \cite{yann07.3}.
(a) The single-particle density [$n({\bf r})$], (b) the corresponding CPD
$\left[ P({\bf r},{\bf r}_{0}) \right]$ [see Eq.\ (\ref{cpds})] in a 3D plot, and (c) the CPD
in a contour plot.
The red dots in the CPD plots indicate the reference (fixed) point ${\bf r}_0$. I note that
the single-particle density in (a) is circularly symmetric. However, the CPDs [(b) and (c)]
reveal a lower intrinsic point-group symmetry that can be denoted as a (1,5) nested polygonal
ring, with one boson at the center of the trap and 5 bosons in the periphery forming a regular
pentagon. Lengths in units of $\Lambda=\sqrt{\hbar/(M\omega_0)}$, with $M$ being the mass of
the scalar bosons. The vertical scales are in arbitrary units.
Reprinted figure with permission from Ref.\ \cite{yann07.3},
Copyright (2007) by the American Physical Society.
}
\label{pra_2007}
\end{figure}
\smallskip

\subsubsection{\small{{\bf Commentary 1.}} Unrestricted HF versus restricted HF: Some history
and areas of application.}
\label{comm1}
The HF method has a history of applications to natural atoms in conjunction with the
imposition of spherical symmetry, i.e., in its RHF version. This is due to the very strong
attraction of the central atomic nucleus and the small size of the atom, which has the
result that the effects of symmetry breaking are rather small and can be neglected
\cite{kohn00,nait21}. This state of affairs contrasts sharply with the case of
artificial atoms which can easily exhibit symmetry-breaking effects (see, e.g., Fig.\
\ref{He-qd-ener}) due to the high degree of variability in their physical parameters (e.g.,
the size of the nanosystem). Natural atoms exhibit physics related to symmetry breaking
and formation of WMs only in exotic cases, like the case of high angular momentum doubly
excited He and alkaline-earth atoms \cite{kell80,berr89,eich25}. Physics associated with
UHF symmetry-broken solutions were studied in the context of quantum chemistry for the case
of natural molecules (in particular, of dissociating or stretched molecules, e.g., the
stretched Hydrogen molecule, see the Coulson-Fisher transition point \cite{coul49}).
In this respect, the title ``Unrestricted Hartree-Fock theory and its applications to
molecules and chemical reactions'' of the paper \cite{fuku81} by Fukutome is suggestive.

\subsection{Importance of the full configuration interaction method}
\label{fci}
\smallskip
   
An alternative approach for studying the emergence of the crystalline regime in a finite
system is the full configuration interaction
\cite{shav98,szabo_book,yann07,ront06,yann15,erca21.2};
a guide to related background literature is given in Appendix \ref{a1}.

In the CI method, one writes the many-body wave function 
$\Phi^{\rm CI}_N ({\bf r}_1, {\bf r}_2, \ldots , {\bf r}_N)$ as a linear
superposition of Slater determinants 
$\Psi^N({\bf r}_1, {\bf r}_2, \ldots , {\bf r}_N)$ that span the many-body
Hilbert space and are constructed out of the single-particle 
{\it spin-orbitals\/} 
\begin{equation}
\chi_j (x,y) = \varphi_j (x,y) \alpha, \mbox{~~~if~~~} 1 \leq j \leq K,
\label{chi1}
\end{equation}
and 
\begin{equation}
\chi_j (x,y) = \varphi_{j-K} (x,y) \beta, \mbox{~~~if~~~} K < j \leq 2K, 
\label{chi2}
\end{equation}
where $\alpha (\beta)$ denote up (down) spins, and the spatial orbitals
$\varphi_j(x,y)$\footnote{
The electrons (in the conduction band) or the holes (in the valence band) in the semiconductor
QDs are described in the present review according to the continuum approximation (effective
mass approach, referred to also as envelope theory) \cite{lutt55,copp07}.}
constitute a complete single-particle basis. Namely
\begin{equation}
\Phi^{\rm CI}_{N,q} ({\bf r}_1, \ldots , {\bf r}_N) = 
\sum_I C_I^q \Psi^N_I({\bf r}_1, \ldots , {\bf r}_N),
\label{mbwf}
\end{equation}
where 
\begin{equation}
\Psi^N_I = \frac{1}{\sqrt{N!}}
\left\vert
\begin{array}{ccc}
\chi_{j_1}({\bf r}_1) & \dots & \chi_{j_N}({\bf r}_1) \\
\vdots & \ddots & \vdots \\
\chi_{j_1}({\bf r}_N) & \dots & \chi_{j_N}({\bf r}_N) \\
\end{array}
\right\vert,
\label{detexd1}
\end{equation}
and the master index $I$ enumerates the number of arrangements $\{j_1,j_2,\ldots,j_N\}$
under the condition that $1 \leq j_1 < j_2 <\ldots < j_N \leq 2K$. $q=2, 3, \ldots$
counts the excited states, with $q=1$ designating the ground state.
With the help of the expansion (\ref{mbwf}), the many-body Schr\"{o}dinger equation 
\begin{equation}
{\cal H} \Phi^{\rm CI}_{N,q} = E^{\rm CI}_{N,q} \Phi^{\rm CI}_{N,q}
\label{mbsch}
\end{equation}
transforms into a matrix diagonalization problem, which yields the coefficients $C_I^q$ and
the eigenenergies $E^{\rm CI}_{N,q}$. Because the resulting matrix is sparse, its numerical
diagonalization is implemented by employing the well known ARPACK solver \cite{arpack}. 

The importance of the FCI rests with the fact that upon convergence this method provides, both
quantitatively and qualitatively, the exact solution of the many-body Schr\"odinger equation.
In the hierarchical chart of Fig.\ \ref{hiera}, it represents the ultimate goal and
it is often referred to simply as the exact-diagonalization (EXD) method.

Like the projected many-body wave functions, the FCI wave functions preserve all the
symmetries of the governing Hamiltonian. Consequently, for a highly symmetric potential
confinement (e.g., a circular one in the case of a planar QD), the intrinsic
(or hidden) point-group symmetry associated with particle localization and Wigner 
molecule formation does not appear explicitly in the single-particle densities,
but it is recognizable through an inspection of the CPDs [the FCI wave function
$\Phi^{\rm FCI}({\bf r}_1,{\bf r}_2, \ldots, {\bf r}_N)$ is then used in Eqs.\ (\ref{cpds})
and (\ref{sponcpd})]. In addition, the hidden symmetry is identifiable via characteristic
trends in the numerically computed excitation spectra.
When numerically feasible, the FCI results provide a definitive answer
and as such they serve as a test to the results obtained through approximation methods (e.g.,
the above two-step method). Note that the underlying physics of Wigner-molecule formation
(for both fermionic and bosonic particles) is less obvious when analyzed with the 
exact-diagonalization method compared to the two-step approach. Consequently,
the significance of using CPDs as a tool for probing the many-body wave functions
cannot be overestimated; see, e.g., the contrast between single-particle density and
CPDs in Fig.\ \ref{pra_2007}.

From the above, it is apparent that both methods, i.e., the two-step method of
symmetry breaking and symmetry restoration at the HF mean-field level 
and the FCI one, complement each other, and it is in this spirit that they will be used
in this review.

Before leaving this Section, it is fitting to mention and make a brief comment on the
recently devised \cite{carl17} neural-quantum-state (NQS) and machine-learning variational
approaches, which are also being applied to strongly correlated continuum fermion problems.
Indeed, several publications \cite{wang22,cass23,bern24,liqi25,zakl26} demonstrated
explicitly that such methods can account for the formation of Wigner molecules in strongly
correlated finite systems. Such variational methodologies,
however, yield in most cases symmetry-broken
(static) WMs which do not preserve the symmetries of the many-body Hamiltonian; see, e.g.,
Fig.\ 3 in Ref.\ \cite{zakl26} and Fig.\ 9 in Ref.\ \cite{bern24}. Thus, unlike the FCI,
they cannot be trusted to describe the subtle interplay between symmetry-preserving
RWMs and symmetry-broken static WMs (see Sec.\ \ref{mqd} for an experimental verification
of this interplay). Note that a proposal to restore broken symmetries in the context of the
NQS approach has also been reported in Ref.\ \cite{szab23}. Note further that traditional
Monte Carlo approaches exhibit a similar propensity for yielding static WMs in place of the
exact RWMs; see, e.g., Fig.\ 3 in Ref.\ \cite{boni08}.

\section{Theoretical results and experimental observations establishing the physics of
Wigner molecules under field-free conditions}
\label{resff}
\smallskip

\subsection{The case of two electrons at zero magnetic field confined in a semiconductor
2D parabolic quantum dot: A first application of the two-step methodology}
\label{2eqd}
\smallskip

In this section, to describe the WM formation in a parabolic QD, I employ
the so-called two-step approach, which involves a first step of symmetry breaking at the
UHF single-determinantal level and a subsequent step of symmetry restoration using Projection
Techniques (PTs). As illustrated in Fig.\ \ref{hiera}, the two-step method yields
multi-determinantal wave functions. (In the process,
I introduce the minimum needed mathematical formalism.\footnote{
For an exposition of the full mathematical theory of projection techniques, based on
group theoretical concepts, see Sec.\ 3 of Ref.\ \cite{shei21}.})
PTs have been employed previously in Quantum Chemistry \cite{lowd55} for the restoration
of the total spin of a natural molecule and in Nuclear Physics \cite{peie57,rs_book,shei21}
for implementing the restoration of the 3D total angular momentum of deformed open-shell
nuclei (space rotational symmetry including the total spin in the $J-J$ coupling scheme).
The extension of such methodologies to circular single quantum dots (SQDs) requires the
simultaneous restoration of both the spin and the 2D angular-momentum symmetries within the
framework of the $L-S$ coupling scheme, as it will be elaborated below.
 
For simplicity and conceptual clarity, I consider here two interacting 
electrons in a SQD (artificial helium, He-QD).\footnote{
For the case of a two-electron 2D lateral double quantum dot under a perpendicular magnetic
field $B$ (artificial hydrogen quantum dot molecule, H$_2$-QDM), which heralded
\cite{loss98,burk99}
the intensive effort to construct solid-state qubits as assembly units of a solid-state
quantum computer, see Ref.\ \cite{yann02.2}.}
In this case, the Hamiltonian for two Coulomb-repelling electrons restricted to move
on a plane is given by,
\begin{equation}
{\cal H} = H({\bf r}_1)+H({\bf r}_2)+e^2/\kappa r_{12}~,
\label{ham}
\end{equation}
where $r_{12}=|\br_1-\br_2|$ and $\kappa$ is the effective dielectric constant of the
semiconductor material. $H({\bf r})$ is the one-body Hamiltonian for a single electron
confined inside a parabolic potential, 
\begin{equation}
H(\br)=T + \frac{1}{2} m^* \omega^2_{0} (x^2 + y^2).
\label{hspho}
\end{equation}
The kinetic energy is $T={\bf p}^2/(2m^*)$, with $m^*$ being the effective mass. 
The single-particle levels and eigenfunctions of $H$ in Eq.\ (\ref{hspho}) are
those of the well-known 2D isotropic harmonic oscillator.

In the first step of the procedure, the Schr\"odinger equation asoociated with the
Hamiltonian in Eq.\ (\ref{ham}) is solved \cite{yann99,yann00.2,yann02} in the
(symmetry-breaking) spin-and-space unrestricted Hartree-Fock (sS-UHF) approximation.
The symmetry-preserving restricted Hartree-Fock (RHF) is also considered for comparison.
For the precise parameters, I use $\hbar \omega_0 =5$ meV and $m^*=0.067 m_e$ (GaAs).
However, the dielectric constant $\kappa$ is varied in order to control the Wigner
parameter $R_W$, which expresses the strength of the Coulomb repulsion relative to the
zero-point quantum kinetic energy. The variation of $R_W$ enables the study of the whole 
range of electronic correlations, from the regime of weak correlations ($R_W < 1$) to
that of strong correlations ($R_W>>1$). As an example, see Fig.\ \ref{He-qd-ener} for the
variation of the total energies at the various levels of the successive approximations.  

Next I focus on the case with $\kappa=8$ ($R_W=2.39$). The RHF solution for the singlet
ground state consists of the same circularly symmetric spatial orbital that is doubly
occupied. As a result, the corresponding charge density is also circularly symmetric
[see Fig.\ \ref{succapp}(a)]. The sS-UHF solution, however, consists of two different
spatial orbitals for the two different up and down spins. The associated spin orbitals
are localized according to an antipodal arrangement within the left and right halves of
the dot [see Fig.\ \ref{succapp}(b1)], and as a result the sS-UHF charge density displays
the two-hump shape in Fig.\ \ref{succapp}(b2), which clearly breaks the circular symmetry.
Note that the sS-UHF charge density plotted in Fig.\ \ref{succapp}(b2) is aligned along
the $x$-axis (azimuthal angle $\gamma=0$). However, depending on the initial input density
in the self-consistent cycles, the sS-UHF yields densities that can be oriented at any
arbitrary $\gamma$. Namely, the broken symmetry UHF solution consists of a manifold of
degenerate wave functions parametrized by the azimuthal angle $\gamma$. This manifold
plays a central role in the approach of symmetry restoration, as will be discussed below.

The sS-UHF determinant for $\gamma=0$, which describes the broken-symmetry ``singlet'' (see
below) ground state of the He-QD, is written in compact notation as follows (henceforth the
prefix sS in subscripts will be dropped)
\begin{equation}
|\Psi_{\rm{UHF}}(1,2)\rangle = 
| u(1) \bar{v}(2) \rangle /\sqrt{2},
\label{det}
\end{equation}
with $u(1) \equiv u({\bf r}_1) \alpha(1)$ and
$\bar{v}(2) \equiv v({\bf r}_2) \beta(2)$, where $u({\bf r})$ and $v({\bf r})$ 
are the $1s$-type (left) and $1s^\prime$-type (right) localized spatial orbitals of 
the sS-UHF solution, and $\alpha$ and $\beta$ denote the up and down spins, 
respectively. For the zero magnetic-field case, these orbitals are displayed 
in Fig.\ \ref{succapp}(b1).\footnote{
Similar localized orbitals (which are complex 
functions) can appear also in the $B \neq 0$ case \cite{yann02,yann02.2}.}

$|\Psi_{\rm{UHF}}(1,2)\rangle$ is an eigenstate of the $z$-projection of the 
total spin, $\hat{{\bf S}} = \hat{{\bf s}}_1 + \hat{{\bf s}}_2$, with an
$S_z=0$ eigenvalue. However, as the quotation marks in ``singlet'' above indicate, it is
not an eigenstate of the square, $\hat{{\bf S}}^2$, of the total spin. Starting with the
determinant $|\Psi_{\rm{UHF}}(1,2)\rangle$, one can construct a proper singlet eigenstate
of $\hat{{\bf S}}^2$ (with zero eigenvalue, $S=0$) by applying the projection operator
$P_0 \equiv (1 - \varpi_{12})/2$ \cite{lowd55,pauncz},   
where the operator $\varpi_{12}$ interchanges the spins of the two electrons.

Thus the ensuing projected wave function, 
\begin{equation}
|\Phi(1,2)\rangle \equiv P_0 |\Psi_{\rm{UHF}}(1,2)\rangle \propto  
| u(1) \bar{v}(2) \rangle - \;| \bar{u}(1) v(2) \rangle,
\label{prj0}
\end{equation}
provides a proper singlet state of the two localized electrons. Note that,
in contrast to the single-determinantal wave functions of the RHF and sS-UHF 
approaches, the wave function (\ref{prj0}) is a linear superposition of two Slater
determinants, and thus it constitutes a post-Hartree-Fock improvement beyond the
mean-field approximation. I note further that the mirror symmetry (i.e., the parity)
about the $y$-axis is automatically restored in the projected wave function together
with the total-spin symmetry.

Eq.\ (\ref{prj0}) has the form of a Heitler-London (HL) wave function \cite{heit27},
familiar from the case of the natural H$_2$ molecule. However, unlike the original HL
scheme which uses the orbitals $\phi_L({\bf r})$ and $\phi_R({\bf r})$ of the separated 
(left and right) atoms, expression (\ref{prj0}) employs the sS-UHF orbitals which are
self-consistently optimized for any value of $R_W$; thus the term generalized Heitler-London
(GHL) is used to describe the wave function $|\Phi(1,2)\rangle$ in Eq.\ (\ref{prj0}). 
Note further that the projection method described here belongs to a class of projection
techniques referred to as variation before projection (VBP). More accurate quantitative
results can be obtained via a generalization which employs a variation after 
projection (VAP) (see, e.g., ch. 11.4.2 in Ref.\ \cite{rs_book}). The VAP will be
utilized only in a few cases in this review. Instead, the FCI method will be
mostly used to obtain the exact wave functions.  

The energy of a projected state can be calculated \cite{rs_book,shei21} using the
expression,
\begin{equation}
E_{\rm{PRJ}} = \left. 
\langle \Psi_{\rm{UHF}}|{\cal H} {\cal O} |\Psi_{\rm{UHF}}\rangle
\right/ \langle \Psi_{\rm{UHF}}|{\cal O}|\Psi_{\rm{UHF}}\rangle,
\label{epr}
\end{equation}
where ${\cal H}$ is the many-body Hamiltonian of the system and ${\cal O}$ denotes a
general projection operator (with the property ${\cal O}^2={\cal O}$) which commutes
with ${\cal H}$. 

Using the spin-projection operator $P_0$ in place of ${\cal O}$, one gets for 
the total energy, $E^{\rm{s}}_{\rm{GHL}}$, of the singlet GHL state,
\begin{equation}
E^{\rm{s}}_{\rm{GHL}}={\cal N}^2_{\rm{s}}
[ H_{uu}+H_{vv} +
 S_{uv}H_{vu} + S_{vu}H_{uv}+J_{uv} + K_{uv}],
\label{engvb}
\end{equation}
where $H_{uu}$, $H_{uv}$, $H_{vu}$, and $H_{vv}$ are the matrix elements of the one-body 
Hamiltonian $H$ in Eq.\ (\ref{ham}), and $J_{uv}$ and $K_{uv}$ are the direct and 
exchange matrix elements of $e^2/\kappa r_{12}$. $S_{uv}$ is the overlap integral of the
$u({\bf r})$ and $v({\bf r})$ spatial orbitals, 
\begin{equation}
S_{uv}= \int u^*({\bf r})v({\bf r}) d{\bf r},
\label{suv}
\end{equation}
and the normalization parameter is given by
\begin{equation}
{\cal N}^2_{\rm{s}} = 1/(1 + S_{uv}S_{vu}).
\label{nuv}
\end{equation}

At zero magnetic field, the electron spatial orbitals are real functions.

For the triplet state with $S_z=\pm 1$, the projected wave function coincides 
with the original HF determinant, so that the corresponding energies in all
three approximation levels are equal, i.e., 
$E^{\rm{t}}_{\rm{GHL}}=E^{\rm{t}}_{\rm{RHF}}=
E^{\rm{t}}_{\rm{UHF}}$.
 
Focussing on the singlet state, one can restore both the spin and angular momentum
symmetries successively, namely, one can generate appropriate 
projected wave functions,
\begin{equation}
|\Phi_L^{S=0} (1,2)\rangle \equiv {\cal O} |\Psi_{\rm{UHF}}(1,2)\rangle~,
\label{psisam}
\end{equation}
by applying the product operator,\footnote{
For a triplet state, the sS-UHF determinant with $S_z= \pm 1$
conserves the total spin, and thus the application of ${\cal P}_L$ alone is
sufficient.}
\begin{equation}
{\cal O} \equiv {\cal P}_L P_0~,
\label{cpr}
\end{equation}
\label{both} 
\noindent
where the spin-projection operator $P_0$ generates the two-determinant singlet 
GHL wave function (\ref{prj0}), as previously elaborated. The angular-momentum projection
operator ${\cal P}_L$ produces a {\it series\/} of multideterminantal wave functions
exhibiting good (i.e., integer) total angular momenta $L$. Indeed, for every given $L$ value,
upon application on the singlet GHL wave function, ${\cal P}_L$ yields a linear superposition
of an infinite number of azimuthally rotated GHL wave functions, organized in a manifold of
degenerate wave functions in analogy with the manifold of the sS-UHF azimuthally degenerate
solutions. ${\cal P}_L$ is given\footnote{
The corresponding formula \cite{peie57} for the 3D angular-momentum projection
employs the Wigner functions ${\cal D}^L_{MK}(\Omega)$ (see also ch. 11.4.6 in 
Ref.\ \cite{rs_book} and Ref.\ \cite{shei21}).}
by \cite{yann07,shei21},
\begin{equation}
2 \pi {\cal P}_L \equiv \int_0^{2 \pi} 
d\gamma \exp[-i \gamma (\hat{L}-L)]~,
\label{amp}
\end{equation} 
where $\hbar L$ are the eigenvalues of the total angular momentum and 
$\hat{L}=\sum_{i=1}^N \hat{l}_i$ is the corresponding {\it operator\/}.
In this Section, I focus on the ground state\footnote{
The family of projected wave functions (\ref{psisam}) describes all the
lowest-energy (yrast band \cite{yann00}) states with good angular momentum
$L=2, 4, ...$, in addition to the ground state ($L=0$).
The yrast-band states with odd values, $L=$ 1, 3, 5, ..., are generated
via a projection of the triplet state.}
~of the system with $L=0$.

It is instructive to examine the restructuring of the singlet-state (ground-state)
electron densities (EDs) associated with the successive approximations,
RHF, sS-UHF, spin projection (SP), and combined spin and angular momentum 
projection (S\&AMP). For $\kappa=8$, these EDs are displayed in Fig.\ \ref{succapp}. 
Because the exact solution for a pair of electrons confined in a parabolic potential
is easily obtainable through the method of separation in center-of-mass (CM) and relative
(rm) coordinates \cite{yann00}, I also display the ED of the exact
ground state in Fig.\ \ref{succapp}(e). The EDs of the initial RHF [Fig.\
\ref{succapp}(a)] and the last S\&AMP [Fig.\ \ref{succapp}(d)] 
approximations are circularly symmetric. In  conntrast, those of the two 
intermediate steps, namely, the sS-UHF and SP, do break the circular 
symmetry. The ED transformations described above illustrate graphically the essence of
the term ``symmetry restoration'' and the fact that the mean-field broken-symmetry
solution is associated with the rotating ({\it intrinsic\/}) frame of reference of
the Wigner molecule. Note that the electron density of the S\&AMP step
shows a characteristic flattening at the top which distinguishes it from the more 
Gaussian-like RHF one. On the other hand, although not identical [the exact ED in Fig.\
\ref{succapp}(e) is slightly flatter at the top], the S\&AMP ED closely resembles the
exact one. Further, I remark that the SP electron density
exhibits a shallower depression in the middle compared to the sS-UHF ED,
in keeping with the fact that the covalent bonding reinforces the probability for
finding an electron in-between the individual dots.
 
Substituting the projection operator (\ref{cpr}) in Eq.\ (\ref{epr}),
one obtains for the energy $E_{\rm{S\&AMP}}$ of the fully projected ground 
state of the two-electron parabolic QD,
\begin{equation}
E_{\rm{S\&AMP}} = \left. { \int_0^{2\pi} h(\gamma) d\gamma } \right/
{ \int_0^{2\pi} n(\gamma) d\gamma},
\label{eproj}
\end{equation}
with
\begin{equation}
h(\gamma) = 
H_{us}S_{vt}+ H_{ut}S_{vs}+ H_{vt}S_{us}+H_{vs}S_{ut}+ 
 V_{uvst} + V_{uvts}, 
\label{hgam}
\end{equation}
and
\begin{equation}
n(\gamma)= S_{us}S_{vt}+S_{ut}S_{vs}.
\label{ngam}
\end{equation}
$s({\bf r})$ and $t({\bf r})$ are the $u({\bf r})$ and $v({\bf r})$
sS-UHF space orbitals rotated by an angle $\gamma$, respectively. 
$V_{uvst}$ and $V_{uvts}$ are two-body matrix elements of the Coulomb 
repulsion.\footnote{
$V_{uvst} \equiv (e^2/\kappa) \int d{\bf r}_1 \int d{\bf r}_2
u^*({\bf r}_1) v^*({\bf r}_2) (1/r_{12}) s({\bf r}_1) t({\bf r}_2)$.
In Eq. (\ref{engvb}) $J_{uv} = V_{uvuv}$ and $K_{uv} = V{uvvu}$.}

To further probe how well expression (\ref{eproj}) describes the ground-state
energy of the He-QD, I display in Fig.\ \ref{He-qd-ener} the energy deviations
of the four successive approximations (i.e., RHF, sS-UHF, SP, and S\&AMP) from the 
exact ground-state energy, as a function of the dielectric constant $\kappa$ or the
Wigner parameter $R_W$.

As aforementioned, the correlation energy $E_{\rm{corr}}$ coincides with the top curve
in Fig.\ \ref{He-qd-ener} (denoted as RHF). In addition, Fig.\ \ref{He-qd-ener} reveals
the existence of two correlation regimes. Namely, one regime is that of weak correlations,
where the three lower curves collapse onto the RHF one; naturally, this regime corresponds
to the normal Fermi liquid/Bohr-model-type atom (associated with a 2D variant of the
shell-model-based Aufbau principle) and it extends to $\kappa \rightarrow \infty$
$(R_W = 0)$. The onset of the second regime of strong correlations and of symmetry-broken
UHF solutions (WM regime) occurs at $\kappa = 16.5$ $(R_W = 1.16)$ and extends to
$\kappa=0$ ($R_W \rightarrow \infty$).

\begin{figure}[t]
\centering
\includegraphics[width=0.7\textwidth]{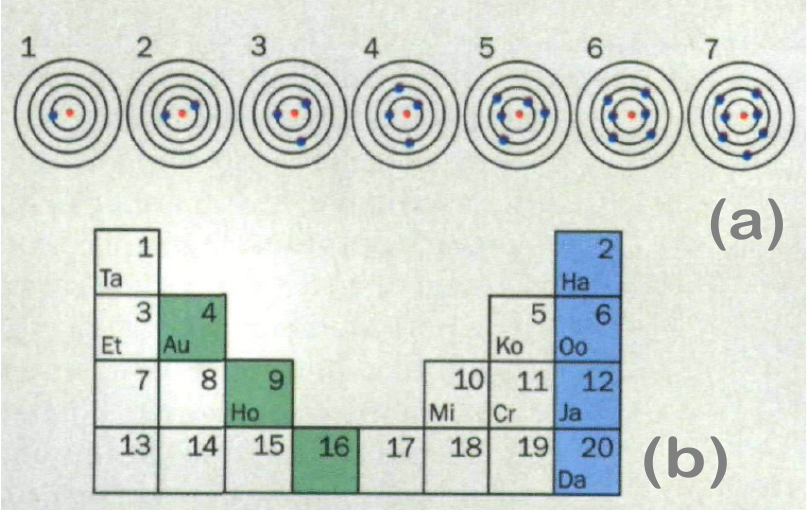}\\
~~~~~\\
\caption{
The Bohr-type 2D artificial atom, as conceptualized in Ref.\ \cite{kouw98}.
(a) The addition of single electrons to the quantum dot can be pictured
in terms of circular orbits. The first shell can contain two electrons, the second
can contain four and so on. This makes it possible to formulate a periodic
table for these artificial two-dimensional atoms. (b) Full shells correspond to
the magic numbers $N = 2,6,12,20$ and so on, while half-filled shells ($N = 4,9,16,$
etc) correspond to maximum spin states. (The elements are named after team members
from NTT and Delft.)
Figure originally published in Ref.\ \cite{kouw98}, Copyright (1998) by the IOP Publishing.
Reused with permission. All rights reserved.
}
\label{pertabl}
\end{figure}

Despite the theoretical evidence above, I note by passing that, paradoxically, 
the independent-model/Bohr-type 2D atom was advanced as the sole governing
principle for the many-body problem of semiconductor QDs in the early years of the
field, as illustrated by the 2D periodic table proposed in Ref.\ \cite{kouw98} and
reprinted in Fig.\ \ref{pertabl}.

For the parabolic QD investigated in this Section, the calculated percentage,
$E_{\rm corr}/E_{\rm exact}$, of the correlation energy relative to the exact one is 6.8\%
and 49.5\% for $\kappa=16.5$ and $\kappa=1$, respectively. These values are considerably
higher than the correlation-energy values encountered in natural atoms and, for this
reason, 2D QDs represent a characteristic class of strongly correlated systems.
In this context, the RHF error in Fig.\ \ref{He-qd-ener} grows exponentially for
stronger correlations (smaller $\kappa$ or larger $R_W$). However, it is remarkable that
the S\&AMP energy (with all symmetries restored) converges to the exact result for
smaller $\kappa$.

The exact ground-state energy is 19.80 meV for $\kappa=8$ ($R_W=2.39$) and
51.83 meV for $\kappa=1$ ($R_W=19.09$), corresponding to a relative error of
4\% and 0.7\% for the S\&AMP energy, respectively.\footnote{
For $\kappa=8$, the fraction of correlation energy captured
by the successive approximations is: sS-UHF 44.4\%, SP 57.2\%, and S\&AMP
73.1\%; the corresponding values for $\kappa=1$ are:
94.5\%, 94.5\%, and 98.5\%.}
~Namely, while the energy deviations are largest in the neighborhood of the bifurcation
point (see Fig.\ \ref{He-qd-ener}), the VBP energies approach the exact ones as
$R_W \rightarrow \infty$ (i.e., the regime of strong correlations). Furthermore, this
trend extends\footnote{
For a similar convergence of the broken symmetry UHF energies with the exact ones as a
function of an applied magnetic field (as $B \rightarrow \infty$), see Ref.\
\cite{szaf04.2}.}
to the sS-UHF energies as well (see again Fig.\ \ref{He-qd-ener}).

It is interesting to note that the SP curve collapses onto the sS-UHF one for
$\kappa \leq 3.2$. This behavior points to an intermediate regime (in the range
$3.2 \leq \kappa \leq 16.5$) between the Fermi-liquid one and that of strongly
formed WMs. In the latter regime ($\kappa \leq 3.2$), the overlap ($S_{uv}$) between
the antipodally localized electronic space orbitals is vanishingly small, so that
the sS-UHF energy is not effectively lowered by the restoration of the total spin.
\smallskip

\subsubsection{\small{{\bf Commentary 2.}} The projected wave function as a superposition of
Slater determinants.}
\label{comm2}
The $\exp[-i \gamma \hat{L}]$ kernel in the definition of the 2D angular momentum
projection operator ${\cal P}_L$ [see Eq.\ (\ref{amp})] rotates the initial sS-UHF Slater
determinant by an azimutal angle $\gamma$, i.e., the action of ${\cal P}_L$ is
equivalent to generating the following (continuous) superposition of Slater determinants
\begin{equation}
|\Phi_{\rm PRJ} \rangle = \int d\gamma C(\gamma) |\Psi_{\rm UHF} (\gamma) \rangle,
\label{cprj}  
\end{equation}  
where the expansion coefficients are given by $C(\gamma)=\exp(i\gamma I)$ as a consequence
of symmetry arguments. Indeed, they coincide \cite{yann03.2,cotton,hamermesh}
with the characters of the continuous cyclic group $C_\infty$.

Alternatively, one can derive the coefficients $C(\gamma)$ (and numerically calculate
the ground-state energy) by formulating a continuous configuration interaction
approach corresponding to the expansion (\ref{cprj}). Utilizing the Gross-Pitaevskii
symmetry-broken solution (which is a simple Hartree product), this was done in
Ref.\ \cite{alon04} for the case of attractive boson condensates in a ring.

Note that the formulation of a many-body Schr\"odinger equation within a restricted
continuously parameter-dependent Hilbert space, which includes the degenerate UHF
manifold of states at the lowest order, was introduced in the context of Nuclear Physics
in the 1950's by John A. Wheeler and his Ph.D. students David L. Hill \cite{hill53} and
James J. Griffin \cite{grif57}. The set of parameters specifying the Hilbert space are
referred to as generator coordinates and the approach is known as the generator coordinate
method (GCM) \cite{hill53,grif57,rs_book,yann07,shei21}.
It belongs to the family of nonorthogonal configuration interaction methods.
It is apparent that the restoration of broken symmetries via projection techniques can
be viewed as a special case of the generator coordinate method.

Note further that another method closely related to the GCM is the resonating HF
approach \cite{fuku88}; for an application to the case of a 2D square quantum dot,
see Ref.\ \cite{okun09}.

\subsection{The case of two electrons at zero magnetic field confined in a semiconductor 
2D parabolic quantum dot: Exact solution and rovibrational spectra}
\label{2eqdex}
\smallskip

Inspired by the key role that spectroscopy played in shaping our understanding of the
electronic structure of natural atoms and the development of quantum physics, this Section
will review (see Ref.\ \cite{yann00}) the exactly solvable excitation spectrum in two
dimensions associated with two electrons (2e) confined in a parabolic QD. This case 
represents a paradigmatic three-body problem comprised of the two electrons (symbolized
below as $X$s) and the (infinitely heavy) confining quantum dot (symbolized as $Y$). 
In particular, by analyzing the anatomy of the exact wave functions with the help of
CPDs, in conjunction with the identification of regular patterns in the excitation spectrum,
it was shown that this spectrum reflects a universal collective behavior associated with
electron localization and formation of an effective linear trimeric molecule $XYX$.
Indeed, it was found \cite{yann00} that the excitation spectrum of the two-electron QD
exhibits for a weak parabolic confinement (i.e., small $\omega_0$ yielding a large $R_W$)
a well-developed, separable ro-vibrational pattern
which is akin to the characteristic spectrum of natural ``near-rigid'' triatomic
molecules (i.e., molecules with stretching and bending vibrational frequencies much
higher than the rotational one). For stronger confinements (i.e., large 
$\omega_0$), the spectrum transforms to that of a ``floppy'' 
triatomic molecule, converging finally to the independent-particle picture 
associated with a circular central mean field. 

The Hamiltonian for a 2e QD, with a parabolic confinement of frequency $\omega_0$,
is given by
\begin{equation}
H= \sum_{i=1,2} \frac{{\bf p}^2_i}{2m^*} + \frac{e^2}{\kappa |{\bf r}_1 - {\bf r}_2|}
+{\textstyle\frac{1}{2} } m^* \omega_0^2 \sum_{i=1,2} {\bf r}_i^2,
\label{h2e}
\end{equation}
where $\kappa$ is the dielectric constant and $m^*$ is the effective electron mass.
The corresponding Schr\"odinger equation is separable
in the center-of-mass (CM) and relative-motion (rm) coordinates.
As a result, the energy eigenvalues can be written as
$E_{NM,nm}=E^{\rm CM}_{NM} + \varepsilon^{\rm rm}(n,|m|)$,
where $E^{\rm CM}_{NM}=\hbar \omega_0 (2N+|M|+1)$ with the $N$ and $M$
being the CM radial and azimuthal quantum numbers, respectively. 
$\varepsilon^{\rm rm}(n,|m|)$ are the eigenvalues of the one-dimensional
Schr\"{o}dinger equation \cite{taut94,garc98},
\begin{equation}
\frac{\partial^2 \Xi}{\partial u^2} + 
\{ \frac{-m^2+1/4}{u^2}-u^2-\frac{R_W \sqrt{2}}{u}+
\frac{\varepsilon}{\hbar \omega_0 /2} \} \Xi =0,
\label{relmot}
\end{equation}
where $\Xi(u)/\sqrt{u}$ is the radial part of the rm wave function
$\Xi(u)e^{im\theta}/\sqrt{u}$, with $m$ being the angular momentum quantum number
associated with the rm azimuthal angle $\theta$. $n$ will denote the number of radial nodes
of $\Xi(u)$ and $u=|{\bf u}_1-{\bf u}_2|$, with ${\bf u}_i={\bf r}_i/l_0 \sqrt{2}$ $(i=1,2)$
being dimensionless electronic coordinates and $l_0=(\hbar/m^* \omega_0)^{1/2}$, which
specifies the spatial extent of the lowest-state wave function of a single electron. $R_W$
is the Wigner parameter previously defined in Eq.\ (\ref{rw}).

The exact spatial wave function of the two-electron QD is the product of the separate
CM and rm wave functions and is denoted as $\Phi_{NM,nm}({\bf u}_1,{\bf u}_2)$, whereas
the spatial two-electron density is given by 
$W_{NM,nm}({\bf u}_1,{\bf u}_2)=|\Phi_{NM,nm}({\bf u}_1,{\bf u}_2)|^2$.
Then the usual (isotropic) pair-correlation function (PCF) is defined as
\begin{equation}
G(v)=2\pi \int \int \delta({\bf u}_1-{\bf u}_2-{\bf v}) 
W({\bf u}_1,{\bf u}_2) d{\bf u}_1 d{\bf u}_2,
\label{pair2}
\end{equation}
whereas the (anisotropic) conditional probability distribution (CPD) for locating one 
electron at position ${\bf v}$, given that the other is at position ${\bf v}_0$, is
expressed as
\cite{yann00.2,yann00}
\begin{equation}
{\cal P}({\bf v}|{\bf u}_2={\bf v}_0)=
\frac{W({\bf v},{\bf u}_2={\bf v}_0)}
{\int d{\bf u}_1 W({\bf u}_1,{\bf u}_2={\bf v}_0)},
\label{cpd2}
\end{equation}

\begin{figure}
\includegraphics[width=1.8\textwidth,left]{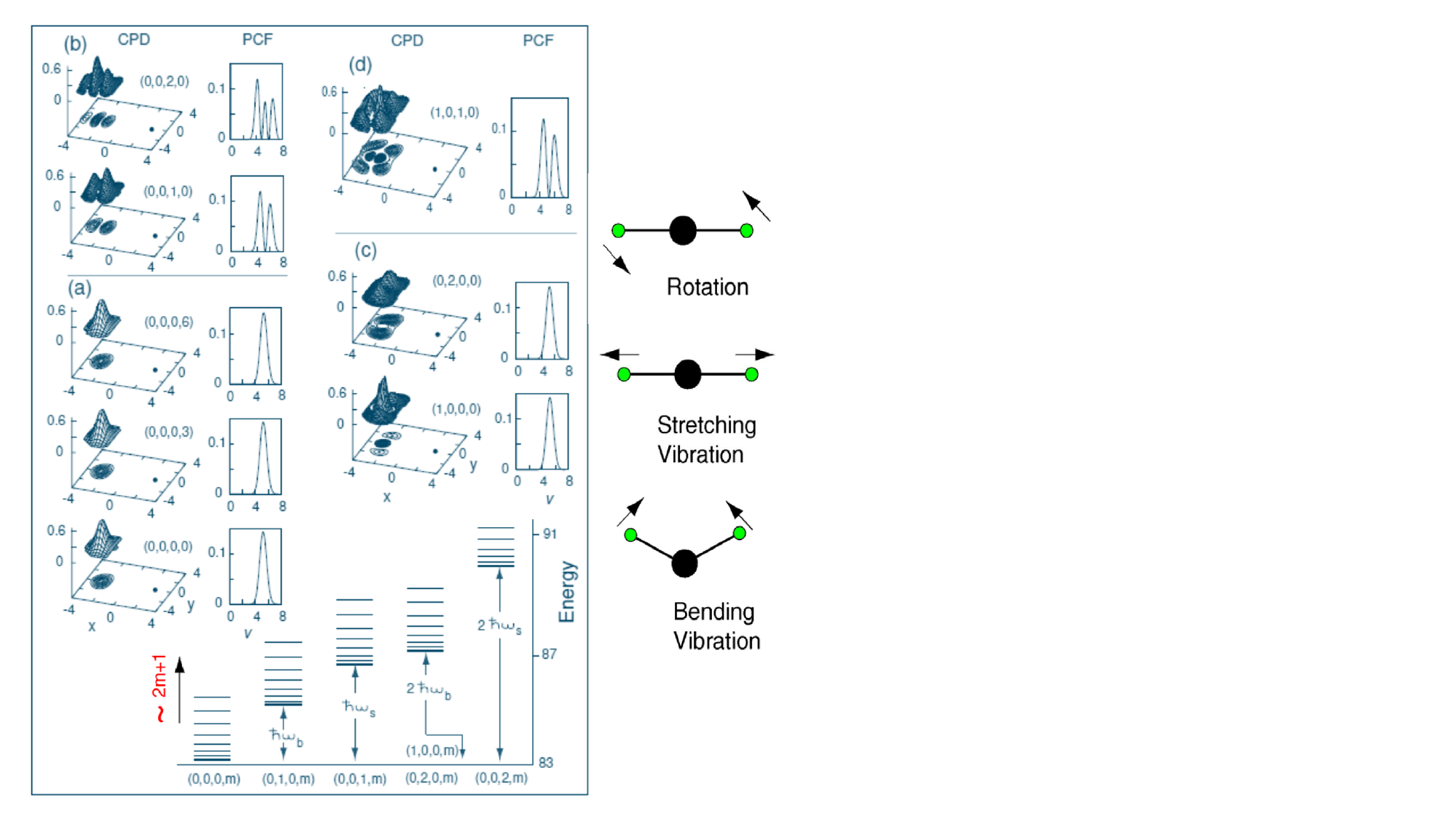}
\caption{
Exact results for a 2eQD with $R_W=200$. (Bottom of figure) Energy spectrum. (a)-(d)
Conditional probability distributions (CPDs) and pair-correlation functions (PCFs)
displayed respectively on the left side and the right side of each subplot.
For each excitation band (resembling a tower of states), the associated quantum numbers
$(N_0,M_0,n_0,m)$ are displayed at the bottom, with $m=0,1,2,...$. The pair of levels with
$m=0$ and $m=1$ cannot be resolved on the energy scale of the figure and appear as thicker
lines. Only a few of the low lying rotational and vibrational states are shown, but the
rovibrational collective behavior extends to higher excitations. The CPDs and PCFs are
labeled according to the set of the quantum numbers of the associated states. See the text
for a discussion of the rules embodied in the spectrum and for the definition and physical
meaning of the CPDs and PCFs. The solid dot in each of the CPD subplots denotes the point
${\bf v}_0=(d_0,0)$, with $d_0=2.6$ being half of the electron separation extracted from 
the PCF of the ground state (0,0,0,0). All lengths $x,y,v$ and $d_0$ are given in
dimensionless units of $l_0 \sqrt{2}$, and the energies are in units of $\hbar \omega_0/2$.
The collective modes of a near-rigid symmetric natural triatomic molecule are also shown
schematically on the right side of the figure.
Reprinted figure with permission from Ref.\ \cite{yann00},
Copyright (2000) by the American Physical Society.
}
\label{2espec}
\end{figure}

\noindent
where the $M,N,n,m$ indices of $W$ (as well as of $G$ and ${\cal P}$) have 
been omitted. Note that the exact electron density \cite{yann00.2}
\begin{equation}
n({\bf v}) =
\int \int \sum_{i=1}^2  \delta({\bf u}_i-{\bf v}) W({\bf u}_1,{\bf u}_2) d{\bf u}_1 d{\bf u}_2
\label{spd}
\end{equation}
is circularly symmetric; see, e.g., Fig.\ \ref{succapp}(e)

Using the above, Ref.\ \cite{yann00} determined the solutions for the 2e-QD energy
spectra and corresponding wave functions for values of $R_W=200$, 20 and 3. I analyze
first the $R_W=200$ case whose spectrum and selected PCFs and CPDs are displayed in Fig.\
\ref{2espec}. For such a large value of $R_W$ (extremely strong correlations approaching
the classical regime), the spectrum of the two-electron QD (displayed at the bottom of
Fig.\ \ref{2espec}) exhibits the following three well-established trends:
(i) for all bands $(N_0,M_0,n_0,m)$, with $m=0,1,2,... $, but with constant values 
$N_0,M_0$ and $n_0$ (the subscript ``0'' denotes a number held constant within
a given sequence), the energy gap between two adjacent rotational levels $m$ and $m+1$
{\it increases linearly \/} in proportion to $2m+1$. Furthermore, the bands
$(N_0, \pm M_0, n_0, \pm m)$ are degenerate. (ii) the sequences $(0,M_0,0,m)$ and
$(N_0,0,0,m)$ are associated with center-of-mass excitations having $M_0$ and $2N_0$ 
vibrational quanta (phonons) of energy $\hbar \omega_0$, respectively.
(iii) the lowest-in-energy levels of the sequences $(0,0,n_0,m)$ exhibit a 1D
harmonic-oscillator spectrum $(n_0+1/2) \hbar \omega_s$.
Note that the energy levels are spin singlets or triplets for even or odd values of
$m$, respectively.

These three ``spectral rules'' above specify a well-developed and separable 
ro-vibrational spectrum exhibiting collective rotations, as well as 
stretching and bending vibrations familiar from the case of natural triatomic linear
molecules, like carbon dioxide (e.g., CO$_2$). Indeed, after neglecting an overall 
constant term, the above three rules can be summed up as,
\begin{equation}
E_{NM,nm}=C m^2~ + (n+1/2) \hbar \omega_s + (2N+|M|+1) \hbar \omega_b,
\label{spec}
\end{equation}
where the rotational constant $C \approx 0.037$ (all energies are given in dimensionless
units of $\hbar \omega_0/2$), the stretching-vibration phonon has an energy
of $\hbar \omega_s \approx 3.50$, and the bending-vibration phonon coincides with that
of the CM motion, i.e., $\hbar \omega_b = \hbar \omega_0=2$. Note that the rotational energy
is proportional to $m^2$, as it is fitting for 2D rotations, unlike the case 
of natural triatomic molecules where the rotational energy varies as $l(l+1)$, with
$l$ being the quantum number associated with the 3D angular momentum. Observe also that
the bending vibration can carry by itself an angular momentum $\hbar M$, and thus the
the total angular momentum is $\hbar (M+m)$.
 
Further insights into the collective nature of the energy spectrum displayed at the
bottom of Fig.\ \ref{2espec} can be gained by inspecting the CPDs and PCFs corresponding
to selected states of the rotational bands $(N_0,M_0,n_0,m)$ (the CPDs are plotted to the
left of the PCFs. Note that the PCFs are always circularly symmetric). The special band
$(0,0,0,m)$, being {\it purely\/} rotational without vibrational excitations, can be
referred to as the ``yrast'' band, in analogy with the customary terminology from the
spectroscopy of rotational spectra in atomic nuclei \cite{bm}.

In Fig.\ \ref{2espec}(a), I display the CPDs and PCFs for three selected states of 
the yrast band, i.e., the (0,0,0,0), the (0,0,0,3), and the (0,0,0,6). The 
corresponding PCFs are all alike and centered around $2d_0=5.2$, which
implies that the two electrons keep apart from each other at a distance
$2d_0$. The formation of an electron Wigner molecule is definitively demonstrated
by the corresponding CPDs [plotted in the left column with
${\bf v_0}=(d_0,0)$; the point ${\bf v}_0$ is denoted by a solid dot]. Indeed,
the CPDs show that the two electrons occupy at all instances diametrically opposite
positions, forming effectively a rotating linear molecule $XYX$
with two symmetric bonds ($X-Y$ and $Y-X$) of equal length, $d_0$. Furthermore,
one sees that all three CPDs in panel (a) are practically identical, in spite of
the fact that the rm angular momentum changes substantially from $m=0$ to $m=6$. This
constancy of the bond lengths, irrespective of the magnitude of the rotational energy,
fittingly characterizes the Wigner molecule as a near-rigid rotor. 

Next, one can inquire about the rotational bands $(0,0,1,m)$ and $(0,0,2,m)$, which are
built upon one- and two-phonon excitations of the stretching vibrational mode. As confirmed
by extensive numerical calculations, the PCFs and the CPDs associated with these bands
share with those of the yrast band the property that they remain practically unchanged 
(at least for the states displayed in Fig.\ \ref{2espec}) as a function
of $m$. Consequently, it suffices to study the bottom states, i.e., those with $m=0$,
(0,0,1,0) and (0,0,2,0), whose corresponding PCFs and CPDs are displayed in Fig.\
\ref{2espec}(b). The PCFs in  Fig.\ \ref{2espec}(b) reveal the presence of internal
excitations with one and two nodes in the relative motion, but they yield no other
information pertaining to the Wigner molecule. On the other hand, the corresponding CPDs
in Fig.\ \ref{2espec}(b) immediately demonstrate that these excitations belong to 
the stretching vibrational mode of the $XYX$ molecule {\it along\/} the 
interelectron axis.

By inspecting the CPDs in Fig.\ \ref{2espec}(c), one can further show that 
the two degenerate rotational bands $(0,2,0,m)$ and $(1,0,0,m)$ are built
upon the lowest two-phonon bending vibrational excitations. Again, it is sufficient
to consider the two lowest-in-energy states of these bands, namely the $(1,0,0,0)$ 
and the $(0,2,0,0)$. It is recognized that both the CPDs describe bending vibrational 
excitations of the $XYX$ molecule which are {\it perpendicular\/} to the 
interelectron axis, with the CPD of the $(1,0,0,0)$ level having one node and that of
the $(0,2,0,0)$ state having no nodes. I note that the corresponding PCFs in Fig.\
\ref{2espec}(c) (plotted on the right of the CPDs) fail to reveal this internal structure.
 
Finally, Fig.\ \ref{2espec}(d) displays the CPD and corresponding PCF of the bottom level
(i.e., with $m=0$) of the rotational band $(1,0,1,m)$ (not included in the spectrum shown in
Fig.\ \ref{2espec}), which is built upon more complex phonon excitations of mixed
stretching and bending character. It is easily seen that the CPD represents combined
vibrations of the Wigner molecule along the interelectron axis (one excited stretching-mode
phonon), as well as perpendicularly to this axis (two excited bending-mode phonons).
Again, one can see that, in contrast to the CPD which enables detailed probing of the
intrinsic architecture of the excited states, the information extracted from the
corresponding PCF may be very limited.

\begin{figure}[t]
\centering
\includegraphics[width=10.0cm]{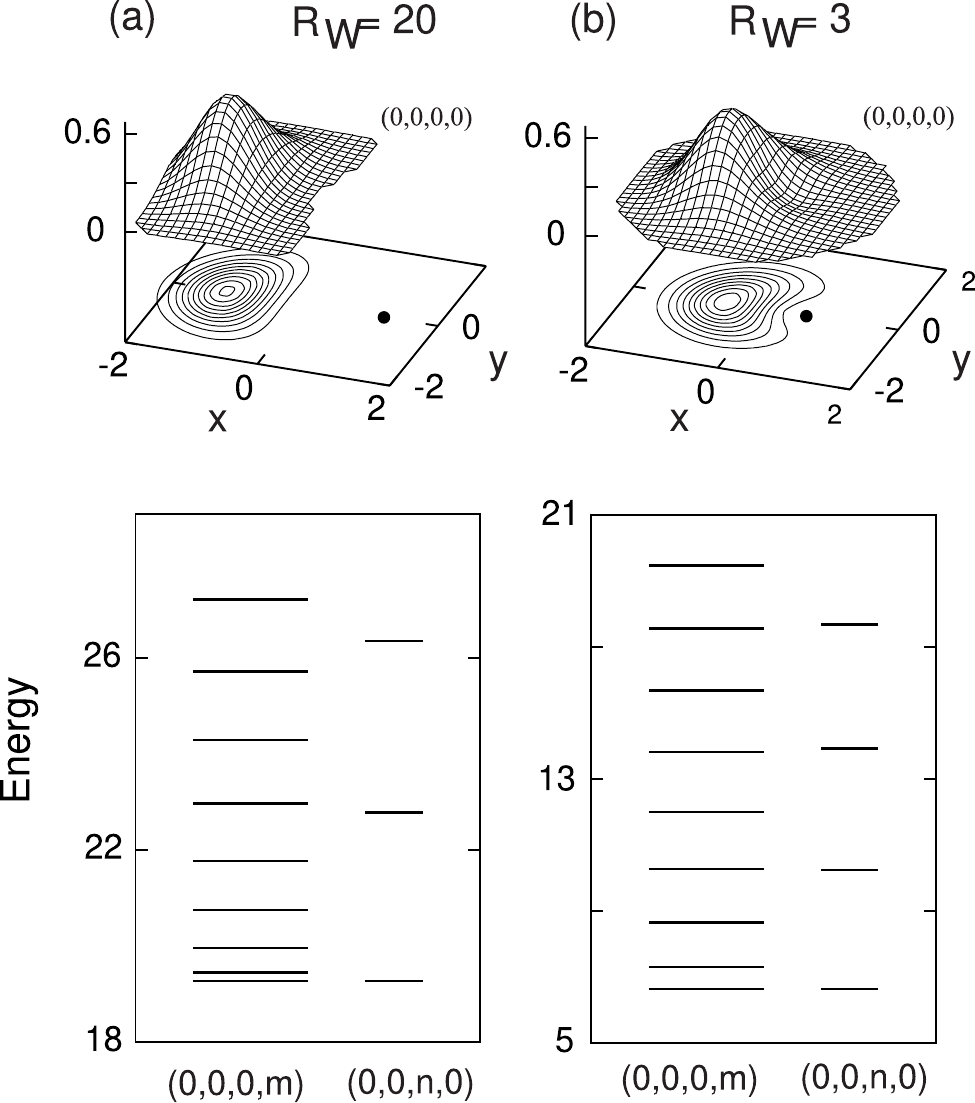}
\caption{
Exact results for a 2eQD with (a) $R_W=20$, and (b) $R_W=3$. (Bottom row)
Energy spectra for the $(0,0,0,m)$ yrast bands (with $m=0,1,2,...$) and the $(0,0,n,0)$
stretching vibrational bands (with $n=0,1,2,...$).
(Top row) CPDs of the ground states $(0,0,0,0)$.
The solid dot in each of the CPD panels represents the fixed point ${\bf v}_0=(d_0,0)$,
where $d_0$ is half of the electron separation. Energies are given in units of
$\hbar \omega_0/2$ and the distances in units of $l_0 \sqrt{2}$. As an example, for
$R_W=3$ and assuming GaAs material parameters ($m^*=0.067m_e$, $\kappa=12.9$),
one has $\hbar \omega_0=1.2$ meV and $l_0 \sqrt{2}=43.54$ nm. 
Reprinted figure with permission from Ref.\ \cite{yann00},
Copyright (2000) by the American Physical Society.
}
\label{rw20and3}
\end{figure}

The near-rigidity of the Wigner molecule, which is well established for
$R_W=200$, naturally weakens as $R_W$ decreases. Then the XYX molecule starts
showing a growing degree of ``floppiness''. The increasing floppiness becomes visible in
the yrast band, which, beginning with
the higher levels, gradually deviates from the spectral rule (i) discussed
above, and ultimately becomes unrecognizable as a rotational band. An illustration
of this trend is presented in the lower subplot of Fig.\ \ref{rw20and3}(a) which displays
the $(0,0,0,m)$ yrast band for $R_W=20$. Specifically, one can recognize that only the
lowest four energy levels follow approximately rule (i), whereas the higher ones start
to develop a constant energy gap between adjacent levels (this energy gap approaches
slowly the energy spacing $\hbar \omega_0$ of the harmonic confinement).

In the $R_W=3$ case, one can hardly recognize any rigid-rotor-like rotational sequence in
the energy levels of the yrast band [the $(0,0,0,m)$ band plotted at the bottom subplot of
Fig.\ \ref{rw20and3}(b)].
Indeed, the ratio between the $(m=0)-$to$-(m=1)$ and the $(m=1)-$to$-(m=2)$ energy spacings
is substantially different from 3/1, whereas the spacing between higher levels approaches
quickly the value 2 of the external confinement. Furthermore, I note that the stretching
vibrations tend to better preserve a constant spacing between the bottom energy levels of
the bands $(0,0,n_0,m)$ [these levels were grouped in a vibrational band $(0,0,n,0)$ and
are plotted on the right-hand-side of the lower subplots in Figs.\ \ref{rw20and3}(a) and
\ref{rw20and3}(b)].

However, in spite of the floppiness exhibited by the excitation spectra in Fig.\
\ref{rw20and3}, the (singlet) ground-state of the 2e QD for both $R_W=20$ and $R_W=3$
is drastically different from the 1$s^2$ closed-shell orbital configuration expected
from the independent-particle \cite{kast93,kouw98} and RHF pictures. Rather, as
revealed by the corresponding CPDs [see panels in the top row of Fig.\ \ref{rw20and3}],
in both these cases of smaller $R_W$'s, the ground state is still within the regime of
formation of rather well-developed XYX electron molecules, but with progressively
smaller bond lengths.

To recap this Section, the remarkable emergence of ro-vibrational excitations for
parabolically confined 2e QDs, under a vanishing magnetic field, provides direct
theoretical evidence for the formation of symmetry-preserving rotating Wigner molecules
in QDs, with their degree of rigidity controlled by the parameter $R_W$.
Such RWMs had been predicted \cite{yann00,yann00.2,yann07} to be a general
feature\footnote{
For a discussion of excitations of rotating Wigner molecules in fully polarized
parabolic QDs in high magnetic fields, see also Ref.\ \cite{maks96}.}
of the physics of few- and many-electron QDs for $R_W > 1$. Their direct
experimental observation was proposed \cite{yann00.2} to be feasible through
the controlled pinning of the collective rotation as early as 2000. Recently, this
experimental program was realized \cite{crom24} successfully for the case of a few
charge carriers (both electrons and holes) in triagonal QDs forming in TMD moir\'e
superlattices, which have emerged as 2D platforms with unprecedented flexibility.

Lastly, I note that the results in this Section invite, in a compelling way, a
discussion for the parallels and differences between the symmetry breaking theory
in finite systems and the concept of spontaneous symmetry breaking
(either in the thermodynamic limit or the limit of large mass in chemistry (associated
with the Born-Oppenheimer approximation); see, e.g., Refs.\ \cite{ande72,pwa_book}).
This discussion is postponed for later; see Commentary 4 (Sec.\ \ref{comm4}).
\smallskip

\begin{figure}[t]
\centering
\includegraphics[width=8.0cm]{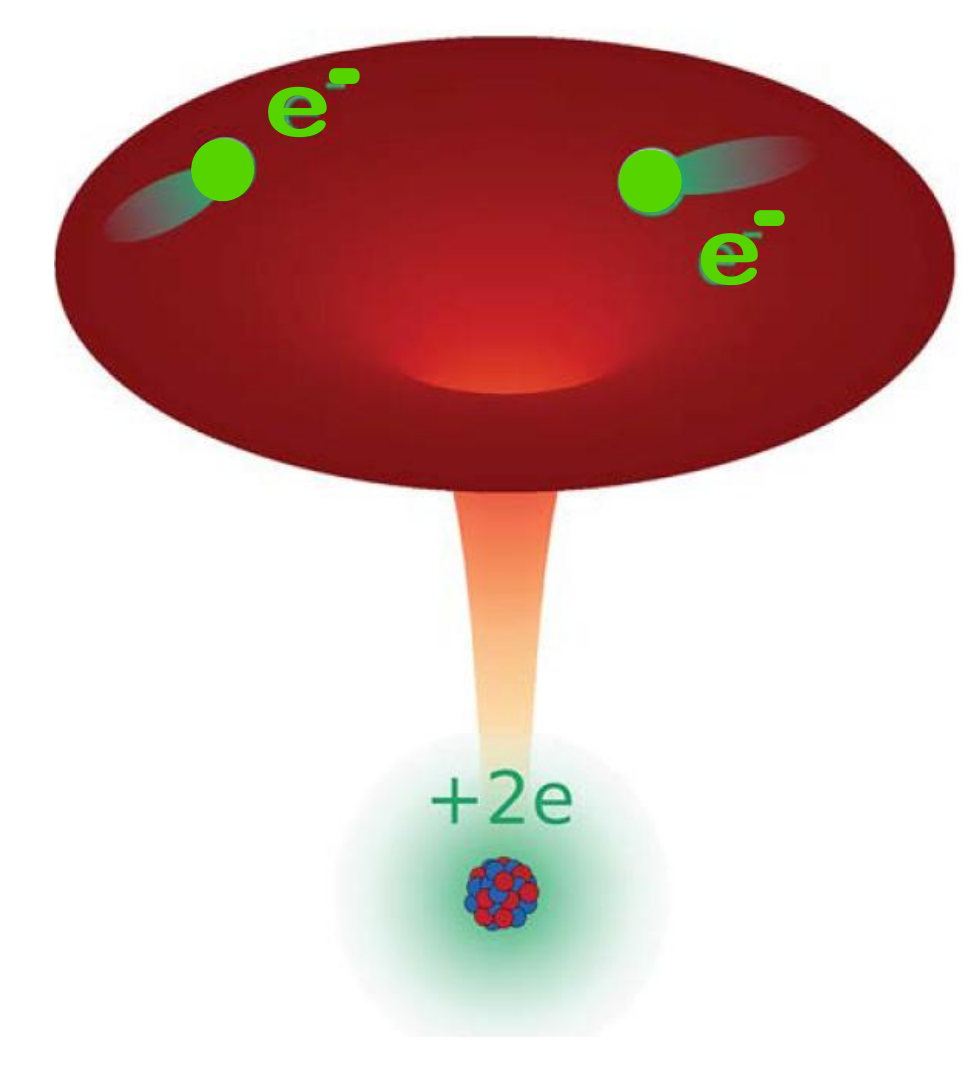}
\caption{The graphical abstract in Ref.\ \cite{gene21} illustrating the funneling shape
of the Coulombic attractive potential of the nucleus and the localization of the two
electrons in highly excited Rydberg states, due to a sharp inflation of the available
volume to the electrons.
Reprinted figure with permission from Ref.\ \cite{gene21},
Copyright (2020) by Informa UK Limited, trading as Taylor \& Francis Group.
}
\label{Cb_funnel}
\end{figure}

\subsubsection{ {\small {\bf Commentary 3.} } Discussion about doubly excited natural
Helium atoms.}
\label{comm3}
As mentioned in passing earlier [see Commentary 1, Sec.\ \ref{comm1}], highly excited
natural atoms can be seen as exotic cases that strongly deviate from the physics of the
Aufbau principle. Such a drastic change in behavior is due to the funneling shape of the
Coulombic attractive potential of the atomic nucleus. In fact, for the atomic ground states
and low-energy excitations, the available volume is small (radius of natural atoms $< 1$ \AA)
and thus the effective $R_W \ll 1$. On the contrary, for highly excited states the electrons
take advantage of a much larger volume by occupying Rydberg states (and resonances
in the continuum), resulting in an effective $R_W \gg 1$ and to electron localization;
see the schematic illustration in Fig.\ \ref{Cb_funnel} which serves as the graphical
abstract of Ref.\ \cite{gene21}. From this, it
follows that natural atoms with a highly excited cluster of a few strongly-correlated
electrons should exhibit spectral signatures analogous to the rovibrational spectrum
discussed above for the case of the 2D parabolic QD at large $R_W$.

Indeed, it is remarkable that the experimentally measured spectra of the doubly-excited
natural He atom and H$^-$ anion have been empirically interpreted, as early as in the
1980's, as rovibrational spectra associated with a near-rigid eZe linear molecule
\cite{kell78,kell80,lin07}, consisting of the nucleus (Z) at the center and two localized
antipodal electrons. Other doubly-excited atomic two-electron systems, which exhibit
rovibrational spectra, include the cations Li$^+$ and Be$^{2+}$, as well as neutral
alkaline-earth elements like Be and Sr \cite{gene21,eich25}.
Triply-excited natural atoms, like Li, and quadruply-excited atoms
have been also found to exhibit relevant 3D rovibrational spectra associated with
geometrical equilibrium configurations of electrons treated as classical
point charges \cite{lin99,mads01,mads02,mads03,mads05}. It is noteworthy that methods
reminiscent of the 2D symmetry-restoration ones elaborated in this review were successfully
employed to describe the corresponding non-rigid 3D rotors \cite{mads01,mads02,mads03,iwai89}.

\begin{figure}[t]
\centering
\includegraphics[width=6.0cm]{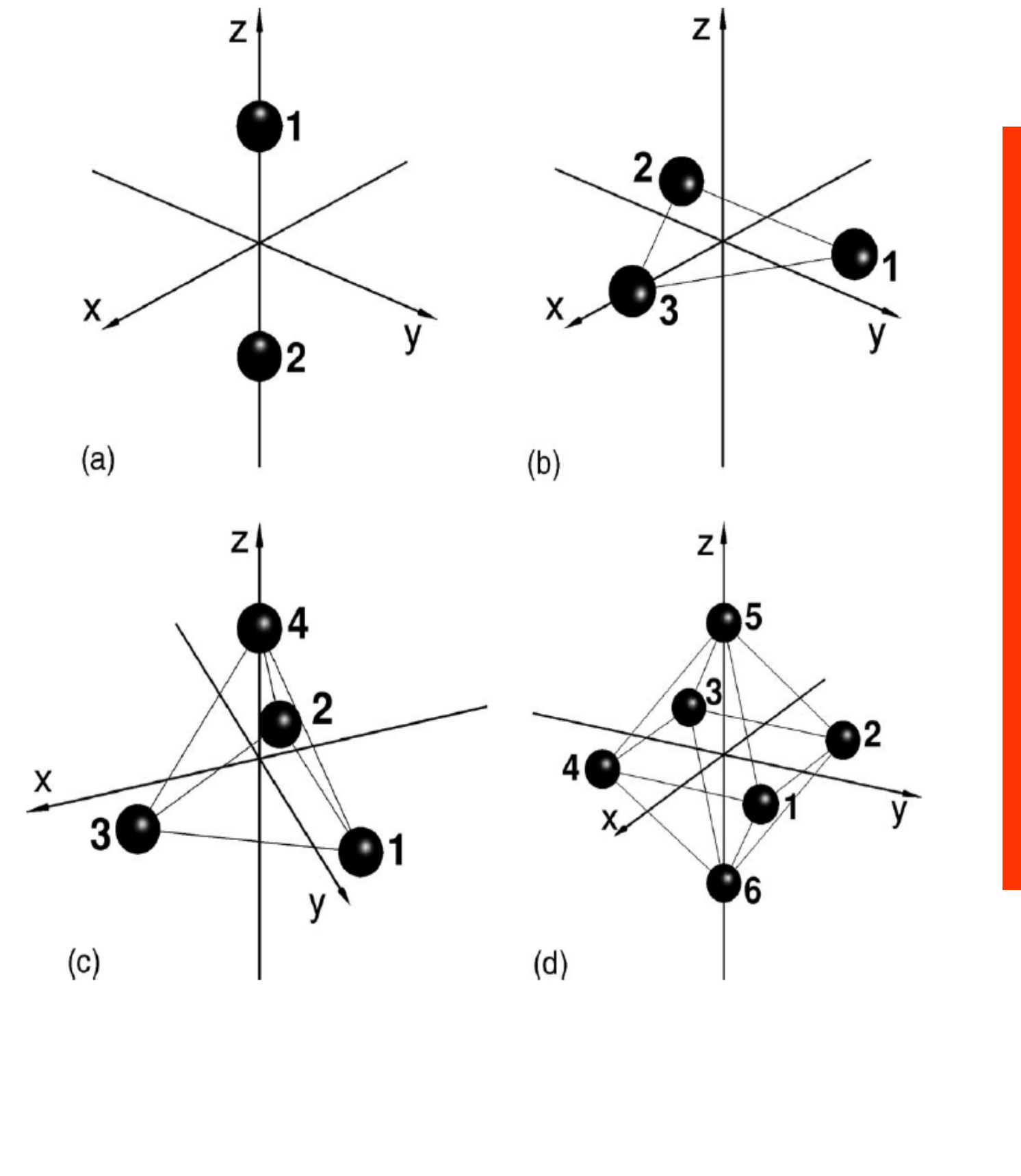} \hspace{0.3cm}
\includegraphics[width=7.0cm]{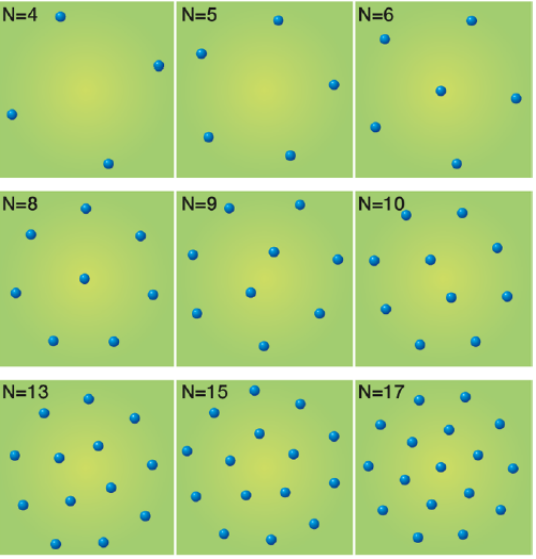}
\caption{(left) Energy-favorable 3D geometrical configurations for (a)
two-, (b) three-, (c) four- and (d) six-valence-electron atoms. The
nuclei are imagined to be in the respective origins. The orientations
of the coordinate systems are chosen in order to visualize as clearly
as possible the three-dimensional structures.
Reprinted figure with permission from Ref.\ \cite{mads05}, Copyright (2005)
by the American Physical Society.
\label{Cb_3D}
}
\caption{(right)
Classical Wigner molecules showing the arrangement of electrons (treated as point charges)
in a circular island for several occupation numbers. The electrons form nested polygonal
ring structures as more electrons are added. At $N=15$ a single inner ring of five electrons
is formed, but at $N=17$ the ring structure comprises three rings.
Adapted figure from Ref.\ \cite{peet09}, Copyright (2009) by the American Physical
Society/Alan Stonebraker. 
\label{Cb_2D} 
}
\end{figure}

Fig.\ \ref{Cb_3D} displays these classical 3D equilibrium configurations for
$N=2-4$ and $N=6$ electrons. They are of course polyhedral in shape, and they differ
from the classical 2D equilibrium configurations which are nested polygonal rings
\cite{bolt93,kong02,peet09} (see an example in Fig.\ \ref{Cb_2D}; for a complete
table of the 2D ground-state and metastable structures, $(n_1,n_2,\ldots,n_r)$,
for $N=2-40$, see Ref.\ \cite{kong02}). I mention again here that the broken-symmetry
UHF solutions in the case of parabolic QDs follow \cite{yann99,yann06.2,yann07,shei21} these
classical nested-polygonal patterns. 

Among the quantum-mechanical theoretical approaches for doubly-excited natural atoms,
of interest here is the molecular orbital (MO) model introduced by J.M. Feagin and J.S.
Briggs \cite{feag86,feag88}. In its essence, this model is an inversion of the BOA for
the H$_2^+$ molecule,\footnote{
For a broader view in this direction and a connection to other finite systems,
see Ref.\ \cite{sala17}.} 
with the molecular internuclear axis being replaced by the
interelectronic axis defined by the two electrons and the H$_2^+$ electronic coordinate
being replaced by the position of the nucleus with respect to the center of mass of the
two electrons \cite{feag88}. This inverted model, where the electrons behave like the
heavy nuclei (they are assumed to be localized) and the nuclei behave like electrons
is reminiscent of Wigner's conception of the WC as an ``inverted alkali metal''
\cite{wign38}. It is natural then to consider that the justification of the MO model
derives from a large effective $R_W$ parameter due to the inflation of the volume
available to the electrons; see Fig.\ \ref{Cb_funnel}.

\begin{figure}[t]
\centering
\includegraphics[width=12.0cm]{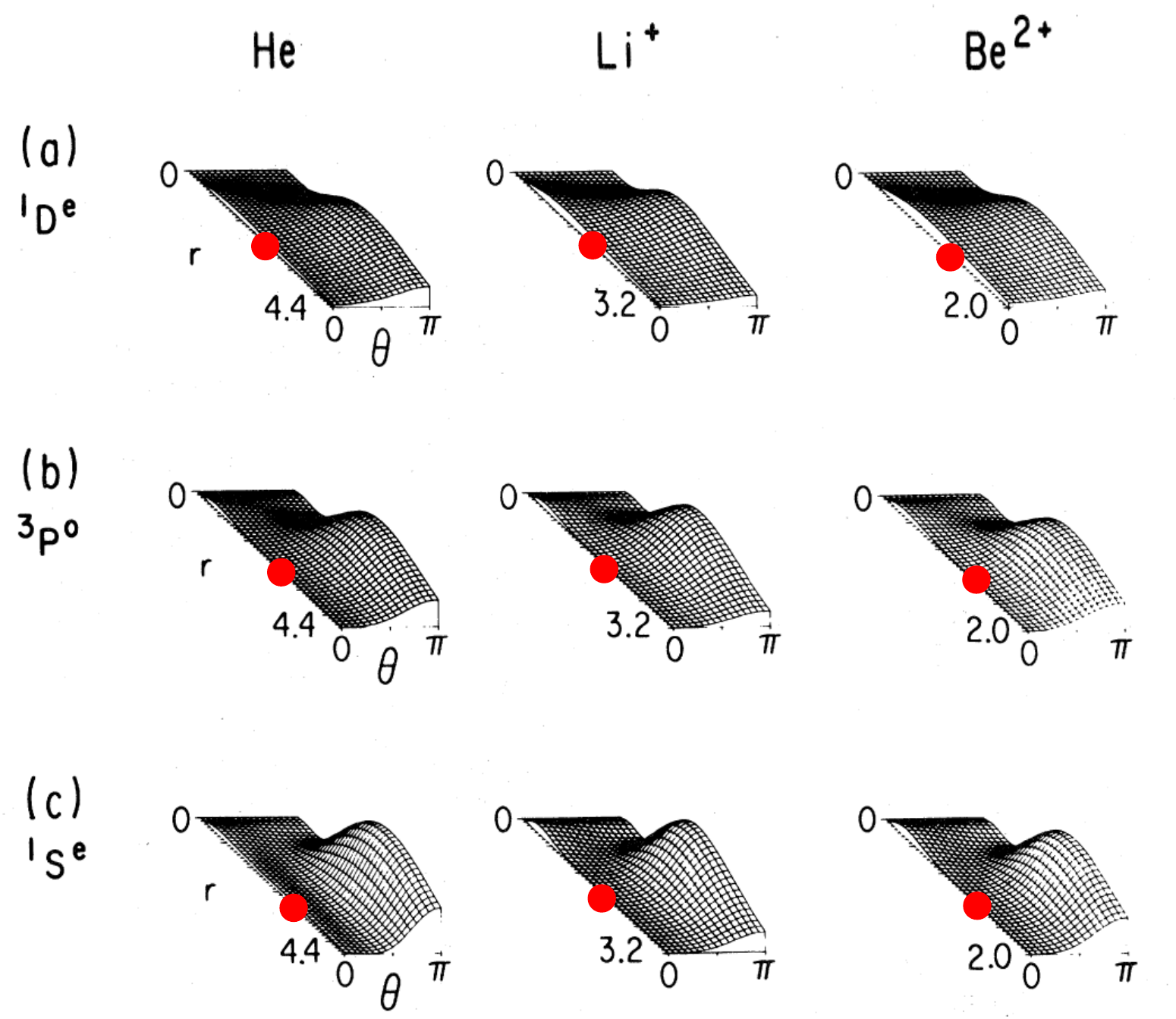}
\caption{CPDs $\rho(r_1,\theta_{12}|r_2=\alpha)$ for rotor series states
$^1S^e$, $^3P^o$, and $^1D^e$. Nuclear charge $Z=2$ (He), 3 (Li$^+$), and 4 (Be$^{2+}$).
Radial coordinates have been scaled to facilitate comparison of wave-function
shapes. The fized point, $r_2=\alpha$, has its most probable value in each case;
it is represented by a red solid dot. The probability of finding the second electron
is maximal at a point antipodal to the red dot (i.e., for $\theta_{12}=\pi$).
Reprinted (and slightly modified) figure with permission from Ref.\ \cite{ezra83},
Copyright (1983) by the American Physical Society.
}
\label{ezra_3D}
\end{figure}

The method of configuration interaction has also played a prominent role in this
field \cite{ezra83,berr89,gene21,eich25}. As mentioned earlier, the Wigner molecular
configurations can been extracted from the CI wave functions with the help of
the conditional probability distributions. An example is given in Fig.\ \ref{ezra_3D};
it illustrates the formation of a 2e antipodal WM associated with yrast (purely rotor)
states (as indicated) for the isoelectronic series He, Li$^+$, and Be$^{2+}$.
For CPDs, analogous to those in Fig.\ \ref{2espec}, illustrating vibational modes of
the 2e antipodal WM in He, Li$^+$, and Be$^{2+}$, see other figures in Refs.\
\cite{ezra83,berr89}.

\subsection{A few electrons at zero magnetic field confined in a semiconductor 2D
parabolic quantum dot: The limit of the rigid rotor employing the two-step method
of symmetry breaking and subsequent symmetry restoration}
\label{qdrr}
\smallskip

In general, at the first step of the two-step method, the localized broken symmetry
orbitals are determined numerically via a selfconsistent solution of the
Pople-Nesbet UHF equations \cite{yann99,yann02.2,yann03.2,yann07}. 
An efficient alternative, however, is to approximate these orbitals by
appropriate analytical expressions, and specifically by displaced Gaussian
functions\footnote{
Wigner \cite{wign38} was the first to discuss, as a quantal correction to the point
charges forming the WC lattice, the appropriateness of displaced Gaussians for
describing the orbitals of the localized electrons.}
\cite{yann02.2,will02}. Namely, for an electron localized at $\bcr_j$,
$j=1,2,\ldots,N$, one uses the spatial orbital 
\begin{equation}
u(\br,\bcr_j) = \frac{1}{\sqrt{\pi} \lambda}
\exp \left( -\frac{|\br-\bcr_j|^2}{2\lambda^2} \right).
\label{uhfo1}
\end{equation}

The positions $\bcr_j$ are taken according to the configuration of nested
regular polygons\footnote{
A single regular polygon with $N$ vertices is denoted as $(0,N)$; a regular polygon
with $N-1$ vertices and one electron at the center is denoted as $(1,N-1)$.}  
of $N$ classical point charges inside an external parabolic confinement of
frequency $\omega_0$ \cite{bolt93,kong02}. Then one proceeds to construct the
UHF determinant $\Psi^{\rm UHF}(\br_1,\ldots,\br_N)$ out of the orbitals
in Eq.\ (\ref{uhfo1}). Assuming fully spin-polarized electrons, correlated
many-body states with good total angular momenta $L$ can be extracted from
the UHF determinant using the projection formalism discussed in Sec.\ \ref{2eqd}.
In particular, only the operator ${\cal P}_L$ in Eq.\ (\ref{amp}) is needed here.
Then, setting ${\cal O}={\cal P}_L$ in the general expression defined by Eq.\
(\ref{epr}), the projected energies are given by
\begin{equation}
E_{\rm PRJ} (L) = \left. { \int_0^{2\pi} h(\gamma) e^{i \gamma L}
d\gamma } \right/ { \int_0^{2\pi} n(\gamma) e^{i \gamma L} d\gamma},
\label{eproj2}
\end{equation}
with
$h(\gamma) = 
\langle \Psi^{\rm UHF}(0) | {\cal H} | \Psi^{\rm UHF}(\gamma) \rangle$
and
$n(\gamma) = 
\langle \Psi^{\rm UHF}(0) | \Psi^{\rm UHF}(\gamma) \rangle,$
where $\Psi^{\rm UHF}(\gamma)$ is the original UHF determinant rotated by an
azimuthal angle $\gamma$ and ${\cal H}$ is the many body Hamiltonian (including
the external confinement and the Coulomb two-body repulsion).
Note that the UHF energies are simply given by $E_{\rm UHF} = h(0)/n(0)$.

At $B=0$, it is advantageous to minimize (for each $L$) the projected energy
[Eq.\ (\ref{eproj2})] by allowing the width $\lambda$ and the positions $\bcr_j$'s
of the displaced Gaussians to vary. For large $R_W$'s and for all cases of $N=2-5$
electrons, the calculated projected energies of the states in the yrast band can be
approximated by
\begin{equation}
E_{\rm PRJ}(L) \approx E_{\rm PRJ}(0) + C_R (\hbar L)^2,
\label{erigid}
\end{equation}
where the rotational coefficient $C_R$ is essentially a constant whose value 
is very close to that of the classical rigid-rotor (denoted by $C_R^{\rm cl}$).
Namely, $C_R \approx C_R^{\rm cl} = 1 /(2 {\cal J}(r_{\rm cl}))$,
where ${\cal J}(r_{\rm cl}) = N m^* r_{\rm cl}^2$ is the moment of inertia of
a {\it pinned\/} (also referred to as {\it static\/}) classical Wigner molecule,
i.e., of $N$ point-like electrons in their $(0,N)$ regular-polygon equilibrium
configuration inside a parabolic confinement of frequency $\omega_0$.
$r_{\rm cl} =l_0 R_W^{1/3} (S_N/4)^{1/3}$ is the (equilibrium) radius of this
classical WM, with $l_0$ being the oscillator length defined in Eq.\ (\ref{l0})
and $S_N= \sum_{j=2}^{N} \left( \sin[(j-1)\pi /N] \right)^{-1}$.

\begin{table}[t]
\caption{\label{b0}
Projected total energies $E_{\rm PRJ}(L)$ at $B=0$ and $R_W=200$ associated
with yrast states for $N=5$ electrons [(0,5) configuration].
$ \protect\widetilde{f} \equiv C_R/C_R^{\rm cl}$ (see text).
The electrons were assumed to be fully polarized, and thus, due to symmetry
constraints \cite{yann03.2}, only projected wave functions with $L=jN$, $j \protect\in
\protect\mathbb{Z}$, are non-vanishing. Energies in units of $\hbar \omega_0$.
Reprinted table with permission from Ref.\ \cite{yann04.2},
Copyright (2004) by the American Physical Society.
}
\begin{tabular}{ccc|ccc}
$L$ & $E_{\rm PRJ}$ & $\protect\tilde{f}$ & 
$L$ & $E_{\rm PRJ} $ & $\protect\tilde{f}$\\ \hline
  0  & 323.3070   &        & 25  & 324.7657 & 0.988 \\
  5  & 323.3656   & 0.992  & 30  & 325.4033 & 0.986 \\
 10  & 323.5414   & 0.992  & 35  & 326.1537 & 0.983 \\
 15  & 323.8338   & 0.991  & 40  & 327.0153 & 0.981 \\
 20  & 324.2422   & 0.989  & 45  & 327.9866 & 0.978 \\                   
\end{tabular}
\end{table}

An detailed example of the calculations above is offered in Table \ref{b0}.
Specifically, Table \ref{b0} lists the calculated $E_{\rm PRJ}(L)$ [see Eq.\
(\ref{eproj2})] values for $N=5$ and for the large value of $R_W=200$. The fact that
$\widetilde{f} \equiv C_R/C_R^{\rm cl} \approx 1$ for all $L \leq 45$ illustrates
that projection techniques can capture RWM's quasiclassical
limit of a rigid rotor. I note that the rigid-rotor spectrum was also produced in Sec.\
\ref{2eqdex} by probing the exact solution of the two-electron Schr\"odinger equation.   
Of course, for smaller values of $R_W$, the rigidity of the RWM is progressively 
reduced, as illustrated for $R_W=20$ and $R_W=3$ in Sec.\ \ref{2eqdex} using the exact
wave functions, or for $R_W=2.39$ in Sec.\ \ref{2eqd} using the two-step method.

I note that, for $R_W=200$, $E_{\rm PRJ} (L=0) - N \hbar \omega_0$ in Table \ref{b0} is
very close to the classical electrostatic value of point-like electrons [in their
$(0,N)$ equilibrium configuration for $N=2-5$] inside a parabolic confinement of
frequency $\omega_0$, namely to
\begin{equation}
E_{\rm cl}(N)=
(3/8) (2 R_W)^{2/3} N S_N^{2/3} \hbar \omega_0.
\label{eclst}
\end{equation}

\subsubsection{\small{{\bf Commentary 4.}} Comparison with the classical rotating dimer and
introduction of the concept of Emergent Symmetry Breaking.}
\label{comm4}
In the preceding two sections, examples were provided demonstrating that
the rigid-rotor semi-classical regime can emerge as an exact (Sec.\ \ref{2eqdex}),
as well as an approximate (Sec.\ \ref{qdrr}), solution of the many-body
Schr\"odinger equation in the case of few-electron circular QDs and in the
limit of large $R_W$.

It is notable that the rotational coefficients $C_R$ in Eq.\ (\ref{erigid}),
extracted from the approximate two-step treatment, is directly related to the
classical moment of inertia (see above). But also, the rotational constant
$C \approx 0.037\;(\hbar \omega_0/2)$
in Eq.\ (\ref{spec}), derived from the exact solution, is expressable through
the associated classical moment of inertia of two point-like electrons. In fact,
$C \approx C^{\rm cl}=\hbar^2/(2m^*d_0^2) = 0.03698\; (\hbar \omega_0/2)$; $d_0=2.6$
$(l_0\sqrt{2})$ is half of the separation between the electrons as reported in the caption
of Fig.\ \ref{2espec}. Beyond the rotational coefficients and the moments of inertia,
a quasi-classical behavior emerges, in addition, relative to the vibrational modes
contained in Eq.\ (\ref{spec}). Before proceeding with the details, one needs to
establish the description of the vibrational modes of the semi-classical 2e Wigner
molecule.

To this effect, one starts by considering the electronic polar coordinates
$(r_1,\theta_1)$ and $(r_2,\theta_2)$ for a 2D parabolic 2e-QD system.
Then letting $s=r_1+r_2$, $t=(r_1-r_2)/2$, and defining $\theta_{12}=\theta_1-\theta_2$
to be the interelectronic azimuthal angle, the
{\it classical\/} equilibrium point for the two antipodal electrons
is $s_0^3=2e^2/(\kappa m^* \omega_0^2),\; t_0=0,\; \theta_{12,0}=\pi$. For $\theta_{12}$
in the neighborhood of $\pi$, $s$ is associated with the symmetric stretch, $t$ with the
antisymmetric stretch, and $\theta_{12}$ is associated with the transverse bending
perpendicular to the interelectron axis. The classical total potential (parabolic
confinement plus Coulombic repulsion), expanded to leading order around the classical
equilibrium point (for a similar analysis applied to the natural He atom, see Ref.\
\cite{kell80}), is given by:
\begin{equation}
V^{\rm cl}(s_0+\delta s, \delta t , \delta q)=
3e^2/(2\kappa s_0) +1.5 \mu \omega_0^2 \delta s^2 + 0.5 {\cal M} \omega_0^2
(\delta q^2 + \delta t^2),
\label{clexp}
\end{equation}
where $\mu$ is the 2e reduced mass and
${\cal M}=2m^*$. $\delta q = s_0(\pi-\theta_{12})/4$, being the displacement of the CM
of the two electrons perpendicular to the interelectron axis, is associated with the
bending vibration. It is straighforward to see that the frequency of the classical
symmetric stretch equals $\sqrt{3} \omega_0$. The bending mode and the antisymmetric
stretch are degenerate with frequency $\omega_0$, and together they give rise to a
two-dimensional vibration which coincides with the 2e CM motion and carries angular
momentum. Because the associated CPDs exhibit bending-like behavior [see Fig.\
\ref{2espec}(c)], this compound mode is treated here as the bending mode.

Upon quantization of the classical vibrations, the corresponding energy quantum
for the symmetric stretch is $\hbar \omega^{\rm cl}_s= 2\sqrt{3}=3.464$ (in units of
$\hbar \omega_0/2$), which is very close to the value $\hbar \omega_s=3.50$ determined
using the exact solution (see Sec.\ \ref{2eqdex}). Obviously, the energy quantum,
$\hbar \omega_b^{\rm cl}$, associated with the classical bending vibration [see last
term in Eq.\ (\ref{clexp})], equals exactly $\hbar \omega_0$, in full agreement
with the findings in Sec.\ \ref{2eqdex}.

The finding that the semi-classical near-rigid-rotor regime can emerge for a finite
system in the limit of large $R_W$, and without the prerequisite of an explicit
symmetry breaking, is quite remarkable. It disagrees strongly with the widely-accepted
theory of spontaneous symmetry breaking proposed in Ref.\ \cite{pwa_book}
in the context of condensed-matter physics, but also used in high-energy particle
physics \cite{wein_book}, which
focuses on an explicit initial stage of symmetry breaking that can emerge only at the
{\it singular\/} thermodynamic (infinite volume) limit \cite{weze19}, followed by a
subsequent quantization of the associated classical gapless collective excitations
(referred to as Nambu-Goldstone modes \cite{weze19}). Instead of being explicit, as
mentioned earlier, the symmetry breaking associated with the exact solutions (or their
symmetry-restored approximations) of the many-body Schr\"odinger equation for a finite
system is hidden, but its presence can be revealed by an analysis of the many-body
correlations. To emphasize this non-trivial difference, I adopt here the
term\footnote{
This term was introduced in Ref.\ \cite{yann07} and was further adopted in
Refs.\ \cite{pape15,pape25}. As mentioned earlier, the term ``obscure symmetry
breaking'' has also been proposed \cite{tasa94,tasaki_book}.}
``emergent symmetry breaking'' (ESB) for this situation pertinent to finite systems.

In addition to Ref.\ \cite{yann07}, the realization that an improved theory of
symmetry breaking (beyond the traditional infinite-volume mathematical formulation
of SSB \cite{pwa_book,zumi69,zumi69.2,wein_book}, referred to henceforth as
{\it formal\/} SSB), can be developed for the case
of finite systems has been discussed recently in the field of mathematical physics
\cite{land20,land13,land17,tasa94,tasaki_book,fras16} and effective field theories
\cite{pape15,pape25}.

As shown in the analysis presented in Secs.\ \ref{2eqdex} and \ref{qdrr}, the emergent
symmetry breaking occurs physically before\footnote{
This point has also been stressed in Refs.\ \cite{land20,land17,tasa94,tasaki_book,fras16}
using discrete (lattice) model spin Hamiltonians. Furthermore, it has been recently
realized that exact diagonalization studies on finite-lattice spin clusters can reveal the
progressive development of a tower of states and other main properties of symmetry
breaking in magnetic materials, associated earlier with the formal SSB, without the need
to enforce the thermodynamic limit \cite{lauc16,kats23}.}
any mathematically singular limit (e.g., the infinite-volume thermodynamic limt
$N \rightarrow \infty$ or the classical limit $\hbar \rightarrow 0$) is enforced.
In particular, the physical spectra derived for large $R_W$ in Secs.\ \ref{2eqdex}
and \ref{qdrr} exhibit the following properties \cite{pwa_book} that are considered
essential consequences of SSB: (i) emergence of a tower of quasidegenerate rotational
states (see Fig.\ \ref{2espec}), (ii) rigidity, (iii) emergence of vibrational modes,
which are gapped, but correspond to the massless Nambu-Goldstone
or massive Higgs bosons\footnote{
The Higgs mode is analogous to the plasmon resonance and appears in the presence of
long-range interactions \cite{pwa_book}.}
associated with the thermodynamic limit of formal SSB \cite{pwa_book,weze19},
and (iv) universality of the spectra because they are organized
according to symmmetry arguments alone; e.g., the strong analogies between the molecular
spectra of the parabolic 2e-QD and those \cite{levine_book} of the natural XYX linear
triatomic molecules are easily recognizable.

Another important property of emergent symmetry breaking is that of autonomy, namely
that the space patterns of the broken-symmetry state are not predetermined \cite{ande94}
by the environmental factor that triggers its emergence. An experimental observation of
this property will be presented in the following section (i.e., Sec.\ \ref{mqd} and
Commentary 6, Sec.\ \ref{comm6}).

As a last paragraph in this Section, it needs to be stressed that the property of
rigidity emerges for large $R_W$ in the absence of an applied magnetic field. For large
$B$, when the single-particle spectrum is organized in infinitely-degenerate Landau
levels, the emergent symmetry breaking is associated with formation of a
{\it hyper-floppy\/} rotor; see Sec.\ \ref{fe2step} below. This hyper-floppiness
underlies the exotic behavior of the topological FQHE states.

\begin{figure}[t]
 \centering 
\includegraphics[width=\textwidth]{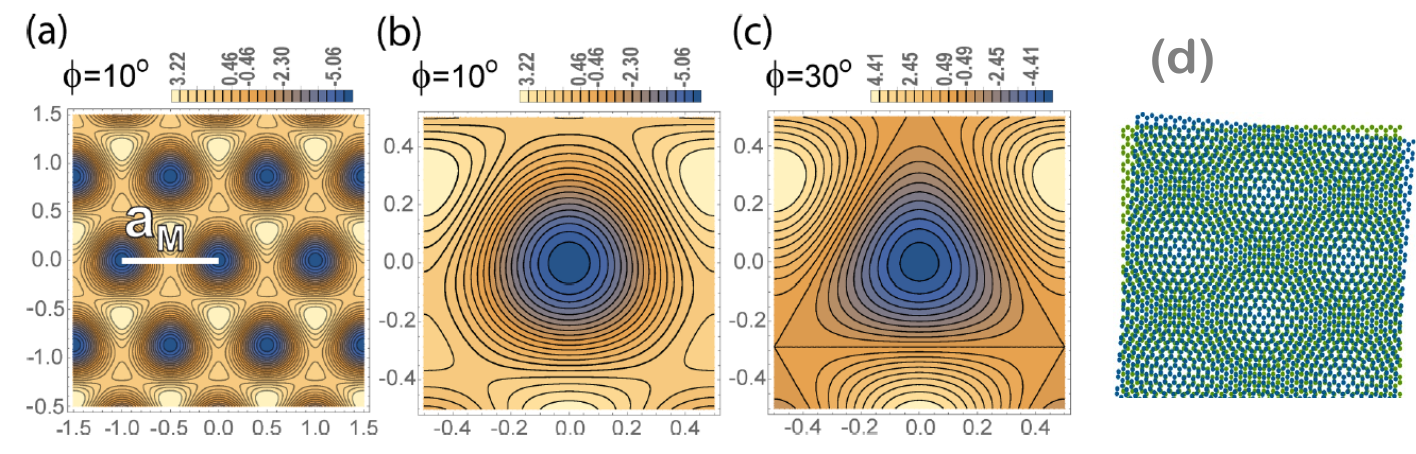}
\caption{
(a-c) Plot of the moir\'{e}-superlattice potential given by Eq.\ (\ref{mpot}).
(a) Broader view of the periodic potential structure for an angle of $\phi=10^\circ$.
(b) Potential of the isolated moir\'{e} QD for $\phi=10^\circ$. (c) Potential of
the isolated moir\'{e} QD for $\phi=30^\circ$. Lengths in units of the moir\'{e}
lattice constant $a_M$. Potential contours in units of $v_0$.
Note the change of the length scale in (b) and (c) compared to (a).
(d) Schematic plot illustrating the formation of a moir\'e superlattice by twisting
two hexagonal monolayers.
Reprinted figure with permission from Ref.\ \cite{yann23},
Copyright (2023) by the American Physical Society.
}
\label{potm}
\end{figure}

\subsection{Interplay of hidden and explicit symmetry breaking, and the
``flea on the elephant'' paradigm: Wigner molecule formation at zero magnetic field
in triangular QDs created in TMD moir\'e superlattices.}
\label{mqd}
\smallskip

In a significant development, most recent investigations
\cite{yann23,redd23,yann24,yann24.2,crom24,hu25,ke25,liqi25,khal25,liu26,chiu26,chak26}
have expanded the WM portfolio to include the newly emerging field of transition-metal
dichalcogenide (TMD) moir\'e materials and superlattices. TMD moir\'e superlattices are 
highly valued because of their potential for fundamental-science discoveries and for
their promise for advancing the applications of quantum-information devices, including for
solving the scalability challenge in quantum computer architectures. Particularly exciting
is the experimental Ref.\
\cite{crom24}, which reported STM pictures of the charge densities in arrays of moir\'e
periodic potential pockets demonstrating the interplay of {\it sliding\/}\footnote{
The use of the term ``sliding'' WM here instead of ``rotating'' WM will be explained in
detail in Sec.\ \ref{swmn4}.}
(symmetry-preserving) and {\it pinned\/} (symmetry-broken) WMs consisting of holes and
for the fillings $\nu=2-4$ (i.e., with $N=2-4$ particles per moir\'e pocket). 

In this section, I uncover, with the use of FCI calculations, the ubiquitous formation
of WMs in this novel class of van-der-Waals vertically stacked heterostructures. These
artificial architectures form moir\'{e} superlattices (with large, tunable lattice
constants, of the order of 10 nm) when TMD layers are vertically stacked with a small
twist angle or lattice mismatch.
Examples of early studies of such moir\'e architectures can be found in the following
references that form part of a developing literature on twisted homo-bilayer (e.g.,
MoS$_2$, WS$_2$), or hetero-bilayer (e.g., WSe$_2$/WS$_2$,
MoSe$_2$/WSe$_2$) TMD constructions \cite{manz17,kaxi20,macd18,fu20,ange21}.

\subsubsection{Many-body Hamiltonian (including crystal field from surrounding
moir\'e pockets).}
Using FCI, Refs.\ \cite{yann23,yann24.2} did investigate\footnote{
Using FCI, Ref.\ \cite{redd23} investigated an isolated MQD with $N=3$ holes,
while Ref.\cite{yann24} investigated $N=4$ and $N=6$ holes in a double MQD.}  
a few-fermion ($N<7$, electrons
or holes) moir\'{e} quantum dots (MQDs) \cite{feen18,zeng22,song22} formed at the upper
layer of integer-filled doped bilayer TMDs \cite{mak21,feld22}. It was found that the
periodic potential pockets (see Fig.\ \ref{potm}) of the moir\'e superlattice, which
confine these fermions and define the MQDs, are closely approximated by the following
expression \cite{macd18,ange21,fu20}
\begin{equation}
V(\br) = -2 v_0 \sum_{i=1}^3 \cos(\bG_i \cdot \br + \phi),
\label{mpot}
\end{equation}      
where $\bG_i=[ (4\pi/\sqrt{3}a_M) ( \sin(2\pi i/3), \cos(2\pi i/3) ) ]$ are the moir\'{e}
reciprocal lattice vectors. The materials specific parameters of $V(\br)$ in Eq.\
(\ref{mpot}) are the potential-depth controlling factor $v_0$, the moir\'{e} lattice
constant $a_M$, and the angle $\phi$. ($v_0$ can also be experimentally controlled through
voltage biasing.) $a_M$ is typically of the order of 10 nm, which is substantially larger
than the lattice constant of a TMD single-layer material (typically a few \AA). The
parameter $\phi$ controls the strength of the trigonal shape ($C_3$ anisotropy) in a given
MQD potential pocket; see Fig.\ \ref{potm} for variations of the shape of $V(\br)$ with
$\phi$. Focusing on a single moir\'e pocket, the details of this trigonal $C_3$ anisotropy
can be seen by Taylor expanding $V(\br)$ in Eq.\ (\ref{mpot}) in powers of the polar
coordinate $r=|\br|$ and defining an approximate confining potential, $V_{\rm MQD}(\br)$,
for a single MQD as follows:
\begin{equation}
V_{\rm MQD}(\br) \equiv V(\br) + 6 v_0 \cos(\phi)  \approx m^* \omega_0^2 r^2/2 +
\cc \sin(3 \theta) r^3,
\label{vexp}
\end{equation}
with $m^*\omega_0^2=16 \pi^2 v_0 \cos(\phi)/a_M^2$, and
$\cc=16 \pi^3 v_0 \sin(\phi)/( 3\sqrt{3}a_M^3)$; $m^*$ is the effective mass and the
expansion of $V(\br)$ can be restricted to the terms up to $r^3$. $(r, \theta)$ are the
polar coordinates of the position vector $\br$.

\begin{figure}[t]
\includegraphics[width=0.495\textwidth]{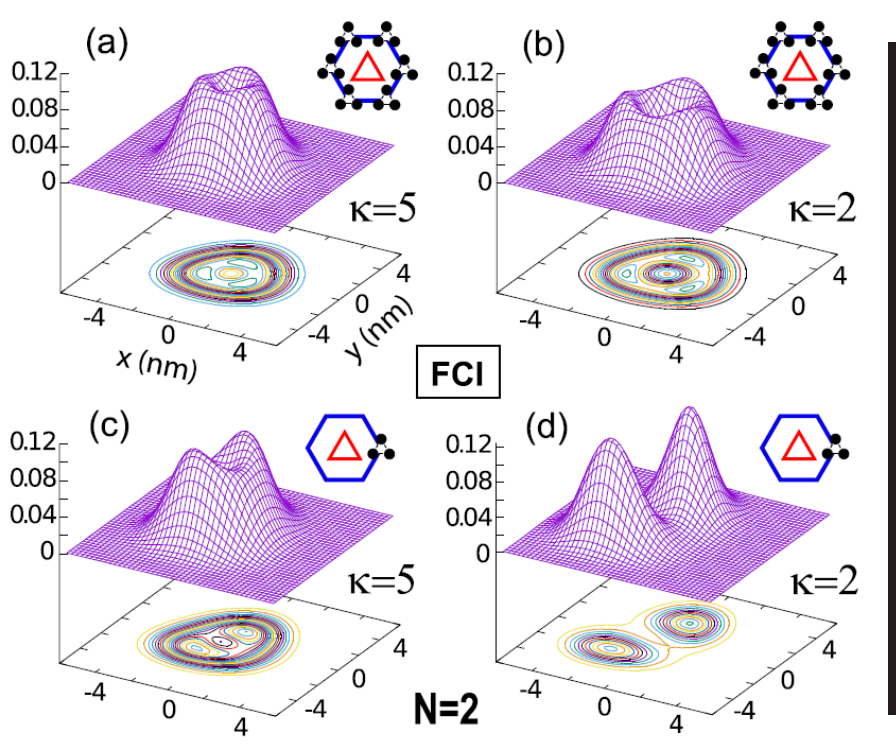} \hspace{0.005\textwidth}
\includegraphics[width=0.495\textwidth]{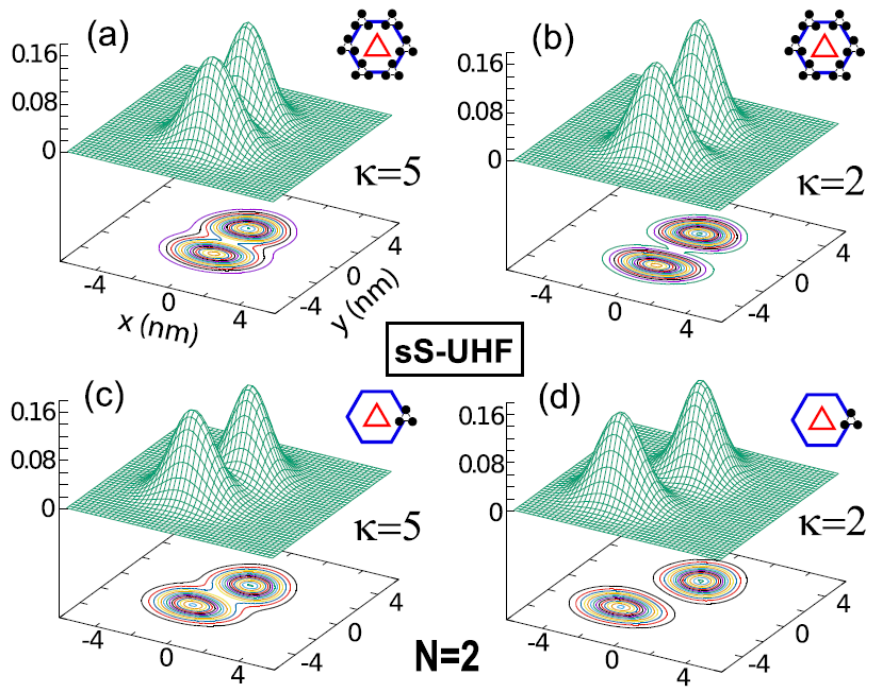}
 \caption{
(left) FCI charge densities for the ground state (a singlet, $S=S_z=0$) of $N=2$
holes in the central MQD (red triangle in the schematics) taking into account
the crystal-field effect arising from the charges that are present in the surrounding six
moir\'e pockets (see blue hexagons, with radius $a_M$, in the schematics). (a) and (c)
$\kappa=5$ in agreement with actual experiments which investigate the properties of TMD
moir\'e superlattices in an hBN environment. For the sake of comparison, (b) and (d) were
calculated with $\kappa=2$, which results in a stronger Coulombic repulsion.
In (a) and (b), each one of the six surrounding MQDs contains $Q=Ne$ charge carriers (see
schematics and the text). In (c) and (d), only one of the surrounding MQDs (the
one to the right, see schematics) is chosen to contain $Q=Ne$ charge carriers; the
remaining surrounding MQDs are empty.
The charge-carrier distributions in each surrounding MQD have been rendered by three point
charges of magnitude $Q/3$ placed at the vertices of a small equilateral triangle 
with radius $a_M/6$ (see schematics). The remaining parameters were chosen to
reproduce the experimental values \cite{crom24}): effective mass
$m^*=0.90 m_e$, moir\'e lattice constant $a_M=9.8$ nm, depth of a moir\'e pocket
$v_0=10.3$ eV, and $\phi=20^\circ$. The charge densities are given in units of 1/nm$^2$.
Reprinted figure with permission from Ref.\ \cite{yann24.2},
Copyright (2024) by the American Physical Society.
\label{cf1}
}
\caption{
(right) Charge densities of the sS-UHF ``singlet'' ground state (with $S_z=0$) for $N=2$
holes in the central MQD (red triangle in the schematics) taking into account
the crystal-field effect arising from the charges that are present in the surrounding six
moir\'e pockets (see blue hexagons, with radius $a_M$, in the schematics). The organization
of the information and the panels in this figure parallels that in Fig.\ \ref{cf1}. 
Reprinted figure with permission from Ref.\ \cite{yann24.2},
Copyright (2024) by the American Physical Society.
\label{cf2}
}
\end{figure}
  
The effective full many-body Hamiltonian, including crystal field contributions,
for a given MQD embedded in the moir\'e superlattice is given by
\begin{equation}
H_{\rm FMB} = H_{\rm MQD} + H_{\rm CF},
\label{fmbh}
\end{equation}
where
\begin{equation}
H_{\rm MQD} = \sum_{i=1}^N \left\{ \frac{{\bf p}_i^2}{2 m^*} +V_{\rm MQD}(\br_i) \right\} +
\sum_{i<j}^N \frac{e^2}{\kappa |\br_i-\br_j|},
\label{hmqd}
\end{equation}
with $\kappa$ being the dielectric constant, and
\begin{equation}
H_{\rm CF} = \sum_{i=1}^N \sum_{j=1}^{3{\cal M}} 
\frac{eQ}{3\kappa |\br_i - {\bf R}_j|} 
\label{vcf}
\end{equation}
is the crystal-field potential from the surrounding moir\'e pockets. ${\cal M}$ denotes
the number of surrounding moir\'e pockets populated with charge carriers, and the
${\bf R}_j$'s are the positions of the point charges that mimic the charge carrier
distributions. In each one of the surrounding moir\'e pockets, three point charges are
located at the vertices of an equilateral triangle with a total charge $Q=Ne$, thus
respecting the $C_3$ triangular symmetry; see, e.g., the solid dots in the schematics
included in Figs.\ \ref{cf1}(a) and (b), as well as Figs.\ \ref{cf2}(a) and (b). Note that
$e > 0$ for holes.

\subsubsection{Results for filling factor $\nu=2$.}
The FCI and sS-UHF CDs for a moir\'e TMD bilayer superlattice (e.g., homo-bilayer WS$_2$
\cite{crom24}) at the integer filling factor $\nu=2$ were computed for the full many-body
Hamiltonian (\ref{fmbh}); they are displayed in Figs.\ \ref{cf1} and \ref{cf2}, respectively.
The equilateral three-point-charge arrangements that generate the crystal-field effect
associated with the first six surrounding moir\'e pockets are sketched in the schematics
displaced in the right upper corner of each panel; see the figure captions for the
parameters used and for a detailed description. The total charge in each equilateral
arrangement is $Q=2e$.

\begin{figure}[t]
\centering
\includegraphics[width=0.9\textwidth]{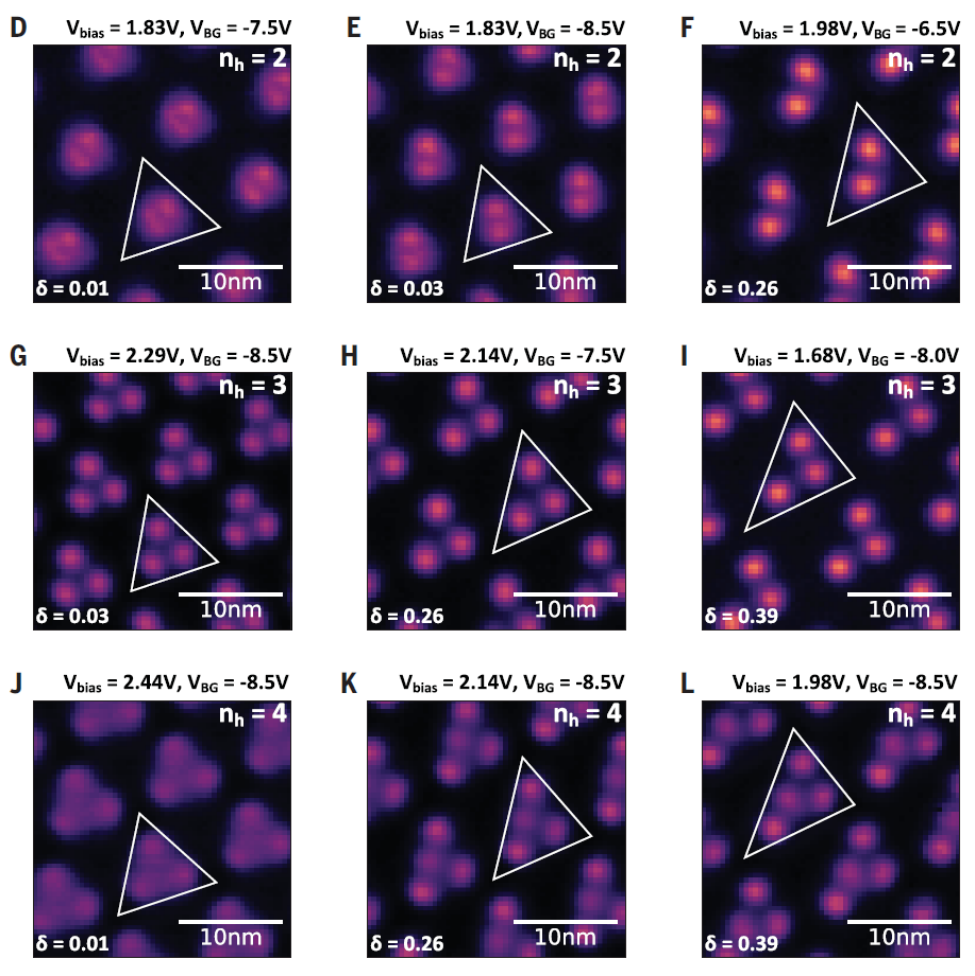}
\caption{
Configuration engineering of Wigner molecular crystal.
(D to F) Evolution of two-hole Wigner molecules as uniaxial strain is increased from (D)
$\delta=0.01$ to (E) $\delta=0.03$, and to (F) $\delta=0.26$.
(G to I). Evolution of three-hole Wigner molecules as uniaxial strain is
increased from (G) $\delta=0.03$ to (H) $\delta=0.26$, and to (I) $\delta=0.39$.
(J to L). Evolution of four-hole Wigner molecules as uniaxial strain is increased from
(J) $\delta=0.01$ to (K) $\delta=0.26$, and to (L) $\delta=0.39$.
The white triangle in each panel labels the potential well contour. T = 5.4 K for
all the above measurements.
The parameter $\delta$ characterizes the strength of the uniaxial strain applied.
Reprinted figure with permission from Ref.\ \cite{crom24},
Copyright (2024) by Science.}
\label{moire_sci1}
\end{figure}

\begin{figure}[t]
\centering
\includegraphics[width=0.6\textwidth]{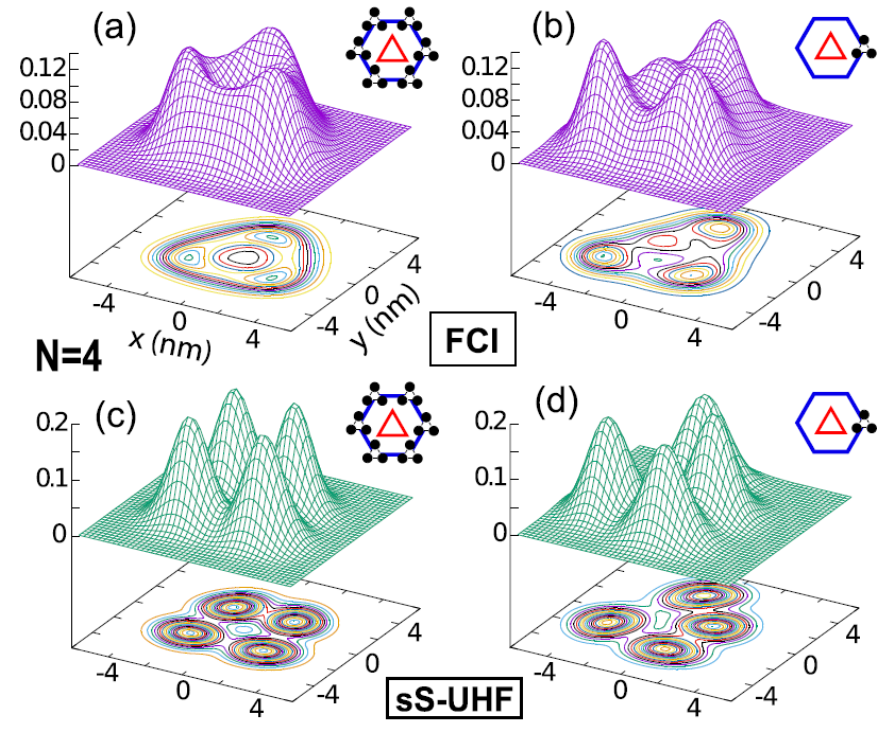}
\caption{
Charge densities of the ground state for $N=4$ holes in the central MQD (red triangle in the
schematics) taking into account the crystal-field effect arising from the charges that are
present in the surrounding six moir\'e pockets (see blue hexagons, with radius $a_M$, in the
schematics). (a) and (b) FCI ground states (with total spin $S=1$ and total-spin projection
$S_z=0$). (c) and (d) sS-UHF ground states (with total spin projection $S_z=0$). $\kappa=5$
in all fours frames.
In (a) and (c), each one of the six surrounding MQDs contains $Q=Ne$ charge carriers (see
schematics). In (b) and (d), only one of the surrounding MQDs (the
one to the right, see schematics) is chosen to contain $Q=Ne$ charge carriers; the
remaining surrounding MQDs are empty. The remaining parameters and notations are the same
as those in Figs.\ \ref{cf1} and \ref{cf2}.
Reprinted figure with permission from Ref.\ \cite{yann24.2},
Copyright (2024) by the American Physical Society.
}
\label{chdn4}
\end{figure}

The FCI CDs [displayed in Figs.\ \ref{cf1}(a) and \ref{cf1}(b)] demonstrate that the
full crystal-field effect (all the first six surrounding moir\'e pockets are involved)
maintains the symmetry-preserving charge densities that are associated with sliding WMs.
Namely, the CDs display the $C_3$ symmetry of the confining potential associated with a
single moir\'e pocket, as was the case with an isolated MQD investigated in Ref.\
\cite{yann23}. I note that the $C_3$ symmetry maintains in the FCI CD [see Fig.\ \ref{cf1}(b)]
even for a stronger Coulomb repulsion ($\kappa=2$ instead of $\kappa=5$, with the latter value
corresponding to the experimental setup). In the context of crystallinity, this
symmetry-preserving performance is counterintuitive, in that the CDs exhibit three weak humps
instead of the naive expectation of a symmetry-breaking double hump associated with the two
Wigner-crystal-type \cite{wign34,wign38,wang21,yazd24} localized charge carriers (with a
single hump per charge carrier). Remarkably, this counterintuitive behavior has been fully
confirmed in the recent experimental findings (see Fig.\ \ref{moire_sci1}(D) in Ref.\
\cite{crom24}) for the {\it unstrained\/} moir\'e sample at filling $\nu=2$.

On the contrary, the CDs in Figs.\ \ref{cf2}(a) and \ref{cf2}(b) show that the sS-UHF fails
to properly describe the unstrained TMD moir\'e supercrystal at filling $\nu=2$.
Indeed, instead of a sliding WM, these sS-UHF CDs represent pinned WMs that exhibit two
strongly-localized humps in agreement with the customary view of a Wigner crystal 
\cite{wign34,wign38,wang21,yazd24}, i.e., one hump per charge carrier. It is worth
stressing again that these sS-UHF CDs do break the $C_3$ symmetry of the MQD confining
potential and thus they strongly disagree with the experimental observations reported in Ref.\
\cite{crom24} for the unstrained case.

Heretofore, the employed crystal field had a $C_6$ symmetry (resulting from the six surrounding
moir\'e pockets when occupied with an equal charge $Q$), which is commensurate with the $C_3$
symmetry of the confinement of the central MQD. A question that arises naturally at this point
pertains to the effect of a crystal field with a symmetry non-commensurate to the $C_3$ symmetry
of an individual MQD. As a first example of a non-commensurability situation, the equilateral
three-point-charge configuration is maintained in one only of the surrounding moir\'e pockets,
i.e., the one situated on the right as indicated in the schematics in Figs.\ \ref{cf1}(c)
\ref{cf1}(d), \ref{cf2}(c), and \ref{cf2}(d).

From the FCI CDs in Figs.\ \ref{cf1}(c) and \ref{cf1}(d), it is clear that this
incommensurate crystal field (denoted as InCF-I below) produces an azimuthally
{\it pinned\/} WM exhibiting two antipodally situated humps. These pinned-WM CDs break the
$C_3$ symmetry of the individual MQD. Moreover, the localization of each charge carrier
is further amplified with stronger Coulombic repulsion; compare the case of $\kappa=5$
[Fig.\ \ref{cf1}(c)] to $\kappa=2$ [Fig.\ \ref{cf1}(d)]. Regarding the corresponding
sS-UHF CDs, it is apparent that, in qualitative agreement with the FCI result, the IcCF-I
CDs result also in pinned two-humped WMs, a stronger charge-carrier localization
notwithstanding; contrast Fig.\ \ref{cf2}(c) to Fig.\ \ref{cf1}(c) and
Fig.\ \ref{cf2}(d) to Fig.\ \ref{cf1}(d). I mention here that the pinning of the WM
that is visible in the IcCF-I FCI CDs agrees with the experimental observations \cite{crom24}
of two-humped WMs\footnote{
See also Fig.\ 4(e) in the most recent experimental Ref.\ \cite{liu26}.}
per moir\'e pocket in the case of {\it strained\/} moir\'e superlattices at a filling
$\nu=2$.

\subsubsection{Results for filling factor $\nu=4$.}
The capability of the crystal-field with the commensurate $C_6$ symmetry to maintain,
and even to bolster, the formation of a quantum sliding WM in the central MQD is
demonstrated in an even more striking way in the case of the filling factor $\nu=4$.
Indeed, within each MQD, Ref.\ \cite{crom24} has documented the observation of CDs with
3 humps ($C_3$ symmetry) in the case of the unstrained lattice [$\delta=0.01$, see Fig.\
\ref{moire_sci1}(J)], but with 4 humps (broken-symmetry pinned WM) in the case of the
strained lattice [$\delta=0.26$, Fig.\ \ref{moire_sci1}(K)] and $\delta=0.39$, Fig.\
\ref{moire_sci1}(L)].
The FCI CDs that incorporate the crystal-field effect agree quite well with the
experimental observations; see the 3-hump, sliding-WM CD in Fig.\ \ref{chdn4}(a)
(corresponding to a crystal field with commensurate $C_6$ symmetry) and the
distorted-4-hump, pinned-WM CD in Fig.\ \ref{chdn4}(b) (case of IcCF-I).
In contrast, the sS-UHF is unable to account for  the
sliding WM, yielding instead a broken-$C_3$-symmetry CD (i.e., a pinned WM) with 4 well
defined humps even for the commensurate crystal-field case; see Fig.\ \ref{chdn4}(c).
In the IcCF-I case, the sS-UHF CD agrees qualitatively with the FCI CD as it describes a
distorted-4-hump, pinned WM; compare Fig.\ \ref{chdn4}(b) and Fig.\ \ref{chdn4}(d). Note,
however, that the localization of the holes is more pronounced in the sS-UHF CD.

\begin{figure}[b]
\centering
\includegraphics[width=0.6\textwidth]{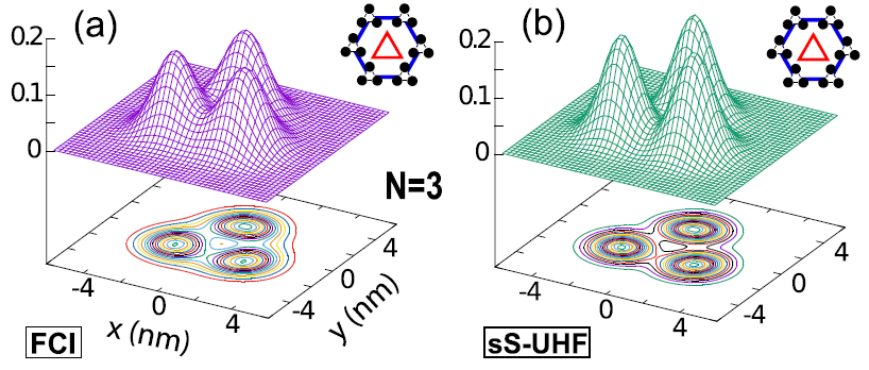}
\caption{
Charge densities of the ground state for $N=3$ holes in the central MQD (red triangle in the
schematics) taking into account the crystal-field effect arising from the charges that are
present in the surrounding six moir\'e pockets (see blue hexagons, with radius $a_M$, in the
schematics). (a) FCI ground state (with total spin $S=1/2$ and total-spin projection
$S_z=1/2$). (b) sS-UHF ground state (with total spin projection $S_z=1/2$). $\kappa=5$ in
both panels.  All other details are the same as those in Figs.\ \ref{cf1}, \ref{cf2}, and
\ref{chdn4}.
Reprinted figure with permission from Ref.\ \cite{yann24.2},
Copyright (2024) by the American Physical Society.
}
\label{chdn3}
\end{figure}

\subsubsection{Results for filling factor $\nu=3$.}
With respect to the cases of $\nu=2$ and $\nu=4$, the case of $\nu=3$ (see Fig.\
\ref{chdn3}) is anomalous, a fact that was already reported in Ref.\ \cite{yann23}
which investigated WM formation in an isolated MQD (i.e., in the absence of any
crystal field arising from the surrounding moir\'e pockets). Specifically, the FCI
CD in the central MQD [see Fig.\ \ref{chdn3}(a)] portrays a 3-hump {\it pinned\/} WM
which, in a straight way, accomodates the $C_3$ symmetry of the confinement.
The harmonization of the intrinsic ($C_3$) and the external ($C_6$) point-group
symmetries prohibits the formation of a sliding WM and yields instead a strongly pinned
WM for both the FCI and sS-UHF calculations. I note further that the sS-UHF CD
for $\nu=3$ is in qualitatively agreement with the FCI one [see Fig.\ \ref{chdn3}(b)],
with the charge-carrier localization, however, being more enhanced in the sS-UHF case.
Again, it is remarkable that the FCI results (including the crystal-field effect) for
$\nu=3$ are in excellent agreement with the experiments; see Figs.\ \ref{moire_sci1}(G to I)
in Ref.\ \cite{crom24} and Fig.\ 4(f) in Ref.\ \cite{liu26}.

\begin{figure}[t]
\centering\includegraphics[width=0.6\textwidth]{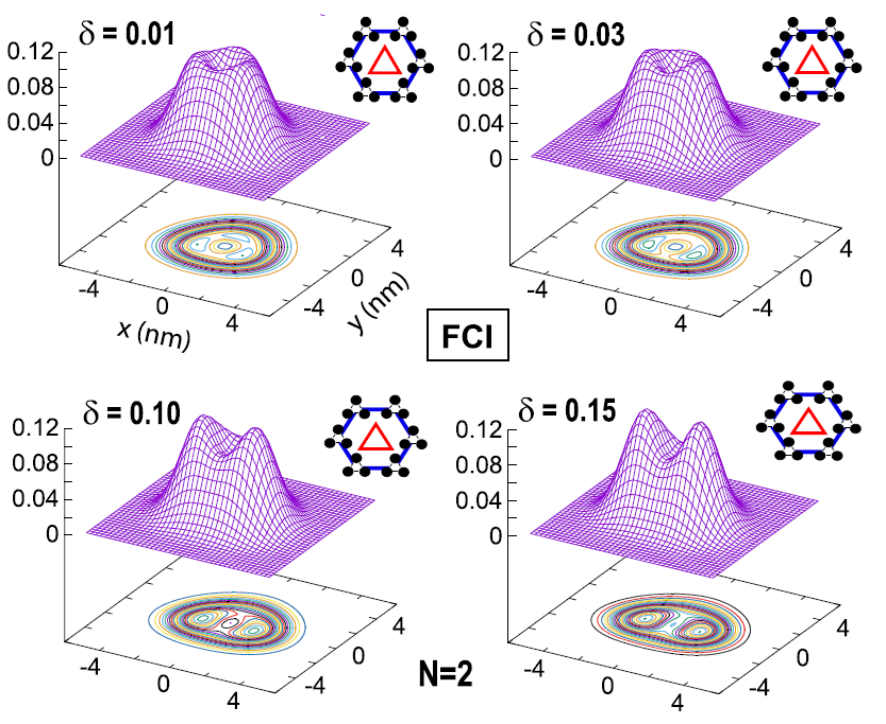}
\caption{
Evolution, as a function of the strain parameter $\delta$, of the FCI charge densities for
$N=2$ holes in the central MQD, including the full crystal-field effect as indicated in the
schematics. The values of $\delta$ are denoted in each panel. The evolution from three weak
peaks to two strong humps is clearly visible in the contour plots in the $x-y$ plane,
as well as in the 3D surfaces. Remaining parameters (in accordance with the experimental
setup \cite{crom24}): effective mass $m^*=0.90 m_e$, moir\'e
lattice constant $a_M=9.8$ nm, depth of a moir\'e potential pocket $v_0=10.3$ eV,
$\phi=20^\circ$, and dielectric constant $\kappa=5.0$. Charge densities are in units of 1/nm$^2$.
Reprinted figure with permission from Ref.\ \cite{yann24.2},
Copyright (2024) by the American Physical Society.
}
\label{strain}
\end{figure}

\subsubsection{Explicit consideration of strain in the FCI calculations.}
Until this point, this Section investigated the effect that the symmetry of the crystal
field has on the formation of Wigner molecules, i.e., sliding versus pinned WMs. Another
contributing factor, implemented in the experimental Ref.\ \cite{crom24}, is the application
of strain to the material. In this context, Fig.\ \ref{strain} displays in some detail the 
evolution of FCI CDs for $N=2$ holes as a function of the strain parameter $\delta$. The
remaining parameters are the same as those for the unstrained case ($\delta=0$) in Fig.\
\ref{cf1}(a). The definition of the strain parameter $\delta$ here follows that of Ref.\
\cite{crom24}, i.e., lengths along the $x$-direction are augmented by a factor $(1+\delta/2)$,
whereas lengths along the transverse $y$-direction are reduced by a factor $(1-\delta/2)$.
In particular, Fig.\ \ref{strain} displays FCI CDs for the values $\delta=0.01$, 0.03, 0.1,
and 0.15. It is seen that a well-formed pinned WM is emerging as the value of $\delta$
increases, in close agreement again with the experimental findings [see Fig.\
\ref{moire_sci1}(D to F)].

\begin{figure}[t]
\centering\includegraphics[width=0.8\textwidth]{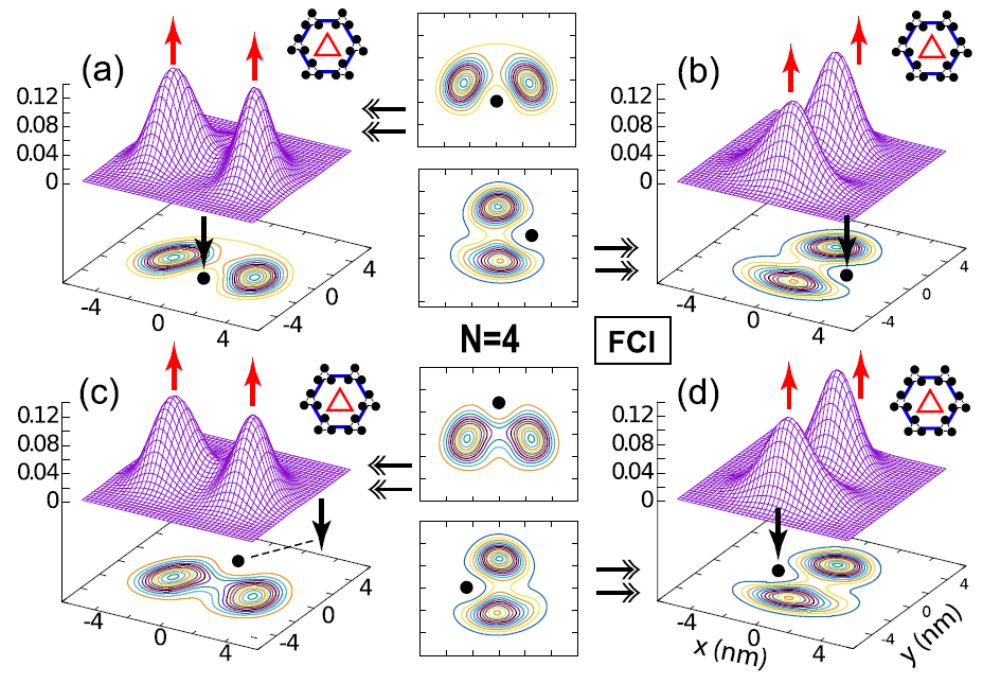}
\caption{
FCI spin-resolved CPDs ($P_{\uparrow\downarrow}$) associated with the ground state (with total spin
$S=1$, $S_z=0$) for $N=4$ holes confined in the central MQD, including the crystal-field effect
created by the charge carriers in the six surrounding moir\'e potential pockets (see blue
hexagon, with radius $a_M$, in the schematics). In all four frames, the dielectric constant was
taken as $\kappa=5$.  The black arrows denote the assumed down spin of the hole at the fixed
point, $\br_0$, which is situated at the following four different positions (marked by a black
solid dot): (a) $\br_0$=(0,-2.09 nm), (b) $\br_0$=(2.4 nm, 0),  (c) $\br_0$=(0, 2.61 nm),
and $\br_0$=(-2.4 nm, 0). The red arrows indicate that the plotted surfaces are associated with 
the distributions of a second spin-up hole. The remaining parameters are the same as those
in Fig.\  \ref{chdn4}(a), which portrays the corresponding charge density. The SR-CPDs are
plotted in units of  1/nm$^4$ and they are normalized to 2/nm$^2$.
Reprinted figure with permission from Ref.\ \cite{yann24.2},
Copyright (2024) by the American Physical Society.
}
\label{cpdn4du}
\end{figure}

\subsubsection{An example of CPDs of the sliding WM at $\nu=4$.}
\label{swmn4}
As demonstrated previously in this Section, the FCI CDs maintain the $C_3$ triangular 
symmetry of the moir\'e potential pocket, even under the influence of the $C_6$ hexagonal
TMD crystal field;  see Figs.\ \ref{cf1}(a),  \ref{cf1}(b) (for $\nu=2$), and Fig.\
\ref{chdn4}(a) (for $\nu=4$).

The $C_3$-symmetry conserving CDs in  Figs.\ \ref{cf1}(a),  \ref{cf1}(b), and
\ref{chdn4}(a) do not suggest by themselves the presence of any underlying Wigner
molecularization. Nevertheless, for these cases, the term "sliding WM" was employed because
the internal architecture of the FCI wave functions is equivalent to a superposition of {\it
unequal\/} static WM configurations with all possible azimuthal orientations. That this is
the case is promptly shown through an examination of the second-order correlations
known as spin-resolved conditional probability distributions [SR-CPDs; see definition in
Eqs.\ (\ref{cpds}) and (\ref{sponcpd}), and description of their physical meaning in the
accompanying text]. Note that, due to this inequality between azimuthal configurations, the
term "sliding" WM is used instead of the term ``rotating'' WM. 

The SR-CPDs that are displayed in Fig.\ \ref{cpdn4du} correspond to the ground state of the
4-hole sliding WM whose charge density was displayed in Fig.\ \ref{chdn4}(a). They belong to
the family $P_{\uparrow\downarrow}$, which can be briefly described as "fix a hole with spin down
and look for another hole with spin up". Of course, one needs to consider also the three
remaining families $P_{\downarrow\downarrow}$, $P_{\uparrow\uparrow}$, and $P_{\downarrow\uparrow}$, which
can be found in Ref.\ \cite{yann24.2}.

Inspection of these four families of SR-CPDs reveals an internal architecture for the sliding
WM consisting of four localized fermionic holes, in an up-down-up-down alternating spin pattern.
Furthermore, in accordance with the reasoning of Ref.\ \cite{yann09}, one can deduce that the
corresponding intrinsic spin eigenfunction is given by the expression
\begin{equation}
\chi^{S=1}_{S_z=0} = ( |\uparrow\downarrow\uparrow\downarrow\rangle -
 |\downarrow\uparrow\downarrow\uparrow\rangle)/\sqrt{2}.
\label{4hspin}
\end{equation}

Unlike the case of a circular confinement (see Secs.\ \ref{2eqd} and \ref{2eqdex},
and also Refs.\ \cite{yann00,yann04,yann07} among many others), the shape of the
CPD surface in Fig.\ \ref{cpdn4du} varies with the azimuthal angle. 
As aforementioned, this variation is due to the fact that the associated
CD in Fig.\ \ref{chdn4}(a) has a triangular doughnut shape.

At this point, a brief comment on the approaches that can bridge the gap between the
broken-symmetry sS-UHF and the symmetry-preserving FCI solutions is pertinent. Indeed, in
the case of a circular confinement, it was shown in Sec.\ \ref{2eqd} that one can use
projection techniques. This amounts to carrying out a mixing (superposition) of the
equivalent (related simply through a rotation) broken-symmetry sS-UHF Slater determinants
over all the azimuthal angles using coefficients given by algebraic expressions that are
supplied by the projection techniques and
other symmetry considerations [see Eq.\ (\ref{cprj}) and related text]. In the case of a
triangular MQD considered in this Section, the mean-field components entering in the mixing
are not equivalent in all the azimuthal angles, and the mixing coefficients $C(\gamma)$ need
to be determined numerically. To this end, one must employ the Griffin-Hill-Wheeler generator
coordinate method \cite{yann07,shei21,grif57,hill53}, or the closely related resonating-HF
approach \cite{fuku88}, which are broader in scope than the symmetry-restoration via
projection-techniques methodology.
The computational implementation of this task for the case of a trigonal
MQD remains to be carried out in future investigations; however, for an application of
the resonating HF to the case of a 2D square quantum dot, see Ref.\ \cite{okun09}.

\subsubsection{{\small{\bf Commentary 5.}} The ``flea on the elephant'' and a new
interpretation of ``spontaneous''.}
\label{comm5}

Previously in this Section, the transition (or crossing) from a symmetry-preserving
Schr\"odinger solution to an explicitly broken-symmetry one due to an additional perturbing
potential term, $V_P$, in the many-body Hamiltonian was demonstrated both experimentally
(see Fig.\ \ref{moire_sci1}) and theoretically (see FCI panels in Figs.\ \ref{cf1},
\ref{chdn4}, and \ref{strain}) for the case of a MQD. Two different families of perturbations
were considered: 1) Adding to $H_{\rm MQD}$ [see Eq.\ (\ref{hmqd})] a crystal-field potential
term from only a single pocket out of the six surrounding moir\'e pockets, and 2) Applying
a uniaxial strain on the full Hamiltonian $H_{\rm FMB}$ [see Eq.\ (\ref{fmbh})]; the strength
of the strain is expressed by the parameter $\delta$.

From the inspection of the figure panels mentioned in the previous paragraph, it follows
that the crossing from symmetry-preserving to symmetry-breaking wave function takes
place within a certain region of values of the parameters (such as $\delta$ or $\kappa$)
that control the strength of the perturbing potential $V_P$. In this context, it is
noteworthy that, experimentally, the effect of the breaking of the $C_3$ symmetry in
the WM charge density could be demonstrated for small values of $\delta$, as small as
$\delta=0.03$ for the filling factor $\nu=2$. At the higher average density associated
with $\nu=4$, a similar effect requires a somewhat larger value of $\delta=0.26$.

These observations suggest that the sensitivity of a finite system to a perturbation
inducing a transitioning to a state of explicit symmetry breaking can be systematically
investigated both experimentally and theoretically. This is apparently a well motivated
project, because a very sharp transition region can provide a finite-size substitute for,
and a reassessment of the meaning of, the adjective ``spontaneous'' in traditional SSB.
In this spirit, it is worth mentioning earlier related literature. Indeed, 
after considering initially a 1D symmetric double well whose ground state has a symmetric
wavefunction (i.e., it is spread out in both wells), G. Jona-Lasinio {\it et al.\/}
\cite{jona81,jona84} succeeded in showing in a mathematically rigorous way that a tiny,
asymmetric perturbation introduces an instability which, spontaneously, can break the
left-right parity symmetry, as long as the kinetic energy in the Schr\"odinger equation
is small, i.e., $\hbar \rightarrow 0$ or $M \rightarrow \infty$ (with $M$ being the
particle mass). Although counterintuitive, this small asymmetry (e.g., a tiny tilt between
the two wells) causes the ground-state wavefunction to become exponentially localized in
one of the wells. Such an effect, where a very small perturbation alters drastically a
much larger system, is often referred to as the ``flea on the elephant'' effect
\cite{simo85,land20,yann20}, a term coined by Ref.\ \cite{simo85}.

Naturally, the single-particle double-well Schr\"odinger equation considered in Refs.\
\cite{jona81,jona84,simo85} is not a many-body problem. A connection to a spin-lattice
many-body problem was presented, however, in Ref.\ \cite{land20} where the quantum
Curie-Weiss Hamiltonian was mapped onto the 1D Schr\"odinger operator describing a
particle in a symmetric double-well potential. Additionally, it was shown \cite{land20}
that the abrupt (spontaneous-like) breaking of the parity symmetry is due to the mixing of
the two lowest-in-energy states of the double well when they become quasidegenerate. This
situation corresponds to a small tower of states in the sense invoked by Anderson
\cite{pwa_book}.

Most importantly, note that the deep analogy between the ``flea on the elephant'' concept and
the symmetry breaking associated with large values of $R_W$ derives from the limit of
smallness of the quantum kinetic energy, either in absolute terms (original flea on the
elephant) or in a relative way (large $R_W$).

\subsubsection{{\small{\bf Commentary 6.}} The concept of autonomy in symmetry breaking.}
\label{comm6}

In this Section, specific examples were presented of explicit symmetry breaking in actual
finite physical systems under the influence of a perturbation term denoted as $V_P$. Two
families of $V_P$ were demonstrated, namely, an applied incommensurate electric field
or the shape-induced deformation of a strained potential trap; see, e.g., Figs.\
\ref{cf1}(c), \ref{cf1}(d), \ref{chdn4}(b), Fig.\ \ref{strain} for computational FCI
results and Figs.\ \ref{moire_sci1}(D to F) and (J to L) for corresponding experimental
observations.

These examples demonstrate that the property of autonomy is an essential aspect of
emergent symmetry breaking. Namely, although the symmetries embodied in $V_P$ are
reflected to some extent through an overall deformation, the defining space pattern
of the broken-symmetry state (i.e., the pinned Wigner molecule with one localized
CD hump per particle) is predetermined by the intrinsic many-body correlations, which are
revealed via inspection of the CPDs. Naturally, these corelations are present in the
symmetry-preserving state before the application of the external agent that triggers the
symmetry breaking. This behavior parallels closely the concept of autonomy as described
in Ref.\ \cite{ande94} in the framework of self-organization.

\begin{figure}[t]
\centering\includegraphics[width=0.9\textwidth]{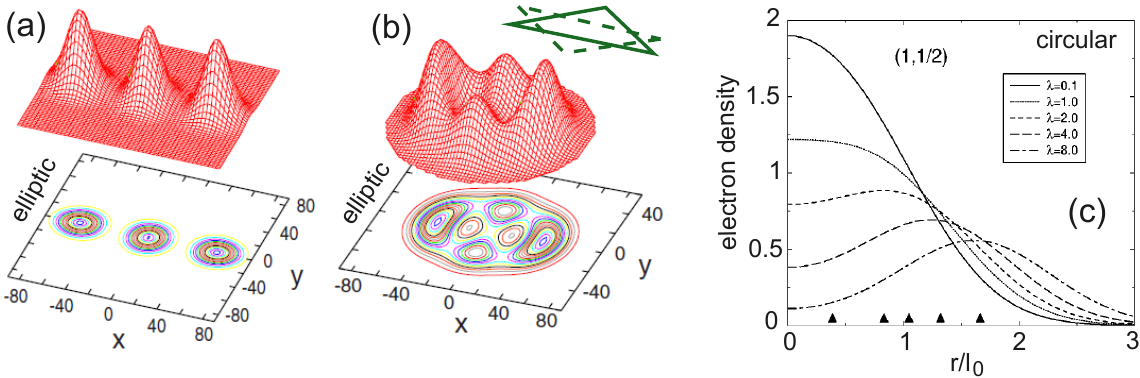}
\caption{
FCI total charge densities (spin-up plus spin-down) at zero magnetic field ($B=0$)
for the ground state of $N=3$ electrons in a 2D quantum dot.
(a) The ($S=1/2,S_z=1/2$) ground state at $\kappa=1.0$ (strong interelectron repulsion)
for an anisotropic QD with parameters $\omega_x=3.137$ meV, $\omega_y=6.274$ meV
($\eta=\omega_x/\omega_y=1/2$, strong anisotropy), and $m^*=0.067m_e$ (GaAs).
(b) The (1/2,1/2) ground state at $\kappa=1.0$ for an anisotropic quantum dot
with parameters $\omega_x=4.23$ meV, $\omega_y=5.84$ meV ($\eta=0.724$, intermediate
anisotropy), and $m^*=0.070m_e$ (GaAs). The schematic in (b) is a visual guide of
the two interlocking virtual isosceles triangles underlying the configuration of
the charge density due to the parity conservation along the $y$-axis. In (a) and (b),
the lengths are in nanometers and the electron densities are in arbitrary units.
Reprinted figure panels with permission from Ref.\ \cite{yann07.2},
Copyright (2007) by the American Physical Society.
(c) Total charge density for the ground state ($L=1,S=S_z=1/2$) of a circular QD
at different $\lambda$. Solid triangles show the positions of the radius $r_{\rm cl}$
of the classical WM at corresponding values of $\lambda$ ($\lambda$ coincides with
$R_W$). Reprinted figure panel with permission from Ref.\ \cite{mikh02.2},
Copyright (2002) by the American Physical Society.
}
\label{n3sqd1}
\end{figure}

\subsection{Extensive results for three particles across different platforms.}
\label{3ferm}

WM formation in the case of three fermions has been studied in various platforms and
confinement potentials and for different species of particles. The corresponding theoretical 
and experimental results are of central importance for the basic physics of
emergent symmetry breaking and WM formation, as well as for the applied-physics aspects
related to the fabrication of solid-state qubits. A particular aspect in this respect is
the interplay between the symmetry breaking and symmetry restoration of parity, which
involves the superposition of only two quasi-degenerate states. In this section, I will
present an anthology of results from relevant publications
\cite{akma99,mikh02.2,yann03.2,yann07.2,burk13,nogu14,yann16,corr21,kim21,
yann22,yann22.2,kim23,kim25}.

\subsubsection{Platform I: Three electrons in a 2D semiconductor quantum dot.}
\label{3epli}

As a first platform, I consider an elliptic or circular QD modeled by the following
in general anisotropic-oscillator potential confinement:
\begin{equation}
V(x,y) = \frac{1}{2} m^* (\omega_x^2 x^2+ \omega_y^2 y^2).
\label{vell}
\end{equation}
Naturally, the elliptic QD reduces to a circular parabolic one for
$\omega_x=\omega_y=\omega_0$. The ratio $\eta=\omega_x/\omega_y$ characterizes the
degree of anisotropy, and it will be referred to thereafter as the anisotropy
parameter. Results will be presented for three cases: $\eta=1/2$ (strongly anisotropic).
$\eta=0.724$ (slightly anisotropic), and $\eta=1$ (circular).

When the confining potential is sufficiently anisotropic, explicit symmetry
breaking and charge localization may be reflected directly in the single-particle
electron densities. Indeed, autonomous electron localization is visible in Figs.\
\ref{n3sqd1}(a) and \ref{n3sqd1}(b), which display the charge densities for the
($S=1/2$, $S_z=1/2$) ground state of $N=3$
electrons in an anisotropic quantum dot with $\eta=1/2$ and $\eta=0.724$,
respectively. For strong anisotropy ($\eta=1/2$), Fig.\ \ref{n3sqd1}(a) shows that
a three-hump pinned WM is formed whose density is shaped linearly. Note that the
three electron humps are sharply defined due to the strong interelectron
repulsion\footnote{
Since the average frequency, $\widetilde{\omega}_0=\sqrt{(\omega_x^2 + \omega_y^2)/2}$,
remains approximately constant (i.e., $\hbar \widetilde{\omega}_0 \approx 5.0$ meV)
for the two anisotropy cases described in this Section, one can calculate (using
$\widetilde{\omega}_0$) an effective value for $R_W \approx 19.10$.
}
($\kappa=1.0$).
For the intermediate anisotropy ($\eta=0.724$), Fig.\ \ref{n3sqd1}(b) displays an
electron density exhibiting pronounced peaks whose locations form a clearly defined
diamond; this indicates the presence of an underlying two-triangle intrinsic
configuration\footnote{
See also Ref.\ \cite{szaf04}, which reported a two-triangle configuration for $N=3$
electrons at high $B$ and for the case of the ($S=3/2$,$S_z=3/2$) fully
spin-polarized state.},
schematically drawn as part of Fig.\ \ref{n3sqd1}(b).

In the case of the circular ($\eta=1$) QD, one expects a hidden symmetry breaking.
In this case, the charge density is circularly symmetric and the total angular
momentum $L$ is a good quantum number. Consequently, Fig.\ \ref{n3sqd1}(c) displays
the radial dependence only for the state $(L=1,S=1/2,S_z=1/2)$\footnote{
This state is the ground state for $R_W < 4.343$, but it becomes the first excited
for $R_W > 4.343$ (with the $L=0$, $S=3/2$ state being the ground state
\cite{mikh02.2}), a behavior that is consistent with the limit of a rigid rotor for
$R_W \rightarrow \infty$.},  
which allows for portraying graphically the evolution of the charge density as a
function of the Wigner parameter\footnote{
The parameter $\lambda=l_0/a_B$ (used in Ref.\ \cite{mikh02.2}), where
$a_B=\kappa \hbar^2/(e^2 m^*)$ is the effective Bohr radius, coincides with $R_W$.}.
Note the depletion of the charge density
at the origin as $R_W$ changes from zero (non-interacting limit) to 8.0, which is a
reflection of the formation of a rotating WM with an intrinsic (0,3)
equilateral-triangle configuration.

Further information regarding the properties of the three-electron WMs discussed in
this Section can be extracted from the corresponding spin-resolved CPDs. Before
proceeding with the inspection of the CPDs, however, it is instructive to briefly
review the possible spin-dependent wave functions for three {\it localized\/} spatial
orbitals. In particular, I focus on the case with a total spin projection $S_z=1/2$,
when the most general three-orbital wave function is given by the superposition of
three Slater determinants, i.e., by the expression 
\begin{equation} 
{\cal W}(S_z=\mbox{$\frac12$})=
a |\uparrow \downarrow  
\uparrow \;\rangle +
b |\uparrow \uparrow 
\downarrow \;\rangle +
c |\downarrow \uparrow 
\uparrow \;\rangle,
\label{wfspabc}
\end{equation}
with the normalization $a^2 + b^2 + c^2=1$, where the positions of the arrows indicate
the localized spatial orbitals. The arrows themselves in Eq.\ (\ref{wfspabc}) indicate 
individual spin projections.

The general states (\ref{wfspabc}) are a superposition of three Slater 
determinants and have attracted a lot of attention in the mathematical theory
of entanglement. Indeed, they represent a prototypical class of three-qubit
entangled states known as $W$-states \cite{woot00}. For general coefficients
$a$, $b$, and $c$, the states (\ref{wfspabc}) are not eigenfunctions of the
square ${\hat{\bf S}}^2$ of the total spin (while the numerical FCI wave-function
solutions of the many-body Schr\"odinger equation are\footnote{
This does not necessarily apply in the case of degeneracies.}
good eigenfunctions of ${\hat{\bf S}}^2$). 
However, the special values of $a$, $b$, and $c$ which yield good total-spin
quantum numbers are well known \cite{lida06,pauncz}. Specifically, adopting the
notation ${\cal W} (S,S_z;i)$ (where the index $i$ is employed in case of a 
total-spin degeneracy \cite{pauncz}), one has
\begin{equation}
{\cal W} (\mbox{$\frac32$},\mbox{$\frac12$}) = 
\big( | \uparrow
 \downarrow
 \uparrow \;\rangle +
| \uparrow
 \uparrow
 \downarrow \;\rangle+
| \downarrow
 \uparrow
 \uparrow \;\rangle \big)/\sqrt{3} 
\label{wf3e3212}
\end{equation}
(i.e., $a=b=c=1/\sqrt{3}$),
\begin{equation}
{\cal W} (\mbox{$\frac12$},\mbox{$\frac12$};1) =
\big( 2 | \uparrow
 \downarrow
 \uparrow \;\rangle 
- | \uparrow
 \uparrow
 \downarrow \;\rangle
- | \downarrow
 \uparrow
 \uparrow \;\rangle \big)/\sqrt{6} 
\label{wf3e12121}
\end{equation}
(i.e., $a=2/\sqrt{6}$, $b=c=-1/\sqrt{6}$), 
\begin{equation}
{\cal W} (\mbox{$\frac12$},\mbox{$\frac12$};2) =
\big( | \uparrow
 \uparrow
 \downarrow \;\rangle
-| \downarrow
 \uparrow
 \uparrow \;\rangle \big)/\sqrt{2}
\label{wf3e12122}
\end{equation}
(i.e., $a=0$, $b=1/\sqrt{2}$, $c=-1/\sqrt{2}$).

For completeness, the much simpler expression for three fully spin-polarized
localized electrons (which is not an entangled $W$-state) is also listed, namely
\begin{equation}
{\cal W} (\mbox{$\frac32$},\mbox{$\frac32$}) = 
| \uparrow
 \uparrow
 \uparrow \;\rangle.
\label{wf3e3232}
\end{equation}

In the case of a circular QD, the $(S=1/2,S_z=1/2)$ spin-preserving three-electron
wave functions must embody the equilateral triangular configuration and thus they
have the form
\begin{equation}
\Theta(\mbox{$\frac12$},\mbox{$\frac12$};1)=
\big( | \uparrow
 \downarrow
 \uparrow \;\rangle 
+ e^{2\pi i/3} | \uparrow
 \uparrow
 \downarrow \;\rangle
+ e^{-2\pi i/3}| \downarrow
 \uparrow
 \uparrow \;\rangle \big)/\sqrt{3} 
\label{wf3e12121c}
\end{equation}
and
\begin{equation}
\Theta (\mbox{$\frac12$},\mbox{$\frac12$};2)=
\big( | \uparrow
 \downarrow
 \uparrow \;\rangle 
+ e^{-2\pi i/3} | \uparrow
 \uparrow
 \downarrow \;\rangle
+ e^{2\pi i/3}| \downarrow
 \uparrow
 \uparrow \;\rangle \big)/\sqrt{3}. 
\label{wf3e12122c}
\end{equation}

Note that, within a phase factor, Eqs.\ (\ref{wf3e12121}) and (\ref{wf3e12122})
can be produced by adding and subtracting Eqs.\ (\ref{wf3e12121c}) and
(\ref{wf3e12122c}), and then dividing by $\sqrt{2}$ to enforce normalization.

\begin{figure}[t]
\includegraphics[width=0.48\textwidth]{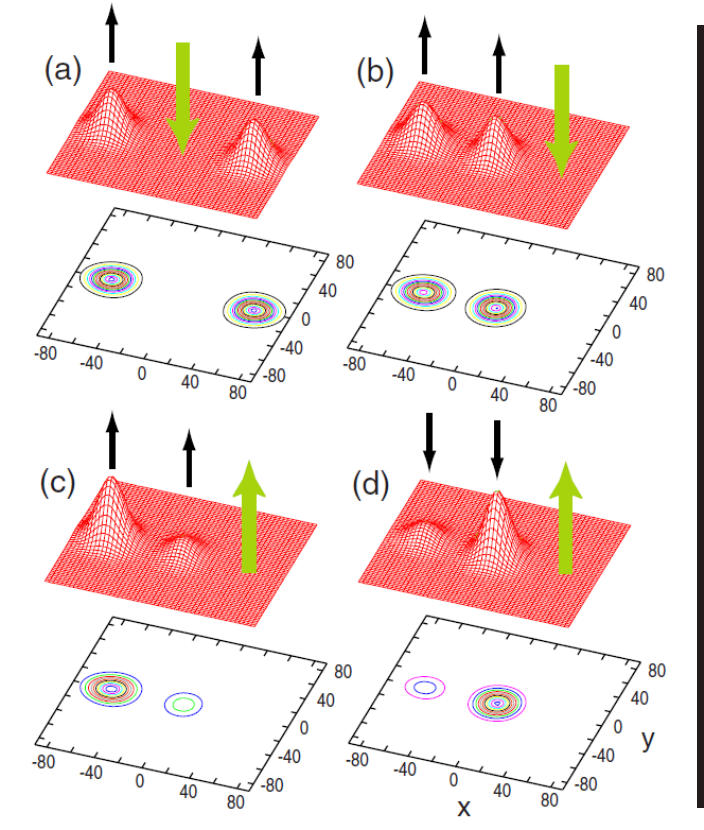}
\includegraphics[width=0.5\textwidth]{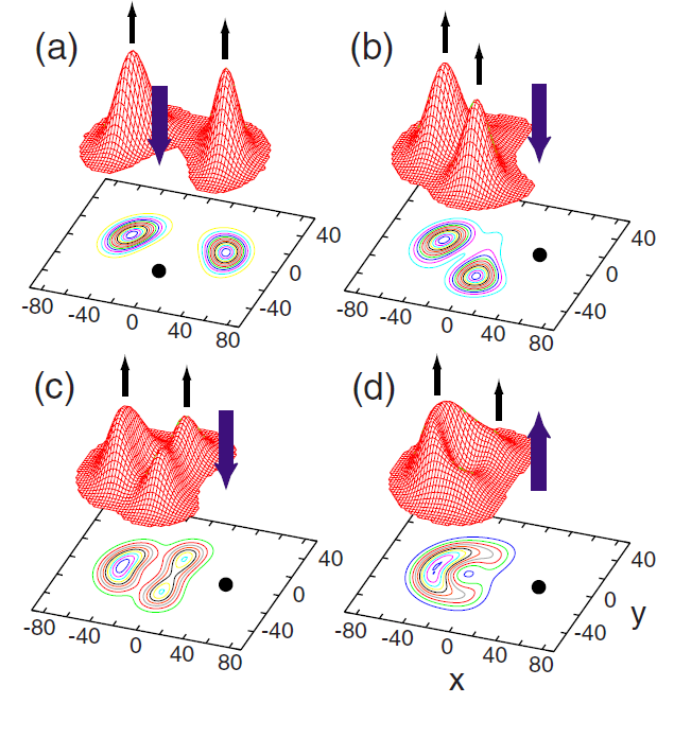}
  \caption{
(Left) Spin-resolved conditional probability distributions for the ($S=1/2,S_z=1/2$)
ground state of $N=3$ electrons in an anisotropic dot at zero magnetic field 
$(B=0)$ with parameters $\hbar\omega_x=3.137$ meV, $\hbar\omega_y=6.274$ meV 
$(\eta=1/2)$, $m^*=0.067m_e$ and $\kappa=1$ (effective $R_W \approx 19.10$) [for the
corresponding electron density, see Fig.\ \ref{n3sqd1}(a)].
The heavy green arrow indicates the location of the fixed electron
at ${\bf r}_0$ [see Eq.\ (\ref{sponcpd})], with the indicated spin projection
$\sigma_0$, i.e., up ($\uparrow$) or down ($\downarrow$).
(a) $\uparrow \downarrow$ CPD with the fixed spin-down electron located at the center.  
(b) $\uparrow \downarrow$ CPD with the fixed spin-down electron located on the right. 
(c) $\uparrow \uparrow$ CPD with the fixed spin-up electron located on the right. 
(d) $\downarrow \uparrow$ CPD with the fixed spin-up electron located on the right. 
The spin of the fixed electron is denoted by a thick arrow (green online).
Lengths in nanometers. The vertical axes are in arbitrary units, but the
scale is the same for all four panels.
Reprinted figure with permission from Ref.\ \cite{yann07.2},
Copyright (2007) by the American Physical Society.
\label{cpdn3li}
}
\caption{
(Right) Spin-resolved conditional probability distributions for the ($S=1/2,S_z=1/2$)
ground state of $N=3$ electrons in an anisotropic dot at $B=0$ with parameters 
$\hbar\omega_x=4.23$ meV, $\hbar\omega_y=5.84$ meV $(\eta=0.724)$, $m^*=0.070m_e$
and $\kappa=1$.
(a) $\uparrow \downarrow$ CPD with the fixed spin-down electron located on the
$y$-axis at (0,-20) (solid dot).
(b) $\uparrow \downarrow$ CPD with the fixed spin-down electron located off 
center at (40,11) (solid dot).
(c) $\uparrow \downarrow$ CPD with the fixed spin-down electron located on the
$x$-axis at (43,0) (solid dot).
(d) $\uparrow \uparrow$ CPD with the fixed spin-ip electron located on the
$x$-axis at (43,0) (solid dot).
The spin of the fixed electron is denoted by a thick blue arrow.
Lengths in nanometers. The vertical axes are in arbitrary units, but the
scale is the same for all panels in this figure.
Reprinted figure with permission from Ref.\ \cite{yann07.2},
Copyright (2007) by the American Physical Society.
\label{cpdn3di}
}
\end{figure}

In Fig.\ \ref{cpdn3li}, several spin-resolved CPDs are displayed that are associated
with the FCI ground state at vanishing magnetic field and strong anisotropy $\eta=1/2$.
Although the FCI expansion in Eq.\ (\ref{mbwf}) consists of a large
number of Slater determinants built from delocalized harmonic-oscillator orbitals,
the CPD patterns in Fig.\ \ref{cpdn3li} reveal an intrinsic structure similar to that
of the wave function ${\cal W}(\mbox{$\frac12$},\mbox{$\frac12$};1)$ in Eq.\
(\ref{wf3e12121}), which is made out of only three localized spatial orbitals.
In particular, when one requires that the fixed electron has a down spin and is
located at the center of the quantum dot, the spin-up electrons are
located on the left and right with equal weights [Fig.\ \ref{cpdn3li}(a)].
Keeping the down-spin direction, but moving the fixed electron to the
right, reveals that the spin-up electrons are located on the left and
the center with equal weights [Fig.\ \ref{cpdn3li}(b)]. Considering
[see Fig.\ \ref{cpdn3li}(c)] a spin-up direction for the fixed electron
and placing it on the right reveals that the remaining spin-up electron
is distributed on the left and the center of the quantum dot with unequal
weights: approximately 4 (left) to 1 (center) following the square of the
coefficients in front of the Slater determinants 
$| \uparrow  \downarrow
 \uparrow \;\rangle$ ($a=2/\sqrt{6}$) and
$| \downarrow  \uparrow
 \uparrow \;\rangle$ ($c=1/\sqrt{6}$) in the expression (\ref{wf3e12121}). 
Likewise [see Fig.\ \ref{cpdn3li}(d)], considering a spin-up direction for
the fixed electron and placing it on the right reveals that the spin-down
electron is distributed on the left and the center of the quantum dot with
unequal weights: approximately 1 (left) to 4 (center), in agreement with the
weights of the contributing Slater determinants in Eq.\ (\ref{wf3e12121}).

The detailed interlocking of the two resonating triangular configurations in the
charge density displayed in Fig.\ \ref{n3sqd1}(b) (elliptic QD with $\eta=0.724$,
$\kappa=1$, and $m^*=0.070m_e$) is further revealed in the FCI spin-resolved CPDs
that are displayed in Fig.\ \ref{cpdn3di}. From the CPDs in Figs.\ \ref{cpdn3di}(a) and
\ref{cpdn3di}(b), it can be seen that one triangle is defined by the 
points ${\bf R}_1 \approx (0,-20)$ nm, ${\bf R}_2 \approx (-43,10)$ nm, and
${\bf R}_3 \approx (43,10)$ nm, while its mirror (the second one) is formed by
the points ${\bf R}_1^\prime \approx (0,20)$ nm, 
${\bf R}_2^\prime \approx (-43,-10)$ nm, and
${\bf R}_3^\prime \approx (43,-10)$ nm. The $\uparrow \downarrow$ [Fig.\ 
\ref{cpdn3di}(c)] and $\uparrow \uparrow$ [Fig.\ \ref{cpdn3di}(d)]
CPDS with the fixed electron on the right at (43,0) nm are similar to those
in Figs.\ \ref{cpdn3li}(b) and \ref{cpdn3di}(c), respectively, with the 
importance difference that the central hump develops clearly to a double one. 
The analysis above indicates that each triangular component is associated with
a wave function of the form ${\cal W} (\mbox{$\frac12$},\mbox{$\frac12$};1)$ that
was given in Eq.\ (\ref{wf3e12121}). Naturally, the emergence of the regime of a
linear arrangement versus that of a two-triangle one depends on both the
strengths of the anisotropy and of the interaction.

\begin{figure}[t]
\centering\includegraphics[width=0.8\textwidth]{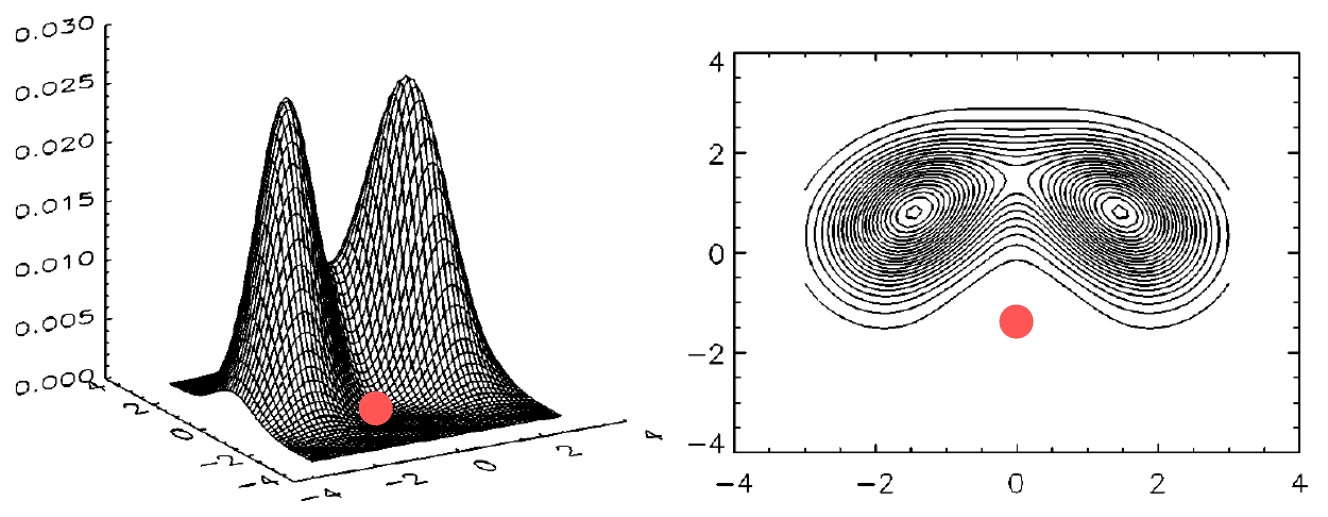}
\caption{
The three coinciding FCI CPDs $P_{\uparrow\uparrow}$, $P_{\downarrow\uparrow}$, and
$P_{\uparrow\downarrow}$ associated with $N=3$ electrons in a circular QD at $R_W=8.0$, and
for the first excited state with total angular momentum $L=1$, $S=1/2$, and $S_z=1/2$.
The fixed point (denoted as a red solid dot) was placed at
$\br_0=(0,-R_{\rm cl})=(0,-1.66 l_0)$.
Part of figures reprinted with permission from Ref.\ \cite{mikh02.2}, Copyright (2002)
by the American Physical Society.
}
\label{n3mikh}
\end{figure}

Turning to the case of $N=3$ electrons in a circular QD, Fig.\ \ref{n3mikh} displays, as
presented in Ref.\ \cite{mikh02.2}, the three coinciding FCI CPDs $P_{\uparrow\uparrow}$,
$P_{\downarrow\uparrow}$, and $P_{\uparrow\downarrow}$ for the state with total angular momentum
$L=1$, $S=1/2$, and $S_z=1/2$ and for the rather large value of $R_W=8 \gg 1$. They
reveal that the intrinsic configuration of the three-electron rotating WM for $R_W=8$
is that of an equilateral triangle. Furthermore, the coincidence of the three CPDs
corresponds to a wave function of the form described by either 
$\Theta(\mbox{$\frac12$},\mbox{$\frac12$};1)$ [see Eq.\ (\ref{wf3e12121c})] or
$\Theta(\mbox{$\frac12$},\mbox{$\frac12$};2)$ [see Eq.\ (\ref{wf3e12122c})]
The fact that the total angular momentum belongs to the sequence $L=3k+1$, $k=0,\pm 1,\pm 2,
\ldots$ enforces $\Theta(\mbox{$\frac12$},\mbox{$\frac12$};1)$ as the proper choice. Indeed,
as shown in Section VI A in Ref.\ \cite{yann03.2}, each one of the expressions
(\ref{wf3e12121c}), (\ref{wf3e12122c}), and (\ref{wf3e3232}) is associated, respectively,
with the following sequences of magic angular momenta $L=3k+1$, $L=3k+2$, and $L=3k$,
$k=0,\pm 1,\pm 2,\ldots$, which are thus coupled to specific total-spin values. This behavior
follows from symmetry requirements \cite{ruan95,maks96,yann03.2}. In the case of a high
magnetic field leading to formation of Landau levels, such magic angular momenta underlie the
emergence of the fractional quantum Hall effect (see Sec.\ \ref{fcillln7fp} below).

\begin{figure}[t]
\centering\includegraphics[width=0.45\textwidth]{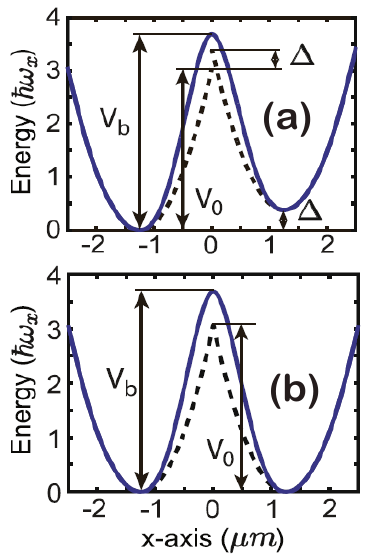}
\caption{
Examples of two-center-oscillator potentials along the $x$-direction illustrating the
function of the smooth neck. (a) Tilted double well ($\Delta>0$). (b) Symmetric double well
($\Delta=0$). The parameters used correspond to the cases of three $^6$Li neutral atoms in
Figs.\ \ref{n3_njp}(f and II) and \ref{n3_njp}(n and VI ) below.
$\hbar \omega_{x1}=\hbar \omega_{x2}=\hbar \omega_x=6.6$ kHz, $V_b=24.30\;
\mbox{kHz}=3.682\hbar \omega_x$, $d=2.5$ $\mu$m. $\Delta=2.5\hbar \omega_y=0.379\hbar\omega_x$
($\hbar \omega_y=1$ kHz) in (a) and $\Delta=0$ in (b). $V_0$ denotes the interwell barrier
(from the left side) for the pure two-parabola confinement without a smooth neck (dashed curve).
When $\Delta \neq 0$, the dashed curve is discontinuous at $x=0$; this discontinuity is overcome
with the use of the smooth neck.
Figure reprinted with permission from Ref.\ \cite{yann16}, Copyright (2016)
by the IOP Publishing Ltd and Deutsche Physikalische Gesellschaft.
}

\label{tcopot}
\end{figure}

\subsubsection{Platform II: Three $^6${\rm Li} ultracold atoms in a 2D double-well trap.}
\label{3aplii}

In this section, I focus on a double-well trap which consists of two needle-like wells
in a parallel arrangement\footnote{
For the case of a double-well trap with a linear (in series) arrangement of the two wells,
as well as for the case of $N=4$ ultracold fermionic atoms, see Ref.\ \cite{yann16}.}
(DWPA). The DWPA trap cannot be described solely along the $x$-direction, because it
involves shape variation along the $y$ coordinate, as well.
To treat this case, one employs a many-body Hamiltonian for $N$ fermions of the form 
\begin{equation}
{\cal H}_{\rm MB} ({\bf r}_i,{\bf r}_j) =\sum_{i=1}^{N} H(i) +
\sum_{i=1}^{N} \sum_{j>i}^{N} g_{xy} \delta({\bf r}_i-{\bf r}_j)~,
\label{mbhd}
\end{equation}
where ${\bf r}_i-{\bf r}_j$ denotes the relative vector between the $i$ and $j$ fermions
(e.g., the $^6$Li atoms). The Hamiltonian ${\cal H}_{\rm MB}$ in Eq.\ (\ref{mbhd}) is the sum
of a one-body part $H(i)$, which implements the needle-like shape of each well, and the
two-particle contact interaction.

The confining potential [$V(x,y)$ in $H(i)$], which describes a double well
(DW), is constructed from a two-center-oscillator (TCO) model \cite{yann09,yann15} that
incorporates a smooth neck along the $x$-direction. Along the
$x$-direction, this TCO model accomodates independent variations of the
separation $d$ and the barrier height $V_b$ between the two wells; see Fig.\
\ref{tcopot}. Along the $y$-direction, the confining potential has the shape of that of a
single harmonic oscillator of frequency $\hbar \omega_y$. The frequencies $\hbar \omega_{x1}$
(of the left well),  $\hbar \omega_{x2}$ (of the right well), and $\hbar \omega_y$ (along the
$y$-direction) can also be varied independently; here I choose $\hbar \omega_{x1}=
\hbar \omega_{x2}=\hbar \omega_x$. In the DWPA case, the needle-like shape of each
individual well is reproduced when $\hbar \omega_x \gg \hbar \omega_y$. 
Finally, the TCO incorporates a tilt $\Delta$ which accounts for the elevation mismatch
between the left and right wells.
Fig.\ \ref{tcopot} illustrates the TCO confining potentials\footnote{
For the defining $H_{\rm TCO}$ Hamiltonian, see Sec.\ \ref{3epliv} below.}   
along the $x$ coordinate used to derive the results (displayed in the next Fig.\ \ref{n3_njp})
for three $^6$Li ultracold atoms in symmetric ($\Delta=0$) and tilted ($\Delta>0$) double wells.

\begin{figure}[t]
\centering\includegraphics[width=0.72\textwidth]{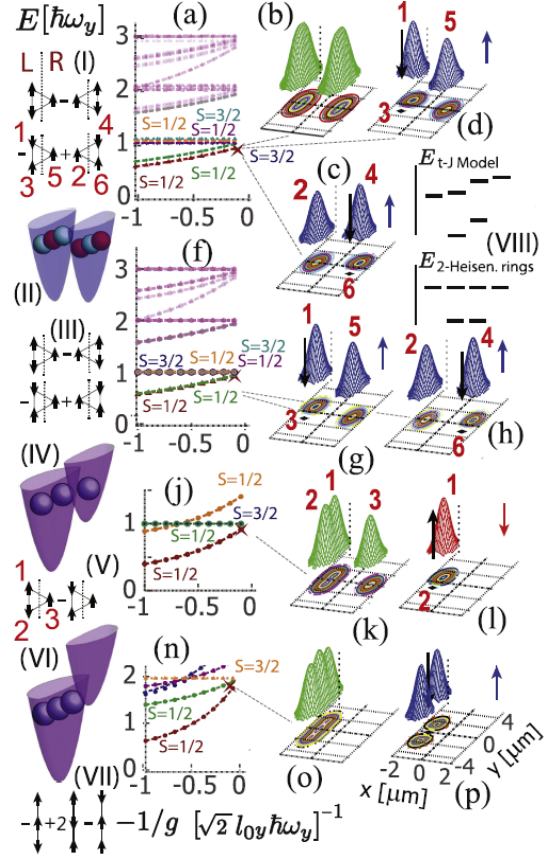}
\caption{
FCI results for $N=3$ {\it strongly-repelling\/} $^6$Li atoms (fermionic) in a double-well
potential confinement, which implements a {\it parallel\/} arrangement (DWPA) of the two
quasi-1D traps. Keeping the interwell separation and the TCO confinement frequencies constant
at $d=2.5$ $\mu$m, $\hbar \omega_x=6.6$ kHz, and $\hbar \omega_y=1$ kHz, the following
quantities are plotted for various values of the tilt, $\Delta$, and the interwell barrier,
$V_b$: Energy versus $-1/g$ spectra, SPDs (green surfaces), and spin-resolved CPDs.
$g$ here is the 1D contact-interaction strength along the $y$ direction \cite{yann15}.
In the SR-CPDs, the red (blue) surfaces denote a second spin-down (spin-up) $^6$Li atom
when a first spin-up (spin-down) atom is located at the fixed point (marked by a diamond
and a thicker, as well as longer, black arrow).
(a,b,c,d) $\Delta=0$ [see schematic in (II)] and $V_b=11.14$ kHz (lower barrier). 
(f,g,h) $\Delta=0$ [see schematic in (II)] and $V_b=24.30$ kHz (high barrier). 
(j,k,l) $\Delta=0.5 \hbar \omega_y$ [see schematic in (IV)] and $V_b=24.30$ kHz.  
(n,o,p) $\Delta=2.5 \hbar \omega_y$ [see schematic in (VI)] and $V_b=24.30$ kHz.
The inserts (I),(III),(V), and (VII) depict schematically the spin functions for the
ground states with $S=1/2$ [see spectra in (a), (f), (j), and (n), respectively].  
The SPDs and CPDs in (b,c,d,g,h,k,l,o,p) are associated with the $S=S_z=1/2$ FCI
ground states [brown curve in the corresponding energy spectra (a,f,j,n)] at
the point (marked by an asterisk) $-1/g=-0.1 (\sqrt{2}l_{0y} \hbar \omega_y )^{-1}$.
The fixed point in the spin-resolved CPDs is located at 
${\bf r}_0=(+1.3$ $\mu$m,$-1.1$ $\mu$m) in (c,h),
${\bf r}_0=(-1.3$ $\mu$m,$-1.1$ $\mu$m) in (d,g,l), and 
${\bf r}_0=(-1.3$ $\mu$m,0) in (p).
Insert (VIII) depicts schematically the degeneracies (lifting of degeneracies) in the two
uncoupled (coupled) Heisenberg rings (in accordance with the $t$-$J$ model), associated with
the energy spectra in (a,f). The ground-state total energies of the non-interacting limiting
cases are 11.14 $\hbar \omega_y$ in (a), 12.62 $\hbar \omega_y$ in (f), 13.11 $\hbar \omega_y$ 
in (j), and  13.62 $\hbar \omega_y$ in (n); they were subtracted from the energy scale
in the corresponding spectra.
Part of figure reprinted with permission from Ref.\ \cite{yann16}, Copyright (2016)
by the IOP Publishing Ltd and Deutsche Physikalische Gesellschaft.
}
\label{n3_njp}
\end{figure}

Fig.\ \ref{n3_njp} presents FCI results for the case of $N=3$ {\it strongly-repelling\/}
ultracold $^6$Li atoms (which are fermions) in a DWPA trap. Specifically, energy versus
$-1/g$ spectra, single-particle densities\footnote{
The term ``single-particle density'' is used for neutral particles in place of
``charge density''.}  
(SPDs), and spin-resolved CPDs are displayed at a given interwell separation $d=2.5$
$\mu$m of the two quasi-1D traps, as a function of the
tilt, $\Delta$, and the interwell barrier, $V_b$. The spectra are shown for
$-1/(\sqrt{2}l_{0y} \hbar \omega_{y} ) \leq -1/g \leq 0$, which includes the range of
strong interparticle contact repulsion. Note that the parameter $g$ here expresses the
1D $s$-wave scattering along the long $y$-direction of the needle-like wells; it relates
to the parameter $g_{xy}$ in Eq.\ (\ref{mbhd}) as follows
\begin{equation}
g = g_{xy} \int_{-\infty}^\infty dx [W(x)]^4,
\label{g1d}
\end{equation}
where $W(x)$ is the lowest-in-energy single-particle state of the TCO Hamiltonian in
the $x$-direction.

A salient feature of the four energy
spectra in Fig.\ \ref{n3_njp} is the appearance of distinct bands consisting of three
low-energy states for the case of tilted wells, or of six low-energy states for the
symmetric case (zero-tilt),\footnote{
See Figs.\ \ref{n3_njp}(j) and (IV) for a moderate tilt, $\Delta=0.5 \hbar \omega_{y}$,
as well as Figs.\ \ref{n3_njp}(n) and (VI) for a strong tilt,
$\Delta=2.5 \hbar \omega_{y}$. For the zero-tilt case, $\Delta=0$, see Figs.\
\ref{n3_njp}(a), (f), and (II).}
as the interaction strength moves closer to infinity (i.e., as
$-1/g \rightarrow -0$). At the point $-1/g=0$, the three or six states participating
in these bands become degenerate.

\begin{figure}[t]
\centering\includegraphics[width=0.7\textwidth]{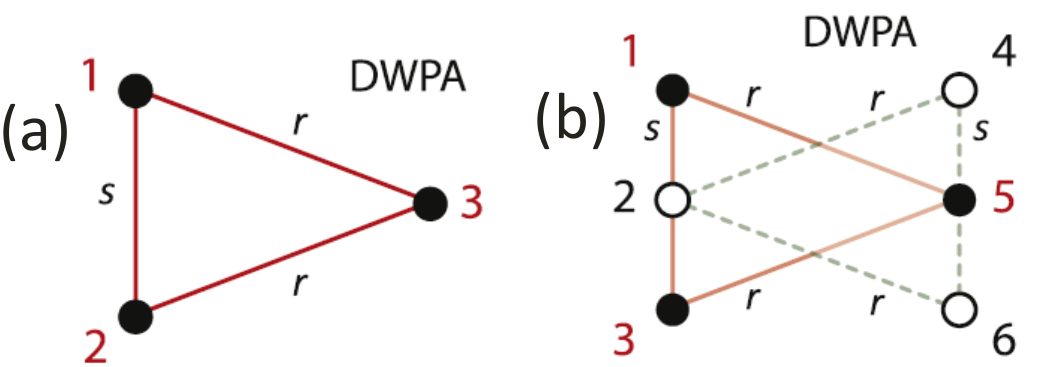}
\caption{
(a) Schematic associated with the three-site Heisenberg model for a DWPA trap in 
the case of tilted wells [see Figs.\ \ref{n3_njp}(IV) and (VI)].
(b) Schematic associated with the six-site $t$-$J$ model for a DWPA trap in the case
of symmetric wells [$\Delta=0$, see Fig.\ \ref{n3_njp}(II)].
Part of figure reprinted with permission from Ref.\ \cite{yann16}, Copyright (2016)
by the IOP Publishing Ltd and Deutsche Physikalische Gesellschaft.}
\label{heisch}
\end{figure}

The cases of asymmetric wells considered in Fig.\ \ref{n3_njp} are amenable to
straightforward modelling using pure Heisenberg models. The moderate tilt [Fig.\
\ref{n3_njp}(IV), $\Delta=0.5 \hbar \omega_y$] generates a ground state with a
$(n_L=2,n_R=1)$ configuration of the atoms (two $^6$Li atoms in the left well and one
$^6$Li atom in the right well, which is tilted upwards). In addition, this configuration
displays the shape of a pinned isosceles-triangle WM [see the SPD in Fig.\ \ref{n3_njp}(k)].
The associated FCI energy spectrum [Fig.\ \ref{n3_njp}(j)] features a three-member
lowest-energy band with a characteristic 1-2 degeneracy pattern.
The total-spin, FCI-derived multiplicities in this band are ${\cal G}(N=3,S=1/2)=2$ and
${\cal G}(N=3,S=3/2)=1$,\footnote{
The same total-spin multiplicities were explicitly demonstrated in Sec.\ \ref{3epli}
for the case of three electrons in a single QD.}  
in agreement with the group-theoretical branching diagram for three fermions
(discussed in Ref.\ \cite{pauncz} and in Appendix A of Ref.\ \cite{yann16}). 
This FCI energy spectrum and the related spin-resolved CPDs [see, e.g., the SR-CPD in Fig.\
\ref{n3_njp}(l)] are recreated by a 3-site Heisenberg-ring Hamiltonian
\begin{equation}
{\cal H}_H^{\rm trg}=J_{12}  {\bf S}_1 {\bf \cdot S}_2 + J_{13} ({\bf S}_1 {\bf \cdot S}_3 +
{\bf S}_2 {\bf \cdot S}_3) - J_{12}/4 - J_{13}/2,
\label{heitrg}   
\end{equation}
where $J_{12}=s$ and $J_{13}=J_{23}=r$; for the numbering convention of the three Heisenberg
sites, consult the schematic in Fig.\ \ref{heisch}(a). In the case of a high
barrier $V_b$ ($r=0$), the eigenenergies of ${\cal H}_H^{\rm trg}$ are given by
${\cal E}_1\;(S=3/2)={\cal E}_2 \; (S=1/2)=0$ and ${\cal E}_3 \; (S=1/2)=-s$, and they
reproduce the 1-2 FCI degeneracy pattern in Fig.\ \ref{n3_njp}(j). 
The FCI-calulated SR-CPDs agree also fully with the eigenvectors of the
${\cal H}_H^{\rm trg}$ Hamiltonian, which coincide\footnote{
For a complete detailed description concerning the solutions (eigenenergies and
eigenvectors) of the Heisenberg Hamiltonian (\ref{heitrg})], as well as the $t$-$J$
one [Eq.\ (\ref{htj})], see Ref.\ \cite{yann16}.}
(for $S_z=1/2$) with the expressions
${\cal W} (\mbox{$\frac32$},\mbox{$\frac12$})$,
${\cal W} (\mbox{$\frac12$},\mbox{$\frac12$};1)$, and
${\cal W} (\mbox{$\frac12$},\mbox{$\frac12$};2)$ in Section \ref{3epli}
[see Eqs.\ (\ref{wf3e3212}), (\ref{wf3e12121}), and (\ref{wf3e12122}), respectively].

For example, the ground-state SR-CPD, ${\cal P}_{\downarrow\uparrow}$, in Fig.\
\ref{n3_njp}(l) [at the point $-1/g=-0.1 (\sqrt{2}l_{0y} \hbar \omega_y )^{-1}$] conforms
with the 3-fermion spin eigenfunction ${\cal W} (\mbox{$\frac12$},\mbox{$\frac12$};2)$
[see Eq.\ (\ref{wf3e12122})]. The corresponding FCI-extracted spin function is
schematically portrayed in Fig.\ \ref{n3_njp}(V); it also agrees with the eigenvector
solutions\footnote{
See previous footnote.}
of the Heisenberg Hamiltonian (\ref{heitrg}).

A larger tilt, $\Delta=2.5 \hbar \omega_y$, generates a ($n_L=3,n_R=0$) FCI ground state,
representing a pinned linear WM [see the SPD in Fig.\ \ref{n3_njp}(o)]. 
The FCI spectrum [Fig.\ \ref{n3_njp}(n)] and the associated SR-CPDs [see, e.g.,
Fig.\ \ref{n3_njp}(p)] are reproduced by a 3-site open-linear-chain Heisenberg Hamiltonian,
obtainable from Eq.\ (\ref{heitrg}) by setting $J_{12}=s=0$. This Hamiltonian yields three
different eigenenergies ${\cal E}_1\; (S=3/2)=0$, ${\cal E}_2\; (S=1/2)=-3r/2$, and
${\cal E}_3\; (S=1/2)=-r/2$, in agreement with the three-member FCI band. The ground-state
FCI-extracted spin function is schematically depicted in Fig.\ \ref{n3_njp}(VII),
and it agrees with the expression ${\cal W} (\mbox{$\frac12$},\mbox{$\frac12$};1)$
[see Eq.\ (\ref{wf3e12121})], which also happens to be an eigenvector of the
Heisenberg Hamiltonian (\ref{heitrg}) when $s=0$ [see Fig.\ \ref{heisch}(a)].

A more complex behavior, opening ingress to the physics of breaking and restoration of
the parity symmetry, is exhibited by the symmetric DWPA cases ($\Delta=0$) for $N=3$
shown in Fig.\ \ref{n3_njp}. In fact, the FCI spectra in Figs.\ \ref{n3_njp}(a)
and (f) consist of a six-member lowest-energy band, which includes four $S=1/2$ states, and
two $S=3/2$ states, i.e., twice as many as in the case of tilted wells [Figs.\
\ref{n3_njp}(j) and \ref{n3_njp}(n)]. For the higher barrier [Fig.\ \ref{n3_njp}(f)],
a 2-4 degeneracy develops, which represents a doubling of the 1-2 degeneracy
pattern in Fig.\ \ref{n3_njp}(j). This doubling is due to the preservation of parity in
the FCI calculation, which requiress consideration of a second triangle (246)
that mirrors the original (135) one; see the schematics in Fig.\
\ref{n3_njp}(I), \ref{n3_njp}(III), and in Fig.\ \ref{heisch}(b), and also the two sets
of differently colored spheres in Fig.\ \ref{n3_njp}(II). The emergence of these
triangular mirror arrangements is reflected in the SR-CPDs displayed in Figs.\
\ref{n3_njp}(c) and (d) for the lower-barrier symmetric double-well case and in Figs.\
\ref{n3_njp}(g) and (h) for the higher-barrier (zero-tunneling) case. This situation
can be rationalized as involving six sites altogether, with the 3 localized fermionic atoms
following either the (135) triangular configuration, or the (246) one [see Fig.\
\ref{heisch}(b)], with 2 atoms in one well and the third atom in the other well; in each
case, the unoccupied (empty) sites may be referred to as ``holes''. This rationalization
generates a physical picture of a 3-atom WM that resonates between the two interlocking
triangles.

To model the FCI results for the symmetric-well case ($\Delta=0$) discussed above, one
must go beyond the simple Heisenberg Hamiltonian in Eq.\ (\ref{heitrg}). Indeed, Ref.\
\cite{yann16} found that an approach similar to the so-called $t$-$J$ model makes possible the
reproduction of all the salient features revealed by the FCI calculations. Motivated by
the $t$-$J$ model for extended systems \cite{auerbook,dago94}, a finite $t$-$J$-type Hamiltonian
based on the schematic in Fig.\ \ref{heisch}(b) can be expressed as 
\begin{equation}
 {\cal H}_{tJ}^{\Delta=0} = {\cal H}_H^{\rm trg}(135)(\{J\}) + {\cal H}_H^{\rm trg}(246)(\{J\})
 + {\cal H}_c(\{t\}),
\label{htj}
\end{equation}
where $H_c$ is the coupling (due to tunneling) between the two Heisenberg Hamiltonians,
${\cal H}_H^{\rm trg}(135)(\{J\})$ and ${\cal H}_H^{\rm trg}(246)(\{J\})$,
associated with the sites (135) and (246); see the upper left and lower right
$3 \times 3$ blocks around the diagonal of the matrix in Eq.\ (\ref{tjmat}). $H_c$,
represented by the two off-diagonal blocks in Eq.\ (\ref{tjmat}), is defined by the matrix
elements $(\alpha 0 \alpha 0 \beta 0 | {\cal H}_c | 0 \alpha 0 \alpha 0 \beta )=
(\alpha 0 \alpha 0 \beta 0 | {\cal H}_c | 0 \alpha 0 \beta 0 \alpha )=t$, and
$(\alpha 0 \alpha 0 \beta 0 | {\cal H}_c | 0 \beta 0 \alpha 0 \alpha )=t_2$,
where the ``0'' indicates an empty site; e.g., $\alpha 0 \alpha 0 \beta 0$ corresponds to
a state where sites 1, 3, and 5 are occupied and 2, 4, and 6 are empty [for the site
arrangements, see Fig.\ \ref{n3_njp}(I) and Fig.\ \ref{heisch}(b)].

The Hamiltonian ${\cal H}_{tJ}^{\Delta=0}$ is equivalent to a six-by-six matrix,
~~~~~~~~\\
{\small
\begin{equation}
{\cal H}_{tJ}^{\Delta=0} = \left(
\begin{array}{ccc|ccc}
-r      & r/2        & r/2       & t_2      & t         & t         \\
r/2     & -s/2-r/2   & s/2       & t        & t         & t_2       \\
r/2     & s/2        & -s/2-r/2  & t        & t_2       & t         \\ \hline
t_2     & t          & t         & -r       & r/2       & r/2       \\
t       & t          & t_2       & r/2      & -s/2-r/2  & s/2       \\
t       & t_2        & t         & r/2      & s/2       & -s/2-r/2  \\
\end{array}
\right)
\begin{array}{l}
\alpha 0 \alpha 0 \beta  0  \\
\alpha 0 \beta  0 \alpha 0  \\
\beta  0 \alpha 0 \alpha 0  \\
0 \beta 0 \alpha 0 \alpha   \\
0 \alpha 0 \alpha 0 \beta   \\
0 \alpha 0 \beta 0 \alpha   \\
\end{array}
.
\label{tjmat}
\end{equation}
}
~~~~~~~\\
The $t$-$J$ Hamiltonian matrix in Eq.\ (\ref{tjmat}) can generate a very rich behavior. Here,
I will limit the analysis to the case of a large interwell barrier $V_b$ (i.e., for $r=0$),
which is also the case for both the FCI spectra in Figs.\ \ref{n3_njp}(a) and \ref{n3_njp}(f).
Under these conditions, the eigenvalues (${\cal E}_i$) and the unnormalized eigenvectors
(${\cal V}_i$) of the Hamiltonian matrix in Eq.\ (\ref{tjmat}) are:
\begin{equation}
{\cal E}_1=-s + t - t_2,\;\; S=1/2, 
\label{e4_1}
\end{equation}
\begin{equation}
{\cal E}_2=-s - t + t_2,\;\; S=1/2, 
\label{4_2}
\end{equation}
\begin{equation}
{\cal E}_3=t - t_2,\;\; S=1/2, 
\label{e4_3}
\end{equation}
\begin{equation}
{\cal E}_4=-t + t_2,\;\; S=1/2,
\label{e4_4}
\end{equation}
\begin{equation}
{\cal E}_5=2 t + t_2,\;\; S=3/2, 
\label{e4_5}
\end{equation}
\begin{equation}
{\cal E}_6=-2 t - t_2,\;\; S=3/2,
\label{e4_6}
\end{equation}

\noindent 
and
\begin{equation}
{\cal V}_1=
\{0, -1, 1, 0, -1, 1 \}^T,\;\;S=1/2,
\label{v4_1}
\end{equation}
\begin{equation}
{\cal V}_2=
\{0, 1, -1, 0, -1, 1 \}^T,\;\;S=1/2,
\label{v4_2}
\end{equation}
\begin{equation}
{\cal V}_3=
\{ 2, -1, -1, -2, 1, 1 \}^T,\;\;S=1/2,
\label{v4_3}
\end{equation}
\begin{equation}
{\cal V}_4=
\{-2, 1, 1, -2, 1, 1 \}^T,\;\;S=1/2,
\label{v4_4}
\end{equation}
\begin{equation}
{\cal V}_5=
\{1, 1, 1, 1, 1, 1 \}^T,\;\; S=3/2,
\label{v4_5}
\end{equation}
\begin{equation}
{\cal V}_6=
\{-1, -1, -1, 1, 1, 1 \}^T,\;\; S=3/2,
\label{v4_6}
\end{equation}
where the eigenvectors are expanded in the basis defined by the rightmost column
in Eq.\ (\ref{tjmat}).

Note that the energy difference between the two lowest-in-energy states is
$\delta t_{\rm gap}=|{\cal E}_2-{\cal E}_1|=2 |t-t_2|$. A sufficiently strong external
perturbation $V_P$ (see Sec.\ \ref{intro_sb} and Commentary 5, Sec.\ \ref{comm5}) can mix
the two states ${\cal V}_2$ and ${\cal V}_1$, and generate the pure single-triangle
states, indexed as (135) and (246), respectively, as follows:
\begin{eqnarray}
{\cal G}_1 & \propto & {\cal V}_1+{\cal V}_2 \nonumber \\
{\cal G}_2 & \propto & {\cal V}_1-{\cal V}_2.
\label{diga}
\end{eqnarray}

Then removing $V_P$ will result in an oscillatory motion between the ${\cal G}_2$
and the ${\cal G}_1$ triangles, with a period inversely proportional to
$\delta t_{\rm gap}$. When $\delta t_{\rm gap}$ is small, but finite [see, e.g., the
spectrum in Fig.\ \ref{n3_njp}(a)], this oscillation may be experimentally 
detectable (see the case of NH$_3$ in Ref.\ \cite{jona86}). For a near-vanishing
$\delta t_{\rm gap}$ [see, e.g., the spectrum in Fig.\ \ref{n3_njp}(f)], however, the
system will appear to be locked to either the ${\cal G}_2$ or the ${\cal G}_1$
triangular configuration. This behavior has actually been observed in pyramidal and
enantiomeric natural molecules, and it was discussed within an 1D double-well model
(directly connected to the ``flea on the elephant'' concept) in Ref.\ \cite{jona86}.
Examples analyzed in Ref.\ \cite{jona86} were the oscillating NH$_3$ (ammonia)
molecule and the phosphine (PH$_3$) and arsine (AsH$_3$) molecules that appear
frozen in a single pyramidal structure.

Going beyond Ref.\ \cite{jona86}, the
parity-breaking treatment in this Section is fully microscopic. In this context,
by analogy, inversion-asymmetric molecules, like NH$_3$, PH$_3$, and AsH$_3$, respresent
exact nonstationary solutions of the many-body Schr\"odinger equation, and thus they
should be considered as an example of {\it weak\/} emergence (i.e., of reductionism,
see Sec.\ \ref{intro_emerg}), in contrast to Anderson's \cite{ande72}
{\it strong\/}-emergence suggestion.

\subsubsection{{\small{\bf Commentary 7.}} Jahn-Teller effects in natural polyatomic
molecules and polymorphism in quantum materials.}
\label{comm7}

It needs to be pointed out, however, that Sections \ref{3epli} and \ref{3aplii},
as well as Ref.\ \cite{jona86} in the mathematical modeling,
considered a single species of particles, unlike the case of natural polyatomic
molecules, which consist of both electrons and nuclear ions. Consideration of both
these species in quantum chemistry has produced the celebrated theory of the Jahn-Teller
(JTE) and pseudo-Jahn-Teller (PJTE) effects, responsible for generating multiple
degenerate nuclear conformations \cite{bersbook,bers21,jahn37}, with the pair of the
inverted pyramids of NH$_3$, PH$_3$, and AsH$_3$ serving as an example. It is then
natural that a PJTE-based understanding of the finite-system SSB (i.e., the ESB and its
experimental detection) in these molecules has also been advanced \cite{bers21}, which
parallels the flea-on-the-elephant interpretation, as presented here. I note further that
the JTE assumes as self-evident the Wigner molecularization of nuclear ions, which is
implicit in the Born-Oppenheimer approximation; instead, it addresses the symmetry
breaking of a highly symmetric configuration of the nuclear framework towards a lower
point-group symmetry, which is driven by a lowering of the total energy.

Further, worth mentioning is the use of projection-operator techniques to describe the
{\it dynamic\/} Jahn-Teller effect in natural polyatomic molecules in the case of
substantial tunneling between equivalently distorted energy-minimum configurations of the
adiabatic potential energy surface \cite{dunn92,dunn12}. Naturally, as mentioned earlier,
due to the very large masses of the ionic cores, the explicit symmetry-broken wave-packet
state localized within a single minimum can be observed when tunneling is fully suppressed
\cite{bers21}.

Finally, note a recent development that investigates the metal-to-insulator question in
quantum materials by exploring, in DFT calculations, effects arising from the polymorphism
associated with symmetry breaking of the crystal lattice, including Jahn-Teller-type
distortions \cite{zung22,zung25,zung25.2}.

\begin{figure}[t]
\includegraphics[width=0.3\textwidth]{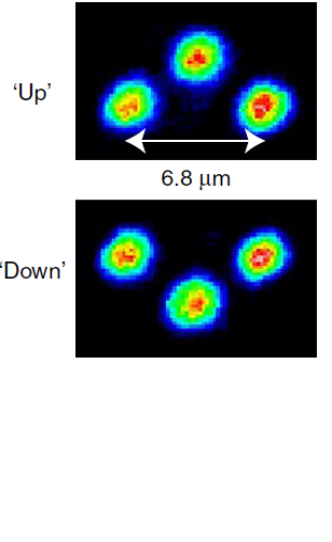}
\includegraphics[width=0.7\textwidth]{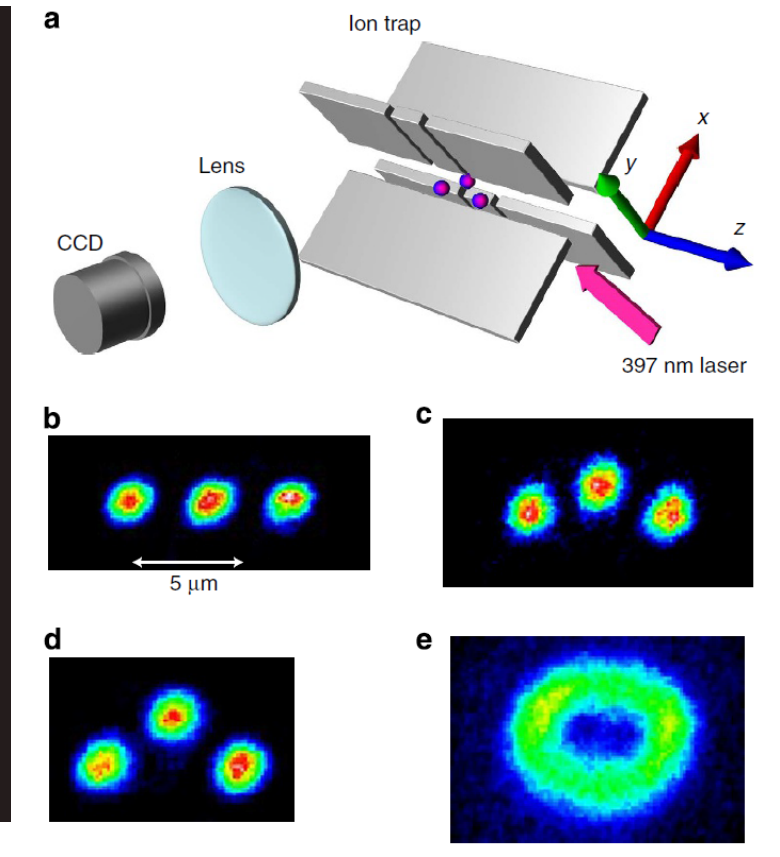}
\caption{
(Left) Illustrative fluorescence images of the two inversion-symmetric isosceles
triangular WM structures.
Reprinted part of figure with permission from Ref.\ \cite{nogu14},
Copyright (2014) by the Macmillan Publishers Limited.
\label{n3itexp1}
}  
\caption{
(Right) (a) The relative geometry of the ion trap and the imaging system. Images were
acquired of three $^{40}${\rm Ca}$^+$ ions under various conditions: (b) under strong
elliptic-anisotropy conditions ($\omega_x=2\pi \times 2.1$ MHz, $\eta=0.533$), (c)
under slightly stronger than intermediate anisotropy  ($\omega_x=2\pi \times 1.65$ MHz,
$\eta=0.678$), (d) under intermediate anisotropy conditions ($\omega_x=2\pi \times
1.523$ MHz, $\eta=0.735$) and (e) near the full-isotropy point ($\omega_x=
2\pi \times 1.119$ MHz, $\eta=1-\epsilon$). At all instances, the confinement along the
$z$ direction was kept constant at $\omega_z=2\pi \times 1.119$ MHz. $\epsilon$ here
denotes a very small quantity.
Reprinted figure with permission from Ref.\ \cite{nogu14},
Copyright (2014) by the Macmillan Publishers Limited.
\label{n3itexp2}
}
\end{figure}

\subsubsection{Platform III: Experimental results for three $^{40}${\rm Ca}$^+$
laser-cooled ions in a harmonic linear Paul trap.}
\label{3ipliii}

Through an analysis of the FCI solutions of the many-body Schr\"odinger equation,
the two previous Sections \ref{3epli} and \ref{3aplii} identified the basic
configurations of a three-fermion Wigner molecule in elliptic and double-well 2D
potential traps. These configurations include: a pinned-WM linear arrangement in a single
strongly-deformed elliptic trap, the interlocking of two isosceles triangles in a
double-well or in a single trap of intermediate deformation (stationary solution),
periodic oscillations between these two isosceles triangles (nonstationary solutions),
and the appearance of ring-like, circularly-symmetric and continuous SP densities
(associated with rotating WMs) in the case of circular traps.

A remarkable experimental realization of these WM configurations was implemented 
\cite{nogu14} by encapsulating three $^{40}$Ca$^+$ ions in a single linear Paul trap,
capable of generating a variety of harmonic confinements. Moreover, the frequencies
of the associated harmonic confining potentials along the $x$ and $z$ axes were
comparable, whereas the frequency along the third $y$-axis was much larger. As a
result the trap was effectively a 2D trap in the $x$-$z$ plane. 

Fig.\ \ref{n3itexp1} displays illustrative fluorescence images of the three $^{40}$Ca$^+$
ions in two inversion-symmetric isosceles-triangle WM configurations, denoted as 'up'
and 'down.' The 'up' and 'down' configurations realize the nonstationary solutions
(but with a long lifetime) of the many-body Schr\"odinger equation in analogy with
the pair ${\cal G}_1$ and ${\cal G}_2$ in Eqs.\ (\ref{diga}) above.

\begin{figure}[t]
\centering\includegraphics[width=0.95\textwidth]{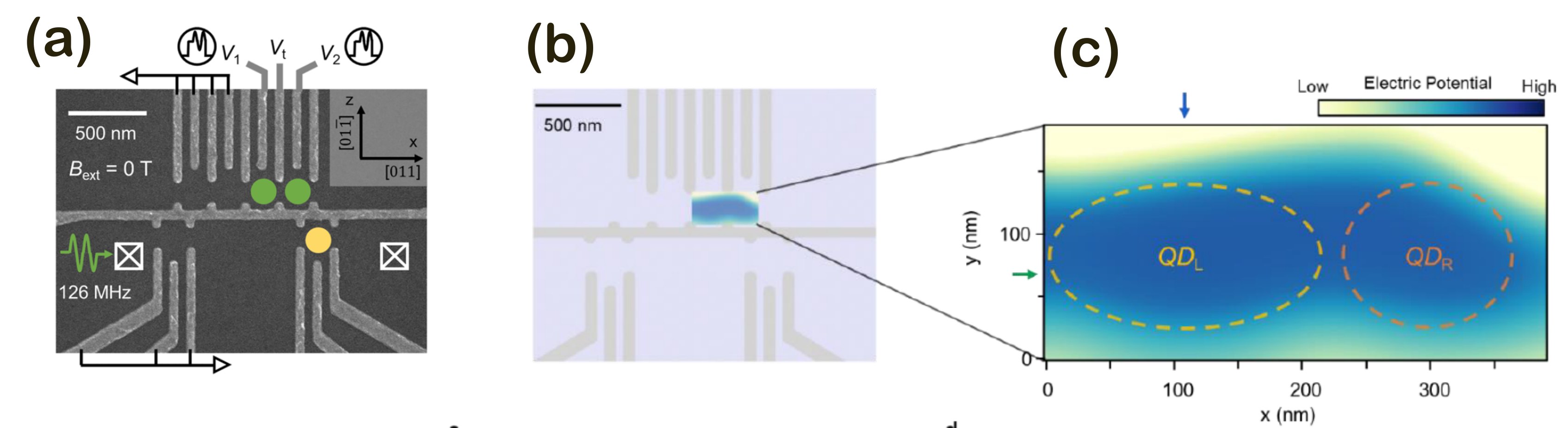}
\caption{
(a) Scanning electron microscope image of the hybrid qubit device similar to the one
used in the experiment of Ref.\ \cite{kim21}. Green (Yellow) solid circles: Double
(Single) quantum dot used to form a hybrid qubit (charge sensor).
(b) The gate geometry of the actual QD device used in Ref.\ \cite{kim21}, scaled for the
purpose of electrostatic simulation. The electric potential near the double-QD site is
simulated usign the COMSOL Multiphysics software, with the dc-voltages employed in
the experiment. The semi-classical electron numbers inside the double
quantum dot (integrals of the Thomas-Fermi electron density over the quantum dot area)
resulting from this gate-voltage set were confirmed to be ($n_L=2,n_R=1$).
(c) Spatial distribution of the confinement potential near the QD sites. Dashed lines
denote the expected position of the QDs with the left (right) QD being distributed over
an oval-shaped (circular-disk) area.
Reprinted parts of figures with permission from Ref.\ \cite{kim21},
Copyright (2021) by the American Chemical Society.
}
\label{kim_dev}
\end{figure}

  Fig.\ \ref{n3itexp2} displays the evolution of the three-ion 'up' configuration as a
function of the trap anisotropy $\eta=\omega_z/\omega_x$. For $\eta=0.533$ (strong
elliptic anisotropy), a pinned linear WM is displayed in Fig.\ \ref{n3itexp2}(b),
in analogy with the three-electron case in Fig.\ \ref{n3sqd1}(a) and the $^6$Li
three-atom case in Fig.\ \ref{n3_njp}(o). For the two intermediate anisotropies in
Fig.\ \ref{n3itexp2}(c) (with $\eta=0.678$) and in Fig.\ \ref{n3itexp2}(d) (with
$\eta=0.735$), pinned WMs are formed exhibiting an isosceles triangular
configuration. Note that for the smaller anisotropy ($\eta=0.735$) the isosceles
triangle is taller with a shorter basis. Similar isosceles triangular WM structures
were discussed previously in the context of FCI calculations; see Fig.\ \ref{n3sqd1}(b)
(three electrons) and Fig.\ \ref{n3_njp}(k) (three $^6$Li atoms). More spectacular is
the case near the circular-trap point which is displayed in Fig.\ \ref{n3itexp2}(e).
In this instance, a {\it rotating\/} WM emerges with a
near-circular ring-like density. This behavior is analogous to that of the ring-like
charge density of three strongly interacting electrons [with $R_W=\lambda=8$, see Fig.\
\ref{n3sqd1}(c)].\footnote{
As an interpretation, Ref.\ \cite{nogu14} employed a model of two inversion-symmetric
{\it equilateral\/} triangular structures inscribed within the same circle, which are
tunnel-coupled by executing clockwise and counter-clockwise $\pi/6$ rotations. This
interpretation is not applicable to the images in Figs.\ \ref{n3itexp1} and
\ref{n3itexp2}, because they do not exhibit any equilateral-triangle configuration.}  
Note that the quantal regime near the ground state was reached \cite{nogu14} by applying
successively the Doppler, sideband, and adiabatic cooling protocols, with the lowest
achieved temperature being 40 nK.

\begin{figure}[t]
\centering\includegraphics[width=0.9\textwidth]{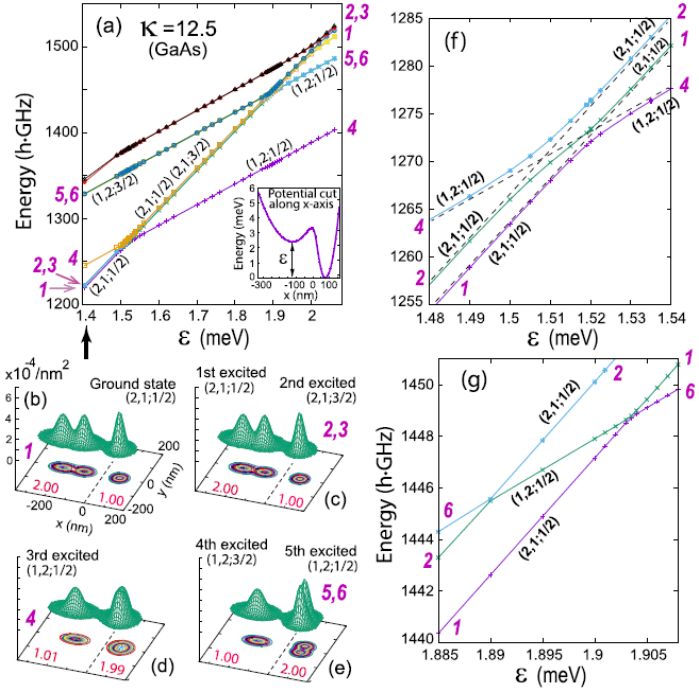}
\caption{
(a) FCI Low-energy spectrum for the three-electron GaAs ($m^*=0.067m_e$, $\kappa=12.5$)
double dot, with parameters corresponding to the experimental device in Ref.\ \cite{kim21}.
The arrow indicates the value of the detuning $\epsilon$ at which the FCI total charge
densities were calculated. 
(b)$-$(e) FCI total electron densities for the ground and first five excited states. 
(f) and (g) Magnification of the neighborhoods of the FCI avoided crossings appearing
in (a). Only the $S=1/2$ states, relevant to the hybrid qubit, are shown.
The notation $(n_L, n_R; S)$ denotes the left electron occupation, the right electron
occupation, and the total spin, respectively. The values of $(n_L, n_R)$ are in agreement
with the FCI-calculated ones that are highlighted in red. For all panels displaying
densities, the scales of all three axes ($x$, $y$, and $z$) are as in (b). The electron
densities are normalized to $N=3$, i.e., the total number of electrons.
Reprinted figure with permission from Ref.\ \cite{yann22},
Copyright (2022) by IOP Publishing Ltd.
}
\label{n3e_dwli}
\end{figure}

\subsubsection{Platform IV: Three electrons in a semiconductor double quantum dot: FCI
calculations associated with an experimentally realized solid-state hybrid qubit.}
\label{3epliv}

A recent progression of experimental papers \cite{kim21,kim23,kim25} is exploring the
formation of Wigner molecules in double-dot semiconductor devices with three electrons,
which can operate as solid-state qubits. Such 3e qubits were introduced in Refs.\
\cite{copp12,copp14} and are often referred to as hybrid qubits.
  
In this Section, motivated by the recent advances
\cite{kim21,kim23,kim25,corr21,cao16,copp17} in the fabrication and operation
protocols of hybrid double-quantum-dot (HDQD) qubits, I review salient features
concerning the many-body spectra and wave functions of three electrons in an
asymmetric two-dimensional double-well external confinement, modeled by a
two-center-oscillator (TCO) potential \cite{yann99,yann02,yann09}. In particular,
these salient features demonstrate the defining role that WM formation play in
shaping the spectra (including the key finding of a pair of left-right
electron-occupancy-dependent avoided crossings) of semiconductor qubits. To this
end, FCI calculations were carried out for the case of a three-electron
double-dot GaAs qubit with parameters comparable to those of the device in Ref.\
\cite{kim21}; see Fig.\ \ref{kim_dev}.

Note that earlier fabricated GaAs QDs \cite{kouw97,kouw01} were characterized by harmonic
confinements with frequencies $\hbar \omega_0 \geq 3$ meV [with $R_W < 1.97$; see Eq.\
(\ref{rw})], which correspond to a range of smaller QD sizes that did not favor
the observation of the WMs in the addition energies at low magnetic fields.\footnote{
Later Ref.\ \cite{nish06} (with the same authors) demonstrated that the non-interacting
Darwin-Fock theory is not sufficient to explain in their entirety the addition spectra for
these smaller QDs. In this context, careful FCI calculations were employed, which confirmed 
the presence of WMs for $\nu \leq 1$ (beyond the maximum density droplet) even in the case
of these intermediate-size, smaller GaAs QDs.}
The strong WM signatures observed in the much larger anisotropic GaAs double dots of Refs.\
\cite{kim21,kim23,kim25}, as well as in the Si/SiGe or Ge/SiGe dots of Refs.\
\cite{corr21,bens26} and in the Germanium QDs of Refs.\ \cite{loss24,yang25}, 
heralded the exploration of heretofore untapped potentialities in the fabrication and
control of QD qubits, a goal that Refs.\ \cite{yann22,yann22.2} aimed to facilitate
from a theory perspective.

The relevant many-body Hamiltonian (for $N$ confined electrons) has the form
\begin{equation}
{\cal H}_{\rm MB} =\sum_{i=1}^{N} H_{\rm TCO}(i) +
\sum_{i=1}^{N} \sum_{j>i}^{N} \frac{e^2}{\kappa |{\bf r}_i-{\bf r}_j|},
\label{mbhd2}
\end{equation}
where $\kappa$ is the dielectric constant of the semiconductor material.

The one-body $H_{\rm TCO}$ \cite{yann99,yann02,yann09} is given by 
\begin{equation}
H_{\rm TCO}=\frac{{\bf p}^2}{2 m^*} + \frac{1}{2} m^* \omega^2_y y^2
+ \frac{1}{2} m^* \omega^2_{x k} x^{\prime 2}_k + V_{\rm neck}(x^\prime_k ) +h_k,
\label{hsp}
\end{equation}
where $x_k^\prime=x-x_k$ with $k=1$ for $x<0$ (left well) and $k=2$ for $x>0$ (right well), 
and the $h_k$'s control the relative depth of the two wells, with the
detuning\footnote{
The detuning $\varepsilon$ is referred to as tilt, $\Delta$, in the literature of trapped
ultracold atoms.}  
defined as $\varepsilon=h_1-h_2$. $y$ denotes the coordinate perpendicular to the
interdot axis ($x$). The most general shapes described by $H_{\rm TCO}$ are
two semiellipses connected by a smooth neck [$V_{\rm neck}(x^\prime_k )$]. $x_1 <
0$ and $x_2 > 0$ are the centers of these semiellipses, $d=x_2-x_1$
is the interdot distance, and $m^*$ is the effective electron mass.

For the smooth neck, the following expression is employed:
\begin{equation}
V_{\rm neck}(x^\prime_k ) = \frac{1}{2} m^* \omega^2_{x k} 
\Big[ {\cal C}_k x^{\prime 3}_k + {\cal D}_k x^{\prime 4}_k \Big] \theta(|x|-|x_k|),
\label{vneck}
\end{equation}
where $\theta(u)=0$ for $u>0$ and $\theta(u)=1$ for $u<0$.
The four parameters ${\cal C}_k$ and ${\cal D}_k$ ($k=1$ or $k=2$) are defined as
follows:
${\cal C}_k= (2-4\epsilon_k^b)/x_k$
and
${\cal D}_k=(1-3\epsilon_k^b)/x_k^2$, 
where the two $\epsilon_1$ and $\epsilon_2$ constants control the height of the interdot
barrier $V_{b}$, because they are given by $\epsilon_k^b=(V_{b}-h_k)/V_{0k}$ with
$V_{0k}=m \omega_{x k}^2 x_k^2/2$; $V_b$ is measured from the zero point of the energy scale.
I note that measured from the bottom of the left ($k=1$) or right ($k=2$) well the
effective interdot barrier is $V_{b}-h_k$.

This one-body $H_{\rm TCO}$ has the advantage of incorporating a smooth interdot
barrier $V_b$, which can be varied independently of the interdot separation $d$;
for an illustration see the inset of Fig.\ \ref{n3e_dwli}(a) (as well as Fig.\
\ref{tcopot}). The GaAs asymmetric double-dot described in Ref.\ \cite{kim21} is
modeled by the following parameters that enter in the TCO Hamiltonian:
The left dot is elliptic with frequencies
$\hbar\omega_{x1}=0.413567~{\rm meV}=100$ h$\cdot$GHz (long $x$-axis) and
$\hbar\omega_{y1}=1.22~{\rm meV}=294.9945$ h$\cdot$GHz  
(short $y$-axis), whereas the right dot is circular with
$\hbar\omega_{x2}=\hbar\omega_{y2}=1.22~{\rm meV}=294.9945$ h$\cdot$GHz (1 h$\cdot$
GHz $=4.13567$ $\mu$eV). The left dot is located at $x_1=-120$ nm, and the right dot
is located at $x_2=75$ nm. The detuning parameter is defined as $\varepsilon=h_1-h_2$,
where $h_1$ and $h_2$ are the chemical potentials of the left and right dot, respectively.
The interdot barrier from the bottom of the right dot is set to $V_b-h_2=3.3123$ meV
$=800.91$ h$\cdot$GHz. Finally, the effective electron mass and the dielectric constant
for GaAs are $m^*=0.067 m_e$ and $\kappa=12.5$, respectively.

Fig.\ \ref{n3e_dwli} summarizes the main results of the FCI calculation, with the
three-part notation $(n_L, n_R; S)$ denoting the left-well electron occupation, the
right-well electron occupation, and the total spin, respectively ($S=1/2$ or $S=3/2$
for three electrons). Fig.\ \ref{n3e_dwli}(a) displays, in the range 
$1.40 \mbox{~meV} \leq \varepsilon \leq 2.1 \mbox{~meV}$ of detunings, the low-energy
spectrum for the GaAs case ($m^*=0.067m_e$ and $\kappa=12.5$). The $(2,1;S)$ states with
two electrons in the {\it left\/} well, as well as  the $(1,2;S)$ states with two
electrons in the {\it right\/} well, are prominently present. In contrast, states with
all three electrons either in the left or right well, designated as $(3,0;S)$ or $(0,3;S)$,
respectively, are absent. The computationally established feature that only the six
$(2,1;S)$ and $(1,2;S)$ states constitute the lowest-energy spectrum of the GaAs double
dot is an essential prerequisite for the realization of the hybrid qubit, which relies
only \cite{copp12,copp14,kim21,cao17,kim23,kim25} on the
four $(2,1;1/2)$ and $(1,2;1/2)$ states. This feature is brought about by the formation
of Wigner molecules resulting from a rather large value of $R_W=5.31$ \cite{yann22,yann22.2}
(for the left well) due to the large-size and strongly asymmetric double dot [see the
definition of the Wigner parameter $R_W$ in Eq.\ (\ref{rw})].

In Fig.\ \ref{n3e_dwli}(a), note the successive numbering of the lowest six states at
$\varepsilon=1.4$ meV, starting from the ground state (\#1) and moving upwards to the
first five excited ones (\#2 to \#6). Outside from the immediate neighborhood of an
avoided crossing, these energy curves are straight lines, which allows for the extension
of the same numbering for all values of the detuning in the window range used in Figs.\
\ref{n3e_dwli}(a), (f), and (g).

The spectrum in Fig.\ \ref{n3e_dwli}(a) invites further remarks, because of
quasi-degeneracies between the states \#2,\#3, and \#5,\#6, as well as the small
energy gap ($\sim$ 3 h$\cdot$GHz) between state \#1 and the quasi-degenerate pair
(\#2,\#3). I stress that the states \#1 and \#2 have two electrons in the {\it left}
well and total spin $S=1/2$, and thus they are denoted as $(2,1;1/2)$, whereas state
\#3 has two electrons in the left well, but a total spin of $S=3/2$ [thus denoted as
$(2,1;3/2)$]. On the contrary, states \#4, \#6 (with $S=1/2$), and \#5 (with
$S=3/2$) have two electrons in the {\it right} well and thus they are denoted as
$(1,2;S)$. A main feature of this six-state spectrum in Fig.\ \ref{n3e_dwli}(a) is that
the curves \#1, \#2, and \#3 form one band of parallel lines, whereas the curves \#4, \#5,
and \#6 form a second band of parallel lines, and the two bands intersect at two avoided
crossings.

A central feature of the spectrum in Fig.\ \ref{n3e_dwli}(a) is the small energy gap
between the two $S=1/2$ states \#1 and \#2 ($\sim$ 3 h$\cdot$GHz), which contrasts
with the larger gap between the other two $S=1/2$ states \#4 and \#5 ($\sim$ 82
h$\cdot$GHz). This behavior follows from the WM physics associated with two different
values of the Wigner parameter (i.e., $R_W=5.31$ and $R_W=3.09$ for the left and right
QD, respectively), as well as with the influence of anisotropy \cite{urie21}. Most
importantly, this behavior agrees with the experimental findings of Refs.\
\cite{kim21,kim23,kim25}. Furthermore, the strong quenching of the energy gaps
compared to their non-interacting values [here $\hbar\omega_{x1} \sim 100$ h$\cdot$GHz
(left QD) and $\hbar\omega_{x2} \sim 295$ h$\cdot$GHz (right QD)] has been considered
as an experimental signature of WM formation in a variety of nanosystems
\cite{peck13,corr21,kim21,loss24}.

Further insights into the properties of the GaAs HDQD qubit can be gained through an
inspection of the FCI electron densities. Illustrative cases, calculated at the detuning
value of $\varepsilon=1.4$ [see arrow in Fig.\ \ref{n3e_dwli}(a)], are displayed in Figs.\
\ref{n3e_dwli}(b)-(e) for the ground and first five excited states. The red numbers 
indicate the left-well and right-well electron occupations as calculated
with the FCI method. The electron densities deviate strongly from those expected from
an independent-particle model. Indeed the formation of a strong 2e WM in the left well
and of a weaker 2e WM in the right well is clearly seen through the presence of a
double hump in all six cases.

Turning to the analysis of the two avoided crossings [see Figs.\ \ref{n3e_dwli}(a), (f),
and (g)], I note that their position and asymmetric anatomy play an essential role in the
operation\footnote{
The qubit is initialized in the ground-state on line \#4 (tuned to the far right of the
left crossing) in Fig.\ \ref{n3e_dwli}(a). After detuning and laser-pulse-induced
jumping to state \#2 [at the left crossing, Fig.\ \ref{n3e_dwli}(f)], readout is achieved
via increased detuning, moving along state \#1 and through the right avoided crossing to
the state \#6 [Fig.\ \ref{n3e_dwli}(g)].}
of the hybrid qubit \cite{kim21}. 
Fig.\ \ref{n3e_dwli}(f) and Fig.\ \ref{n3e_dwli}(g) display magnifications of the
neighborhoods of the left and right FCI avoided crossings, respectively, which are
present in the spectrum of the GaAs double dot [Fig.\ \ref{n3e_dwli}(a)]. Only the
$S=1/2$ states are shown, because the $S=3/2$ states are not relevant for the workings
of three-electron spin qubits \cite{kim21,copp12.2,vinc00,burk17}.

The left avoided crossing (in the neighborhood of 1.49 meV $<~\varepsilon~<$
1.54 meV) is shaped through the interaction of the three curves \#1, \#2, and \#4 [the
same numbering is kept for all the curves here as in Fig.\ \ref{n3e_dwli}(a)]. On the
other hand, the curves \#1, \#2, and \#6 participate in the formation of the right avoided
crossing in the neighborhood of 1.885 meV $<~\varepsilon~<$ 1.908 meV. I note that,
according to the FCI calculation, the two avoided crossings are separated by a detuning
interval of $\Delta \varepsilon \sim 400$ $\mu$eV, which is in agreement with the
experimentally determined value for the hybrid-qubit device in Ref.\ \cite{kim21}.

The continuous lines in Figs.\ \ref{n3e_dwli}(f) and (g) represent {\it adiabatic\/}
paths, which the system follows for slow time variations of the detuning variable.
For fast time changes of the detuning, or due to an external laser pulse, the
system can instead jump over the energy gap from one adiabatic curve to another,
following the {\it diabatic\/} paths designated explicitly with dashed lines in Fig.\
\ref{n3e_dwli}(f). These jumps develop according to the Landau-Zener-St\"{u}ckelberg-Majorana
\cite{copp12,cao13,burk13,kim25} dynamical interference theory, and they are an
integral part of the operation of the hybrid qubit.

\subsubsection{{\small{\bf Commentary 8.}} The properties of autonomy and universality
of Wigner molecularization, again.}
\label{comm8}

In Sect.\ \ref{3ferm}, Wigner-molecule formation for three mutually repelling
particles was analyzed in a variety of nanosystems of different size and constituents,
belonging to a broad range of physics subfields, i.e., electrons in semiconductor single
and double quantum dots, neutral atoms in ultracold optical traps, and atomic ions in
an electric Paul trap. The dimensions of these systems ranged from nm to $\mu$m, whereas
the corresponding masses differed by six-orders of magnitude [from 0.067$m_e$ (GaAs) to
73440$m_e$ ($^{40}$Ca$^+$)]. In addition, WM formation was demonstrated for both the
Coulombic and contact repulsive interactions. Thus, these studies offer another prominent
illustration of the universality of the physics of Wigner molecularization and of the
autonomy of the underlying process of emergent symmetry breaking.

\section{Wigner molecularization under applied magnetic fields and in rotating traps}
\label{wmmagrot}
\smallskip

The many-body problem regarding a few 2D fermionic or bosonic particles under an
applied magnetic field $B$ (or in ultracold traps rotating with angular frequency
$\Omega$) occupies a special place in modern-physics research investigations. Indeed, both
the celebrated Laughlin wave function \cite{laug83,laug99} and the composite-fermion (CF)
\cite{jain89,jainbook} modelings of the fractional quantum Hall effect \cite{pranbook}
(FQHE, associated with fully polarized electrons) do claim a loose connection to exact
solutions of the MBSE in the lowest Landau level. The emergence of magic angular momenta
\cite{maks90,ruan95} in assemblies of a few particles is another illustration of
unconventional physical behavior under high $B$ or large $\Omega$. Most recently, a
nontraditional FQHE-physics regime has been experimentally realized \cite{joch24} with
a few neutral and {\it spinful\/} ultracold fermionic atoms in a rapidly rotating trap.

This section will review how this $B$- or $\Omega$-dependent few-body problem is
primarily connected to the formation of rotating Wigner molecules, and for both fully
polarized and spinful fermionic particles, as well as scalar bosons. Specifically,
some preliminary material is discussed in Secs.\ \ref{simi} (equivalence between $B$ and
$\Omega$) and \ref{dfll} (Darwin-Fock oscillator). Sec.\ \ref{resmagqd} presents theoretical
and experimental results concerning Wigner molecularization in 2D semiconductor QDs under
an applied perpendicular magnetic field $B$, with the considered $B$ values ranging from
$B=0$ to values in the neighborhood immediately before the LLL limit. 
For the LLL case (that forms under rapid rotation, $\Omega = \omega_0$, or high magnetic
fields, $B \rightarrow \infty$), see Secs.\ \ref{lllrot} (neutral atoms) and \ref{lllfp}
(fully polarized electrons), respectively.

In the case of bulk incompressible Hall fluids, the thermodynamic FQHE physics
\cite{stor99,tsui99} has been interpreted using approximate wave functions in the spirit of
the Jastrow-Laughlin \cite{laug83,halp83,laug90,moor91,laug99,girv99,simo20} and
composite-fermion \cite{jain89,jainbook,jain20,yang20} approaches. These two lines of research
have generated a large body of literature and are regarded as the standard approach on the
subject in the case of macroscopic samples. Detailed description of the bulk FQHE literature is
beyond the scope of this review, which focuses on planar (so-called disc-geometry) few-body LLL
droplets; for a most recent experimental realization of such LLL droplets, see Refs.\
\cite{grus24,joch24,grus23}.

\subsection{Similarity between magnetic fields and rotations, and the corresponding
many-body Hamiltonian.}
\label{simi}
\smallskip

The many-body Hamiltonian that governs the physics of $N$ electrons confined in a 
planar QD and interacting via a Coulomb repulsion is given by
\begin{equation}
H^B_{\rm MB} =\sum_{i=1}^N H(i) +
\sum_{i=1}^N \sum_{j>i}^N \frac{e^2}{\kappa r_{ij}},
\label{mbhb}
\end{equation}
where $r_{ij}=|{\bf r}_i - {\bf r}_j|$ and $\kappa$ is the dielectric constant of the
material. The single-particle Hamiltonian in a perpendicular external magnetic field $B$ is
written as
\begin{equation}
H=\frac{({\bf p}-e{\bf A}/c)^2}{2m^*} + V(x,y) +
\frac{g^* \mu_B}{\hbar} {\bf B \cdot s}.
\label{hspb}
\end{equation}
In Eq.\ (\ref{hspb}), the potential confinement is represented by $V(x,y)$ and $m^*$ is
the effective electron mass, while the vector potential ${\bf A}$ is given in the
symmetric gauge by
\begin{equation}
{\bf A}({\bf r})=\frac{1}{2}{\bf B} \times {\bf r} =\frac{1}{2}(-By,Bx,0).
\label{vectp}
\end{equation}
The last term in Eq.\ (\ref{hspb}) is the Zeeman interaction, with $g^*$
being the effective Land\'{e} factor, $\mu_B$ the Bohr magneton, and ${\bf s}$
the spin of an individual electron. The confining potential $V(x,y)$ can assume
various parametrizations which can model a single circular or elliptic quantum dot,
or a multiwell QD.
 
\begin{figure}[t]
\centering\includegraphics[width=0.6\textwidth]{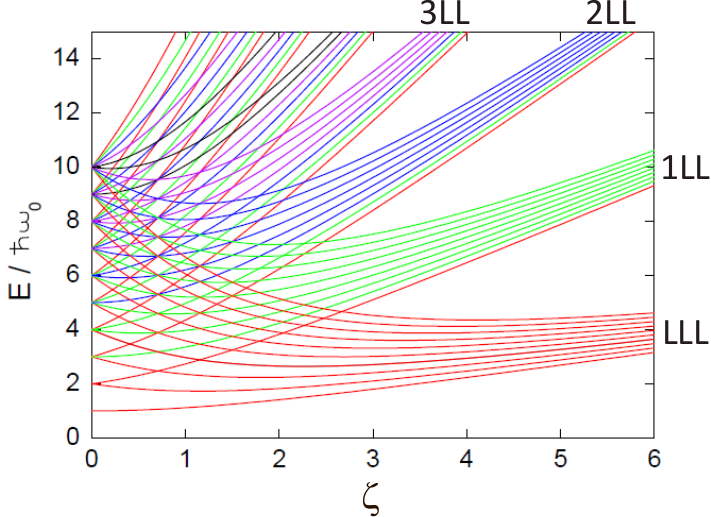}
\caption{
The $B$-dependent eigenenergies of the Darwin-Fock Hamiltonian in Eq.\ (\ref{hb}) as a
function of $\zeta_B = \omega_c/\omega_0$, where $\omega_c$ is the cyclotron frequency
and $\omega_0$ is the frequency specifying the 2D-harmonic confinement at $B=0$. A
given color specifies orbitals with the same number of radial nodes $n$, i.e., red
$\rightarrow$ $n=0$, green $\rightarrow$ $n=1$, blue $\rightarrow$ $n=2$, magenta
$\rightarrow$ $n=3$.
Reprinted figure with permission from Ref.\ \cite{yann07},
Copyright (2007) by IOP Publishing Ltd.
}
\label{fdb}
\end{figure}
\begin{figure}[t]
\centering\includegraphics[width=0.6\textwidth]{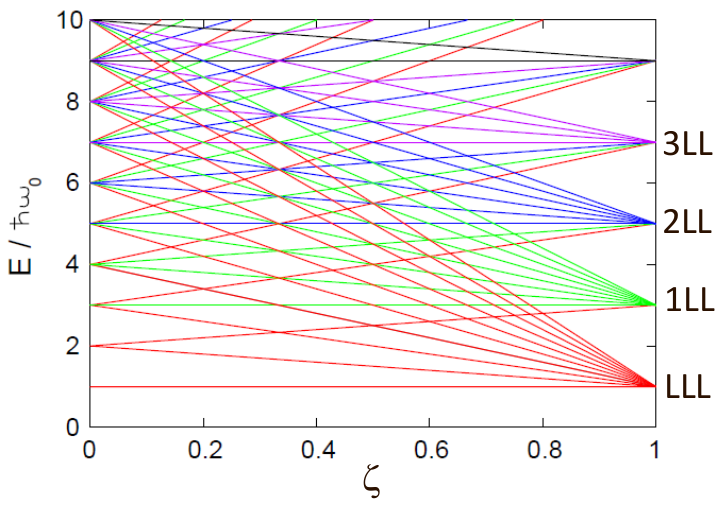}
\caption{
The single-particle energy levels of the Darwin-Fock Hamiltonian in Eq.\ (\ref{hom})
(2D harmonic oscillator rotating with angular frequency $\Omega$) as a function of
$\zeta_\Omega=\Omega/\omega_0$, where $\omega_0$ is the frequency associated with the
harmonic confinement. A given color specifies orbitals with the same number of radial
nodes. The encoding of colors is the same as in Fig.\ \ref{fdb}.
Reprinted figure with permission from Ref.\ \cite{yann07},
Copyright (2007) by IOP Publishing Ltd.
}
\label{fdrot}
\end{figure}

In the case of $N$ neutral atoms or atomic ions (of mass $M$) in an ultracold
2D harmonic trap (of frequency $\omega_0$), rotating with angular frequency
$\Omega$ (around an axis $\hat{\bf z}$), the many-body Hamiltonian can be written
in the form,
\begin{equation}
H^\Omega_{\rm MB}=\sum_{i=1}^N \left\{
\frac{\left({\bf p}_{i}-M\Omega\hat{\bf z} \times 
{\bf r}_{i}\right)^{2}}{2 M}+\frac{M}{2}(\omega_0^2-\Omega^2) {\bf r}_i^2
\right\}  
+\sum_{i < j}^{N} v({\bf r}_{i}-{\bf r}_{j}), 
\label{mbhrot}
\end{equation}
with the two-body potential $v({\bf r}_{i}-{\bf r}_{j})$ being the long-range
Coulomb or the short-range contact interaction. 
The kinetic term of $H^\Omega_{\rm MB}$ is formally equivalent to that (in the symmetric
gauge) of a 2D electron under a constant perpendicular magnetic field $B$ [see
Eqs.\ (\ref{mbhb}) and (\ref{hspb})], if one recognizes the correspondence that the cyclotron
frequency $\omega_c=eB/(m_ec)\rightarrow 2\Omega$.

\subsection{The Darwin-Fock 2D isotropic oscillators and the formation of Landau
levels.}
\label{dfll}
\smallskip

The main (diamagnetic\footnote{
The diamagnetic effect on the spectra and single-particle orbitals contrasts with
the paramagnetism associated with the Zeeman term in the Hamiltonian. Diamagnetism
is prominent in 2D quantum dots due to the much larger size compared to the
natural atoms.})
effect leading to the formation of the Landau levels (LLs), compared to the case of
a vanishing applied magnetic field, or a non-rotating trap, results
from the single-particle part in the many-body Hamiltonians (\ref{mbhb}) and
(\ref{mbhrot}). 
In the case of a parabolic external confinement, $V(x,y)=m^*\omega_0^2 \br^2/2$, the
corresponding 2D Hamiltonians, $H^B_{\rm DF}$ and $H^{\rm rot}_{\rm DF}$, are known as
Darwin-Fock isotropic oscillators, after the names of the authors of the two original
papers \cite{darw31,fock28} on this subject. 

This Section offers an outline of the Darwin-Fock single-particle spectra and
wave functions.

\subsubsection{Two-dimensional isotropic oscillator for electrons in a perpendicular
magnetic field.}
\label{dfb}

In this case, with the help of Eq.\ (\ref{vectp}), the Darwin-Fock Hamiltonian for
an electron can be written in the form
\begin{equation}
H^B_{\rm DF}=\frac{{\bf p}^2}{2m^*} - \frac{1}{2} \omega_c \hat{l}
+ \frac{1}{2}m^*\tilde{\omega}^2 {\bf r}^2,
\label{hb}
\end{equation}
where $\hat{l}=-i \hbar (x \partial/\partial y - y \partial/\partial x)$
is the angular momentum operator in the $z$ direction of an individual electron,
$\omega_c = eB/(m^*c)$ is the usual cyclotron frequency, and $\tilde{\omega} = 
\sqrt{\omega_0^2 + \omega_c^2 /4}$ is the modified frequency of the effective
parabolic confinement.

The $B$-dependent eigenfunctions of the Hamiltonian (\ref{hb}) have the same functional
form as those of a 2D harmonic oscillator at zero magnetic field, but with the modified
frequency $\tilde{\omega}$ defined above, i.e., in polar coordinates, one has
\begin{equation}
\phi_{n,l}(\rho,\theta)={\cal N}_{n,l} \rho^{|l|} e^{-\rho^2/2} 
e^{i l \theta} L_n^{|l|} (\rho^2),
\label{hbwf}
\end{equation}
with $\rho = r/\tilde{l}$ and the characteristic length 
$\tilde{l}=\sqrt{ \hbar/(m^* \tilde{\omega}) }$. In Eq.\ (\ref{hbwf}),
the index $n$ denotes the number of radial nodes, and the symbol $l$ (without
any tilde or hat) stands for the angular-momentum quantum numbers;
the $L_n^{|l|}$'s are associated Laguerre polynomials.

The $B$-dependent eigenenergies of the Hamiltonian (\ref{hb}) are given by
\begin{equation}
\frac{E_{n,l}}{\hbar \omega_0} = (2n+|l|+1) \sqrt{1+\frac{\zeta_B^2}{4}}
-\frac{l}{2}\zeta_B,
\label{dfeb}
\end{equation}
with $\zeta_B=\omega_c/\omega_0$; the corresponding energy spectrum is displayed
in Fig.\ \ref{fdb}.

In the limit of $\zeta_B \rightarrow \infty$, one can neglect the external
confinement, and the energy spectrum in Eq.\ (\ref{dfeb}) reduces to
that of the celebrated Landau levels, namely,
\begin{equation}
E_{\cal M} = \hbar \omega_c ({\cal M} + \frac{1}{2}),
\label{ebll}
\end{equation}
where ${\cal M}=n+(|l|-l)/2$ is the Landau-level index. It is apparent that the LLs
are infinitely degenerate. In particular, the lowest Landau level (LLL) comprises all
the nodeless ($n=0$) orbitals with non-negative angular momenta $l=0, 1, 2, \ldots$.

\subsubsection{The two-dimensional rotating harmonic oscillator.}
\label{dfrot}

In the case of a rotating isotropic oscillator, instead of the expression
(\ref{hb}), one has the following single-particle Hamiltonian:
\begin{equation}
H^\Omega_{\rm DF}=\frac{{\bf p}^2}{2M} - \Omega \hat{l}
+ \frac{1}{2}M \omega_0^2 {\bf r}^2,
\label{hom}
\end{equation}
where $M$ denotes the mass of the trapped particle (e.g., a bosonic or fermionic
atom, or an atomic ion); as aforementioned, $\Omega$ denotes the rotational frequency. 

From a comparison of the second terms in Eqs.\ (\ref{hb}) and (\ref{hom}), one 
verifies again the correspondence $\Omega \rightarrow \omega_c/2$. 

Note that, unlike the case of the $B$-dependent Darwin-Fock oscillator, the 
rotation does not generate an effective-confinement frequency different from that 
of the original external confinement [compare the third terms between Eq.\ (\ref{hb})
and Eq.\ (\ref{hom})]. As a result, the eigenfunctions of the Hamiltonian (\ref{hom})
are given by the same functional expression displayed in Eq.\ (\ref{hbwf}), but with
$\rho=r/l_0$, where the characteristic length is $l_0=\sqrt{\hbar/(M \omega_0)}$.

The single-particle energy spectrum associated with the Hamiltonian (\ref{hom})
is displayed in Fig.\ \ref{fdrot}, and the mathematical expression for the
corresponding eigenenergies is given by
\begin{equation}
\frac{E_{n,l}}{\hbar \omega_0} = (2n+|l|+1) -l \zeta_\Omega,
\label{dfeom}
\end{equation}
with $\zeta_\Omega=\Omega/\omega_0$. 

For  $\zeta_\Omega = 1$, the energy spectrum in Eq.\ (\ref{dfeom}) describes formation
of rotational Landau levels, i.e.,
\begin{equation}
E_{\cal M} = 2 \hbar \omega_0 ({\cal M} + \frac{1}{2}),
\label{eomll}
\end{equation}
where ${\cal M}=n+(|l|-l)/2$ is the index of the Landau level.

As is the case with the $B$-dependent Darwin-Fock oscillator, the rotational LLs are 
infinitely degenerate, and the lowest Landau level (with ${\cal M}=0$) consists of all
the nodeless single-particle levels ($n=0$) with angular momenta $l \geq 0$.
However, in contrast to the magnetic-field case where the energy gap, $\hbar \omega_c$,
between the Landau levels does depend on $B$, the energy gap between the rotational Landau
levels is independent of $\Omega$ and equals a constant value of $2\hbar\omega_0$.

\begin{figure}[t]
\centering\includegraphics[width=0.8\textwidth]{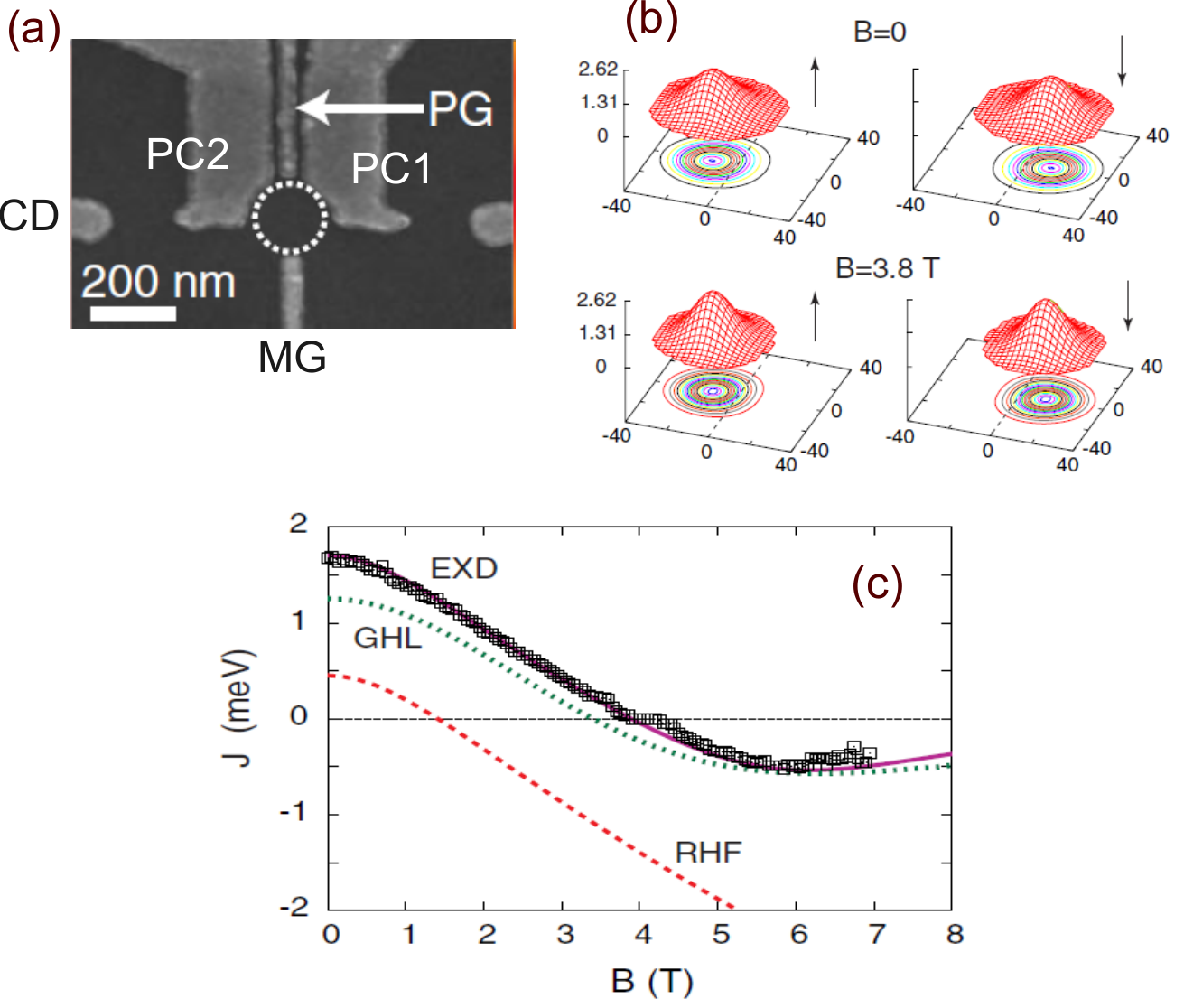}
\caption{
(a) Scanning electron micrograph of the two-electron quantum dot with integrated charge
readout fabricated by electron-beam lithography on a parabolic quantum well based on
GaAs/Al$_x$Ga$_{1-x}$As heterostructures. The quantum well material has been designed and
measured to have a Land\'e factor $g^* \sim 0$ [see Eq.\ (\ref{hspb})].
The QD is formed by applying negative voltages to the gates PCI, PC2, PG and MG.
(b) Single-particle UHF orbitals (modulus square) that are used in the construction of the
GHL wave function. Lengths in nm and orbital densities in $10^{-3}$ nm$^{-2}$. Arrows
indicate up and down spins.
(c) Comparison of $J(B)$ calculated with different methods and the experimental results
(open squares). Solid line: EXD. Dotted line: GHL. Dashed line: RHF.
For the parameters used in the calculation to model the anisotropic QD, see text.
Reprinted figures with permission from Ref.\ \cite{yann07.4}, Copyright (2007) by World
Scientific Publishing Company.
}
\label{ihnf}
\end{figure}

\subsection{Theoretical results and experimental observations related to Wigner molecules
in semiconductor QDs under an applied magnetic field in the range from vanishing to fields
near the LLL.}
\label{resmagqd}
\smallskip

\subsubsection{$B$-dependence of the excitation spectrum of two correlated electrons in an
elliptic lateral quantum dot.}
\label{2eqdb}

Understanding the quantum dot helium (He-QD), a man-made two-electron system, was a hallmark
for enabling the design and control of more complex interacting quantum systems such as
those required for the implementation of quantum information processing schemes. Refs.\
\cite{yann06,yann07.4} presented a study of quantum dot Helium fabricated on a parabolic
quantum well [see Fig.\ \ref{ihnf}(a)] which went beyond earlier work \cite{kouw97.2} by
detecting higher lying excited states as a function of magnetic fields and by presenting
evidence for the importance of correlation effects associated with Wigner molecularization.

In order to interpret the measured excitation spectra in detail, an exact diagonalization
(i.e., FCI) and an approximate (generalized Heitler-London, GHL, see Sec.\ \ref{2eqd})
microscopic treatments for two electrons in a single elliptic QD were used.
The elliptic shape of the QD is implemented due to the plunger [see Fig.\ \ref{ihnf}(a)].
The two-electron Hamiltonian that corresponds to the experimental device is given by Eq.\
(\ref{ham}), but with the following two modifications:
(1) $H({\bf r})$ is the single-particle Hamiltonian for an electron in an elliptic
potential confinement, $H=T + (m^*/2) (\omega^2_x x^2 + \omega^2_y y^2)$,
with the kinetic energy being $T=( {\bf p}-e{\bf A}/c )^2 /(2m^*)$ and ${\bf A}(\br)=
0.5(-Bx, By, 0)$ and (2) the Coulombic term has a prefactor $\gamma$ which accounts for
the reduction of the Coulomb strength due to
the thickness (along the $z$ direction) of the electronic layer and for any screening effects
arising from the gate electrons.    
The specific parameters used in the EXD calculation were as follows: $m^*=0.070m_e$ (GaAs),
$\kappa=12.5$ (GaAs), $\hbar \omega_x= 4.23 $ meV, $\hbar \omega_y= 5.84 $ meV, and
$\gamma=0.862$.

\begin{figure}[t]
\centering\includegraphics[width=0.99\textwidth]{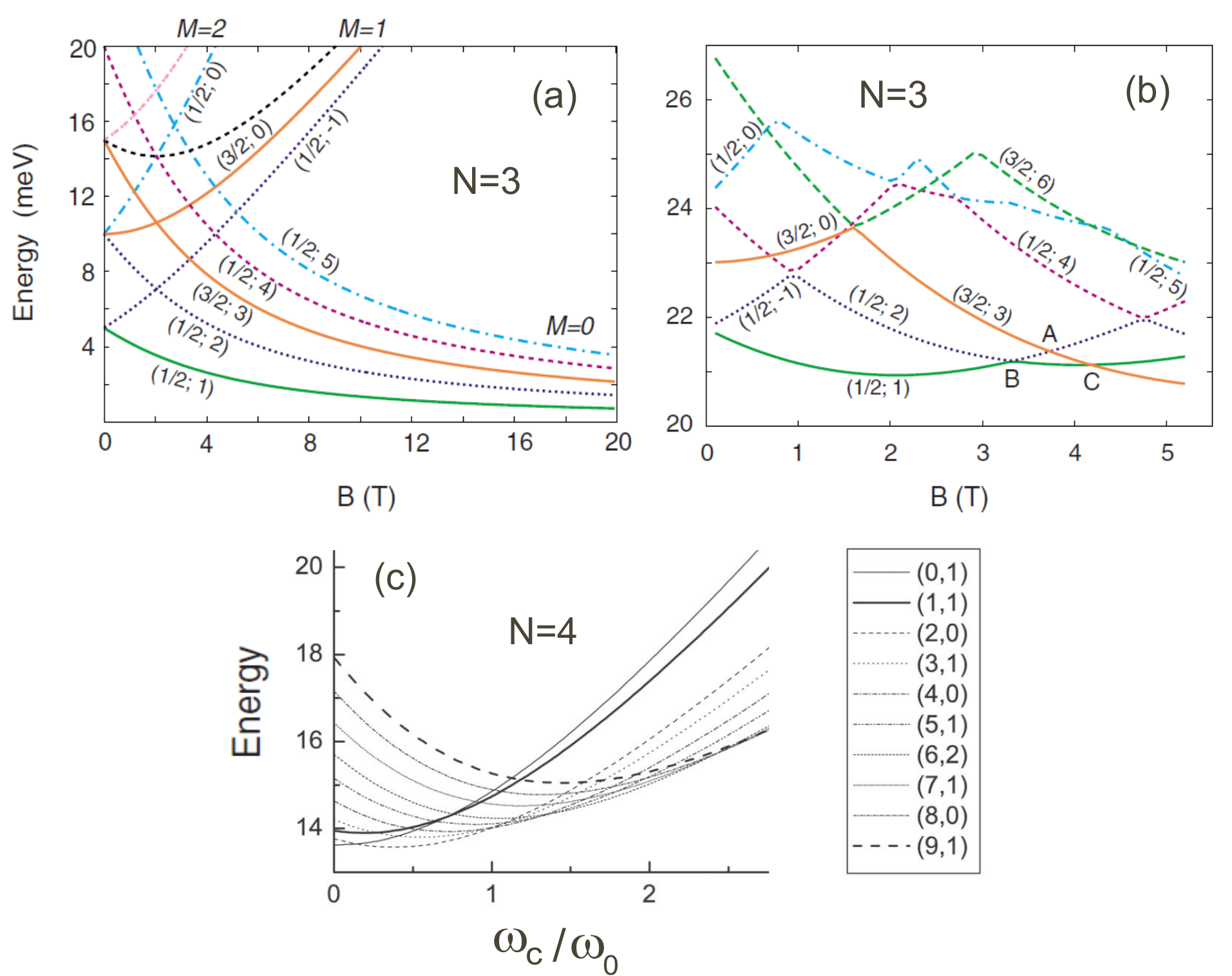}
\caption{
(a) FCI ground-state and excitation energy spectra [referenced to $3\hbar \sqrt{\omega_0^2
+ \omega_c^2/4}$, with $\omega_0=\sqrt{(\omega_x^2+\omega_y^2)/2}$] as a function of the
magnetic field for $N=3$ {\it noninteracting\/} electrons in a circular quantum dot
($\eta=\omega_x/\omega_y=1$). Parameters: external confinement $\omega_x=\omega_y=5$ meV, 
dielectric constant $\kappa = \infty$, effective mass $m^*=0.067m_e$, and effective
Land\'e coefficient $g^*=0$. The labels $(S;L)$ denote the quantum numbers for the total
spin and the total angular momentum. Different forming Landau bands are denoted by the
different ${\cal M}$ values. The $S_z$ indices are not indicated, since the multiplets
$(S=1/2,S_z)$ and $(S=3/2,S_z)$ are degenerate in energy when $g^*=0$.
Reprinted figure with permission from Ref.\ \cite{yann07.2},
Copyright (2007) by the American Physical Society.
(b) FCI ground-state and excitation energy spectra [referenced to $3\hbar \sqrt{\omega_0^2
+ \omega_c^2/4}$, with $\omega_0=\sqrt{(\omega_x^2+\omega_y^2)/2}$] as a function of $B$
for $N=3$ {\it interacting\/} electrons in a circular quantum dot with same parameters
as in (a), except for $\kappa=12.5$.
With regard to the noninteracting case in (a), characteristic crossings between the
energy levels within each Landau level do appear.
Reprinted figure with permission from Ref.\ \cite{yann07.2},
Copyright (2007) by the American Physical Society.
(c) FCI energy spectrum for $N=4$ interacting electrons in a circular quantum dot as a
function of the magnetic field for $R_W=2.0$. The lowest-energy states of angular momentum
up to 9 are shown. The Zeeman energy is included with $g^*=-0.44$. The energy is given in
units of $\hbar\omega_0$ (which specifies the trapping confinement). States are labeled
$(L,S)$, with $L$ the total angular momentum and $S$ the total spin of the state.
Note again the level crossings.
Adapted figure with permission from Ref.\ \cite{szaf03},
Copyright (2003) by the American Physical Society.
}  
\label{bsmallf}
\end{figure}

As demonstrated in detail in Refs.\ \cite{yann06,yann07.4}, these experimental findings
could be quantitatively interpreted through a comparison with the results of FCI
calculations for two electrons confined in an anisotropic harmonic potential specified
by the parameters tabulated above.
All the states observed in the measured spectra (as a function of the magnetic field)
could be unambiguously identified with calculated ground-state and excited states of
the two-electron Hamiltonian described above. In particular, the calculated
magnetic-field-dependent energy splitting, $J_{\rm EXD}(B) = E^t_{\rm EXD}(B) - E^s_{\rm EXD}(B)$,
between the two lowest singlet $(s)$ and triplet $(t)$ states is found to be in remarkable
agreement with the experiment [see Fig.\ \ref{ihnf}(c)].

The importance of the hidden breaking of the parity symmetry along the $x$-axis in the
two-electron wave function can be assessed by comparing the measured $J(B)$ with that
calculated within the GHL and RHF approximations (see Sec.\ \ref{2eqd}). To facilitate
the comparisons, the calculated $J_{\rm GHL}(B)$ and $J_{\rm RHF}(B)$ curves are plotted also
in Fig.\ \ref{ihnf}(c), along with the EXD result and the experimental measurements.
The RHF scheme is appealing, because it minimizes the total energy using a single Slater
determinant. From Fig.\ \ref{ihnf}(c), it becomes apparent, however, that the RHF approach,
which requires in addition that the two electrons in the singlet state occupy a common orbital,
is unable to account for the experimental findings. On the contrary, the
GHL method, which permits the two electrons to occupy two spatially separate orbitals,
is a superior approximation. Plotting the two GHL orbitals [see Fig.\ \ref{ihnf}(b)]
for the singlet state clearly demonstrates their significant spatial separation.

The GHL orbitals\footnote{
The GHL orbitals coincide with the UHF ones; see Sec.\ \ref{2eqd}.}  
are displayed in Fig.\ \ref{ihnf}(b) for both the $B = 0$ and $B = 3.8$ T cases. The
spatial shrinking of these orbitals at the higher $B$-value illustrates enhancement of
WM formation towards the classical limit with increasing magnetic field. The asymptotic
convergence of the energies of the singlet and triplet states, [i.e., $J(B) \rightarrow 0$
as $B \rightarrow \infty$] is also a reflection of a process of ``dissociation'' of the 2e
Wigner molecule, since the ground-state energy of two fully spatially separated electrons
(zero overlap) does not depend on the total spin.

\subsubsection{$B$-dependence of the excitation spectrum of three and four correlated
electrons in a circular quantum dot from FCI calculations.}
\label{34eqdb}

That the exact spectra of two interacting electrons in a circular quantum dot as a function
of $B$ exhibit crossings between the constant-$L$ levels within each Landau level was found
rather early \cite{chap92}. Fig.\ \ref{ihnf}(c) demonstrated an analogous property for the
case of two electrons in a deformed QD, as well.

Moving to the cases of $N=3$ and $N=4$ interacting electrons in circular QDs under a magnetic
field, Figs.\ \ref{bsmallf}(b) and (c) demonstrate that this level-crossing behavior persists
for larger number of electrons. The fact that this level crossing is a signature of strong
correlations follows clearly through a comparison with the $B$-dependent spectrum of $N=3$
noninteracting electrons, which exhibits no level crossings within each Landau level [see
Fig.\ \ref{bsmallf}(a)].

\begin{figure}[t]
\centering\includegraphics[width=0.90\textwidth]{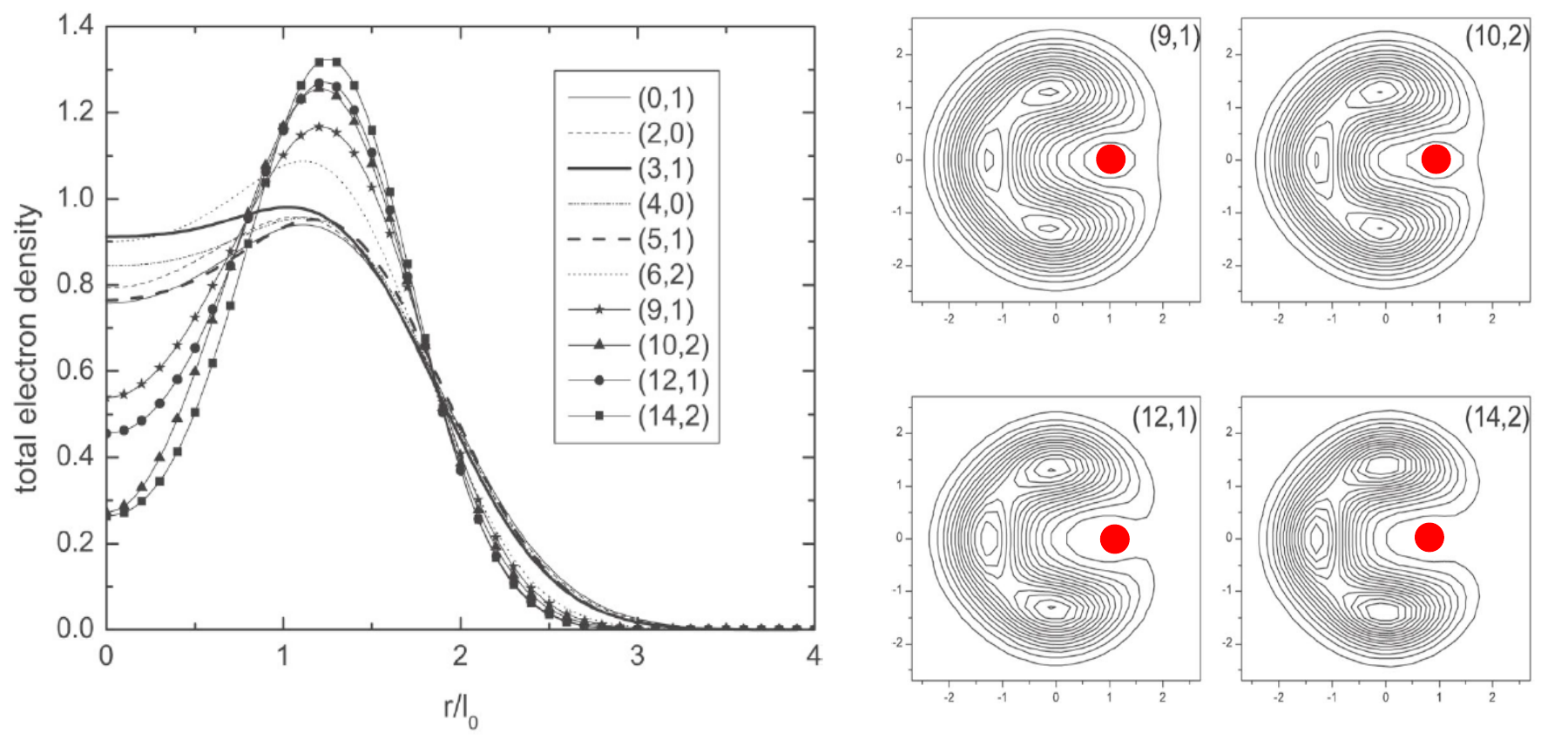}
\caption{
(a) FCI total electron densities in a number of different angular momentum states for $N=4$
interacting electrons in a circular quantum dot at $R_W=2.0$ and different $B$ values.
In concurrence with the level crossings, these states are the ground states at, respectively,
$\omega_c/\omega_0$ equal to 0.0, 0.5, 1.0, 1.1, 1.27, 2.0, 2.7, 3.0, 3.6, and 4.0. States
are labeled $(L,S)$, with $L$ the total angular momentum and $S$ the total spin of the state.
Angular momenta up to 14 are included. These FCI electron densities are rotationally invariant
and they are plotted as a function of the polar coordinate $r$.
(b) Two-dimensional FCI spin-unresolved CPDs for four different angular momentum states of
$N=4$ interacting electrons at $R_W=2.0$. One electron (designated by a red dot) is fixed at
$\br_0=(1.24,0)$. These states become ground states at, respectively, $\omega_c/\omega_0$ equal
to 2.7, 3.0, 3.6, and 4.0. Lengths are in units of $l_0=\sqrt{\hbar/(m^* \omega_0)}$. States
are labeled $(L,S)$, with $L$ the total angular momentum and $S$ the total spin of the state.
Reprinted and adapted figures with permission from Ref.\ \cite{szaf03},
Copyright (2003) by the American Physical Society.
}
\label{szaf4f}
\end{figure}

As a result of the level crossings, states with specific increasing angular momenta become
the ground state with increasing $B$, as exemplified for $N=4$ interacting electrons in the
caption of Fig.\ \ref{szaf4f}(a). These specific $L$'s belong to different magic angular
momenta series \cite{yang07}, each series being associated with a given total angular
momentum $S$. For example, for $S=2$ (fully spin-polarized electrons), one has the momenta
$L=6$, 10, and 14, which belong to the series $4n+2$; see Table I in Ref.\ \cite{yang07}
and the description of the spinful-4e RWM in Sec.\ V of Ref.\ \cite{shi07}.
The formation of RWMs [with a (0,4) square-ring intrinsic configuration] in these ground
states is seen from corresponding charge densities [shown in Fig.\ \ref{szaf4f}(a)] and CPDs
[shown in Fig.\ \ref{szaf4f}(b)]. The enhancement of the RWM formation for the larger
angular momenta is clearly seen in Fig.\ \ref{szaf4f}(a) from the fact that the density
depression at the center of the ring is getting deeper.

\subsubsection{Description of RWMs formed in circular QDs under a magnetic field using
the two-step symmetry-breaking/symmetry-restoration method and their equivalence to
superfloppy rotors.}
\label{fe2step}

In the method of successive hierarchical approximations (see Sec.\ \ref{hieras}), one begins
with a {\it pinned\/} (also referred to as {\it static\/}) Wigner molecule (SWM), described by
an unrestricted Hartree-Fock determinant which violates the circular symmetry. Subsequently,
the {\it rotation\/} of the Wigner molecule is described by a post-Hartree-Fock step of
restoration of the broken circular symmetry via projection techniques. Here, the
symmetry-broken UHF orbitals (first step of the two-step procedure) are approximated by
(parameter free) displaced Gaussian functions; that is, for an electron localized at
${\bf R}_j$ ($Z_j$), one can use the orbital
\begin{equation}
u(z,Z_j) = \frac{1}{\sqrt{\pi} \lambda}
\exp \left( -\frac{|z-Z_j|^2}{2\lambda^2} - i\vartheta(z,Z_j;B) \right),
\label{uhfo}
\end{equation}
with $\lambda = \sqrt{\hbar /m^* \tilde{\omega}}$.
$\tilde{\omega}=\sqrt{\omega_0^2+\omega_c^2/4}$, where $\omega_c=eB/(m^*c)$ is the cyclotron
frequency (defined previously in Sec.\ \ref{simi}), and $\omega_0$ specifies the external
parabolic confinement, as usually. Complex numbers are used to represent the position
variables, so that $z=x+iy$, $Z_j = X_j +i Y_j$. The phase in Eq.\ (\ref{uhfo}) guarantees
gauge invariance in the presence of a perpendicular magnetic field and is given in the
symmetric gauge by $\vartheta(z,Z_j;B) = (x Y_j - y X_j)/2 l_B^2$, with
$l_B = \sqrt{\hbar c/ e B}$. Note that expression (\ref{uhfo}) has the form of a London
atomic orbital, widely used \cite{lond37,popl62,schm88,helg08} in atomic physics to describe
electronic orbitals in a magnetic field.

In the case of a two-dimensional extended system, the $Z_j$'s are arranged in a triangular
lattice \cite{yosh83}. For a finite 2D assembly of $N$ electrons, however, the $Z_j$'s
form \cite{yann04.2} $r$ concentric regular polygons denoted as ($n_1, n_2,...,n_r$),
where $n_1>0$ and $n_r>0$ correspond to the innermost and outermost rings, respectively.
Furthermore, the vertices of these regular polygons coincide with the equilibrium positions
of $N=\sum_{q=1}^r n_q$ classical point charges inside a circular confinement
\cite{kong02}. For a single polygonal formation, the notation $(0,N)$ is mostly employed; it
is then understood that $n_1=N$. 

A {\it single\/} UHF Slater determinant, $|\Psi^{\rm SWM} [z] \rangle$, made out of the
one-electron orbitals $u(z_i,Z_i)$, $i = 1,...,N$, represents the symmetry-broken {\it static\/}
Wigner molecule. RWM correlated and symmetry-preserving many-body wave functions (exhibiting good
total angular momenta $L$) are generated (second step in the two-step approach, see Secs.\
\ref{hieras}, \ref{2eqd}, and \ref{qdrr}) from this UHF determinant by applying appropriate
projection operators. Then, apart from a proportionality constant, the ensuing
rotating-Wigner-molecule states are given by
\cite{yann06.3}
\begin{equation}
|\Phi^{\rm RWM}_L \rangle =  \int_0^{2\pi} ... \int_0^{2\pi}
d\gamma_1 ... d\gamma_r 
|\Psi^{\rm SWM}(\gamma_1, ..., \gamma_r) \rangle
\exp \left( i \sum_{q=1}^r \gamma_q L_q \right).
\label{wfprj}
\end{equation}
The total angular momentum $L=\sum_{q=1}^r L_q$ and $|\Psi^{\rm SWM}[\gamma] \rangle$ is the original
UHF Slater determinant after {\it all the one-electron orbitals of the $q$th ring\/} have been
rotated collectively, (i.e., coherently) by the {\it same\/} azimuthal angle $\gamma_q$. Note
that Eq.\ (\ref{wfprj}) can be written as a product of projection operators ${\cal P}_{L_q}$'s
[see Eq.\ (\ref{amp})] acting on the original Slater determinant
$|\Psi^{\rm SWM}(\gamma_1=0, ..., \gamma_r=0) \rangle$.
In the limit $B \rightarrow \infty$, $\lambda = l_B \sqrt{2}$, and the single-electron
wave function in Eq. (\ref{uhfo}) is contained entirely in the lowest Landau level
\cite{yann06.3} (see Appendix A therein). For a finite $B$, the orbital in Eq.\
(\ref{uhfo}) contains contributions from all the Darwin-Fock levels. At $B=0$, one has
$\lambda=l_0$, and the orbital in Eq. (\ref{uhfo}) reduces to the form displayed in Eq.\
(\ref{uhfo1}). The continuous-configuration-interaction form of the projected wave
functions [i.e., the linear superposition of determimants in Eq.\ (\ref{wfprj})]
implies a highly entangled state; see Sec.\ \ref{comm2}.
The electrons in Eq.\ (\ref{wfprj}) are fully spin polarized.

Due to the point-group symmetries of each polygonal ring of electrons
in the SWM wave function, the total angular momenta $L$ of the rotating
crystalline electron molecule are restricted to the so-called {\it magic} 
angular momenta, i.e.,
\begin{equation}
L_m = L_0 + \sum_{q=1}^r k_q n_q,
\label{lmeq}
\end{equation}
where the $k_q$'s are non-negative integers\footnote{
For multiple rings, see Refs.\ \cite{yann02.2,yann04}. For the simpler cases of $(0,N)$
or $(1,N-1)$ rings, see, e.g., Refs.\ \cite{ruan95} and \cite{maks96}.
}
(when $n_1=1$, $k_1=0$).

The partial angular momenta associated with the $q$th ring,
$L_q$ [see Eq.\ (\ref{wfprj})], are given by
\begin{equation}
L_q = L_{0,q} + k_q n_q,
\label{lmpar}
\end{equation}
where $L_{0,q}=\sum_{i=i_q+1}^{i_q+n_q} (i-1)$ with
$i_q = \sum_{s=1}^{q-1} n_s$ $(i_1=0)$, and
$L_0 = \sum_{q=1}^r L_{0,q}$.

The energy of the RWM state [Eq.\ (\ref{wfprj})] is given \cite{yann06.3,yann07} by
\begin{equation}
E^{\rm RWM}_L = \left. { \int_0^{2\pi} h([\gamma]) e^{i [\gamma] \cdot [L]}
d[\gamma] } \right/%
{ \int_0^{2\pi} n([\gamma]) e^{i [\gamma] \cdot [L]} d[\gamma]},
\label{eproj3}
\end{equation}
with the Hamiltonian and overlap matrix elements $h([\gamma]) =
\langle \Psi^{\rm SWM}([0]) | H | \Psi^{\rm SWM}([\gamma]) \rangle$ and
$n([\gamma]) =
\langle \Psi^{\rm SWM}([0]) | \Psi^{\rm SWM}([\gamma]) \rangle$, 
respectively, and
$[\gamma] \cdot [L] = \sum_{q=1}^r \gamma_q L_q$.
The SWM energies are simply given by $E_{\rm SWM} = h([0])/n([0])$.

\begin{figure}[t]
\centering\includegraphics[width=0.85\textwidth]{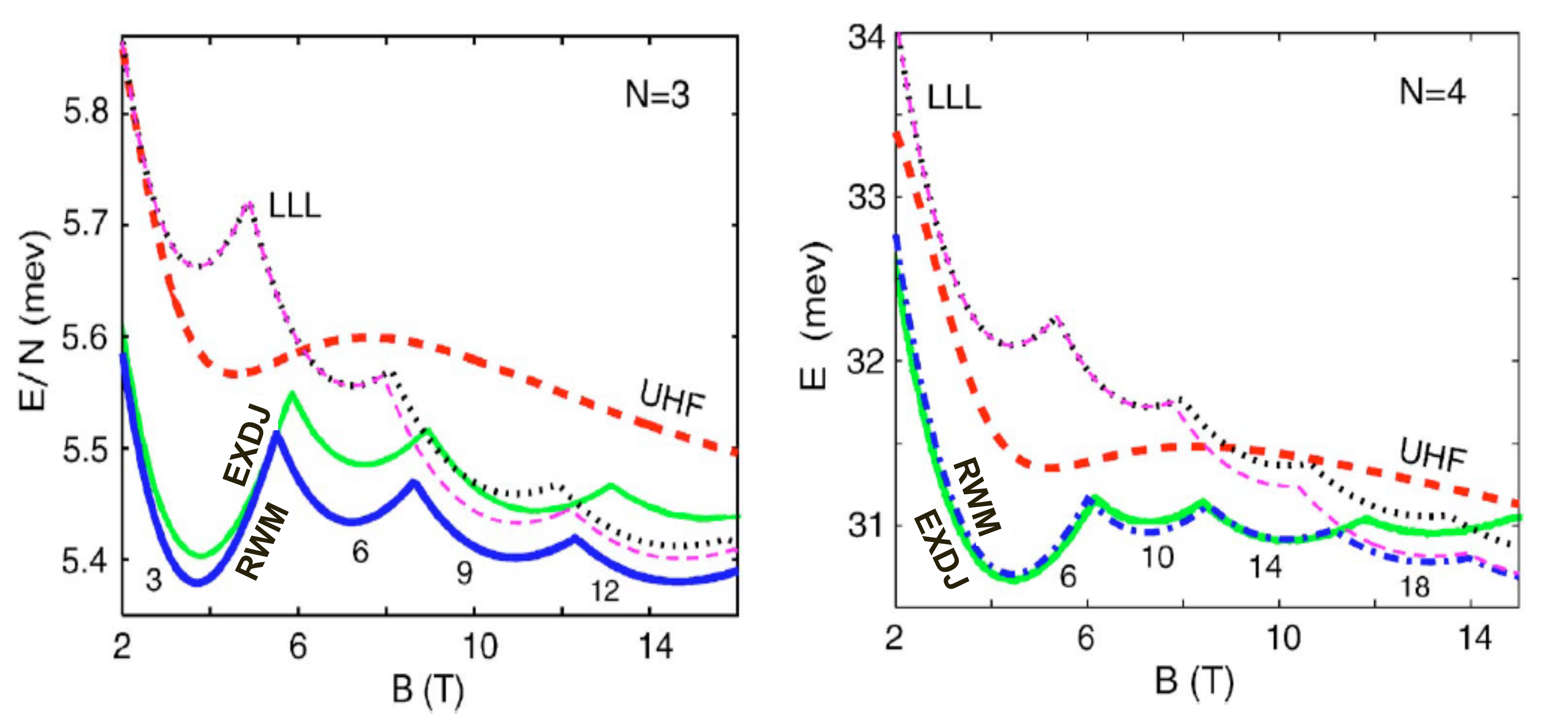}
\caption{
Two-step method versus other-type of calculations.
(a) Ground-state energies (per particle, referenced to $\hbar \protect\tilde{\omega}$)
for $N=3$ fully spin-polarized electrons. The electrons are arranged in a (0,3) structure
in the intrinsic frame of reference. EXDJ calculation from Ref.\ \cite{hawr93}. 
Parameters used: confinement $\hbar \omega_0=3.37$ meV, 
dielectric constant $\kappa=12.4$, effective mass $m^*=0.067 m_e$.
(b) Ground-state energies (referenced to 4$\hbar \protect\tilde{\omega}$)
for $N=4$ fully spin-polarized electrons. The electrons are arranged in a (0,4) configuration
in the intrinsic frame of reference. EXDJ calculation from Ref.\ \cite{ruan99}.
Parameters used: confinement $\hbar \omega_0=3.60$ meV, dielectric constant $\kappa=13.1$,
effective mass $m^*=0.067m_e$.
Both (a) and (b). Thick dashed line (red): broken-symmetry UHF (SWM). Thinner solid line
(green): EXDJ (uses Jacobi variables). Thick solid line (blue) or thick dashed-dotted
line (blue): RWM. Thin dashed line (violet): energies according to the commonly used
Hamiltonian $H^{B,\prime}_{\rm LLL}$ in Eq.\ (\ref{hblll2}) below. Thin dotted line (black):
Similar to the violet line, bur with the interaction contributions calculated within the
LLL using the RWM wave function (see text).
Reprinted panels with permission from Ref.\ \cite{yann06.3},
Copyright (2006) by the American Physical Society.
}
\label{yues1f}
\end{figure}

To check the accuracy of the two-step multi-ring approximation described above, the
$B$-dependent RWM energies according to Eq.\ (\ref{eproj3}) for three [see Fig.\
\ref{yues1f}(a)] and four [see Fig.\ \ref{yues1f}(b)] electrons in an external parabolic
confinement are compared with those from other methods and approximations, including an
exact diagonalization (marked as EXDJ) that separates out the center of mass by utilizing
Jacobi variables \cite{hawr93,ruan99}.

The thick dotted line (red) represents the broken-symmetry UHF approximation (first step of
the two-step method), which naturally is a smooth curve lying above the RWM and EXDJ [solid
line (green)] ones. To further evaluate the accuracy of the two-step method, Fig.\
\ref{yues1f} also displays [thin dashed line (violet)] ground-state energies
calculated with the commonly employed \cite{maks00,yang93,jeon04} approximate Hamiltonian
$H^{B,\prime}_{\rm LLL}$ in Eq.\ (\ref{hblll2}) below, 
which is valid near the LLL limit (i.e., only the nodeless Darwin-Fock states contribute
to the kinetic energy). These energies (marked as LLL) tend to substantially overestimate
the RWM (and EXDJ) energies for lower values of $B$. On the other hand, for higher values
of $B$ ($>$ 12 T), they tend to agree rather well with the RWM ones. A similar behavior is
exhibited also by the curve [dotted line (black)] for which the interaction contributions
are calculated within the LLL using the RWM wave function [i.e., by setting
$\lambda=l_B \sqrt{2}$ in Eq.\ (\ref{uhfo})].

The central message drawn from both Figs.\ \ref{yues1f}(a) and \ref{yues1f}(b) is that the
RWM approach is able to capture the discontinuous ``oscillations'' of the ground-state
(yrast-band) energies as a function of increasing $B$. Moreover, each oscillatory
parabola-like segment is associated with a given magic angular momentum $L_m$ [see Eq.\
(\ref{lmeq})], belonging to the $L_m=3+3m$ series for $N=3$ [hidden (0,3) RWM configuration]
and the $L_m=6+4m$ series for $N=4$ [hidden (0,4) RWM configuration].\footnote{
A similar ``oscillatory'' yrast spectrum [associated with a (0,8) configuration] was also
found for the case of $N=8$ neutral repelling bosons in a rotating toroidal trap as a
function of the reduced rotational frequency $\Omega/\omega_0$ for $R_\delta=50$, as elaborated
in Ref.\ \cite{roma06}.}
Corresponding fractional filling factors are specified by $\nu = N(N-1)/(2L_m)$.

\begin{figure}[t]
\centering\includegraphics[width=0.75\textwidth]{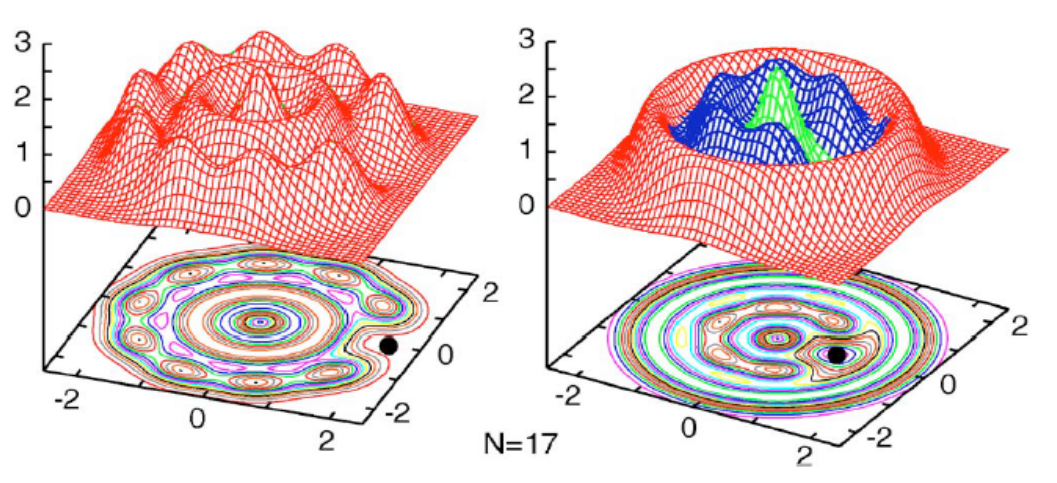}
\caption{
Ground-state conditional probability distributions, CPDs, obtained from the RWM wave 
function for the ground state of $N=17$ electrons at $B=10$ T ($L_m=228$). The electrons
are arranged in a (1,6,10) configuration \cite{kong02}. The fixed point (solid dot) is
placed on the outer ring at $r_0=1.858R_0$ (left frame), and on the inner ring at
$r_0=0.969R_0$ (right frame). Parameters used: confinement $\hbar \omega_0=3.6$ meV, 
dielectric constant $\kappa=13.1$, effective mass $m^*=0.067 m_e$. Lengths in units of
$R_0 =( 2e^2/(\kappa m^* \omega_0^2) )^{1/3}$. The CPDs are given (along the vertical axes)
in arbitrary units, but the same for both panels.
Reprinted figure with permission from Ref.\ \cite{yann06.3},
Copyright (2006) by the American Physical Society.
}
\label{yues2f}
\end{figure}

Fig.\ \ref{yues2f} displays CPDs for the RWM wave function of $N=17$ electrons.
This case has a nontrivial three-ring configuration (1,6,10) \cite{kong02},
which is sufficiently complex to allow generalizations for larger numbers of
particles. The remarkable combined character (partly crystalline and partly liquid
leading to a non-rigid rotational inertia) of the RWM is illustrated in these CPDs.
Indeed, as the two CPDs [reflecting the choice of taking the fixed point, ${\bf r}_0$ in
Eq.\ (\ref{cpds}), on the outer ring (left panel) or the inner ring (right panel)]
demonstrate, the polygonal electron rings rotate {\it independently\/} of each other.
As a result, to an observer situated on the inner ring, the outer ring will appear
as having a uniform density, and vice versa.
The wave functions obtained from
FCI exhibit also the property of independently rotating rings (see e.g., the $N=12$ and
$L=132$ ($\nu=1/2$) case in Fig.\ 11 of Ref.\ \cite{yann06.3}). This is a testimony to
the ability of the RWM wave functions to capture the essential physics of a finite number
of electrons under an applied magnetic field $B$.

As was shown in Sec.\ \ref{qdrr}, the RWM at vanishing magnetic field but high $R_W$
behaves like a rigid rotor with the ground-state (yrast-band) energies varying like $L^2$. 
The independent rotation of the concentric rings illustrated in Fig.\ \ref{yues2f},
as well as the segmented parabola-type ground-state energies as a function of $B$
demonstrated in Fig.\ \ref{yues1f}, correspond to a classical rotor that is hyper-floppy,
i.e., the corresponding moment of inertia depends strongly on the angular momentum $L$. 
Focussing here for simplicity and brevity to a $(0,N)$ ring structure, for high $L$
(and/or $B$), the radius $a$ entering in the moment of inertia ${\cal J}=N m^*a^2$ of the
floppy classical ring is determined by the balance between the classical kinetic energy
and the potential energy due to the harmonic confinement $\tilde{\omega}$. Then
$a \approx \lambda \sqrt{L/N}$, and the yrast-band energies of the hyper-floppy rotor
can be approximated by \cite{maks96,yann04.2}
\begin{equation}
E^{\rm floppy}_{\rm class}(L) \approx \hbar(\tilde{\omega}-\omega_c/2)L + C_V/L^{1/2},
\label{eflop}    
\end{equation}
where the constant $C_V=N^{3/2} S_N e^2/(4 \kappa \lambda)$, with
$S_N= \sum_{j=2}^{N} \left( \sin[(j-1)\pi /N] \right)^{-1}$.

The generalization of Eq.\ (\ref{eflop}) to the case of multiple concentric rings was
derived in Ref.\ \cite{yann06.3}.

\subsection{Neutral spinful fermionic atoms in a rapidly rotating ultracold trap
at the LLL limit.}
\label{lllrot}
\smallskip

This Section will review symmetry-restoration and FCI theoretical investigations
\cite{yann25,yann20} demonstrating the underlying physical picture of rotating
Wigner molecules in the case of neutral spinful fermionic atoms ($^6$Li) in rapidly
rotating traps, i.e., at the $\zeta_\Omega=1$ limiting regime of LL formation
(referred to also as the ``deconfinement limit'' in the experimental Ref.\
\cite{joch24}). In addition, most recent experimental results \cite{joch24} supporting
the RWM physics in the LLL will be discussed.

Near $\zeta_\Omega =1$, and for conditions applicable to the large majority of the
experiments in this field, the rotational Hamiltonian (\ref{mbhrot}) can be simplified
\cite{yann07.3} to the form
\begin{equation}
H^{\Omega,\prime}_{\rm LLL}=N \hbar \omega_0 + \hbar(\omega_0-\Omega) L
+\sum_{i < j}^{N} v({\bf r}_{i}-{\bf r}_{j}),
\label{hlll22}
\end{equation}
with $L$ being the total angular momentum.

Exactly in the LLL, the second term in the right-hand-side of Eq.\ (\ref{hlll22}) is suppressed,
because $\omega_0=\Omega$. Thus the many-body Hamiltonian governing ultracold neutral atoms can
be approximated \cite{popp04,roma06,yann07.3,coop08,hazz08,palm20,yann20,yann21}
with only the contact-interaction term,\footnote{
Based on FCI calculations, RWMs for scalar bosons in the LLL were predicted
in Ref.\ \cite{yann07.3}.}
i.e.,
\begin{equation}
H^{\Omega,\delta}_{\rm LLL} = (g/\Lambda^2) \sum_{i < j}^{N} \delta^2(z_i-z_j),
\label{hlll}
\end{equation}
with $\Lambda=\sqrt{\hbar/(M\omega_0)}$. In Eq.\ (\ref{hlll}), I reverted to the customary
use of complex particle coordinates in the 2D plane, where the position of particle $j$
is denoted as $z_j = x_j + i y_j$, and the square of the particle's distance from the
origin is given by $z_j z^*_j$. This notation will be used in the remaining of this Section.
$g$ is the strength of the repulsive contact interaction.

\begin{figure}[t]
\centering\includegraphics[width=0.9\textwidth]{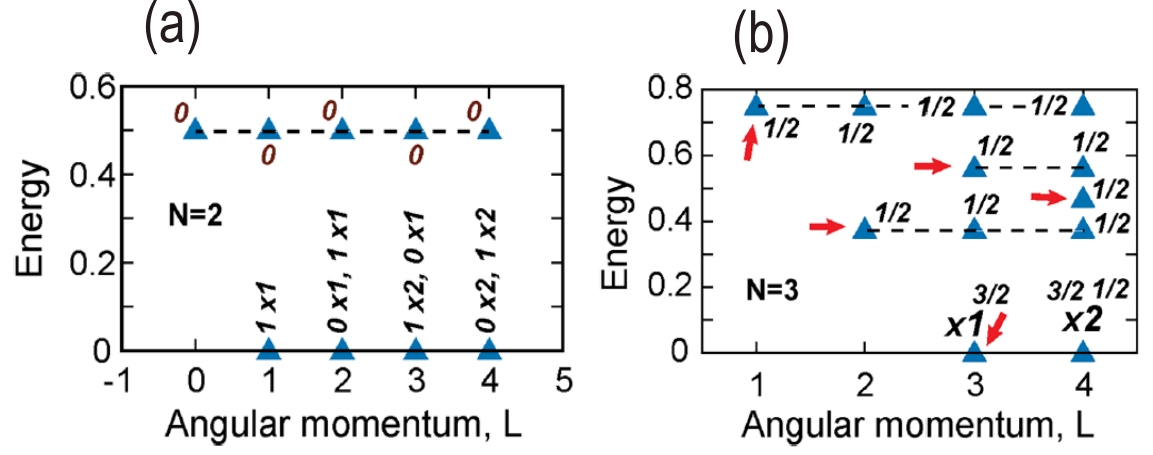}
\caption{
FCI energy spectra for (a) $N=2$ and (b) $N=3$ spinful contact-interacting LLL fermions with
total-spin projection $S_z=0$ and $S_z=1/2$, respectively. In (a), the total spin $S=0$ (singlet)
or $S=1$ (triplet) for each level is given next to the corresponding solid-triangle symbol.
In (b), the total spin $S=1/2$ or $S=3/2$ for a given level is displayed next to the
corresponding solid-triangle marker. In both (a) and (b), the lebel ``x$n$''
indicates an $n$-degenerate zero-interaction-energy state. States that lie on a
horizontal {\it dashed\/} line relate through a center-of-mass translation.
Translationally invariant states for $N=3$ are marked by a red arrow.
For translationally invariant states for $N=2$, see the text. 
Energies are in units of $g/(\pi \Lambda^2)$.
Reprinted figures with permission from Ref.\ \cite{yann25},
Copyright (2025) by the American Physical Society.
}
\label{specLLLdelta}
\end{figure}

\subsubsection{FCI spectra for neutral spinful fermionic atoms in the LLL.}
\label{speclll}
\smallskip

The FCI energy spectra of the $H^{\Omega,\delta}_{\rm LLL}$ Hamiltonian for $N=2$ and $N=3$
spinful fermions, as a function of the total angular momentum $L$, are displayed in Fig.\
\ref{specLLLdelta}. Despite the high degree of complexity involved in these LLL spectra,
Ref.\ \cite{yann25} succeeded in providing exact algebraic expressions for the full
variety of all the displayed states. Given the attention of the current experimental
investigations on assemblies of a few ultracold atoms \cite{grus23,joch24}, the complexity
uncovered theoretically in Ref.\ \cite{yann25} for the $N=3$ $^6$Li atoms, compared to the 
experimentally studied $N=2$ case \cite{grus23,joch24}, identifies the former (i.e., the $N=3$
fermionic system) as a foremost future experimental target.

\subsubsection{Derivation of RWM analytic wave functions for two spinful fermions or
two scalar bosons using the symmetry-restoration approach.}
\label{anan2li6}

To this effect, Ref.\ \cite{yann25} expanded upon similar-in-spirit two-step
derivations regarding the case of any-$N$ fully spin-polarized electrons
\cite{yann02.2} or of scalar bosons \cite{yann10}. The starting step of such derivations
involves a Slater determinant that describes through displaced Gaussians an {\it azimuthally
pinned\/} WM (referred to also as static WM). Then the second step restores the broken
circular symmetry with the help of projection techniques \cite{yann07}; see Sections \ref{2eqd}
and \ref{qdrr} for the cases of a few electrons in QDs in the absence of a magnetic field.

Specifically, according to the two-step method (introduced in Sec.\ \ref{hieras}):

(I) At the {\it first step\/}, the broken-symmetry wave function of a fermionic $N=2$
{\it azimuthally pinned\/} WM can be written as
\begin{equation}
\Psi^{\rm PWM}_{\pm}(z_1,z_2) (\alpha(1)\beta(2) \mp \alpha(2)\beta(1))
\label{pinnwm}
\end{equation}
where $\alpha$, $\beta$ denote a spin up and a spin down, respectively, and the space
part is given by symmetric and antisymmetric combinations
\begin{equation}
\Psi^{\rm PWM}_{\pm}(z_1,z_2)=u(z_1,Z_1)u(z_2,Z_2) \pm u(z_2,Z_1)u(z_1,Z_2), 
\label{psirot}
\end{equation}
with
\begin{equation}
u(z,Z_j) = \frac{1}{\sqrt{\pi}} 
 \exp[-|z-Z_j|^2/2] \exp[-i (xY_j-yX_j)],
\label{gaus}
\end{equation}
being a displaced Gaussian function in the LLL localized at the position
$Z_j$; $z_j=x_j+iy_j$ and $Z_j = X_j+iY_j = R e^{i\phi_j}$; lengths are in units of $\Lambda$.
To proceed further, I impose $\phi_1=0$ and $\phi_2=\pi$, which yields an antipodal
configuration for the two {\it pinned\/} fermionic particles. Unlike the $B=0$ case of the
displaced Gaussian in Eq.\ (\ref{uhfo1}), the phase factor in Eq.\ (\ref{gaus}) is caused by
the gauge invariance associated with the rotation of the trap, in accordance with the
equivalence between $B$ and $\Omega$. For two fermions, $\Psi_{+}$ describes the space
part of a total-spin singlet state and $\Psi_{-}$ that of a triplet state. For two
scalar bosons, only the symmetric space part, $\Psi_{+}$, needs to be considered. 

The localized orbitals $u(z,Z)$ reside fully within the LLL. Indeed, they can be expanded in
an infinite series over the complete set of nodeless Darwin-Fock (see Sec.\ \ref{dfrot})
single-particle wave functions
\begin{equation}
\psi_{l_i}(z) = \frac{ z^{l_i} } { \sqrt{ \pi {l_i}!} } \exp(-zz^*/2),
\label{psilll}
\end{equation}
with $l_i \geq 0$. One obtains (see Appendix A in Ref.\ \cite{yann06.3})
\begin{equation}
u(z,Z)=\sum_{l=0}^{\infty} C_l(Z) \psi_l(z),
\label{uexp}
\end{equation}
with 
\begin{equation}
C_l(Z)=(Z^*)^l \exp(-ZZ^*/2)/\sqrt{l!}
\label{clz}
\end{equation} 
for $Z \neq 0$. Naturally, $C_0(0)=1$ and $C_{l>0}(0)=0$.
Then the following expansion (within a proportionality constant) is obtained
\begin{equation}
\Psi^{\rm PWM}_{\pm}(z_1,z_2) =  
 e^{-R^2} \sum_{l_1=0,l_2=0}^{\infty}
\frac{ (-)^{l_2}R^{l_1+l_2} } {l_1! l_2!}
 (z_1^{l_1} z_2^{l_2} \pm  z_2^{l_1} z_1^{l_2}) |0\rangle.
\label{exppsi}
\end{equation}

In Eq.\ (\ref{exppsi}), the common factor $|0\rangle$ represents the product
$\psi_0(z_1)\psi_0(z_2)$ of Gaussians functions given in Eq.\ (\ref{psilll}). This trivial
factor will be left out from most of the algebra below.

(II) {\it Second step:\/}
The azimuthally pinned wave functions $\Psi^{\rm PWM}_\pm (z_1,z_2)$ do break the rotational
symmetry and thus they do not have good total-angular-momentum,
$\hbar \hat{L}=\hbar \sum_{j=1}^2 \hat{l}_j$, quantum numbers.
However, one can restore \cite{yann02.2,yann02.3,yann04,yann07,shei21} the rotational symmetry
by applying onto $\Psi^{\rm PWM}_\pm (z_1,z_2)$ the projection operator ${\cal P}_L$ specified
in Eq.\ (\ref{amp}). 

When applied onto $\Psi^{\rm PWM}_\pm (z_1,z_2)$, the projection operator ${\cal P}_L$
performs as a Kronecker delta: from the infinite sum in Eq.\ (\ref{exppsi}), it retrieves
only the finite number of terms having a given total angular momentum $L$ (the prefactor
$\hbar$ is hereby overlooked in connection to angular momenta).
The spatial part of the $L$-preserving RWM wave function,
$\Psi^{\rm RWM}_{N=2}(L) ={\cal P}_L \Psi^{\rm PWM}_\pm$
(see Refs.\ \cite{yann02.2,yann02.3,yann04,yann07}), is
found to be (within a proportionality constant)
\begin{equation}
\Psi^{\rm RWM}_{N=2}(L) \propto (z_1-z_2)^L,
\label{rwm2}  
\end{equation}
where an {\it even\/} $L$ is associated with a fermionic spin-singlet or a bosonic scalar
state, while an {\it odd\/} $L$ is associated with a fermionic spin-triplet state.

In addition to being rotationally invariant, the states (\ref{rwm2}) are translationally
invariant (TI), as well. Furthermore, they are also zero-interaction
energy (0IE) states \cite{yann21} and thus they lie along the $x$-axis in the FCI spectra
shown in Fig.\ \ref{specLLLdelta}(a). The states (\ref{rwm2}) represent 
only a small part of the spectrum. Indeed, there are additional 0IE states [see the
degeneracies (indicated as ``x$n$'') in Fig.\ \ref{specLLLdelta}(a)], as well as excitations
with non-zero energy. These additional states can be investigated with the help of the FCI
solutions (see below). As an example, one finds that the second 0IE state at $L=2$ (a
triplet with $S_z=0$) is $\propto (z_1^2-z_2^2)$. (For the remaining states in the
$N=2$ spectrum, see Appendix B in Ref.\ \cite{yann25}.)

\begin{figure}[t]
\centering\includegraphics[width=1.0\textwidth]{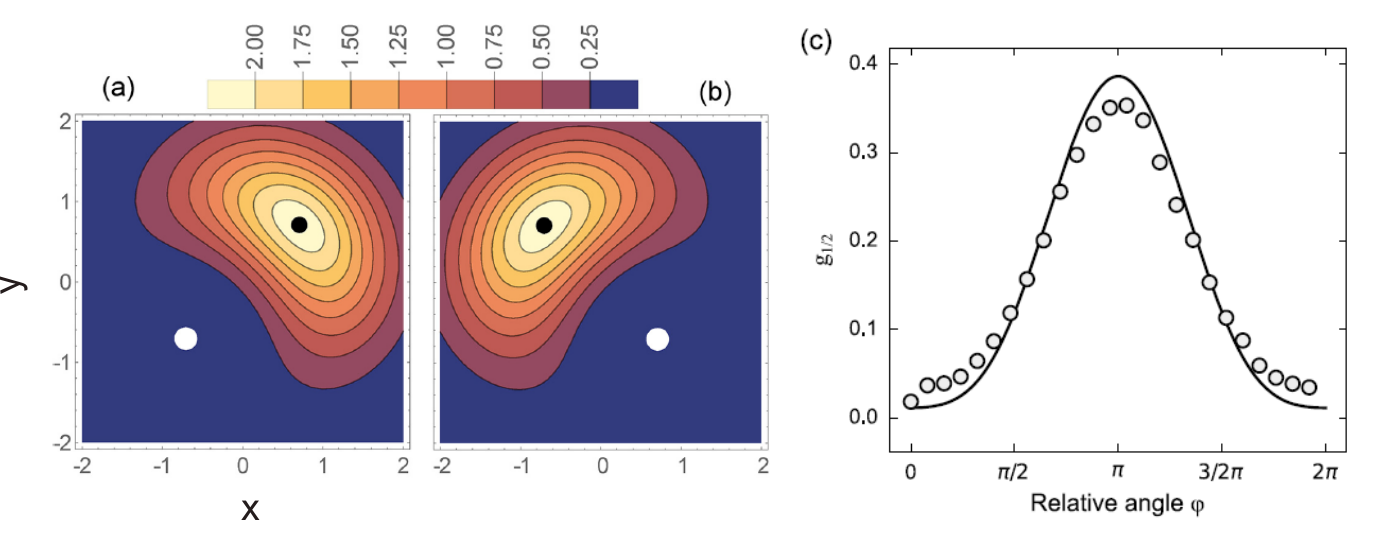}
\caption{
(a) and (b) The two-body correlation distribution ${\cal P} (z0_1; x,y)$ [see Eq.\
(\ref{p2l2n1})] corresponding to the singlet 0IE state [see Eq.\ (\ref{rwm2})] of $N=2$
LLL fermions having $L=2$. The positions of the fixed fermion, spotlighted by white
solid dots, are $z0_1=R e^{ -3\pi i/4}$ in (a) and $z0_1=R e^{-\pi i/4}$ in (b), with $R=1 \Lambda$.
A black solid dot marks the maximum of the distribution. The two dots in each panel are
antipodal. Lengths in units of $\Lambda$. ${\cal P} (z0_1; x,y)$ in units of $\Lambda^4$.
Reprinted figures with permission from Ref.\ \cite{yann25},
Copyright (2025) by the American Physical Society.
(c) Experimental normalized histogram (empty circles) of relative angle second-order
correlations between the spin-up and spin-down fermion ($^6$Li). The solid line is the
theoretical azimuthal-angle correlation function $g_{1/2}(\varphi)$, see text in Ref.\
\cite{joch24}. Error bars of the 95\% confidence interval, determined using a bootstrapping
technique, are smaller than the circle size if not visible.
Reprinted figure with permission from Ref.\ \cite{joch24},
Copyright (2024) by the American Physical Society.
}
\label{2bodycsm}
\end{figure}

Although the SPDs associated with the RWM wave functions are circularly symmetric,
one can uncover the underlying molecular configuration of an RWM by plotting
higher-order correlation distributions \cite{yann00,yann04,yann07,yann07.3}. As a simple
example for $N=2$, I consider here the (unnormalized) spin-unresolved CPDs (second-order
correlations) of the state with $L=2$ given by Eq.\ (\ref{rwm2}), namely
\begin{equation}
  {\cal P} (z0_1; x,y) =  (z_1-z_2)^2 (z^*_1-z^*_2)^2 e^{-\sum_{i=1}^2 z_i z_i^*},
\label{p2l2n1}
\end{equation}
where one fixes one particle at a point $z_1=z0_1$ and inquires about the position,
$z_2=x+i y$, of the second fermion; the asterisk denotes complex conjugation.

Figs.\ \ref{2bodycsm}(a) and \ref{2bodycsm}(b) display the ${\cal P} (z0_1; x,y)$
correlation distribution described above
for two different fixed points, highlighted by white solid dots, at $z0_1=R e^{ -3\pi i/4}$ and
$z0_1=R e^{-\pi i/4}$, respectively, with $R=1.0 \Lambda$. One sees that the black solid dots,
marking the maximum probability for finding the second fermion, are antipodal to the white dots,
corroborating thus the physical picture of a rotating Wigner molecule.

Note further that the states (\ref{rwm2}) coincide with the Jastrow-type ones, which are
customarily referred to \cite{joch24,grus24} as the topological Halperin/Laughlin states
\cite{halp83,laug83}, and which are interpreted as ``special quantum liquids.'' However,
most recent experimental work \cite{joch24} has explicitly measured [through the
correlations in the relative angle $\varphi$, see Fig.\ \ref{2bodycsm}(c)] the antipodal
two-body quantum correlations embodied in the expression (\ref{rwm2}) in agreement with
the theoretical CPDs in Figs.\ \ref{2bodycsm}(a) and (b), confirming thus the RWM
underlying physics.

\subsubsection{Derivation of RWM analytic wave functions for three spinful fermions using
the symmetry-restoration approach.}
\label{anan3li6}

The three spin eigenfunctions $\chi(S,S_z=1/2)$ associated with three fermions in an
equilateral triangular configuration are given by Eqs.\ (\ref{wf3e12121c}),
(\ref{wf3e12122c}), and (\ref{wf3e3212}) in Sec.\ \ref{3epli}. However, for the purpose
of this Section, it is advantageous to rewrite them as follows:
\begin{eqnarray}
& \chi(1/2,1/2;1)=(\cz_{123} + e^{2\pi i/3}\cz_{231} + e^{-2\pi i/3}\cz_{312})/\sqrt{3} \nonumber \\
& \chi(1/2,1/2;2)=(\cz_{123} + e^{-2\pi i/3}\cz_{231} + e^{2\pi i/3}\cz_{312})/\sqrt{3} \nonumber \\
& \chi(3/2,1/2)=(\cz_{123} + \cz_{231} + \cz_{312})/\sqrt{3},
\label{chi3}
\end{eqnarray}
where the spin primitives are
\begin{equation}
\cz_{ijk}=\alpha(i)\alpha(j)\beta(k),
\end{equation}
and the $\alpha$ and $\beta$ denote, as usually, up and down spins,

Straightforward wave functions, $\Phi^{\rm PWM}_{N=3}$, for an {\it azimuthally pinned\/}
three-fermion WM are constructed by using displaced Gaussians $u(z,Z_j)$ [see Eq.\ (\ref{gaus})]
centered at the vertices $Z_j=R e^{2 j \pi i/3}$, $j=0,1,2$ of an equilateral triangle, and
then replacing the spin primitives $\cz_{ijk}$ in the three $\chi$'s in Eq.\ (\ref{chi3})
by the corresponding Slater determinants
\begin{equation}
   {\cal D}_{ijk} = 
   {\rm Det}[ u(z_i,Z_1)\alpha(i), u(z_j,Z_2)\alpha(j), u(z_k,Z_3)\beta(k) ]/ \sqrt{6},
\label{dijk}
\end{equation}
where the Slater determinants are denoted through a listing of their diagonal elements.

\begin{figure}[t]
\centering\includegraphics[width=0.7\textwidth]{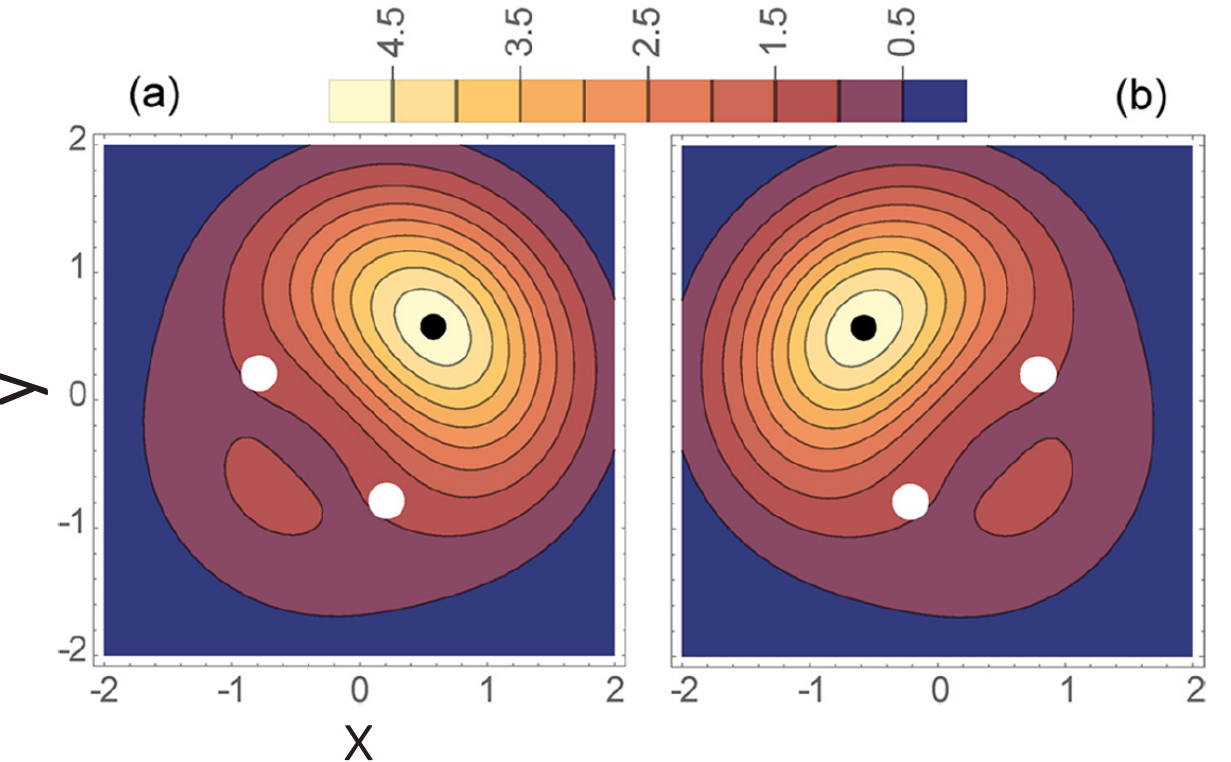}
\caption{
The three-particle correlation distribution ${\cal P} (z0_1,z0_2; x,y)$
[see Eq.\ (\ref{p3l2n1})] associated with the TI ground state [see Eq.\ (\ref{rwm3l2n1})]
of $N=3$ LLL fermions having angular momentum $L=2$ and spin ($S=S_z=1/2$). The positions
of the two fixed fermions, spotlighted by solid white dots, were chosen as follows: (a)
($z0_1=R e^{ 11\pi i/12}$, $z0_2=R e^{- 5\pi i/12}$) and (b) ($z0_1=R e^{17\pi i/12}$,
$z0_2=R e^{\pi i/12}$), with $R=0.816 \Lambda$. In each panel, a black solid dot marks
the maximum of the distribution. The three dots in each panel are located at the
vertices of an equilateral triangle. Lengths in units of
$\Lambda$. ${\cal P} (z0_1,z0_2; x,y)$ in units of $\Lambda^4$.
Reprinted figures with permission from Ref.\ \cite{yann25},
Copyright (2025) by the American Physical Society.
}
\label{3bodycsm}
\end{figure}

To obtain the {\it rotating\/} WM wave functions, $\Phi^{\rm RWM}_{ijk}(L)$, one
subsequently expands the Slater determinants, as well as the displaced Gaussians
according to Eqs.\ (\ref{uexp}) and (\ref{clz}), and then applies the operator
${\cal P}_L$ [Eq.\ (\ref{amp})] and carries out the angular momentum projection 
by extracting the coefficients of the powers $R^{l_1+l_2+l_3}$ for a given
$L=l_1+l_2+l_3$. Using symbolic-algebra scripts (e.g., MATHEMATICA \cite{math22})
to sum the terms in these coefficients, one can derive the following compact formula:
\begin{equation}
\Phi^{\rm RWM}_{N=3}(L)=\sum_{ijk} \Psi^{\rm RWM}_{ijk}(L) \cz_{ijk}.
\label{rwm3}  
\end{equation}
The symbolic summation index ``$ijk$'' in Eq.\ (\ref{rwm3}) runs over the three cyclic
permutations \{1,2,3\}, \{2,3,1\}, and \{3,1,2\}, whereas the space parts are given by
\begin{equation}
\Psi^{\rm RWM}_{ijk}(L) \propto i[(z_{i+j-k} - i z_{i-j})^L - (z_{i+j-k} + i z_{i-j})^L],
\label{psi3}
\end{equation}
with $z_{i+j-k}=\sqrt{2/3}\big((z_i+z_j)/2-z_k\big)$ and $z_{i-j}=(z_i-z_j)/\sqrt{2}$ being
three-particle Jacobi coordinates.

Simple examples for the $N=3$ case are given by the state for $L=1$ and the two states for
$L=2$, whose FCI energies are displayed in Fig.\ \ref{specLLLdelta}(b). From the FCI analysis
(presented in detail below), one finds that the state with $S=S_z=1/2$ and $L=1$
equals 
\begin{equation}
\Phi^{\rm RWM}_{N=3}(L=1) \propto \sum_{ijk} (z_i-z_j) \cz_{ijk}.
\label{rwm3l1}
\end{equation}

Furthermore, from the FCI analysis (see below again), one also finds that the ground state
(with $S=S_z=1/2$ and $L=2$) is equal to
\begin{equation}
\Phi^{\rm RWM}_{N=3} (L=2) \propto   \sum_{ijk} \Psi^{\rm RWM}_{ijk} (L=2) \cz_{ijk},
\label{rwm3l2n1}
\end{equation}
where the space part is given by [see Eq.\ (\ref{psi3})]
\begin{equation}
\Psi^{\rm RWM}_{ijk} (L=2)=(z_i-z_j) (z_i+z_j-2 z_k).
\label{psi3l2n1}
\end{equation}

At this point, it is instructive to investigate the spin-unresolved three-body
correlations\footnote{Compared to the CPD, this is the next higher-order correlation
function.}
of $\Phi^{\rm RWM}_{N=3}(L=2)$, which are given by
\begin{equation}
{\cal P} (z0_1,z0_2; x,y) =  
\sum_{ijk} \Psi^{\rm RWM}_{ijk}(L=2) \Psi^{\rm RWM*}_{ijk}(L=2) e^{-\sum_{i=1}^3 z_i z_i^*}.
\label{p3l2n1}  
\end{equation}
In ${\cal P} (z0_1,z0_2; x,y)$, one fixes two fermions at points $z0_1$ and $z0_2$ and
inquires about the position ($z_3=x+i y$) of the third fermion.

Fig.\ \ref{3bodycsm} displays the ${\cal P} (z0_1,z0_2; x,y)$ distribution for two different
pairs of fixed fermions, highlighted by white solid dots, placed at (a)
($z0_1=R e^{ 11\pi i/12}$, $z0_2=R e^{- 5\pi i/12}$) and (b) ($z0_1=R e^{17\pi i/12}$,
$z0_2=R e^{\pi i/12}$). It is immediately recognized that the black solid dots, which
designate the location of maximum probability for finding the third fermion, coordinate with
the white dots to form equilateral triangles when $R=0.816 \Lambda$, thus confirming again the
underlying physics of a rotating Wigner molecule. (In this context, recall
that the corresponding charge density is circularly symmetric.) 

Regarding the excited state for $N=3$ with $S=S_z=1/2$ and $L=2$, the FCI analysis
shows that it equals
\begin{equation}
 \sum_{ijk} (z_i-z_j)(z_1+z_2+z_3) \cz_{ijk} \propto 
 \Phi^{\rm RWM}_{N=3} (L=1) z_{\rm c.m.}^{N=3},
\label{rwm3l2n2}
\end{equation}
where $z_{\rm c.m.}=\sum_{i=1}^N z_i/N$ denotes the center of mass.

Note that the factor $z_{\rm c.m.}$ is a structural earmark of the states on any horizontal
dashed line in the LLL spectrum. Indeed, each integer polynomial associated with these states
at a given $L$ [see Fig.\ \ref{specLLLdelta}(b)] equals that of the previous state at $L-1$
times the factor $z_{\rm c..m.}$, apart from the original states that are translationally invariant
(marked by a red arrow). This factor $z_{\rm c.m.}$ represents center-of-mass vibrations
which in the LLL case degrade to simple translations. 

\begin{table}[t]
\caption{\label{tn2l5n1}
FCI numerical coefficients, $c_{\rm CI}(I)$, in the FCI expansion of the 0IE LLL
state, and the equivalent transcribed algebraic ones, $c_{\rm alg}(I)$,
for $N=2$ fermions with $L=5$ and ($S=1$, $S_z=0$).
The Slater determinants ${\cal D}_I$ are identified through the set of single-particle
angular momenta $(l_1,l_2)$ associated with up ($l_1$) and down ($l_2$) spins.
Reprinted table with permission from Ref.\ \cite{yann25},
Copyright (2025) by the American Physical Society.
}
\begin{tabular}{rcrc}
$I$ & $c_{\rm CI} (I)$ & $c_{\rm alg} (I)$ &
$(l_1\uparrow,l_2\downarrow)$  \\ \hline
1 & 0.176777   & $ \sqrt{1/32}   $ & (0,5) \\
2 & -0.395285  & $ -\sqrt{5/32}  $ & (1,4) \\
3 & 0.559017   & $ \sqrt{10/32}  $ & (2,3) \\
4 & -0.559017  & $ -\sqrt{10/32} $ & (3,2) \\
5 & 0.395285   & $  \sqrt{5/32}  $ & (4,1) \\
6 & -0.176777  & $ -\sqrt{1/32}  $ & (5,0) \\
\end{tabular}
\end{table}
\begin{table}[b]
\caption{\label{tn3l2n1}
FCI numerical coefficients, $c_{\rm CI}(I)$, in the CI expansion of the relative
LLL {\it ground\/} state, and the equivalent transcribed algebraic ones,
$c_{\rm alg}(I)$, for $N=3$ fermions with $L=2$ and $S=S_z=1/2$.
The spinful-fermion Slater determinants ${\cal D}_I$ are identified through the set
of single-particle angular momenta and spins, $(l_1\uparrow,l_2\uparrow,l_3\downarrow)$.
Reprinted table with permission from Ref.\ \cite{yann25},
Copyright (2025) by the American Physical Society.
}
\begin{tabular}{rcrc}
$I$ & $c_{\rm CI} (I)$ & $c_{\rm alg} (I)$ &
$(l_1\uparrow,l_2\uparrow,l_3\downarrow)$  \\ \hline
1 & -0.816496  & $ -\sqrt{2/3}  $ & (0,1,1) \\
2 &  0.577350  & $  1/\sqrt{3}  $ & (0,2,0) \\
\end{tabular}
\end{table}

\subsubsection{Extracting the analytic wave functions for a few spinful fermions
from the FCI numerical computations.}
\label{fcian}

The exact FCI wave functions are represented by the expression
\begin{equation}
\Phi^{\rm CI} (z_1\sigma_1, \ldots , z_N\sigma_N) =
\sum_I c_{\rm CI}(I) {\cal D}_I(z_1\sigma_1, \ldots , z_N\sigma_N),
\label{phici}
\end{equation}
with the basis Slater determinants that span the Hilbert space being 
\begin{equation}
{\cal D}_I = {\rm Det}({\cal S})/ \sqrt{N!}, 
\label{detexd}
\end{equation}
where the elements of the matrix ${\cal S}$ are 
${\cal S}_{s,r}=\psi_{l_r}(z_s)\sigma_{l_r}(s)$, $r,s=1,\ldots,N$, with the LLL
single-particle space orbitals being given by Eq.\ (\ref{psilll}); they are specified
by the angular momentum index $l_r$, whereas $\sigma$ can be either an up ($\alpha$) or a down
($\beta$) spin. The master capital index $I$ counts the number of ordered arrangements (lists)
$\{j_1,j_2,\ldots,j_N\}$ obeying the rule that $1 \leq j_1 < j_2 <\ldots < j_N \leq K$;
$K \in \mathbb{N}$. The pertinent point here is that the FCI coefficients $c_{\rm CI}(I)$
are numerical. 

As a next step, one recasts the FCI wave functions $\Phi^{\rm CI}$ in Eq.\ (\ref{phici}) as 
\begin{equation}
\Phi^{\rm CI}_{\rm alg} (z_1\sigma_1, \ldots , z_N\sigma_N) =
\sum_I c_{\rm alg}(I) {\cal D}_I(z_1\sigma_1, \ldots , z_N\sigma_N),
\label{phialg1}
\end{equation}
where the introduction of the subscript ``alg'' reflects the fact that, using a symbolic
computer language code, one can obtain algebraic coefficients $c_{\rm alg}$ by performing
a transcription $c_{\rm CI}(I) \rightarrow c_{\rm alg}(I)$.
In this way, the FCI computer-assisted calculations, processed subsequently in Ref.\
\cite{yann25} through the use of symbolic-language scripts, led up to specific analytical
algebraic many-body wave-functions, $\Phi^{\rm CI}_{\rm alg}$, that are fully equivalent to
the corresponding numerical FCI results.

\begin{table}[t]
\caption{\label{tn3l2n2}
FCI numerical coefficients, $c_{\rm CI}(I)$, in the CI expansion of the LLL
{\it excited\/} state, and the equivalent transcribed algebraic ones,
$c_{\rm alg}(I)$, for $N=3$ fermions with $L=2$ and $S=S_z=1/2$. The spinful-fermion
Slater determinants ${\cal D}_I$ are identified through the set of single-particle
angular momenta and spins, $(l_1\uparrow,l_2\uparrow,l_3\downarrow)$.
Reprinted table with permission from Ref.\ \cite{yann25},
Copyright (2025) by the American Physical Society.
}
\begin{tabular}{rcrc}
$I$ & $c_{\rm CI} (I)$ & $c_{\rm alg} (I)$ &
$(l_1\uparrow,l_2\uparrow,l_3\downarrow)$  \\ \hline
1 & 0.577350 & $ 1/ \sqrt{3}  $ & (0,1,1) \\
2 & 0.816496 & $  \sqrt{2/3}  $ & (0,2,0) \\
\end{tabular}
\end{table}

Examples\footnote{
For a full exposition of more complex cases associated with the spectra in
Fig.\ \ref{specLLLdelta}, see Ref.\ \cite{yann25}.}
of such transcriptions of coefficients are given in Tables \ref{tn2l5n1},
\ref{tn3l2n1} and \ref{tn3l2n2} corresponding to the first 0IE two-fermion state
$\Phi^{\rm CI}_{N=2}(L=5;\;1)$ with $L=5$, as well as to the two three-fermion states
$\Phi^{\rm CI}_{N=3}(L=2;\;1)$ and $\Phi^{\rm CI}_{N=3}(L=2;\;2)$ with $L=2$ [see Fig.\
\ref{specLLLdelta}(b)]. It needs to be stressed again that the compact polynomial
expressions produced through the expansions over Slater determinants, given by
$\Phi^{\rm CI}_{{N=2},{\rm alg}}(L=5;\;1)$, $\Phi^{\rm CI}_{{N=3},{\rm alg}}(L=2;\;1)$, and
$\Phi^{\rm CI}_{{N=3},{\rm alg}}(L=2;\;2)$ according to Eq.\ (\ref{phialg1}), coincide
with the RWM-based polynomial expressions in Eqs.\ (\ref{rwm2}), (\ref{rwm3l2n1}),
and (\ref{rwm3l2n2}), respectively.

\subsection{Fully spin polarized electrons in the LLL.}
\label{lllfp}  
\smallskip

This section reviews the underlying physics of Wigner molecularization in the LLL
for the case of {\it spin polarized\/} electrons under a high magnetic field
($\zeta_B \rightarrow \infty$, see Sec.\ \ref{dfb}). (For the LLL case of scalar bosonic
ultracold atoms in rotating traps ($\zeta_\Omega =1$, Sec.\ \ref{dfrot}), confer
Ref.\ \cite{yann07.3}.)

\subsubsection{RWM analytic wave functions for a few spin-polarized electrons in the LLL.}
\label{rwmanlllfp} 

The derivation of RWM analytic wave functions for any $N$ spin-polarized electrons was
first reported in Refs.\ \cite{yann02.3,yann03,yann04}. This derivation is similar to the
two-step method described in Secs.\ \ref{anan2li6} and \ref{anan3li6} for the two and three
spinful-fermion cases. It is, however, simpler because only the space part of the wave
function needs to be considered.

At the first step, which describes the azimuthally pinned WM, this space part is a single
Slater determinant built out from the space orbitals in Eq.\ (\ref{gaus}), but with lengths
in units of $\sqrt{2}l_B$, where the magnetic length $l_B=\sqrt{\hbar c/(eB)}$.
At the next step, the required rotational symmetry is restored by applying a generalization
of the angular-momentum projection operator in Eq.\ (\ref{amp}) which takes into
consideration the nested-polygonal-ring arrangements $(n_1,n_2,\ldots,n_q)$,
$\sum_{i=1}^q n_i=N$, associated with the classical ground-states, as well as their excited
isomers (see, e.g., Ref.\ \cite{kong02}).

The detailed execution of the two steps outlined in the previous paragraph, and the
ensuing analytic expressions, can be found in Refs.\
\cite{yann02.3,yann03,yann04,yann10,yann11}.
For the purpose of this review, it is sufficient to give the simpler expressions for
$(0,N)$ (a single regular polygon) and $(1,N-1)$ (one electron at the center), i.e., 
\begin{eqnarray}
\Phi_L^{\rm RWM} (0,N) = &&
\sum^{l_1 + \cdot \cdot \cdot +l_N=L}%
_{0 \leq l_1<l_2< \cdot \cdot \cdot <l_N}
\left( \prod_{i=1}^N l_i! \right)^{-1} 
\times \left( \prod_{1 \leq i < j \leq N} 
\sin \left[\frac{\pi}{N}(l_i-l_j)\right] \right)  \nonumber \\
&& \times \; D(l_1,l_2,...,l_N)
\exp(-\sum_{i=1}^N z_i z_i^*/2),
\label{phi1}
\end{eqnarray} 
with 
\begin{equation}
L_m=L_0+Nm, \;\; m=0,1,2,3,..,
\label{lwm0n}
\end{equation}
and
\begin{eqnarray}
\Phi_L^{\rm RWM} (1,N-1) = && 
\sum^{l_2+ \cdot \cdot \cdot +l_N=L}%
_{1 \leq l_2 < l_3 < \cdot\cdot\cdot < l_N}
\left( \prod_{i=2}^N l_i! \right)^{-1} 
\times \left( \prod_{2 \leq i < j \leq N} 
\sin \left[\frac{\pi}{N-1}(l_i-l_j)\right] \right)  \nonumber \\
 \nonumber \\
&& \times \; D(0,l_2,...,l_N)
\exp(-\sum_{i=1}^N z_i z_i^*/2),
\label{phi2}
\end{eqnarray}
with 
\begin{equation}
L_m=L_0+(N-1)m, \;\; m=0,1,2,3,...,
\label{lwm1nm1}
\end{equation}

In Eqs.\ (\ref{phi1}) and (\ref{phi2}), the determinants, $D(l_1,l_2,...,l_N) \equiv 
{\rm Det}[z_1^{l_1},z_2^{l_2}, \cdot \cdot \cdot, z_N^{l_N}]$, are specified through
their diagonal elements. Moreover, $L_0=N(N-1)/2$ is the minimum allowed total angular
momentum for $N$-polarized electrons in the LLL (case of the maximum density droplet).

Note that the RWM wave functions [Eqs.\ (\ref{phi1}) and (\ref{phi2})] vanish
identically for values of the total angular momenta outside the specific $L_m$ values
given by Eqs.\ (\ref{lwm0n}) and (\ref{lwm1nm1}), respectively. These $L_m$ values,
which are directly connected to the point-group symmetries of the nested-polygonal
configurations, are known as ``magic angular momenta'', because the corresponding FCI
ground-state (yrast-state) energies exhibit enhanced stability compared to those of
neighboring $L$'s [when the linear in $L$ confinement Hamiltonian term is considered;
see Eq.\ (\ref{hblll2}) below]. The magic angular momenta in high $B$ were first
reported in Refs.\ \cite{girv83,maks90}.

\begin{figure}[t]
\centering\includegraphics[width=0.7\textwidth]{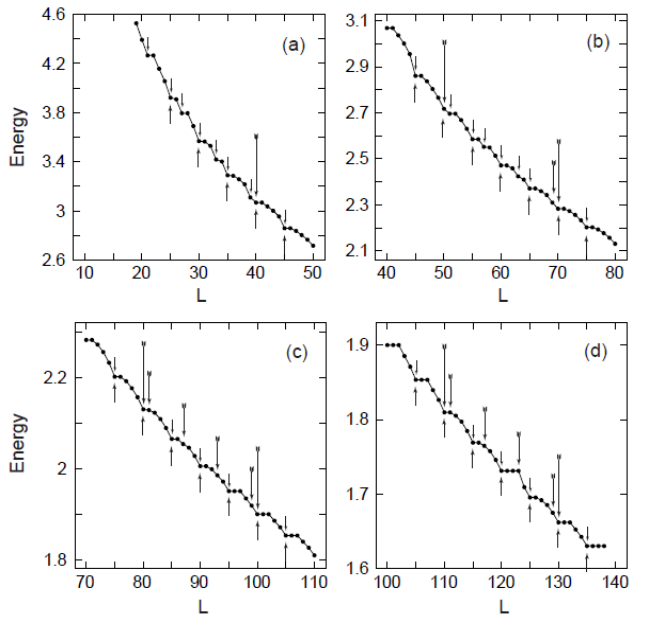}
\caption{
Total interaction energy from FCI calculations as a function of the total
angular momentum $(19 \leq L \leq 140)$ for $N=6$ electrons in the lowest Landau
level [see Hamiltonian (\ref{hblll})].
The upwards pointing arrows indicate the cusp states with magic angular momenta
$L_m=10+5m$, $m=3,4,5,\ldots$, corresponding to the classically most stable (1,5)
intrinsic polygonal-ring arrangement of the rotating Wigner molecule.
The downwards pointing short arrows designate partial successful predictions of the
composite-fermion theory. The medium-size downwards pointing arrows designate false
predictions of the composite-fermion approach that fail to materialize as actual FCI magic
angular momenta. The downward long arrows point to FCI magic angular momenta not predicted
by the composite-fermion approach. The cusp state with angular momentum $L=21$ corresponds
to the (0,6) polygonal-ring isomer.
Energies in units of $e^2/\kappa l_B$, where $\kappa$ is the dielectric constant.
Reprinted figures with permission from Refs.\ \cite{yann03,yann07},
Copyright (2003, 2007) by the American Physical Society.}
\label{n6lllspec}
\end{figure}

\subsubsection{An example of FCI calculations in the LLL: The cases of $N=6$ and $N=7$
fully spin-polarized electrons and comparisons with the RWM, composite-fermion, and
Laughlin wave-function results.}
\label{fcillln7fp} 

Near $\zeta_B \rightarrow \infty$, taking into account the harmonic confinement,
the corresponding $B$-dependemt many-body Hamiltonian (\ref{mbhb}) can be simplified
\cite{yann07.2,yann11,jainbook} to the form
\begin{equation}
H^{B,{\prime}}_{\rm LLL}=N \frac{\hbar \omega_c}{2} +
\hbar(\sqrt{ \omega_0^2+\omega_c^2/4 }-\omega_c/2) L
+\sum_{i=1}^N \sum_{j>i}^N \frac{e^2}{\kappa r_{ij}},
\label{hblll2}
\end{equation}
with $L$ being the total angular momentum. The linear term in $L$ in the Hamiltonian
(\ref{hblll2}) influences only the total energies of the LLL states, but not their
many-body structure, which is determined solely by the interaction term
\begin{equation}
H^B_{\rm LLL} =  \sum_{i=1}^N \sum_{j>i}^N \frac{e^2}{\kappa r_{ij}}.
\label{hblll}
\end{equation}
Indeed the total angular momentum commutes with $H^B_{\rm LLL}$.

Results obtained \cite{yann03,yann04,yann07} with FCI calculations in the LLL for a
few electrons in the disk geometry have revealed discrepancies of the composite-fermion
approach \cite{jainbook} (including the Jastrow-Laughlin (JL) wave function
\cite{laug83,laug99}) in the context of quantum dots under high magnetic fields.

For $N=6$, Fig.\ \ref{n6lllspec} displays (in four frames) the total interaction energy
from FCI computations in the LLL [see the Hamiltonian (\ref{hblll})] as a function of
the total angular momentum $L$ in the range $ 19 \leq L \leq 140 $. One immediately
observes the appearance of cusps, implying states of enhanced stability, at the
magic angular momenta as aforementioned.

An inspection of the total-energy-vs-$L$ plots in Fig.\ \ref{n6lllspec} reveals 
that the CF prediction misses many of the actual magic angular momenta, specified by 
the FCI calculations as those associated with the cusps. Indeed,
the CF theory\footnote{
For an exposition of the details regarding the application of the CF approach
to the six-electron case in the LLL, see Ref.\ \cite{yann03}.}
fails in two aspects: (I) There are FCI magic angular momenta which are
regularly missing from the CF prediction in every interval
${\cal I}_p \leftrightarrow 15(2p-1) \leq L \leq 15(2p+1)$, $p=1,2,3,4$ (associated
with the four panels, respectively). These missing {\it exact\/} magic numbers are
marked by a long downward arrow in the figure. 
(II) There are CF magic numbers that do not correspond to cusps in
the FCI calculations (indicated by the medium-size downward arrows in the figure). 
This reflects the fact that cusps with $L$'s whose difference from $L_0$ 
is divisible by 6 (but not simultaneously by 5) gradually weaken and totally
disappear in the later intervals. The only cusps which survive have the difference
$L-L_0$ divisible by 5, i.e., according to the FCI, the (1,5) intrinsic arrangement
becomes energetically more stable relative to the (0,6) isomer. On the contary, the
CF model predicts the appearance of four magic angular momenta with $L-L_0$ divisible
solely by 6 in every interval ${\cal I}_p$, at $L=30i\mp9$ and $30i\mp3$, $i=1,2,3,\ldots$.
Overall the CF model predicts six false cases (long and medium-size downward arrows)
in every interval ${\cal I}_p$ with $p \geq 3$, compared to only five correct ones [short
downward arrows, see Figs.\ \ref{n6lllspec}(c) and (d)].

\begin{figure}[t]
\centering\includegraphics[width=0.9\textwidth]{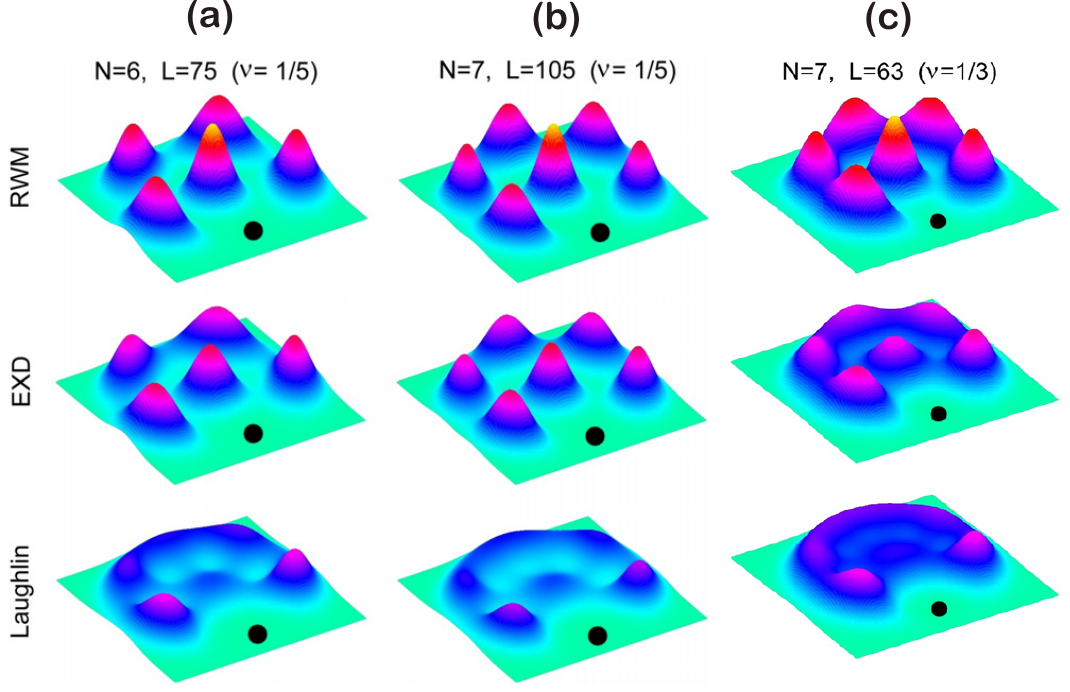}
\caption{
Conditional probability distributions in the LLL for (a) $N=6$ electrons and 
$L=75$ ($\nu=1/5$, left column), (b) $N=7$ electrons and $L=105$ (again 
$\nu=1/5$, middle column), and (c) $N=7$ electrons and $L=63$ ($\nu=1/3$,
right column). Top row: RWM case. Middle row: The case of FCI (indicated as EXD
on the figure). Bottom row: The Jastrow-Laughlin case.
The FCI and RWM wave functions have a pronouned crystalline
character, corresponding to the (1,5) polygonal configuration of the RWM for 
$N=6$, and to the (1,6) polygonal configuration for $N=7$. In contrast, the 
JL wave functions exhibit a characteristic liquid profile that 
depends smoothly on the number $N$ of electrons. The observation (fixed) point 
(identified by a solid dot) is located at $r_0 = 5.431 l_B$ in (a), $r_0=5.883 l_B$
in (b), and $r_0=4.568 l_B$ in (c). Lengths in units of the magnetic length $l_B$.
The CPDs are given in arbitrary units, which are the same for all three panels in a
given column.
Reprinted figures with permission from Refs.\ \cite{yann04,yann07},
Copyright (2004, 2007) by the American Physical Society.
}.
\label{ellllcpds}
\end{figure} 

\begin{figure}[t]
\centering\includegraphics[width=0.5\textwidth]{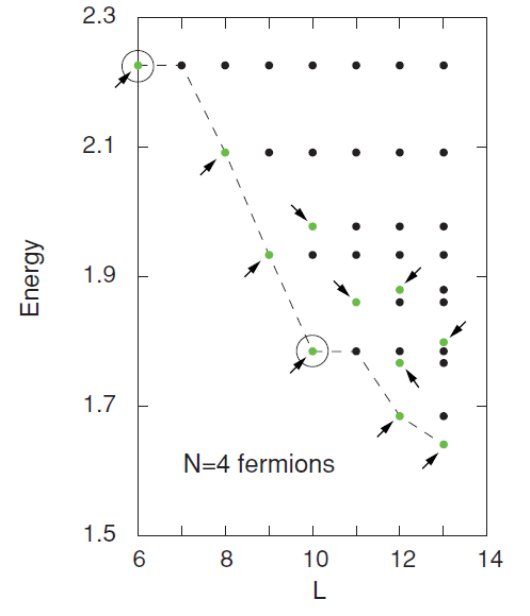}
\caption{
FCI LLL spectra for $N = 4$ spin-polarized electrons calculated using
the Hamiltonian (\ref{hblll}). The green solid dots (marked by arrows)
denote the translationally invariant states. The dark solid dots are the
spurious states (see text). The dashed line denotes the yrast band, while the
cusp states are marked by a circle. Energies in units of $e^2/(\kappa l_B)$.
The number of translationally invariant states is much smaller than the total
number of LLL states.
Reprinted figure with permission from Ref.\ \cite{yann10},
Copyright (2010) by the American Physical Society.
}
\label{n4ellllspec}
\end{figure}

To illustrate the differences between the intrinsic structure of the RWM and
FCI states in the lowest Landau level versus the familiar Jastrow-Laughlin 
ones (which belong also to the composite-fermion family for fractions $1/(2m+1)$,
$m \in \mathbb{N}$), Fig.\ \ref{ellllcpds} displays the CPDs for cusp states
corresponding to two filling factors and for two different sizes, i.e.,
for $N=6$ electrons ($L=75$, $\nu=1/5$, left column) and $N=7$ electrons ($L=105$,
$\nu=1/5$, middle column and $L=63$, $\nu=1/3$, right column). 
In Fig.\ \ref{ellllcpds}, the top row depicts the RWM case; the FCI case is given
by the middle row, while the JL wave functions are displayed in the
bottom row. Note that for a finite number of electrons, $N$, the fractional filling
is connected to the filling factor through $\nu=L_0/L$ \cite{laug83,girv83,yann04,yann07}.

The following conclusions can be drawn from an inspection of Fig.\
\ref{ellllcpds}:
(I) The intrinsic character of the FCI states is unmistakably crystalline with the
corresponding CPDs exhibiting a well developed molecular polygonal configuration [(1,5)
for $N=6$ and (1,6) for $N=7$, with one electron at the center], in agreement with 
the RWM cases. 
(II) For the two displayed fractional fillings 1/5 and 1/3 (and others discussed in
Ref.\ \cite{yann04}), the Jastrow-Laughlin wave functions fail to capture the intrinsic
crystallinity of the FCI states. In contrast, they represent ``liquid-like'' states in
agreement with an analysis that goes back to the original papers \cite{laug83,pranbook}
by Laughlin.

\begin{table}[t] 
\caption{\label{ene_ferm_np4}%
LLL energies of the TI states of four spin-polarized electrons interacting via the
Coulomb repulsion $e^2/(\kappa |z_i-z_j|)$ . Second column: Dimensions of the FCI
($D^{\rm FCI}$) and the nonspurious TI ($D^{\rm TI}$, in parenthesis) Hilbert spaces.
(Note that the FCI space is spanned by uncorrelated determinants of nodeless
Darwin-Fock orbitals.) Last three columns: Energy eigenvalues [in units of
$e^2/(\kappa l_B)$] from the diagonalization of the Coulomb interaction in the TI
subspace spanned by the analytic wave functions $\Phi^{\rm RWM}_{6+4k}(0,4) Q_\lambda^m$
and $\Phi^{\rm RWM}_{6+3k}(1,3) Q_\lambda^m$ (RVWM diagonalization).
Third to sixth columns: the molecular configurations $(n_1,n_2)$ and the
quantum numbers $k$, $\lambda$ and $m$ are indicated within brackets.
There is no nonspurious state with $L=7$. As another option to the FCI diagonalization,
the full FCI spectrum at a given $L$ can be constructed by including, in addition to
the listed TI energy eigenvalues [$D^{\rm TI}(L)$ in number], all the energies associated
with angular momenta smaller than $L$. An integer in square brackets indicates the
energy ordering in the FCI spectrum (including both spurious and TI states), with [1]
denoting an yrast state. Eight decimal digits are displayed, but the energy eigenvalues
from the RVWM diagonalization agree with the corresponding FCI ones within machine
precision.
Reprinted table with permission from Ref.\ \cite{yann10},
Copyright (2010) by the American Physical Society.
}
\begin{tabular}{llllll}
$L$ & $D^{\rm FCI}$($D^{\rm TI}$)& $[(n_1,n_2)\{ k, \lambda, m \}]$ &
\multicolumn{3}{l}{Energy eigenvalues (RVWM diag. or FCI)} \\
\hline
6 & 1(1) & [(0,4)\{0,$\lambda$,0\}] &
2.22725097[1]  &   ~~~~      & ~~~~~ \\
8 & 2(1) & [(0,4)\{0,2,1\}] &
2.09240211[1] &   ~~~~      & ~~~~~~ \\
9 & 3(1) & [(1,3)\{1,$\lambda$,0\}] &
 1.93480798[1] &   ~~~~      & ~~~~ \\
10 & 5(2) & [(0,4)\{1,$\lambda$,0\}] [(0,4)\{0,2,2\}] &
 1.78508849[1]  &  1.97809256[3] & ~~~~ \\
11 & 6(1) & [(1,3)\{1,2,1\}] & 1.86157215[2]  &
  ~~~~      & ~~~~~ \\
12 & 9(3) &  [(0,4)\{1,2,1\}] [(0,4)\{0,2,3\}] [(1,3)\{2,$\lambda$,0\}] &
 1.68518201[1] & 1.76757420[2]  & 1.88068652[5] \\
13 & 11(2) & [(1,3)\{1,2,2\}] [(0,4)\{1,3,1\}]  & 1.64156849[1]  &
 1.79962234[5] & ~~~~ \\
14 & 15(4) & [(0,4)\{2,$\lambda$,0\}] [(0,4)\{1,2,2\}] [(0,4)\{0,2,4\}] &
 1.50065835[1]  & 1.63572496[2]   &  1.72910626[5]  \\
~~~ & ~~~~ & [(1,3)\{2,2,1\}] &
 1.79894008[8] & ~~~~~   & ~~~~~~ \\
15 & 18(3) & [(1,3)\{3,$\lambda$,0\}] [(1,3)\{2,3,1\}] [(1,3)\{1,3,2\}]  &
 1.52704695[2] & 1.62342533[3] & 1.74810279[8] \\
18 & 34(7) & [(0,4)\{3,$\lambda$,0\}] [(0,4)\{2,2,2\}] [(0,4)\{1,2,4\}] &
 1.30572905[1]  &  1.41507954[2]   &  1.43427543[4] \\
~~~~ & ~~~~ & [(0,4)\{0,2,6\}] [(1,3)\{4,$\lambda$,0\}] [(1,3)\{2,2,3\}]  &
 1.50366728[8]  &   1.56527615[11]   & 1.63564655[15] \\
~~~~ & ~~~~ & [(1,3)\{3,3,1\}]  &
 1.68994048[20]  &   ~~~~      & ~~~~~ \\
\end{tabular}
\end{table}

For $N=4$ spin-polarized electrons, one needs to
consider two distinct molecular configurations, i.e., $(0,4)$ and $(1,3)$.
Vibrations with $\lambda \geq 2$ must also be considered. In this case the 
RVM states are not always orthogonal, and the Gram-Schmidt orthogonalization is 
implemented.

\begin{table}[b] 
\caption{\label{exp_coeff}%
$N=4$ LLL electrons with $L=18$: Expansion coefficients in the RVWM basis (labelled by 
the $|i\rangle$'s) for the three lowest-in-energy translationally invariant FCI states
(labelled as [1], [2], [4]; see Table \ref{ene_ferm_np4}). The 4th column gives the
RVWM expansion coefficients of the corresponding JL expression. 
Reprinted table with permission from Ref.\ \cite{yann10},
Copyright (2010) by the American Physical Society.
}
\begin{tabular}{ccccc}
RVWM & FCI[1] & FCI[2] & FCI[4] & JL \\
\hline
$|1\rangle$ & \underline{0.9294} & -0.3430~ & 0.0903 & 0.8403 \\ 
$|2\rangle$ & -0.1188~ & -0.0693~ & \underline{0.8930} & -0.1086~ \\ 
$|3\rangle$ & 0.0067  &  0.0382 & -0.2596~ & 0.0076 \\ 
$|4\rangle$ & 0.0137  &  0.0191 & -0.0968~ & 0.0395 \\ 
$|5\rangle$ & 0.2540  &  \underline{0.8486} & 0.1519 & 0.4029  \\ 
$|6\rangle$ & 0.0211  &  0.0283 & 0.3097 & 0.0616 \\ 
$|7\rangle$ & -0.2387~ & -0.3935~ & 0.0877 & -0.3380~ \\ 
\end{tabular}
\end{table}

\subsubsection{Beyond the cusp states and magic angular momenta: Describing the complete
LLL spectrum of polarized electrons with rovibrational analytic wave functions.}
\label{corrbasisn4fp} 

Using the case of $N=4$ spin-polarized electrons, this Section will illustrate the
introduction of a generalized class of analytic wave functions portraying combined
rotations and vibrations of Wigner molecules (referred to as rovibrating WMs, RVWMs)
which encompasse the complete LLL spectra. To this end, Fig.\ \ref{n4ellllspec}
displays the corresponding FCI spectrum for $N=4$ electrons, i.e., the spectrum
associated with the exact diagonalization of the Hamiltonian (\ref{hblll}). As already
discussed in Sec.\ \ref{anan3li6}, the LLL states fall into two different groups, i.e.,
(i) those that are related through a simple translation (referred to also as
``spurious'' \cite{trug85}), and whose energies remain constant, and (ii) those that are
translationally invariant. Naturally, the spurious states lie on horizontal lines in
Fig.\ \ref{n4ellllspec}.

As aforementioned, the pure RWM analytic wave functions in the LLL [$\Phi^{\rm RWM}_L$,
see, e.g., Eqs.\ (\ref{phi1}) and (\ref{phi2})] are associated exclusively with the cusp
states of the yrast band. The generalized RVWM functions provide a correlated basis that
spans fully the translationally invariant part of the LLL spectra. The RVWM wave functions
for $N=4$ spin-polarized electrons\footnote{
For other numbers of electrons, as well as for bosonis particles, see the expanded
exposition in Refs.\ \cite{yann10,yann11}.}
have the form \cite{yann10,yann11}
\begin{equation}
\Phi^{\rm RVWM}_L=\Phi^{\rm RWM}_{\cal L} Q^m_\lambda,
\label{rvwm}  
\end{equation}
where
\begin{equation}
Q_\lambda = \sum_{i=1}^N (z_i-z_{\rm c.m.})^\lambda,
\label{qlam}
\end{equation}
with $m,\lambda=0,1,2,3,\ldots$, and the total angular momentum
$L={\cal L}+\lambda m$. The center-of-mass variable $z_{\rm c.m.}$ was defined in the line
below Eq.\ (\ref{rwm3l2n2}). The $Q_\lambda$'s represent multipolar vibrational modes
in the LLL Hilbert space \cite{ston92,mott99,pape01,ueda01,yann10,yann11}, and in the form
of expression (\ref{qlam}) they are translationally invariant.

For $N=4$ spin-polarized electrons, two distinct intrinsic molecular configurations need to
be considered, i.e., $(0,4)$ and $(1,3)$. Vibrations with $\lambda \geq 2$ must also be
considered. In this case the RVWM states are not always orthogonal, and the Gram-Schmidt
orthogonalization is implemented.

Of particular interest is the $L=18$ case ($\nu=1/3$) which is considered
\cite{laug83,laug99} as the  prototype of quantum-liquid states. However, in this case, 
Ref.\ \cite{yann10} found (see Table \ref{ene_ferm_np4}) that the exact TI solutions
are linear superpositions of the following seven RVWM states: 
\begin{eqnarray}
&& |1\rangle=\Phi^{\rm RWM}_{18}(0,4),\;\;\;\;\;\;  |2\rangle=\Phi^{\rm RWM}_{14}(0,4) Q_2^2, 
\;\;\;\;\;\;  |3\rangle=\Phi^{\rm RWM}_{10}(0,4) Q_2^4, \nonumber \\
&& |4\rangle=\Phi^{\rm RWM}_{6}(0,4) Q_2^6, \;\;\;\;\;\; |5\rangle=\Phi^{\rm RWM}_{18}(1,3),
\;\;\;\;\;\; |6\rangle=\Phi^{\rm RWM}_{12}(1,3) Q_2^3,\nonumber \\
&& |7\rangle=\Phi^{\rm RWM}_{15}(1,3) Q_3.  ~~~~~
\label{rvmbasisel}
\end{eqnarray}

The expansion coefficients of the three lowest-in-energy translationly invariant
FCI states (labelled [1], [2], [4]; see Table \ref{ene_ferm_np4}) 
in this RVWM basis are listed in Table \ref{exp_coeff}. One sees that for each 
case, one component (underlined) dominates this expansion; this applies 
for both the yrast state (No. [1]) and the two excitations (Nos. [2] and [4]).

The celebrated Jastrow-Laughlin ansatz \cite{laug83,laug90,laug99} ($p=1,2,3,\ldots$)
\begin{equation}
\Phi^{\rm JL}[z] = \prod_{1\leq i<j \leq N} (z_i-z_j)^{2p+1} e^{-\sum_{i=1}^N z_iz_i^*/2}, 
\label{jlwf}
\end{equation}
has been given exclusively an interpretation of a quantum-fluid state
\cite{laug83,jainbook}. However, since the RVWM functions span the TI subspace, it
follows that any TI trial function (including the JL ansatz above and the compact CF
states) can be expanded in the RVWM basis. As an example, I give in Table \ref{exp_coeff}
(4th column) the RVWM expansion of the JL state for $N=4$ electrons and $L=18$. One sees
that, compared to the RVWM yrast state (1st column), the relative weight of the pure 
(0,4) RWM [denoted by $|1\rangle$, see Eq.\ (\ref{rvmbasisel})] is strongly reduced,
while the weights of higher-in-energy vibrational excitations are enhanced. In this
context, the liquid character of the JL states is due to the stronger weight of
the vibrational modes which diminish the granularity of the Wigner molecule.

The presence of vibrational modes in the yrast-band states (relative ground states at
a given $L$) explains also the moderate relaxation of the FCI electron humps compared
to the RWM case (compare the middle and top rows in Fig.\ \ref{ellllcpds}).
This state of affairs is a natural occurrence within the context of a molecular
physical picture. For example, in nuclear physics, it is well known that vibrational
modes participate in the constitution of the correlated ground state, as one improves
the description of the finite system by transitioning from the time-dependent
Hartree-Fock to the RPA approach \cite{rs_book}. Such RPA-based vibrational correlations
are also well known in the context of constructing DFT energy functionals; see, e.g.,
Ref.\ \cite{enge07}.

\subsubsection{ {\small{\bf Commentary 9.}}
A direct analytic comparison between exact, RVWM, and Jastrow-Laughlin wave functions for
three polarized LLL electrons.} 
\label{comm9}

Although unrecognized, the solution of the problem of three spin-polarized electrons in
the LLL using rovibrating Wigner molecular functions, $\Phi^{\rm RVWM}_L$, was presented by
Laughlin in Ref.\ \cite{laug83.2}, which preceeded the well-known paper \cite{laug83} that
proposed the approximate trial Jastrow-type wave function in Eq.\ (\ref{jlwf}). In
accordance with this unidentified fact, Laughlin attested in Ref.\ \cite{laug83} that the
inspiration for the Jastrow-type wave function in Eq.\ (\ref{jlwf}) originated from his
analysis of the exact results in Ref.\ \cite{laug83.2}. However, this process introduced
deviations from the exact solutions, which can be detailed as follows:

Namely, the main result of Ref.\ \cite{laug83.2} [see Eq.\ (18) therein] was the
introduction of a correlated basis consisting of the following set of analytic wave
functions
\begin{equation}
|k,m\rangle \propto
 \left[ \frac{(z_a+iz_b)^{3k}-(z_a-iz_b)^{3k}}{2i} \right]
(z_a^2+z_b^2)^m e^{-(z_a z_a^*+z_b z_b^*)/2},
\label{laug3}
\end{equation}
where 
$z_a=\sqrt{2/3} \big( (z_1+z_2)/2-z_3 \big)$, and $z_b= (z_1-z_2)/\sqrt{2}$ are
three-electron Jacobi variables, and the lengths here are in units of $l_B \sqrt{2}$.

Expression (\ref{laug3}) is precisely of the form $\Phi^{\rm RWM}_{{\cal L}=3k} Q_2^m$
[associated with an intrinsic equilateral triangular ring (0,3)], as can be checked
after transforming back to Cartesian coordinates $z_1$, $z_2$, and $z_3$. Specifically,
using an algebraic computer language, it can be shown that
\begin{equation}
\Phi^{\rm RWM}_{{\cal L}=3k}(0,3) \propto  \left[ \frac{(z_a+iz_b)^{3k}-(z_a-iz_b)^{3k}}{2i}
\right] e^{-(z_a z_a^*+z_b z_b^*)/2}  e^{-3z_{\rm c.m.}z^*_{\rm c.m.}/2},
\label{rwmlaug}
\end{equation}
and
\begin{equation}
Q_2 = z_a^2+z_b^2.
\label{q2laug}
\end{equation}

Note that, according to the properties of the Jacobi coordinates, the center-of-mass
contribution in Eq.\ (\ref{rwmlaug}) separates out.

The discussion above demonstrates that Laughlin's first paper \cite{laug83.2} in 1983
reported a RVWM diagonalization that provided exact solutions for $N=3$ spin-polarized
electrons in the LLL. The follow up paper \cite{laug83}, however, deviated from this
line of research by introducing the JL trial wave function in Eq.\ (\ref{jlwf}); see
Sec.\ 7.3 in Ref.\ \cite{laug90}. The introduction of the zero-parameter JL trial wave
function and of subsequent ad hoc anz\"atze in the same vein (referred to recently as
``wavefunctionology'' \cite{simo20}) led to the overlooking of the connections between
the FQHE and the physics of the rotating and rovibrating WMs. Note that, in the context
of the wavefunctionology approach, specific signatures of the intrinsic crystallinity of
the three-electron LLL problem (reported by Laughlin through the CPDs displayed in Fig.\
7.3 of Ref.\ \cite{laug90}) were not given any further due consideration.

A high accuracy of the JL wave functions was claimed \cite{laug83,laug90} based on
near-unity overlaps with the exact wave functions for $N=3$. However, for larger sizes
$N$, the quality of the overlaps deteriorate rather rapidly, in agreement with a process
known as the van Vleck-Anderson orthogonality catastrophe
\cite{vlec36,ande67,kohn99,deml12,ares18}. Naturally, the RVWM diagonalization is
immune to this orthogonality catastrophe. 

At this point, another difference between the JL description and the RVWM one should be
mentioned, concerning the competition between LLL liquid and Wigner-crystal
\cite{fukuy78,maki83} states. Indeed, the JL and WC states are considered unrelated,
but they compete energetically, with the lowest-in-energy one accepted as the ground state.
Proposing an alternative perspective, the RVWM states express this competition as an
interplay between symmetry-preserving rotating WMs and broken-symmetry WMs, that are
pinned due to weak disorder in the sample \cite{yann04,yann11}. In contrast to the
traditional JL-versus-WC energetics-only approach, which predicted the appearance of
LLL Wigner crystallization only at low fractional fillings $\nu \le 1/5$ \cite{girv84},
the WM description was able to predict the appearance of Wigner crystallization in the
immediate neighborhood of the large fractions $\nu=1/3$ and $\nu=1$, a behavior that was
experimentally observed \cite{tsui10}.

Note that this alternative RVWM view for the liquid-versus-crystal interplay in the LLL is
associated \cite{yann11,yann20} with a superposition of quasi-degenerate rotationally
invariant states, and thus it is akin to the perturbation-driven pinning of sliding WMs
reported (in both experimental and theoretical investigations) for the case of moir\'e
materials, as discussed in Sec.\ \ref{mqd}.  

\begin{figure}[t]
\centering\includegraphics[width=0.99\textwidth]{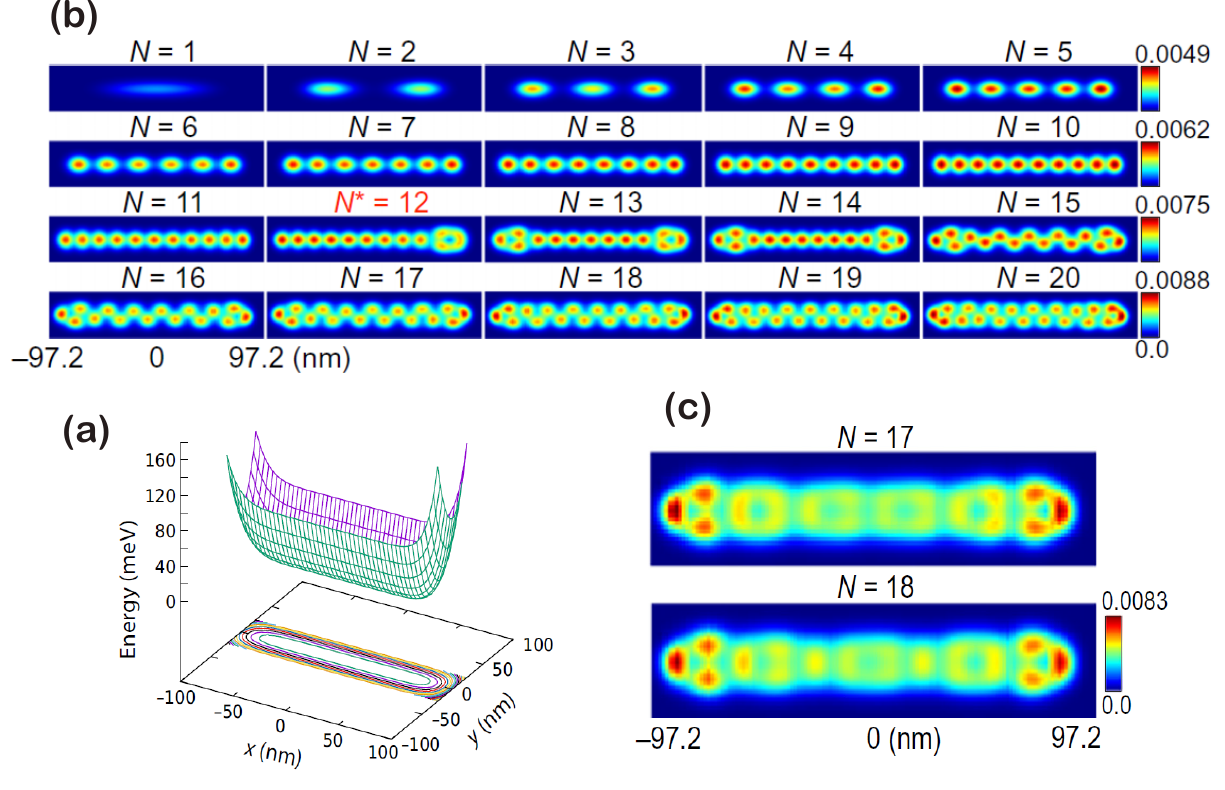}
\caption{
(a) Illustration of the Viking-longboat-type total confining potential $V(x,y)$
employed. The lateral potential, along the $x$-axis is square-type \cite{yann24.3}. 
The transverse harmonic potential, along the $y$-direction corresponds to
$\hbar \omega_y=10$ meV and to an effective electron mass $m^*=0.19 m_e$ (appropriate
for silicon).
(b) UHF ground-state charge densities for $N=1-20$ electrons for a silicon elongated QD
with a transverse confinement of $\hbar \omega_y=10$ meV. Dielectric constant $\kappa=11$,
appropriate for silicon (strong interaction). The CDs exhibit a transition from single-row
Wigner-molecular structures for $N \leq 11$ to a zig-zag chain for $N \geq N^* = 12$ that
initiates via charge accumulation occurring at the two ends of the elongated QD (see
$N=12-14$), with the zig-zag pattern fully developed for $N = 15 - 20$. For larger number
of electrons a three-row zig-zag chain structure emerges (not shown).
(c) Parity-restored (along the $y$-direction) CDs for $N=17$ and 18 fully spin and valley
polarized electrons, and
for same silicon elongated QD. $\kappa=11$ appropriate for silicon (strong interaction).
The symmetry-broken UHF CDs, corresponding to the parity-restored ones are given in (b).
The color bars (on the right) in (b) and (c) indicate the charge-density scale (in units
of 1/nm$^2$).
Reprinted figures with permission from Ref.\ \cite{yann24.3},
Copyright (2024) by the American Physical Society.
}
\label{1delwm}
\end{figure}

\section{Theoretical results and experimental observations associated with
Wigner molecules in one-dimensional systems} 
\label{res1d}
\smallskip

Theoretical investigations regarding Wigner crystallization in strictly one-dimensional or
quasi-1D electronic systems
\cite{schu93,jaur93,haus93,piac04,szaf04.3,meye08,crem11,esco19,vu20,yann24.3,gull26.2}
(including quantum rings \cite{wend96,szaf05,bao06,yann17}, carbon nanotube QDs
\cite{roy12,peck13,shap19}, edge states in graphene dots \cite{wuns08,roma09}),
as well as chains of trapped ions \cite{mori25} in linear \cite{ejte23,paga18} and ring
\cite{haef17} geometries, have a long history. Experimental investigations are represented
for example by Refs.\ \cite{crem11,peck13,shap19,ejte23,paga18,haef17,pepp18,crom24.2}.
This section will review some recent pertinent experimental and theoretical literature.
In the process, salient aspects of Wigner molecularization in finite 1D systems will be
elucidated.

\subsection{Recent theoretical and experimental results in quasi-1D electronic
nanosystems.}
\label{1delsys} 
\smallskip

As aforementioned, quantum-dot qubits are fundamental elements for semiconductor-based
solid-state quantum computing architectures \cite{dzur13,warr22,burk23}. 
A central challenging issue in constructing scalable quantum processors is that of
quantum chip large-scale integration, which would allow transfer of information
between computing qubits while preserving information during transfer. 

To this effect, currently, attention focuses on patterned, gate-controlled elongated
quantum dots (EQDs) \cite{dzur23,kuem24} and quantum interconnects
\cite{awsc21,gull26,gull26.2}, enabling coherent transfer of spins between relatively
distant quantum-dot qubits. 
 
This Section provides fundamental insights regarding the many-body quantum nature
of the electronic states in such patterned, long-distance-coupled silicon EQDs
\cite{yann24.3} (which are also relevant \cite{gull26.2} to silicon interconnects).
Such understanding is imperative for enabling theory-guided fabrication and integration
of these elements into solid-state Si-based quantum information devices. The main
finding is that the wire-like nature of the silicon EQD results in conditions where the
inter-electron repulsion energy dominates over the electron quantal kinetic energy,
leading to formation of pinned multi-chain Wigner Molecules.

Fig.\ \ref{1delwm}(a) portrays the confining potential (used in the theoretical
investigations reported here) that mimics the fabricated EQD (referred to also as
jellybean QD) according to the description given in Ref.\ \cite{dzur23}.
This potential is squarelike along the long lateral side (the $x$-axis) and harmonic
along the $y$-direction with a tight transverse confinement of $\hbar \omega_y=10$ meV.

Fig.\ \ref{1delwm}(b) displays the corresponding UHF CDs for fully spin and valley
polarized $N=1-20$ electrons. These CDs exhibit a high degree of organization with
Wigner-chain-like features. In particular, the single chain transitions to a double
zig-zag chain at the region $N=12-14$, with the double chain starting to form from
the edges.\footnote{This behavior contrasts with that in harmonic confinements along
the lateral $x$-direction, where the zig-zag chain starts forming at the center of
the linear chain, a fact well known from classical calculations in the literature of
trapped heavy ions; see, e.g., Refs.\ \cite{schi93,ejte15,yan16}.}
Note that, based on classical calculations \cite{piac04}, a third row (not shown) is
expected to develop at larger values of $N$.\footnote{
For a wider transverse confinement, $\hbar \omega_y = 2.5$ meV, the UHF transition
from single to zig-zag double row happens at $N=6$, and from a double to a triple
chain at $N=11$; see Fig.\ 3 of Ref.\ \cite{yann24.3}.}

From an inspection of the zig-zag geometries for $N=15-20$ in Fig.\ \ref{1delwm}(b),
it is apparent, as discussed in Sec.\ \ref{metho}, that these UHF solutions do break
the symmetries of the external confinement. In particular, most prominent is the
symmetry breaking of parity along the transverse $y$-direction, which is associated
with a symmetric harmonic confinement. It is to be stressed again that the broken-symmetry
UHF solutions violate a fundamental axiom of quantum mechanics, namely, that the
single-particle CDs must preserve the symmetries of the many-body Hamiltonian. Here,
because the UHF solutions involve the light-mass electrons, one needs to further take
a step in restoring the parity symmetry along the transverse $y$-direction,

To restore the $y$-parity symmetry, one applies on the UHF Slater determinant,
$\Psi_{\rm UHF}$, the projection  operator
\cite{yann07,shei21,paca70}
\begin{equation}
\hat{\Pi}_p=\frac{1}{2} (1+p \cpy),
\label{prt}
\end{equation}
with $p=\pm 1$. $\cpy$ is the many-body parity operator, which inverts about the $x$-axis the 
$y$-coordinates for all electrons, namely, $\cpy=\prod_{i=1}^N \hat{P}^{(i)}_y$.

In a similar way to the superposition of the two mirror triangular configurations in
Sec.\ \ref{3aplii} [see Fig.\ \ref{heisch}(b)], the application of the projection
operator (\ref{prt}) is equivalent to adding or subtracting two mirror wave functions
(often referred to as ``zig'' and ``zag''). The CDs resulting from this restoration are
displayed in Fig.\ \ref{1delwm}(c) for $N=17$ and 18 electrons confined in the potential
depicted in Fig.\ \ref{1delwm}(a). It is apparent that the zig-zag motif gets obliterated
by the parity restoration.

\begin{figure}[t]
\hspace{-1cm} \centering\includegraphics[width=1.05\textwidth]{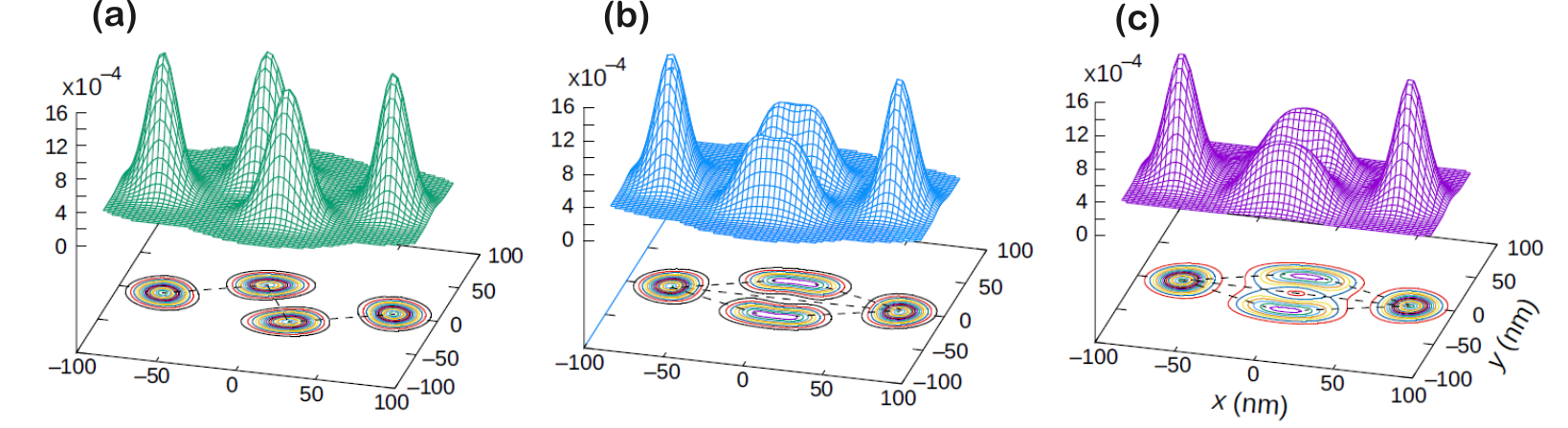}
\caption{
Comparison of (a) symmetry-broken UHF, (b) $y$-parity-restored UHF, and (c) FCI CDs for
the associated ground states of $N=4$ electrons at zero magnetic field in a single elliptic
quantum dot, with an effective mass $m^*=0.067m_e$ and an elliptic  potential confinement
specified by $\hbar \omega_x=3.00$ meV and $\hbar \omega_y=5.64$ meV. The dielectric constant
used was $\kappa=1$. In (a) and (b), $S_z=0$. In (c) $S=S_z=0$. Lenghts in units of nm.
Charge densities in units of 1/nm$^2$. The charge densities are normalized to the total
number of fermions, $N=4$. The dashed lines are a guide to the eye; in (b) and (c), they
delineate two mirror trapezoids.
Reprinted figures with permission from Ref.\ \cite{yann24.3},
Copyright (2024) by the American Physical Society.
}
\label{1delfci}
\end{figure}
  
It is worth comparing the UHF plus symmetry-restoration results with the exact CDs
obtained through an FCI calculation. Such a comparison is depicted in Fig.\ \ref{1delfci}
for the case of an elliptical external two-dimensional potential that confines $N = 4$
conduction electrons for a one-band (no valley present) semiconductor material.

The UHF CD in Fig.\ \ref{1delfci}(a) displays a broken-symmetry zig-zag configuration.
However, restoration of the parity along the $y$-axis in Fig.\ \ref{1delfci}(b) (which
also restores the $x$-parity in this example) produces a configuration of two mirror
trapezoids, which is formed by the spreading out (along the $x$-axis) of the two middle
sharp single-electron humps in panel Fig.\ \ref{1delfci}(a). In this context, note the
reduction in height of the corresponding broad humps in Fig.\ \ref{1delfci}(b). The
parity-restored UHF CD in Fig.\ \ref{1delfci}(b) is in good agreement with the FCI CD
in Fig.\ \ref{1delfci}(c). Naturally, the FCI solution displays a certain degree of
additional relaxation effects in the CD humps.

\begin{figure}[t]
\centering\includegraphics[width=0.60\textwidth]{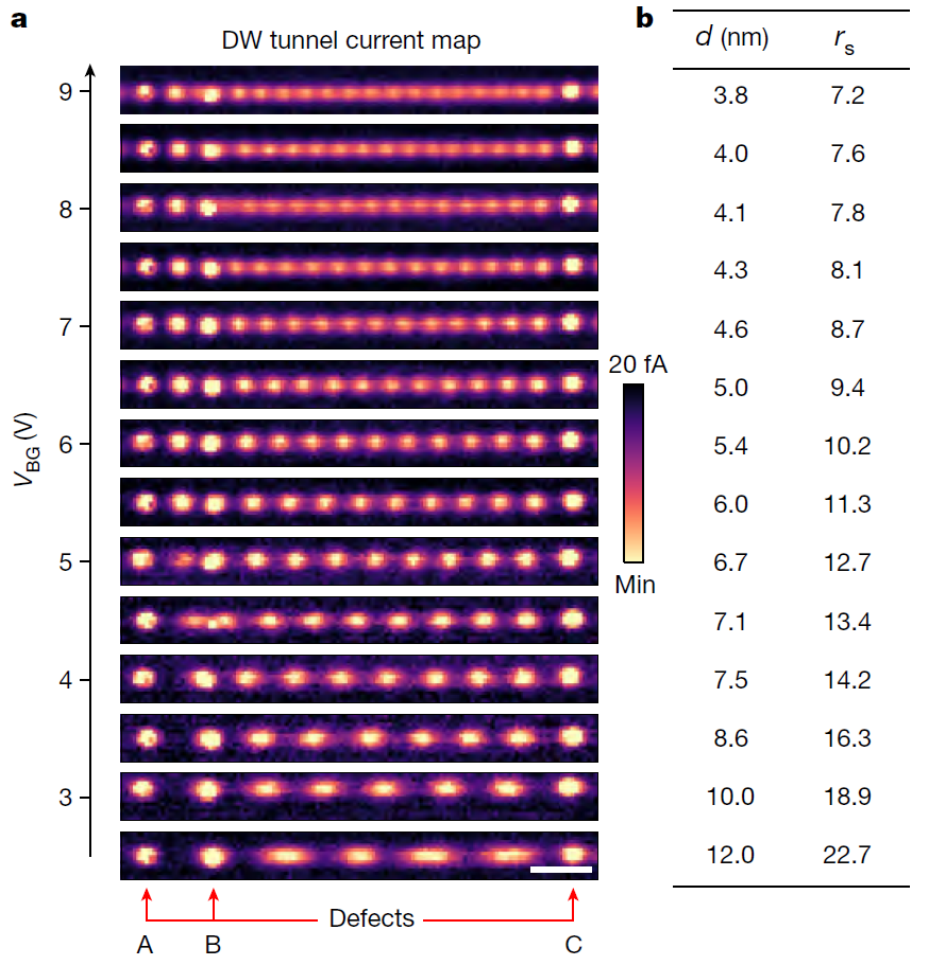}
\caption{
Tunnel current measurement of 1D Wigner crystallization. {\bf a}, Conduction band edge (CBE)
tunnel current maps for the centre DW in a triple-DW group with the back gate voltage,
$V_{\rm BG}$, increasing from 2.5 V to 9.0 V. The sample-tip bias, $V_{\rm bias}$, is selected
in the range of $-0.85$ V $< V_{\rm bias} < -0.30$ V to minimize the tip-sample vacuum level
mismatch. The maps show pinned 1D Wigner molecules in which each bright dot corresponds to
one localized electron. Three pinning defects are labelled with red arrows at the bottom.
{\bf b}, Table of the average electron separations and corresponding values of $r_s$ (here a
quantity similar to $R_W$, see text) for the images shown in {\bf a}. Min, minimum. Scale bar,
10 nm in {\bf a}. 
Reprinted figure with permission from Ref.\ \cite{crom24.2},
Copyright (2024) by Springer Nature Limited.
}
\label{1dwmDW}
\end{figure}

The formation of pinned WM chains in 1D finite-length channels has been verified
experimentally through direct imaging of the charge densities, both in the case of carbon
nanotubes \cite{shap19} and in domain walls (DWs) \cite{crom24.2} arising from differential
uniaxial strain in 2D van der Waals heterosctructures. As an example, Fig.\ \ref{1dwmDW}(a)
presents some indicative STM measurements from a gate-tunable bilayer WS$_2$ device
\cite{crom24.2} that demonstrate WM formation in the range of $N=4-17$ electrons.
The transverse confinement is very tight, and the DW behaves as a strictly 1D system where
the double- and higher-row geometries are missing. Between the defects named as B and C, one
can observe a periodic arrangement of strongly localized electrons with a integer electron
number which varies from 4 to 17 as $V_{\rm BG}$ is increased from 2.5 V to 9.0 V. 

The prominance of the 1D Wigner molecule in the STM images of Fig.\ \ref{1dwmDW}(a)
is a testament of the strong impact that electron-electron interactions have in one
dimension. Ref.\ \cite{crom24.2} employed the dimensionless parameter\footnote{
To avoid confusion, I mention here that the symbol $r_s$ is most often used to indicate
the Wigner-Seitz radius.}
$r_s = d/a_B$ to qualitatively characterize the propensity towards electron localization
where $d$ is the electron separation and $a_B=\kappa \hbar^2/(m^* e^2)$ is the effective
Bohr radius, $m^* = 0.39m_e$ is the effective electron mass obtained from  DFT
calculations, and $\kappa=3.9$ is the effective dielectric constant. The values of $d$
and $r_s$ for the Wigner molecular chain shown in Fig.\ \ref{1dwmDW}(a) are listed in
Fig.\ \ref{1dwmDW}(b). Note that the parameter $r_s$ here is a direct variant of the
Wigner parameter $R_W$ introduced in Sec.\ \ref{intro_rw}. 

Before leaving this Section, it is beneficial to note that a recent large-scale density
matrix renormalization group (DMRG) theoretical study \cite{gull26.2} has further
confirmed the appearance of Wigner molecularization in strictly 1D electronic
nanosystems; for earlier FCI studies, see, e.g., Refs.\ \cite{haus93,szaf04.3}.

\begin{figure}[t]
\centering\includegraphics[width=0.999\textwidth]{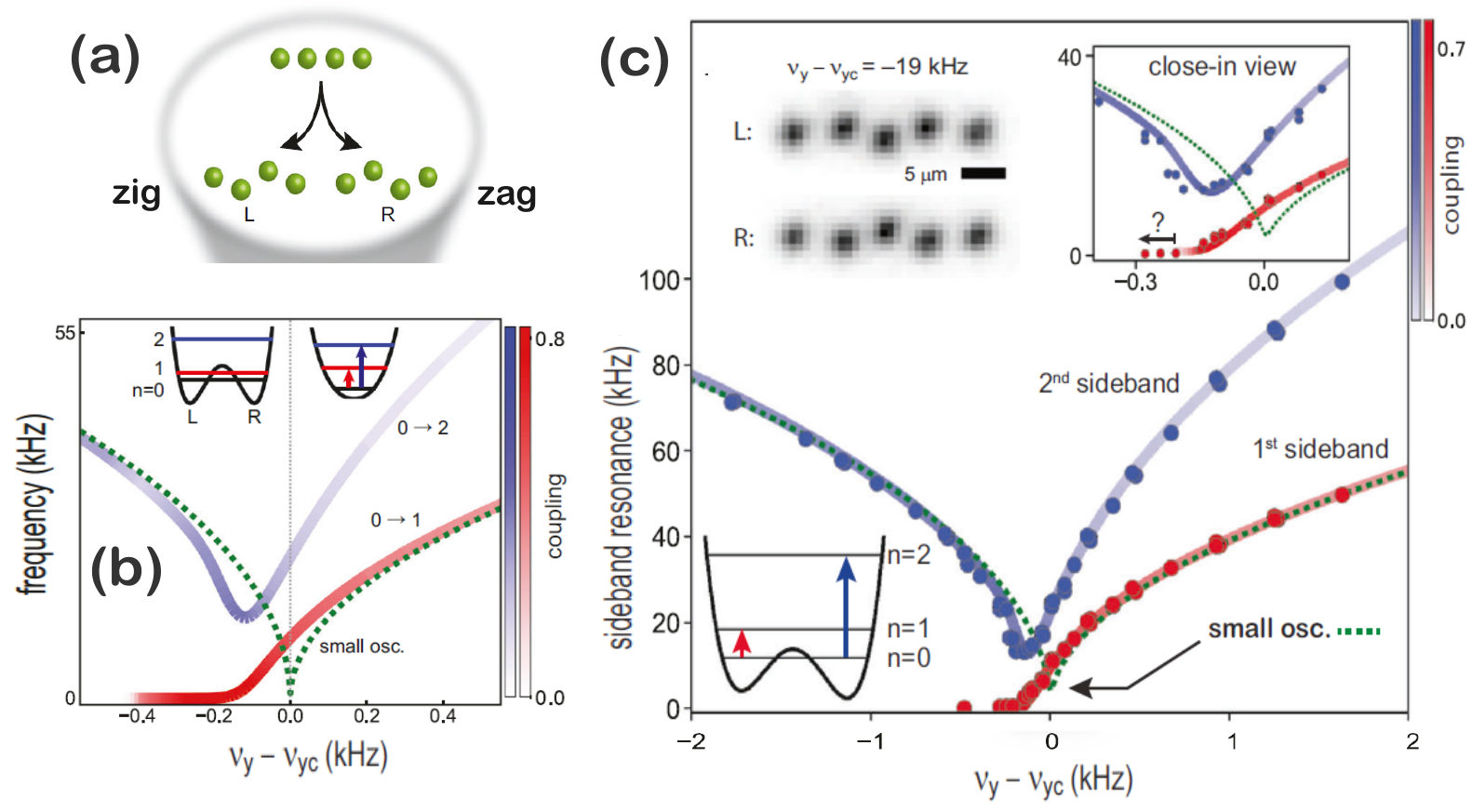}
\caption{
(a) Schematic depiction of the 1D linear to 2D zigzag structural transition for a four-ion
crystal confined in a linear rf Paul trap, consisting of four rod electrodes and two
endcap needles.  The ``zig''(''zag'') structure is denoted as left,L(right,R). 
(b) Theoretically simulated quantum energy-level spectrum for the first two excited states
of the transverse-$y$ zig-and-zag Schr\"odinger-cat entangled mode of a four-ion crystal,
calculated relative to the ground state using the form of the zigzag potentials on either
side of the transition, as shown. Also shown (dotted curve) is the classical small-oscillation
prediction of the broken-symmetry $L$ or $R$ structures. Simulations assume $\nu_{yc} = 740$
kHz for the classical transition point [see schematic in (a)]. $\nu_y$ is the frequency that
specifies the harmonic confinement of the trap in the transverse direction. 
(c) Experimental spectroscopic characterization of the quantum-mechanical linear-zigzag
transition for five $^{171}$Yb$^+$ ions: the frequency of first and second upper Raman
sidebands for the transverse-$y$ zigzag mode as a function of the secular trap frequency
$\nu_y$ relative to the classical critical value $\nu_{yc} \approx 760$ kHz. Solid lines are
quantum energy-level differences of $n=1$ and $n=2$ numbered excited states with respect to
the $n=0$ ground state for the quartic potential defined in Eq.\ (\ref{vpotfield}) (with bias
$|C_1| = 3.3 \times 10^{-7}$). The dotted line shows the classical small-oscillation prediction
for the effective potential in Eq.\ (\ref{vpotpaul}) with equivalent bias.
The left (top tow) inset shows sample images of the two broken-symmetry equilibrium structures
of the five-ion Wigner molecule far from the critical point, where the ``zig'' and ``zag''
superposition is suppressed. The right (top row) inset shows a close-in view of data
acquisition near the critical point.
Adapted figures from Ref.\ \cite{ejte23}, Copyright (2023) under license CC BY 4.0.
}
\label{ejte}
\end{figure}

\subsection{Theoretical and experimental results in finite quasi-1D ion Coulomb
geometries.}  
\label{1diontrap} 
\smallskip

Whether a zig-zag or a symmetry-restored CD (see previous Sec.\ \ref{1delsys}) will be
the actual finding in a quasi-1D system may depend in practice on several additional
factors, including the length of the system, the value of the effective mass, the
presence or not of impurities, and other external perturbing agents. Indeed, the longer
the 1D system, the higher the expectation that the experimental system exhibits a broken
symmetry CD. Interestingly, because of the much larger mass
compared to electrons, the CDs of trapped heavy ions are routinely found experimentally
to exhibit broken-symmetry geometries, including zig-zag chains. Nevertheless, despite
the large ionic mass, the Schr\"odinger cat superposition (or entanglement) of both the
mirror ``zig'' and ``zag'' symmetry-broken configurations is quantum mechanically allowable,
and recently\footnote{
For the realization of quantum mechanical superpositions and entanglement of the
azimuthal configurations of 2D polygonal-ring-like ionic structures, see Sec.\
\ref{3ipliii}.}
it has been experimentally realized \cite{ejte23} for the case of $N=3-5$ ultracold
$^{171}$Yb$^+$ ions in a linear radio-frequency (rf) Paul trap.

Using reduced coordinates and neglecting small perturbative imperfections, the classical
potential felt by the $^{171}$Yb$^+$ ions in the Paul trap can be approximated as an
anisotropic harmonic confinement plus the inter-ion Coulomb repulsion, namely
\begin{equation}
V(\br_1,\ldots,\br_N)=\sum_{i=1}^N {\textstyle\frac{1}{2}} (a_x x_i^2+a_y y_i^2 +z_i^2) +
\sum_{i<j}^N \frac{1}{|\br_i-\br_j|},
\label{vpotpaul}  
\end{equation}  
where the lengths are given in units of $d_z=(q^2/(\kappa M \omega_z^2)^{1/3}$ and the
energies are in units of $q^2/(\kappa d_z)$. The aspect ratio of the effective harmonic
confinement is controlled by $a_{x(y)}=\omega_{x(y)}/\omega_z$, where the frequencies
$\omega_r=2\pi \nu_r$, $r \in (x,y,z)$.

As was discussed in Ref.\ \cite{ejte23}, the linear-to-zigzag (LZ) transition for a pinned
ionic WM is controlled by the aspect ratio of the Paul trap. By applying a dc quadrupole
potential in the transverse trapping plane, the transverse confinement could be modified
only along the principal $y$-axis. The LZ transition is thus pinned to the $y$-$z$ plane.
In analogy with the case of the elongated QD (Sec.\ \ref{1delsys}), for strong transverse
confinement the ions form a linear string along the axial $z$-direction of the linear trap.
At a critical aspect ratio $a_{yc}$, which depends on the number of ions, the
ions undergo a structural transition to a 2D zigzag configuration [Fig.\ \ref{ejte}(a)].
This is consistent with the transition of the effective potential in Eq.\ (\ref{vpotpaul})
from a single well to a double well along the transverse direction, as was explicitly
shown\footnote{
Indeed a Taylor expansion of the effective potential along the transverse direction reveals
the presence of quartic terms, which are responsible for double-well formation, depending
on their strength as a function of the aspect ratio.}
in Ref.\ \cite{retz08}. The classical dynamics of the linear ion chain restricted to 2D can
be described in terms of its $2N$ vibrational modes, with the transverse-$y$ zigzag mode
representing a ``soft mode'' that goes to zero frequency at the LZ critical point; see
dotted line in Fig.\ \ref{ejte}(b).

As further elaborated in Ref.\ \cite{ejte23}, a coupled-mode analysis and adiabatic elimination
of the other vibrational modes related to the potential (\ref{vpotpaul}) generate an effective
field theory for the zigzag mode \cite{retz08}. Accordingly, the underlying dimensionless
potential $U(\varphi)$, up to the 4th order, is given by 
\begin{equation}
U(\varphi)=C_1 \varphi + {\textstyle\frac{1}{2}} C_2 \varphi^2 +
{\textstyle\frac{1}{3}} C_3 \varphi^3 + {\textstyle\frac{1}{4}} C_4 \varphi^4,
\label{vpotfield}
\end{equation}  
where $\varphi$ is the zigzag order parameter (the normal mode coordinate). Again, depending
on the strength of the four $C$ coefficients, $U(\varphi)$ can display a single-well (for
$\nu-\nu_{yc} > 0$) or a double-well (for $\nu-\nu_{yc} < 0$) profile; see the $U(\varphi)$
schematics in the insets of Figs.\ \ref{ejte}(b) and (c).

At the zero temperature limit, the quantum mechanical modeling proceeds by adding a
kinetic energy term to $U(\varphi)$, and then solving the resulting Hamiltonian to obtain
the quantum energy levels for the effective potential across the transition, as shown in
Fig.\ \ref{ejte}(b) for the example of four ions. Near the critical point, $\nu_{yc}$, the
frequency splitting between the parity-conserving ground state, $|0\rangle$, and the first
excited state, $|1\rangle$, deviates from the corresponding classical small oscillation
frequency. Specifically, the level splitting remains finite at the critical point before
approaching zero away from it, as tunnel coupling between the two sides of the double well,
associated with the broken-symmetry states $|L\rangle = (|0\rangle - |1\rangle)/\sqrt{2}$
(``zig'') and $|R\rangle = (|0\rangle + |1\rangle)/\sqrt{2}$ (``zag''), is suppressed by
the increasing barrier.

The parity-preserving spectral behavior (described in the previous paragraph) was
experimentally observed for $N=3-5$ ultracold $^{171}$Yb$^+$ ions, demonstrating thus the
realization of the Schr\"odinger-cat superposition and entanglement of the two mirror ``zig''
and ``zag'' chains; see, e.g., the actual measurements for $N=5$ that are displayed in Fig.\
\ref{ejte}(c).

\subsection{ {\small{\bf Commentary 10.}} Remarks on the similarities between electronic and
ionic Wigner-molecular chains.}
\label{comm10}

The analogies between the physics of Wigner-molecular chains of a few electrons in
elongated QDs (Sec.\ \ref{1delsys}) and that of small Wigner-crystal chains of trapped
ultracold ions in a linear Paul trap (Sec.\ \ref{1diontrap}), in spite of the large
difference in the composition of particles and the energy scales, are indeed most remarkable.
It can be reliably stated that the underlying factor is the all unifying behavior related to
the interplay of symmetry breaking and symmetry restoration. The universality and autonomy
aspects of this behavior have been expounded throughout this review; see, e.g., Sec.\
\ref{comm8}.

Specifically, the symmetry governing the physics of the LZ transition is the parity, which
is a discrete symmetry involving only two mirror states. Specific examples are the
$y$-parity-broken pair of electronic states, $(\Psi_{\rm UHF},{\cpy}\Psi_{\rm UHF})$, of the
EQD [see Eq.\ (\ref{prt})] and the ($|L\rangle$, $|R\rangle$) pair of states relevant to
the LZ ionic transition. A further example was the pair of mirror triangular states
$({\cal G}_1,{\cal G}_2)$ [Eq.\ (\ref{diga}) in Sec.\ \ref{3aplii}] for three ultracold
$^6$Li atoms in a double-well optical trap.

This universal behavior emerges in spite of the dissimilarity in the theoretical treatments
in the literature between the ionic and electronic cases. Indeed, based on the much larger
mass and experimentally observed strong localization of ions, a classical modeling using a
field-theoretical nonlinear potential can be rationalized as a valid approximation. This
nonlinear potential [see Eq.\ (\ref{vpotfield})] is then combined with a kinetic
Hamiltonian term to intoduce quantum corrections. In contrast, the theoretical treatment
of the electronic chains relates directly to the symmetry-restored UHF approximations of
the many-body Schr\"odinger equation, including FCI exact solutions.

In this context,
I note the conceptual equivalence of the electronic tunnel splitting generated by the
non-orthogonality of the two mirror UHF states, $\Phi_{\rm UHF}$ and $\cpy \Phi_{\rm UHF}$,
with the phenomenological ionic one associated with the barrier between the left and right
wells of $U(\varphi)$. I further note that the magnitude of the electronic tunnel splitting
was estimated to be $\sim 0.4$ meV (see Fig.\ 10 in Ref.\ \cite{yann24.3}) compared to
the smallness of the ionic tunnel splitting \cite{ejte23} ($\sim 3$ kHz, 1 h kHz = $4.136
\times 10^{-9}$ meV). The ability to measure such small tunnel splittings is a testament to
the unprecedented experimental advances in the cooling of trapped ionic systems.

\begin{figure}[t]
\centering\includegraphics[width=0.80\textwidth]{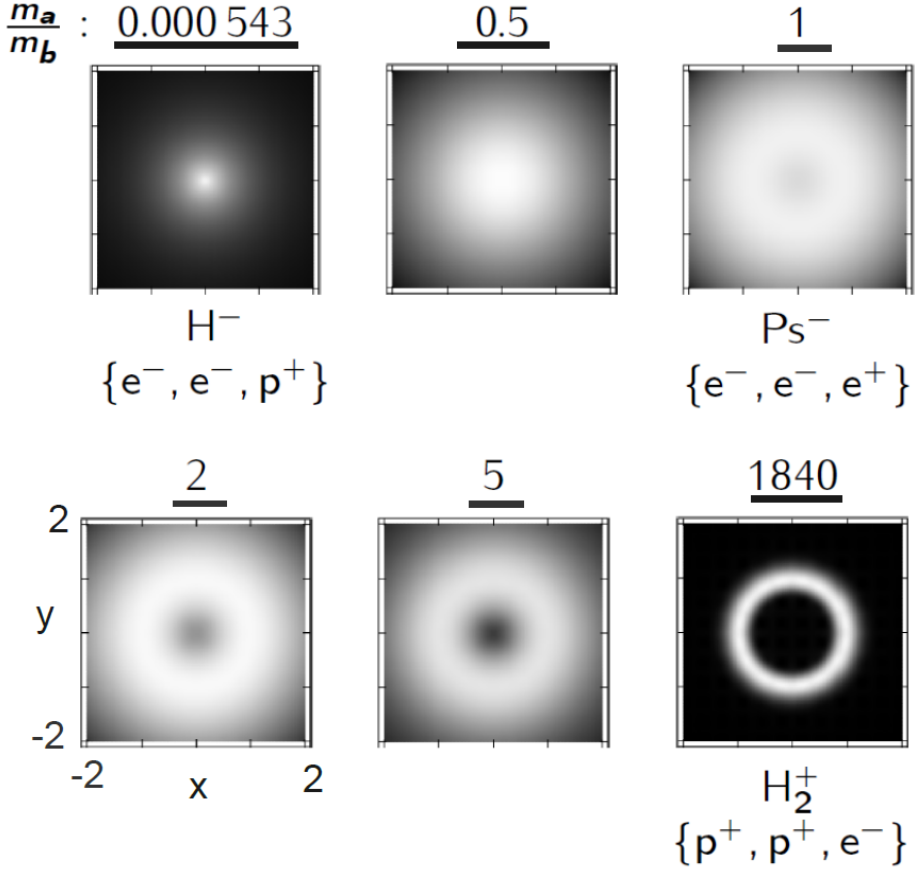}
\caption{
Transition from an atomic- to a molecular-type single-particle density of the a particles
in \{a$^\pm$,a$^\pm$,b$^\mp$\}-type Coulombic systems. In each panel, the cut of the normalized
3D density $\rho_{\rm a}(\br)$ is plotted along the plane defined by $\br=(x,y,0)$ in the range
$(x,y) \in [-2,2]$ $au$. The assembled particles are specified by the ratio $m_{\rm a}/m_{\rm b}$
of the masses; see underlined numbers on top of the panels. The center of mass for all three
particles is the center of each plot. The mass of the proton is $m_p \approx 1840 m_e$. Note
that $1/1840 \approx 0.000543$. Lighter colors indicate larger densities, with white
corresponding to maxima.
Figure adapted with permission from Ref.\ \cite{maty11}, Copyright (2011) by the
American Physical Society.}
\label{matyf} 
\end{figure}

\section{ {\small{\bf Commentary 11.}} The hierarchical approach as a bridge to quantum
chemistry}  
\label{comm11}

The chemical concept of molecular structure (directly related to the BOA) has often been
considered (see, e.g., Ref.\ \cite{hend19}) a prime example of a physical picture that
belongs squarely to the realm of strong emergence. More generally, the issue of how
molecular structure, i.e., the classical ball-and-stick model representing the relative
positions of the constituent nuclei, relates to the probabilistic nature of quantum
mechanics has been debated over several decades. A complete review of this literature
is beyond the scope of this paper, but an overview can be gained from the following recent
Refs.\ \cite{hend19,seif20,scer25,wool25,lomb25}. The goal of this Section is more modest,
i.e., to further enrich this debate by highlighting the analogies with the description of
Wigner molecularization in 2D and 1D artificial nanosystems.

In any case, it seems that, in parallel with many aspects of the hierarchical
methodological scheme presented in this review (see Sec.\ \ref{hieras}), the
center-of-weight of this discussion (concerning the nature of emergence as it relates to
the concept of molecular structure) is shifting \cite{scer25} in favor of weak emergence
and reductionism. Indeed, due to computational advaces and under the broad terms of
``non-Born-Oppenheimer'' (non-BO) and ``pre-Born-Oppenheimer'' (pre-BO) chemistry, renewed
interest has emerged in obtaining highly accurate (including exact FCI) solutions in three
dimensions of the many-body Schr\"odinger equation associated with the full combined
Hamiltonian which incorporates both the electrons and nuclei of the molecule on an equal
quantum footing \cite{maty11,maty11.2,naka11,maty19,agos22,lang24}.

In analogy with the 2D exact FCI description of Wigner molecularization, in the case of
pre-BO and non-BO calculations, information concerning the intrinsic molecular structure
is extracted from the combined (both electrons and nuclei treated on an equal footing)
many-body wave function by analyzing the partial single-particle densities and second-order
correlation functions for the nuclei \cite{maty11,maty19,lang24}. 

Fig.\ \ref{matyf} displays the variation of the same-particle (denoted as ``a'') charge
density, $\rho_{\rm a}(\br)$, along the transition
H$^-$ $\rightarrow$ Ps$^-$ (Positronium anion) $\rightarrow$ H$^+_2$. This transition is
implemented numerically through the variation of the mass ratio $m_{\rm a}/m_{\rm b}$
in a three-particle \{a$^\pm$,a$^\pm$,b$^\mp$\}-type Coulombic system. It is
apparent that the formation of a shell in the charge density, which becomes sharper as the
ratio $m_{\rm a}/m_{\rm b}$ increases, is a signature of an underlying process of Wigner
molecularization for the two ``a'' particles. Naturally, this rotating-WM formation is
associated with an increase of the corresponding effective Wigner parameter, $R_W^{\rm a}$
(see Sec.\ \ref{intro_rw} for the direct relation between the Wigner parameter and the
particle mass).

The symmetry-preserving partial charge density in the last panel of Fig.\ \ref{matyf}
does not reveal the angular correlations of the two protons in H$^+_2$. In analogy again
with the CPDs and PCFs in Fig.\ \ref{2espec}(a), second-order angular correlations were
employed in Refs.\ \cite{maty11.2,maty19} to demonstrate that the two protons exhibit an
antipodal intrinsic (hidden) configuration.

The symmetry-preserving pre-BO and non-BO studies do not usually
address the process of formal SSB, apparently because natural molecules are finite
objects. However, as elaborated in this review, finite Coulombic systems are subject
to symmetry breaking (referred to as ESB) for large values of the Wigner parameter under
a very small external perturbation supplied by the environment. Thus the very large value
of $R_W^{\rm n}$ (resulting from the proton mass $m_p \approx 1840 m_e$) generates
tower-of-states spectra of molecular rotations (compare, e.g., Fig.\ 1(a) in Ref.\
\cite{agos22} to Fig.\ \ref{2espec}), which in turn give rise to explicitly
symmetry-broken configurations of the constituent nuclei in accordance with the ``flea on
the elephant'' effect (see Sec.\ \ref{mqd}).

\begin{figure}[t]
\centering\includegraphics[width=0.80\textwidth]{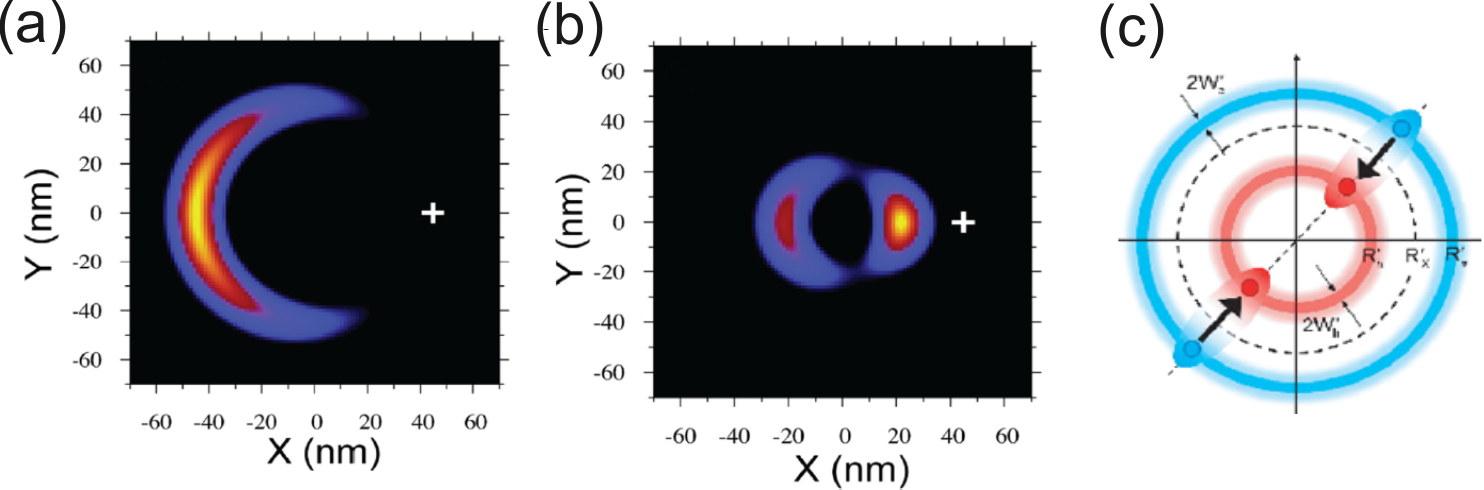}
\caption{
(a) and (b) Two-body CPDs with RWM character for a biexciton confined in a nanoquantum ring
at vanishing magnetic field. (a) The probability distribution of the other electron with
respect to one electron fixed at the position marked by a $+$. (b) The probability
distribution of the two holes for a fixed electron at the same position. (c) Schematic
of the biexciton RWM. The blue dots portray electrons. The red dots portray holes. The
ring radius for the holes is smaller than the radius for the electrons due to the larger mass.
For the same reason, the holes are more localized compared to the electrons.
Figures adapted with permission from Ref.\ \cite{okuy16}, Copyright (2015) by the
American Chemical Society.}
\label{etof}
\end{figure}  

From the discussion above, it follows that the symmetry-broken states are the most probable
physical outcome in the case of nuclei in natural molecules, as well as of bare ions confined
in traps. In this context, I mention again that
the realization of actual symmetry-preserving many-body states, which has
recently been experimentally achieved with trapped ultracold $^{40}${\rm Ca}$^+$ and
$^{171}$Yb$^+$ cations (see Secs.\ \ref{3ipliii} and \ref{1diontrap}, respectively),
constitutes a major advance in demonstrating the universality of the interplay between
symmetry breaking and symmetry restoration in finite systems.

Ref.\ \cite{maty19} contrasted the pre-BO symmetry-preserving, ring-like charge density
for the two protons of H$^+_2$ to the commonly used picture of a classical rotating dumbbell
(see Fig.\ 7 therein).
However, the bridging of these two seemingly inconsistent pictures is achieved with the
two-step quantum formalism of symmetry-breaking and symmetry-restoration via projection
techniques, as previously elaborated; see, e.g., Fig.\ \ref{succapp}, as well as Secs.\
\ref{2eqd} and \ref{qdrr}. Note that the restoration of good total angular momenta and of the
rotational symmetry via projection techniques in the case of natural molecules has been
implemented in recent publications \cite{reih19,reih23,reye25}. 

Further examples of the analogies between the pre-BO/non-BO treatment in natural molecules
and Wigner molecularization in 2D semiconductor nanosystems are provided by the FCI
calculations for assemblies of trapped biexcitons reported in Refs.\
\cite{okuy11,okuy16,okuy18}. A biexciton, usually denoted as ``XX'', is composed of two
electrons and two associated holes and is overall neutral in charge. Thus the FCI
calculations solve the combined Schr\"odinger equation which involves both holes (positive
charges) and electrons (negative charges) on an equal footing. Using CPDs, Refs.\
\cite{okuy11,okuy16,okuy18} showed that both two-electron and two-hole RWMs can be formed
in  nanoquantum rings for a large range of parameters. An example of such CPDs is
displayed in Fig.\ \ref{etof}. Experimentally, the signature of RWM formation consists of
the Aharonov-Bohm period becoming fractional compared to that of uncorrelated charges, as
was qualitatively observed through the application of a perpendicular magnetic field.

\begin{figure}[t]
\centering\includegraphics[width=0.90\textwidth]{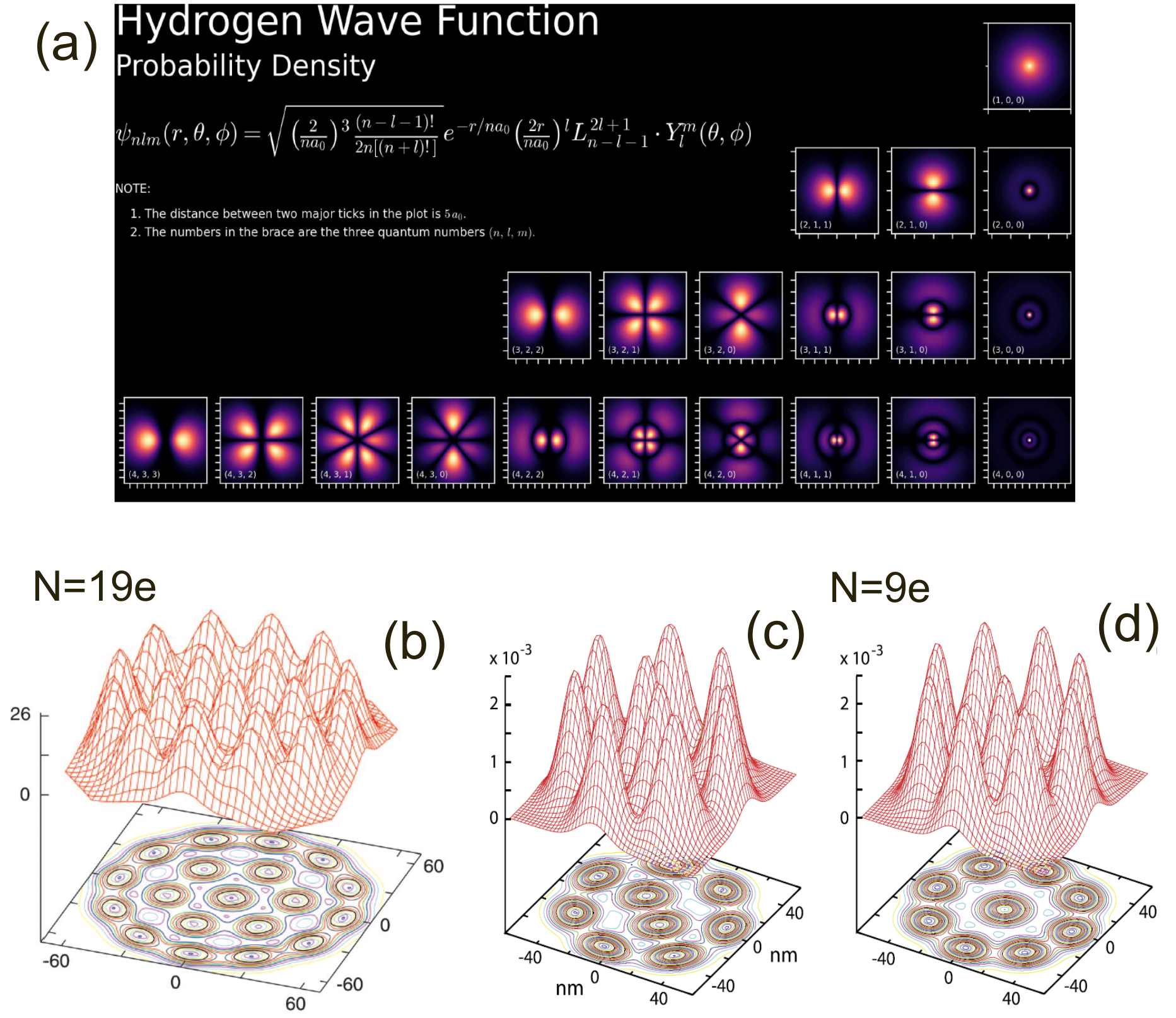}
\caption{
(a) The Hydrogen-atom one-electron orbitals derived from the Schr\"odinger equation in 1926
\cite{schr26} (see a historical outline in Ref.\ \cite{gali26}). The independent-particle
approach utilizes similar delocalized orbitals to build a single Slater determinant as the
many-body wave function. Reprinted figure from Ref.\ \cite{hydr25}.
(b) Mean-field UHF charge density in a 2D parabolic QD for $N=19$ and $S_z = 19/2$, exhibiting
breaking of the circular symmetry at $R_W = 5$ [see Eq.\ (\ref{rw})] and vanishing magnetic
field $B = 0$. The localized orbitals correspond to a WM with a (1,6,12) geometric
configuration of concentric polygonal rings. The choice of the remaining parameters is
$\hbar\omega_0 = 5$ meV, dielectric constant $\kappa=3.8191$, and effective electron mass
$m^* = 0.067m_e$. Lengths along the horizontal $x$- and $y$-axes are in nanometers and the
electron density is in $10^{-4}$ nm$^{-2}$.
Reprinted figure with permission from Ref.\ \cite{yann07},
Copyright (2007) by the Institute of Physics and IOP Publishing.
(c and d) Mean-field UHF charge densities for two WM isomers in a 2D parabolic quantum dot
with $N = 9$ electrons and $S_z = 9/2$, exhibiting breaking of the circular symmetry at
$R_W = 6.365$  and $B = 0$. (c) The (2,7) ground-state WM isomer with total energy 570.0093
meV. (d) The (1,8) first-excited WM isomer with total energy 570.2371 meV. The choice of the
remaining parameters is: parabolic confinement $\hbar\omega_0 = 5$ meV, dielectric constant
$\kappa = 3$, and effective mass of electron $m^* = 0.067m_e$. Lengths are in nanometers and
the electron densities are in nm$^{-2}$.
Reprinted figure with permission from Ref.\ \cite{shei21},
Copyright (2021) by the Institute of Physics and IOP Publishing.
}
\label{schrf}
\end{figure}

\section{ {\small{\bf Commentary 12.}} From de Broglie's waves to Wigner's corpuscular
molecularization}  
\label{comm12}
\smallskip

2026 is the 100-year anniversary of the Schr\"odinger equation \cite{schr26}, which is
being observed worldwide with celebratory meetings \cite{esi26} and commemorative articles
in internet blogs \cite{gali26} and scientific news journals \cite{cen26}. All these
conference talks and articles describe eloquently the historical path that led to the
formulation by Schr\"odinger of his foundational quantum-mechanical equation. The main
point is that Schr\"odinger was influenced in an outright way by de Broglie's work on
the matter waves \cite{brog25} and that his efforts (during the late 1925 and early 1926)
were focused on producing an appropriate wave equation \cite{gali26,mehr87,renn13} able
to explain the spectrum of the Hydrogen atom, and thus transcend Bohr's model of old
quantum mechanics which had failed to account for the spectrum of the He atom.

Soon thereafter, the Schr\"odinger equation was used to describe the two-electron problem
in the Helium atom by Heisenberg \cite{heis26}, Slater \cite{slat27}, and Hylleraas
\cite{hyll29} (see also the review by Bethe and Salpeter \cite{bethe_book}). Hylleraas 
succeeded in carrying out accurate enough calculations on the ground state \cite{hyll29}
and low-energy excitations of the Helium atom \cite{hyll64}, thereby confirming the validity
of the many-body Schr\"odinger equation to describe the natural atoms.

It is thus highly counterintuitive that the same equation, which sprung out of de Broglie's
matter waves and which provided the interpretational tool for the undulatory mechanics of
atomic physics and the aufbau organization of natural elements, may provide solutions which
relate to corpuscular geometric architectures that underlie the process of
Wigner molecularization and Wigner crystallization. Indeed the contrast between the
two-electron results in the book by Bethe and Salpeter \cite{bethe_book} and those
describing a 2e RWM in Sec.\ \ref{2eqdex} cannot be more striking. In this context, Fig.\
\ref{schrf} helps one to visualize further the contrast between the ``undulatory'' [Fig.\
\ref{schrf}(a)] versus the ``corpuscular'' [Fig.\ \ref{schrf}(b to d)] solutions of the
many-body Schr\"odinger equation.

The counterintuitiveness described above accounts also for the fact that the microscopic
theory of quantum Wigner molecularization (based on the Schr\"odinger equation as elaborated
in this review) was developed much later than the publication of the original papers by
Schr\"odinger in 1926, i.e., in the last 25 years, following the entrance of artificial
nanosystems in the scientific and technological stage. Surprising as it may be, the formal
coexistence of these two incongruous aspects (undulatory versus corpuscular) in the same
mathematical equation offers another instance of the famous Wigner pronouncement ``The
Unreasonable Effectiveness of Mathematics in the Natural Sciences'' \cite{wign60}.

\section{Summary and Conclusions}
\label{conc}   

The main focus of this paper is to review the quantum physics of Wigner molecularization
in finite-size, artificial nano- and micro-systems. Such systems encompass a broad diverse
collection consisting of 2D (see Secs.\ \ref{resff} and \ref{wmmagrot}), 1D (see Sec.\
\ref{res1d}), and 3D\footnote{
For theoretical investigations of Wigner molecularization in assemblies of few electrons
in spherical confinements (the socalled harmonium atoms), see Refs.\
\cite{alav04,cios06,cios14,cios17}. For experimental signatures of WM
formation in 3D self-assembled QDs, see Refs.\ \cite{honi14,mint16,mint18}.}
semiconductor QDs and 2D moir\'e QDs in TMD materials (see Sec.\ \ref{mqd}), as well
as neutral trapped ultracold bosonic and fermionic natural atoms (see Secs.\ \ref{3aplii}
and \ref{lllrot}) and ultracold Coulombic ions in radiofrequency traps (see Secs.\
\ref{3ipliii} and \ref{1diontrap}). The semiconductor and TMD QDs comprise either electrons
or holes and are referred to generally as artificial atoms and molecules, a terminology
adopted also here to trapped ultracold neutral atoms and Coulombic ions.

At a preliminary level \cite{lozo90,bolt93,beda94,lozo99,kong02,piac04},
a Wigner molecule is often viewed as a finite small fragment of the broken-symmetry bulk
Wigner crystal governed by classical dynamics, as was considered by Eugene Wigner in 1934
\cite{wign34}. The present paper reviews the WM literature of the last
25 years which established Wigner molecularization as a fully quantum process emerging (in
the regime of strong interparticle repulsive correlations) from the exact solutions of the
many-body Schr\"odinger equation, or related approximations organized in a well defined
hierarchical scheme leading to the exact solutions (see Sec.\ \ref{hieras}). The regime
of strong repulsive correlations arises for large values of the Wigner parameter $R_W$ (or
$R_\delta$) (see Sec.\ \ref{intro_rw}), large values of an applied magnetic field, or rapid
rotation of trapped particles (see Sec.\ \ref{intro_mag}). The classical WMs investigated in
Refs.\ \cite{lozo90,bolt93,beda94,lozo99,kong02,piac04}
(referred to also as Coulomb clusters) emerge for the Schr\"odinger equation in the limit
$R_W \rightarrow\infty$ (when the quantum kinetic energy can be neglected), in conjuction with
the implicit assumption of symmetry breaking.

Beyond the broken-symmetry, geometric point-charge configurations identified in the classical
Coulombic-cluster treatments, the quantal aspects of WMs include the zero-point broadening (see
Sec.\ \ref{qdrr}) of the localized-particle orbitals (considered already by Wigner in Ref.\
\cite{wign38}),\footnote{
This ``blurring'' quantal effect was stressed recently in the experimental Ref.\ \cite{yazd24}.}
and most importantly the preservation of the Hamiltonian symmetries,
which is a fundamental property of the exact solutions of the many-body Schr\"odinger equation.
(see, e.g., Secs.\ \ref{2eqdex} and \ref{34eqdb}).

Wigner molecules that preserve the Hamiltonian symmetries are referred to as rotating WMs
(in circular semiconductor 2D QDs) or sliding WMs (in triangular TMD moir\'e QDs). The
corresponding charge densities do not exhibit explicit symmetry breaking. However, the
rotating or sliding WMs exhibit a hidden or emergent symmetry breaking, which is revealed
in higher-order correlation functions. WMs with explicit symmetry breaking in the charge
densities (referred to as pinned or static) appear when a perturbing potential term is
added in the Hamiltonian. The corresponding solutions of the Schr\"odinger equation are
then superpositions of stationary eigenstates, which in the thermodynamic limit
\cite{pwa_book} or the $R_W \rightarrow \infty$ limit (see Sec.\ \ref{comm4}) form the
so-called Anderson ``tower of states''. Reversely, the symmetry-preserving WMs are
superpositions, in their turn, of broken-symmetry pinned WMs, a fact best understood
through the step of symmetry restoration via projection techniques in the hierarchical
scheme (see Sec.\ \ref{comm2}).

Control of the perturbing potential (including modifications of the confinement) enables the
interplay between symmetry-preserving and broken-symmetry (pinned) WMs, as was theoretically
elaborated in Ref.\ \cite{yann24.2} (see Sec.\ \ref{mqd}). In a pioneering experimental
investigation \cite{crom24}, the transition from a sliding WM (with a
$C_3$-symmetry-preserving charge density, independent of the number, $N$, of trapped holes)
to a pinned WM (with a CD autonomously exhibiting exactly $N$ localized particle humps) as
a function of the applied uniaxial strain on a TMD moir\'e QD has been recently demonstrated
(see Fig.\ \ref{moire_sci1}).
Moreover, a remarkable experiment \cite{nogu14} involving heavy ions demonstrated the
reversible transition from a rotating WM (with a continuous circular CD) consisting of three
$^{40}$Ca$^+$ ultracold ions to a pinned WM (exhibiting three localized ions), as a function
of the ellipticity of the 2D confinement in a radio-frequency linear Paul trap (see Sec.\
\ref{3ipliii}).

Wigner molecularization under an applied magnetic field $B$ or rotation $\Omega$ was reviewed
in Sec.\ \ref{wmmagrot}. An important finding discussed in Secs.\ \ref{lllrot} and \ref{lllfp}
is that the RWMs relate directly to the physics of the fractional quantum Hall effect,
including experimental realizations \cite{joch24} of non-traditional FQHE states of few spinful
neutral ultracold atoms in rapidly rotating traps. 

The controlled transition between symmetry-preserving and broken-symmetry states in
nanosystems points towards a reappraisal of the concept of spontaneous symmetry breaking,
which requires invocation of the singular thermodynamic limit \cite{weze19}. Indeed,
this transition has been shown to be highly abrupt for minute perturbations in the limiting
case of relatively small kinetic energy (a large $R_W$ falls in this category), leading to the
``flea on the elephant'' effect (see Sec.\ \ref{comm5}), which has been explicitly demonstrated
within the theoretical framework of the Schr\"odinger equation. Furthermore, it has become
apparent \cite{yann07,tasaki_book,land20,fras16,pape25} that the attributes of SSB must be
present already in a finite system in the regime of strong correlations, as was explicitly
shown for the process of Wigner molecularization throughout this review (see, e.g., Sec.\
\ref{comm4}). Accordingly, the terms emergent symmetry breaking \cite{yann07,pape15,pape25}
or obscure symmetry breaking \cite{tasa94,tasaki_book} (both reflecting the finiteness of
actual physical systems) have been introduced, with the former preferred in this review.
Two main attributes of ESB (shared with the concept of SSB) are those of autonomy and
universality (see Secs.\ \ref{comm6} and \ref{comm8}).

At a fundamental-physics perspective, the present review establishes how the ESB and Wigner
molecularization are rooted in the properties of the solutions of the many-body Schr\"odinger
equation, a fact that classifies them in the category of weak (epistemological) emergence 
(see Sec.\ \ref{intro_emerg}). In particular, the review reveals an overlooked until now
``arrow of explanation ... traced down to the level of the quantum mechanics of electrons
[and] atomic nuclei'', according to Weinberg's reductionist approach \cite{wein87}.

Just as importantly, beyond the fundamental-physics aspects, the widespread recent experimental
demonstration, that Wigner molecularization plays a central role in shaping the
quantum properties of solid-state qubits and manmade quantum devices and materials (see Sec.\
\ref{intro}), positions the present review to further contribute to future advances in
the emerging field of quantum-information technology. In this context, to be noted is the
great potential for future developments and technological applications of the relevant
computational tools (i.e., symmetry restoration and configuration interaction), following
anticipated rapid advances in High Performance Computing.

\section{Special dedication}
This review, completed in March 2026, is a tribute to the 100-year anniversary of the
Schr\"odinger equation.

\ack{
The author is indebted to his earlier co-authors at Georgia Tech with whom he had the privilege
to collaborate over the past years on this exciting line of research. In particular,
Prof. Uzi Landman, Director of the Center for Computational Materials Science, as well as
Yuesong Li, Ying Li, Igor Romanovsky, Leslie Baksmaty, Benedikt Brandt, and Arnon Goldberg.
The conceptualization and writing of a first-draft outline proceeded while the author was an
active faculty member of the School of Physics, Georgia Institute of Technology. The fruition
of the manuscript in its final form was completed after the author's retirement.
}  

\appendix

\section{Appendix The Configuration Interaction Method}
\label{a1}

The full configuration interaction (FCI) methodology has a long history, starting in
quantum chemistry; see Refs.\ \cite{shav98,szabo_book}. The method was adapted to two
dimensional problems and found extensive applications in the fields of semiconductor
quantum dots
\cite{yann03,mikh02,szaf03,ront06,yann07.2,yann07,yann09,erca21.2,yann22.2,yann22.3},
of the fractional quantum Hall effect \cite{yann04,shi07,yann21,yann25}, moir\'e
TMD quantum dots \cite{yann24.2}, and trapped ultracold atoms \cite{yann15}.

The reader will find a comprehensive exposition of the 2D CI in Appendix B of Ref.\
\cite{yann22.2} and the Supporting Information of Ref.\ \cite{yann15}, where the method
was applied to GaAs double-quantum-dot quantum computer qubits and to untracold fermions
trapped in a double well, respectively.
In the application to moir\'{e} QDs, similar space orbitals,
$\varphi_j(x,y)$, $j=1,2,\ldots,K$, were kept; they were employed in the building of the 
single-particle basis of spin-orbitals used to construct the Slater determinants $\Psi_I^N$,
which span the many-body Hilbert space [see Eq.\ (\ref{mbwf}) in Sec.\ \ref{fci}, the index
$I$ counts the Slater determinants].

Most often, the space orbitals $\varphi_j(x,y)$ are determined as solutions (in Cartesian
coordinates) of a simpler auxiliary Hamiltonian
\begin{equation}
  H_{\rm aux}=\frac{{\bf p}^2}{2m^*} + V_{\rm aux}(x,y),
\label{haux}
\end{equation}
where $V_{\rm aux}(x,y)$ has in general an elliptic (including circular) shape for single
wells or a two-center-oscillator shape (see Secs.\ \ref{3aplii} and \ref{3epliv}) for
double wells, but without the smooth neck. The remaining Hamiltonian
${\cal H}_{\rm MB}-H_{\rm aux}$ contains the interaction terms, the magnetic-field contribution,
and single-particle potential deviations from the shapes in $H_{\rm aux}$. A sparse-matrix
eigensolver \cite{arpack} based on Implicitly Restarted Arnoldi methods to diagonalize the
resulting many-body Hamiltonian matrix is employed. 

The smooth-neck (one-body) and Coulomb and contact-interaction (two-body) matrix elements
required for the sparse-matrix diagonalization are calculated numerically as described in Ref.\
\cite{yann22.2} and \cite{yann15}. Similarly, the matrix elements between the orbitals
$\varphi_i(x,y)$ and $\varphi_j(x,y)$ of the trilobal (one-body) term in the moir\'{e} QD
confinement [see Eq.\ (\ref{mpot})] are also calculated numerically.

In the case of the LLL, the space orbitals $\varphi_j(x,y)$ are given by the corresponding
Darwin-Fock nodeless ones (see Sec. \ref{dfll}) with non-negative angular momenta
$l=0, 1, 2, \ldots$. In this case, The two-body matrix elements of the Coulomb \cite{tsip02},
as well as the contact \cite{pape99,yann25}, interaction are given by closed analytic
expressions.

\section{Appendix The Spin-and-Space Unrestricted Hartree-Fock and Symmetry Restoration}
\label{a2}

Early on in the field of 2D semiconductor QDs, the spin-and-space unrestricted Hartree-Fock
(sS-UHF) was employed in Ref.\ \cite{yann99} to describe formation of Wigner molecules at
the mean-field level. This methodology employs the Pople-Nesbet equations
\cite{szabo_book,yann07}. The sS-UHF WMs are self-consistent solutions of the Pople-Nesbet
equations that are obtained by relaxing both the total-spin and space symmetry requirements
(which are implemented in the restricted Hartree Fock, RHF). For a detailed description of the
Pople-Nesbet equations in the context of three-dimensional natural atoms and molecules, see
Ch.\ 3.8 in Ref.\ \cite{szabo_book}. For a detailed description of the Pople-Nesbet equations
in the context of two-dimensional artificial atoms and semiconductor quantum dots, see Sec.\
2.1 of Ref.\ \cite{yann07}. Convergence of the self-consistent iterations was achieved in all
cases by mixing the input and output charge densities at each iteration step. The convergence
criterion was set to a difference of $10^{-12}$ meV between the input and output total UHF
energies at the same iteration step.    

Note that the book of Szabo and Ostlund \cite{szabo_book} does not describe the
post-Hatree-Fock theory of symmetry restoration. For a detailed description of the theory of
symmetry restoration, see Sec.\ 2.2 in Ref.\ \cite{yann07} and Ref.\ \cite{shei21}.

\section{Appendix Charge Densities from FCI and UHF Wave Functions}
\label{a3}

The FCI single-particle density (also referred to as charge density or electron density in
the case of charged particles) is defined as the expectation value of a one-body operator
as follows:
\begin{equation}
\rho({\bf r}) = \langle \Phi^{\rm FCI}
\vert  \sum_{i=1}^N \delta({\bf r}-{\bf r}_i)
\vert \Phi^{\rm FCI} \rangle,
\label{elden}
\end{equation}
where $\Phi^{\rm FCI}$ denotes the many-body (multi-determinantal) FCI wave function, namely,
\begin{equation}
\Phi^{\rm FCI} ({\bf r}_1, \ldots , {\bf r}_N) =
\sum_I C_I \Psi_I({\bf r}_1, \ldots , {\bf r}_N),
\label{mbwf2}
\end{equation}
with $\Psi_I({\bf r)}$ denoting the Slater determinants that span the many-body Hilbert space. 

For the sS-UHF case, one substitutes $\Phi^{\rm FCI}$ in Eq.\ (\ref{elden}) with the 
single-determinant, $\Psi^{\rm UHF}({\bf r})$, solution of the Pople-Nesbet equations. 
$\Psi^{\rm UHF}({\bf r})$ is built out from the UHF spin-orbitals whose space part has the form:
\begin{equation}
u^\alpha_i=\sum_{\mu=1}^K {\cal C}^\alpha_{\mu i} \varphi_\mu, \;\;\; i=1,\ldots,K,
\end{equation}
and
\begin{equation}
u^\beta_i=\sum_{\mu=1}^K {\cal C}^\beta_{\mu i} \varphi_\mu, \;\;\; i=1,\ldots,K,
\end{equation}
where the expansion coefficients ${\cal C}^\alpha_{\mu i}$ and ${\cal C}^\beta_{\mu i}$ are
solutions of the Pople-Nesbet equations.

\bibliographystyle{my-iopart-num-long}
\bibliography{WM_topical_review}

@article{yann99,
  title = {{Spontaneous Symmetry Breaking in Single and Molecular Quantum Dots}},
  author = {Yannouleas, Constantine and Landman, Uzi},
  journal = {Phys. Rev. Lett.},
  volume = {82},
  issue = {26},
  pages = {5325--5328},
  numpages = {0},
  year = {1999},
  month = {Jun},
  publisher = {American Physical Society},
  doi = {10.1103/PhysRevLett.82.5325},
  url = {https://link.aps.org/doi/10.1103/PhysRevLett.82.5325}
}

@article{yann00,
  title = {{Collective and Independent-Particle Motion in Two-Electron Artificial Atoms}},
  author = {Yannouleas, Constantine and Landman, Uzi},
  journal = {Phys. Rev. Lett.},
  volume = {85},
  issue = {8},
  pages = {1726--1729},
  numpages = {0},
  year = {2000},
  month = {Aug},
  publisher = {American Physical Society},
  doi = {10.1103/PhysRevLett.85.1726},
  url = {https://link.aps.org/doi/10.1103/PhysRevLett.85.1726}
}

@article{yann07,
	doi = {10.1088/0034-4885/70/12/r02},
	url = {https://doi.org/10.1088/0034-4885/70/12/r02},
	year = 2007,
	month = {dec},
	publisher = {{IOP} Publishing},
	volume = {70},
	issue = {12},
	pages = {2067--2148},
	author = {Constantine Yannouleas and Uzi Landman},
	title = {Symmetry breaking and quantum correlations in finite
	  systems: {Studies} of quantum dots and ultracold {Bose} gases and related nuclear and 
          chemical methods},
          journal = {Reports on Progress in Physics},
}

@article{yann22,
	doi = {10.1088/1361-648x/ac5c28},
	url = {https://doi.org/10.1088/1361-648x/ac5c28},
	year = 2022,
	month = {mar},
	publisher = {{IOP} Publishing},
	volume = {34},
	issue = {21},
	pages = {21LT01},
	author = {Constantine Yannouleas and Uzi Landman},
	title = {{Wigner} molecules and hybrid qubits},
	journal = {J. Phys.: Condens. Matter (Letter)}
}

@article{yann22.2,
  title = {Molecular formations and spectra due to electron correlations in three-electron
           hybrid double-well qubits},
  author = {Yannouleas, Constantine and Landman, Uzi},
  journal = {Phys. Rev. B},
  volume = {105},
  issue = {20},
  pages = {205302},
  numpages = {14},
  year = {2022},
  month = {May},
  publisher = {American Physical Society},
  doi = {10.1103/PhysRevB.105.205302},
  url = {https://link.aps.org/doi/10.1103/PhysRevB.105.205302}
}

@Article{mak21,
author={Li, Tingxin
and Zhu, Jiacheng
and Tang, Yanhao
and Watanabe, Kenji
and Taniguchi, Takashi
and Elser, Veit
and Shan, Jie
and Mak, Kin Fai},
title={Charge-order-enhanced capacitance in semiconductor {moir{\'e}} superlattices},
journal={Nature Nanotechnology},
year={2021},
month={Oct},
day={01},
volume={16},
issue={10},
pages={1068-1072},
issn={1748-3395},
doi={10.1038/s41565-021-00955-8},
url={https://doi.org/10.1038/s41565-021-00955-8}
}

@Article{feld22,
   author = {Carlos R. Kometter and Jiachen Yu and Trithep Devakul and Aidan P. Reddy and Yang Zhang and 
     Benjamin A. Foutty and Kenji Watanabe and Takashi Taniguchi and Liang Fuand Benjamin E. Feldman},
      title     = {Hofstadter states and reentrant charge order in a semiconductor {moir{\'e}} lattice},
      journal   = {},
      volume    = {},
      year      = {},
      pages     = {},
      eprint    = {arXiv:2212.05068},
      doi       = {10.48550/arXiv.2212.05068},
      url       = {https://arxiv.org/abs/2212.05068}
}

@article{yann03,
  title = {Two-dimensional quantum dots in high magnetic fields: {Rotating-electron-molecule}
           versus composite-fermion approach},
  author = {Yannouleas, Constantine and Landman, Uzi},
  journal = {Phys. Rev. B},
  volume = {68},
  issue = {3},
  pages = {035326},
  numpages = {11},
  year = {2003},
  month = {Jul},
  publisher = {American Physical Society},
  doi = {10.1103/PhysRevB.68.035326},
  url = {https://link.aps.org/doi/10.1103/PhysRevB.68.035326}
}

@article{yann06.3,
  title = {From a few to many electrons in quantum dots under strong magnetic fields:
           Properties of rotating electron molecules with multiple rings},
  author = {Li, Yuesong and Yannouleas, Constantine and Landman, Uzi},
  journal = {Phys. Rev. B},
  volume = {73},
  issue = {7},
  pages = {075301},
  numpages = {14},
  year = {2006},
  month = {Feb},
  publisher = {American Physical Society},
  doi = {10.1103/PhysRevB.73.075301},
  url = {https://link.aps.org/doi/10.1103/PhysRevB.73.075301}
}

@article{yang07,
  title = {Spin-dependent rotating {Wigner} molecules in quantum dots},
  author = {Dai, Zhensheng and Zhu, Jia-Lin and Yang, Ning and Wang, Yuquan},
  journal = {Phys. Rev. B},
  volume = {76},
  issue = {8},
  pages = {085308},
  numpages = {8},
  year = {2007},
  month = {Aug},
  publisher = {American Physical Society},
  doi = {10.1103/PhysRevB.76.085308},
  url = {https://link.aps.org/doi/10.1103/PhysRevB.76.085308}
}

@article{szaf03,
  title = {Four-electron quantum dot in a magnetic field},
  author = {Tavernier, M. B. and Anisimovas, E. and Peeters, F. M. and Szafran, B.
            and Adamowski, J. and Bednarek, S.},
  journal = {Phys. Rev. B},
  volume = {68},
  issue = {20},
  pages = {205305},
  numpages = {9},
  year = {2003},
  month = {Nov},
  publisher = {American Physical Society},
  doi = {10.1103/PhysRevB.68.205305},
  url = {https://link.aps.org/doi/10.1103/PhysRevB.68.205305}
}

@article{mikh02,
  title = {Two ground-state modifications of quantum-dot beryllium},
  author = {Mikhailov, S. A.},
  journal = {Phys. Rev. B},
  volume = {66},
  issue = {15},
  pages = {153313},
  numpages = {4},
  year = {2002},
  month = {Oct},
  publisher = {American Physical Society},
  doi = {10.1103/PhysRevB.66.153313},
  url = {https://link.aps.org/doi/10.1103/PhysRevB.66.153313}
}

@article{mikh02.2,
  title = {Quantum-dot lithium in zero magnetic field: Electronic properties,
           thermodynamics, and Fermi liquid--Wigner solid crossover in the ground state},
  author = {Mikhailov, S. A.},
  journal = {Phys. Rev. B},
  volume = {65},
  issue = {11},
  pages = {115312},
  numpages = {12},
  year = {2002},
  month = {Feb},
  publisher = {American Physical Society},
  doi = {10.1103/PhysRevB.65.115312},
  url = {https://link.aps.org/doi/10.1103/PhysRevB.65.115312}
}

@article{yann02.2,
	doi = {10.1088/0953-8984/14/34/101},
	url = {https://doi.org/10.1088/0953-8984/14/34/101},
	year = 2002,
	month = {aug},
	publisher = {{IOP} Publishing},
	volume = {14},
	issue = {34},
	pages = {L591--L598},
	author = {Constantine Yannouleas and Uzi Landman},
	title = {Strongly correlated wavefunctions for artificial atoms and molecules},
	journal = {Journal of Physics: Condensed Matter},
}

@article{song22,
author = {Song, Zhigang and Wang, Yu and Zheng, Haimei and Narang, Prineha and Wang,
          Lin-Wang},
title = {Deep Quantum-Dot Arrays in {moir\'{e}} Superlattices of Non-van der {Waals}
         Materials},
journal = {Journal of the American Chemical Society},
volume = {144},
issue = {32},
pages = {14657-14667},
year = {2022},
doi = {10.1021/jacs.2c04390},
URL = {https://doi.org/10.1021/jacs.2c04390},
}

@article{feen18,
author = {Pan, Yi and {F\"{o}lsch}, Stefan and Nie, Yifan and Waters, Dacen and Lin, Yu-Chuan and 
Jariwala, Bhakti and Zhang, Kehao and Cho, Kyeongjae and Robinson, Joshua A. and Feenstra, Randall M.},
title = {Quantum-Confined Electronic States Arising from the {moir\'{e}} Pattern of {MoS$_2$–WSe$_2$} 
Heterobilayers},
journal = {Nano Letters},
volume = {18},
issue = {3},
pages = {1849-1855},
year = {2018},
doi = {10.1021/acs.nanolett.7b05125},
URL = {https://doi.org/10.1021/acs.nanolett.7b05125},
eprint = {https://doi.org/10.1021/acs.nanolett.7b05125},
}

@article{zeng22,
  title = {Strong modulation limit of excitons and trions in {moir\'{e}} materials},
  author = {Zeng, Yongxin and MacDonald, Allan H.},
  journal = {Phys. Rev. B},
  volume = {106},
  issue = {3},
  pages = {035115},
  numpages = {15},
  year = {2022},
  month = {Jul},
  publisher = {American Physical Society},
  doi = {10.1103/PhysRevB.106.035115},
  url = {https://link.aps.org/doi/10.1103/PhysRevB.106.035115}
}

@article{yann09,
  title = {Artificial quantum-dot {Helium} molecules: Electronic spectra,
           spin structures, and {Heisenberg} clusters},
  author = {Li, Ying and Yannouleas, Constantine and Landman, Uzi},
  journal = {Phys. Rev. B},
  volume = {80},
  issue = {4},
  pages = {045326},
  numpages = {17},
  year = {2009},
  month = {Jul},
  publisher = {American Physical Society},
  doi = {10.1103/PhysRevB.80.045326},
  url = {https://link.aps.org/doi/10.1103/PhysRevB.80.045326}
}

@article{shei21,
doi = {10.1088/1361-6471/ac288a},
url = {https://dx.doi.org/10.1088/1361-6471/ac288a},
year = {2021},
month = {nov},
publisher = {IOP Publishing},
volume = {48},
issue = {12},
pages = {123001},
author = {J A Sheikh and J Dobaczewski and P Ring and L M Robledo and C Yannouleas},
title = {Symmetry restoration in mean-field approaches},
journal = {Journal of Physics G: Nuclear and Particle Physics},
}

@article{shav98,
author = {Isaiah Shavitt},
title = {The history and evolution of configuration interaction},
journal = {Molecular Physics},
volume = {94},
issue = {1},
pages = {3-17},
year  = {1998},
publisher = {Taylor & Francis},
doi = {10.1080/002689798168303},
URL = {https://www.tandfonline.com/doi/abs/10.1080/002689798168303}
}

@article{ront06,
author = {Rontani,Massimo  and Cavazzoni,Carlo  and Bellucci,Devis  and Goldoni,Guido},
title = {Full configuration interaction approach to the few-electron problem in
         artificial atoms},
journal = {The Journal of Chemical Physics},
volume = {124},
issue = {12},
pages = {124102},
year = {2006},
doi = {10.1063/1.2179418},
URL = { https://doi.org/10.1063/1.2179418},
}

@article{yann22.3,
  title = {Valleytronic full configuration-interaction approach: Application to the
           excitation spectra of Si double-dot qubits},
  author = {Yannouleas, Constantine and Landman, Uzi},
  journal = {Phys. Rev. B},
  volume = {106},
  issue = {19},
  pages = {195306},
  numpages = {19},
  year = {2022},
  month = {Nov},
  publisher = {American Physical Society},
  doi = {10.1103/PhysRevB.106.195306},
  url = {https://link.aps.org/doi/10.1103/PhysRevB.106.195306}
}

@article{yann16,
author = {Constantine Yannouleas and Benedikt B. Brandt and Uzi Landman},
title = {Ultracold few fermionic atoms in needle-shaped double wells: spin chains and
         resonating spin clusters from microscopic {Hamiltonians} emulated via
	 antiferromagnetic {Heisenberg} and $t-{J}$ models},
journal = {New Journal of Physics},
year = {2016},
month = {jul},
publisher = {{IOP} Publishing},
volume = {18},
issue = {7},
pages = {073018},
doi = {10.1088/1367-2630/18/7/073018},
url = {https://doi.org/10.1088/1367-2630/18/7/073018}
}

@Article{wang21,
author={Li, Hongyuan
and Li, Shaowei
and Regan, Emma C.
and Wang, Danqing
and Zhao, Wenyu
and Kahn, Salman
and Yumigeta, Kentaro
and Blei, Mark
and Taniguchi, Takashi
and Watanabe, Kenji
and Tongay, Sefaattin
and Zettl, Alex
and Crommie, Michael F.
and Wang, Feng},
title={Imaging two-dimensional generalized {Wigner} crystals},
journal={Nature},
year={2021},
month={Sep},
day={01},
volume={597},
issue={7878},
pages={650-654},
issn={1476-4687},
doi={10.1038/s41586-021-03874-9},
url={https://doi.org/10.1038/s41586-021-03874-9}
}

@article{yann04,
  title = {Structural properties of electrons in quantum dots in high magnetic fields:
           Crystalline character of cusp states and excitation spectra},
  author = {Yannouleas, Constantine and Landman, Uzi},
  journal = {Phys. Rev. B},
  volume = {70},
  issue = {23},
  pages = {235319},
  numpages = {9},
  year = {2004},
  month = {Dec},
  publisher = {American Physical Society},
  doi = {10.1103/PhysRevB.70.235319},
  url = {https://link.aps.org/doi/10.1103/PhysRevB.70.235319}
}

@book{szabo_book,
  title     = "{Modern Quantum Chemistry}",
  author    = "A. Szabo and N. S. Ostlund",
  year      = "1989",
  publisher = "McGraw-Hill",
  address   = "New York",
  note      = "{For the Slater-Condon rules, see Chap. 4}"
}

@book{pranbook,
  title     = "{{The Quantum Hall Effect}}",
  editor    = "Richard E. Prange and Steven M. Girvin",
  year      = "1990",
  publisher = "Springer",
  address   = "New York, NY",
  doi = "10.1007/978-1-4612-3350-3",
  url = "https://doi.org/10.1007/978-1-4612-3350-3"
}

@incollection{laug90,
  title = "{Elementary Theory: The Incompressible Quantum Fluid} in",
  booktitle     = "{{The Quantum Hall Effect}}",
  author = "Robert B. Laughlin",
  editor    = "Richard E. Prange and Steven M. Girvin",
  chapter = {7},
  pages = {233-301},
  year      = "1990",
  publisher = "Springer",
  address   = "New York, NY",
  doi = "10.1007/978-1-4612-3350-3_7",
  url = "https://doi.org/10.1007/978-1-4612-3350-3_7"
}

@article{yann06,
  title = {{Excitation Spectrum of Two Correlated Electrons in a Lateral Quantum Dot with
            Negligible Zeeman Splitting}},
  author = {Ellenberger, C. and Ihn, T. and Yannouleas, C. and Landman, U. and Ensslin, K.
            and Driscoll,   D. and Gossard, A. C.},
  journal = {Phys. Rev. Lett.},
  volume = {96},
  issue = {12},
  pages = {126806},
  numpages = {4},
  year = {2006},
  month = {Mar},
  publisher = {American Physical Society},
  doi = {10.1103/PhysRevLett.96.126806},
  url = {https://link.aps.org/doi/10.1103/PhysRevLett.96.126806}
}

@article{kim21,
author = {Jang, Wonjin and Cho, Min-Kyun and Jang, Hyeongyu and Kim, Jehyun and Park, Jaemin and Kim, Gyeonghun and Kang, Byoungwoo and Jung, Hwanchul and Umansky, Vladimir and Kim, Dohun},
title = {{Single-Shot Readout of a Driven Hybrid Qubit in a GaAs Double Quantum Dot}},
journal = {Nano Letters},
volume = {21},
issue = {12},
pages = {4999-5005},
year = {2021},
doi = {10.1021/acs.nanolett.1c00783},
URL =    {https://doi.org/10.1021/acs.nanolett.1c00783},
}

@article{urie21,
  title = {Two-body {Wigner} molecularization in asymmetric quantum dot spin qubits},
    author = {Abadillo-Uriel, Jos\'e C. and Martinez, Biel and Filippone, Michele and
              Niquet, Yann-Michel},
      journal = {Phys. Rev. B},
        volume = {104},
	  issue = {19},
	    pages = {195305},
	      numpages = {17},
	        year = {2021},
		  month = {Nov},
		    publisher = {American Physical Society},
		      doi = {10.1103/PhysRevB.104.195305},
		        url = {https://link.aps.org/doi/10.1103/PhysRevB.104.195305}
			}

@article{corr21,
  title = {Coherent Control and Spectroscopy of a Semiconductor Quantum Dot {Wigner} Molecule},
    author = {Corrigan, J. and Dodson, J. P. and Ercan, H. Ekmel and Abadillo-Uriel, J. C. and Thorgrimsson, Brandur and Knapp, T. J. and Holman, Nathan and McJunkin, Thomas and Neyens, Samuel F. and MacQuarrie, E. R. and Foote, Ryan H. and Edge, L. F. and Friesen, Mark and Coppersmith, S. N. and Eriksson, M. A.},
      journal = {Phys. Rev. Lett.},
        volume = {127},
	  issue = {12},
	    pages = {127701},
	      numpages = {6},
	        year = {2021},
		  month = {Sep},
		    publisher = {American Physical Society},
		      doi = {10.1103/PhysRevLett.127.127701},
		        url = {https://link.aps.org/doi/10.1103/PhysRevLett.127.127701}
			}

@article{erca21.2,
  title = {Strong electron-electron interactions in {Si/SiGe} quantum dots},
    author = {Ercan, H. Ekmel and Coppersmith, S. N. and Friesen, Mark},
      journal = {Phys. Rev. B},
        volume = {104},
	  issue = {23},
	    pages = {235302},
	      numpages = {22},
	        year = {2021},
		  month = {Dec},
		    publisher = {American Physical Society},
		      doi = {10.1103/PhysRevB.104.235302},
		        url = {https://link.aps.org/doi/10.1103/PhysRevB.104.235302}
			}

@Article{Peck13,
author={Pecker, S.
and Kuemmeth, F.
and Secchi, A.
and Rontani, M.
and Ralph, D. C.
and McEuen, P. L.
and Ilani, S.},
title={Observation and spectroscopy of a two-electron {Wigner} molecule in an ultraclean
       carbon nanotube},
journal={Nature Physics},
year={2013},
month={Sep},
day={01},
volume={9},
issue={9},
pages={576-581},
issn={1745-2481},
doi={10.1038/nphys2692},
url={https://doi.org/10.1038/nphys2692}
}

@article{macd18,
  title = {Hubbard Model Physics in Transition Metal Dichalcogenide {moir\'e} Bands},
  author = {Wu, Fengcheng and Lovorn, Timothy and Tutuc, Emanuel and MacDonald, A. H.},
  journal = {Phys. Rev. Lett.},
  volume = {121},
  issue = {2},
  pages = {026402},
  numpages = {5},
  year = {2018},
  month = {Jul},
  publisher = {American Physical Society},
  doi = {10.1103/PhysRevLett.121.026402},
  url = {https://link.aps.org/doi/10.1103/PhysRevLett.121.026402}
}

@article{burk23,
  title = {Semiconductor spin qubits},
  author = {Burkard, Guido and Ladd, Thaddeus D. and Pan, Andrew and Nichol,
            John M. and Petta, Jason R.},
  journal = {Rev. Mod. Phys.},
  volume = {95},
  issue = {2},
  pages = {025003},
  numpages = {58},
  year = {2023},
  month = {Jun},
  publisher = {American Physical Society},
  doi = {10.1103/RevModPhys.95.025003},
  url = {https://link.aps.org/doi/10.1103/RevModPhys.95.025003}
}

@article{yann07.3,
  title = {Rapidly rotating boson molecules with long- or short-range repulsion:
           An exact diagonalization study},
  author = {Baksmaty, Leslie O. and Yannouleas, Constantine and Landman, Uzi},
  journal = {Phys. Rev. A},
  volume = {75},
  issue = {2},
  pages = {023620},
  numpages = {14},
  year = {2007},
  month = {Feb},
  publisher = {American Physical Society},
  doi = {10.1103/PhysRevA.75.023620},
  url = {https://link.aps.org/doi/10.1103/PhysRevA.75.023620}
}

@Article{yann15,
author={Brandt, Benedikt B.
and Yannouleas, Constantine
and Landman, Uzi},
title={Double-Well Ultracold-Fermions Computational Microscopy: Wave-Function Anatomy of
Attractive-Pairing and {Wigner}-Molecule Entanglement and Natural Orbitals},
journal={Nano Letters},
year={2015},
month={Oct},
day={14},
publisher={American Chemical Society},
volume={15},
issue={10},
pages={7105-7111},
issn={1530-6984},
doi={10.1021/acs.nanolett.5b03199},
url={https://doi.org/10.1021/acs.nanolett.5b03199}
}

@article{yann21,
  title = {Exact closed-form analytic wave functions in two dimensions: Contact-interacting
           fermionic spinful ultracold atoms in a rapidly rotating trap},
  author = {Yannouleas, Constantine and Landman, Uzi},
  journal = {Phys. Rev. Research},
  volume = {3},
  issue = {3},
  pages = {L032028},
  numpages = {7},
  year = {2021},
  month = {Jul},
  publisher = {American Physical Society},
  doi = {10.1103/PhysRevResearch.3.L032028},
  url = {https://link.aps.org/doi/10.1103/PhysRevResearch.3.L032028}
}

@article{roma06,
  title = {Bosonic Molecules in Rotating Traps},
  author = {Romanovsky, Igor and Yannouleas, Constantine and Baksmaty, Leslie O. and Landman,
            Uzi},
  journal = {Phys. Rev. Lett.},
  volume = {97},
  issue = {9},
  pages = {090401},
  numpages = {4},
  year = {2006},
  month = {Aug},
  publisher = {American Physical Society},
  doi = {10.1103/PhysRevLett.97.090401},
  url = {https://link.aps.org/doi/10.1103/PhysRevLett.97.090401}
}

@Article{kim23,
author={Jang, Wonjin
and Kim, Jehyun
and Park, Jaemin
and Kim, Gyeonghun
and Cho, Min-Kyun
and Jang, Hyeongyu
and Sim, Sangwoo
and Kang, Byoungwoo
and Jung, Hwanchul
and Umansky, Vladimir
and Kim, Dohun},
title={Wigner-molecularization-enabled dynamic nuclear polarization},
journal={Nature Communications},
year={2023},
month={May},
day={23},
volume={14},
issue={1},
pages={2948},
doi={10.1038/s41467-023-38649-5},
url={https://doi.org/10.1038/s41467-023-38649-5}
}

@article{yann06.2,
    author = {Constantine Yannouleas  and Uzi Landman },
    title = {Electron and boson clusters in confined geometries: Symmetry breaking in
             quantum dots and harmonic traps},
    journal = {Proceedings of the National Academy of Sciences},
    volume = {103},
    issue = {28},
    pages = {10600-10605},
    year = {2006},
    doi = {10.1073/pnas.0509041103},
    URL = {https://www.pnas.org/doi/abs/10.1073/pnas.0509041103},
}

@Inproceedings{kouw97,
  author    = "L. P. Kouwenhoven and C. M. Marcus and P. L. McEuen and S. Tarucha and 
               R. M. Westervelt and N. S. Wingreen",
  title     = "Electron transport in quantum dots",
  booktitle = "Mesoscopic Electron Transport",
  editor    = "{L. L. Sohn, L. P. Kouwenhoven, G. Sch\"{o}n}",
  series    = "",
  year      = "1997",
  pages     = "105--214",
  publisher = "Springer",
  address   = "Berlin"
}

@article{roma09,
  title = {Edge states in graphene quantum dots: Fractional quantum {Hall} effect analogies 
           and differences at zero magnetic field},
  author = {Romanovsky, Igor and Yannouleas, Constantine and Landman, Uzi},
  journal = {Phys. Rev. B},
  volume = {79},
  issue = {7},
  pages = {075311},
  numpages = {12},
  year = {2009},
  month = {Feb},
  publisher = {American Physical Society},
  doi = {10.1103/PhysRevB.79.075311},
  url = {https://link.aps.org/doi/10.1103/PhysRevB.79.075311}
}

@Article{Manz17,
author={Manzeli, Sajedeh
and Ovchinnikov, Dmitry
and Pasquier, Diego
and Yazyev, Oleg V.
and Kis, Andras},
title={2D transition metal dichalcogenides},
journal={Nature Reviews Materials},
year={2017},
month={Jun},
day={13},
volume={2},
number={8},
pages={17033},
issn={2058-8437},
doi={10.1038/natrevmats.2017.33},
url={https://doi.org/10.1038/natrevmats.2017.33}
}

@Article{kaxi20,
author={Carr, Stephen
and Fang, Shiang
and Kaxiras, Efthimios},
title={Electronic-structure methods for twisted {moir{\'e}} layers},
journal={Nature Reviews Materials},
year={2020},
month={Oct},
day={01},
volume={5},
number={10},
pages={748-763},
issn={2058-8437},
doi={10.1038/s41578-020-0214-0},
url={https://doi.org/10.1038/s41578-020-0214-0}
}

@article{fu20,
  title = {{Moir\'{e}} quantum chemistry: Charge transfer in transition metal dichalcogenide superlattices},
  author = {Zhang, Yang and Yuan, Noah F. Q. and Fu, Liang},
  journal = {Phys. Rev. B},
  volume = {102},
  issue = {20},
  pages = {201115},
  numpages = {6},
  year = {2020},
  month = {Nov},
  publisher = {American Physical Society},
  doi = {10.1103/PhysRevB.102.201115},
  url = {https://link.aps.org/doi/10.1103/PhysRevB.102.201115}
}

@article{yann07.2,
  title = {Three-electron anisotropic quantum dots in variable magnetic fields: Exact results
           for excitation spectra, spin structures, and entanglement},
  author = {Li, Yuesong and Yannouleas, Constantine and Landman, Uzi},
  journal = {Phys. Rev. B},
  volume = {76},
  issue = {24},
  pages = {245310},
  numpages = {12},
  year = {2007},
  month = {Dec},
  publisher = {American Physical Society},
  doi = {10.1103/PhysRevB.76.245310},
  url = {https://link.aps.org/doi/10.1103/PhysRevB.76.245310}
}

@article{szaf04,
  title = {Anisotropic quantum dots: Correspondence between quantum and classical {Wigner}
           molecules, parity symmetry, and broken-symmetry states},
  author = {Szafran, B. and Peeters, F. M. and Bednarek, S. and Adamowski, J.},
  journal = {Phys. Rev. B},
  volume = {69},
  issue = {12},
  pages = {125344},
  numpages = {15},
  year = {2004},
  month = {Mar},
  publisher = {American Physical Society},
  doi = {10.1103/PhysRevB.69.125344},
  url = {https://link.aps.org/doi/10.1103/PhysRevB.69.125344}
}

@article{ange21,
author = {Mattia Angeli  and Allan H. MacDonald },
title = {{$\Gamma$}-valley transition metal dichalcogenide {moir\'{e}} bands},
journal = {Proceedings of the National Academy of Sciences},
volume = {118},
number = {10},
pages = {e2021826118},
year = {2021},
doi = {10.1073/pnas.2021826118},
URL = {https://www.pnas.org/doi/abs/10.1073/pnas.2021826118},
}

@article{jain89,
  title = {Composite-fermion approach for the fractional quantum {Hall} effect},
  author = {Jain, J. K.},
  journal = {Phys. Rev. Lett.},
  volume = {63},
  issue = {2},
  pages = {199--202},
  numpages = {0},
  year = {1989},
  month = {Jul},
  publisher = {American Physical Society},
  doi = {10.1103/PhysRevLett.63.199},
  url = {https://link.aps.org/doi/10.1103/PhysRevLett.63.199}
}

@book{jainbook,
  title     = "{Composite Fermions}",
  author    = "J. K. Jain",
  year      = "2007",
  publisher = "Cambridge University Press",
  address   = "Cambridge",
}

@book{arpack,
  title     = "{ARPACK USERS’ GUIDE: Solution of Large-Scale Eigenvalue Problems with
                Implicitly Restarted ARNOLDI Methods}",
  author    = "R. B. Lehoucq and D. C. Sorensen and C. Yang",
  year      = "1998",
  publisher = "SIAM",
  address   = "Philadelphia"
}

@misc{math22,
  author = {{Wolfram Research, Inc.}},
  title = {Mathematica, {V}ersion 13.2},
  url = {https://www.wolfram.com/mathematica},
  note = {{Champaign, IL, 2022}}
}

@article{yann00.2,
  title = {Formation and control of electron molecules in artificial atoms:
           Impurity and magnetic-field effects},
  author = {Yannouleas, Constantine and Landman, Uzi},
  journal = {Phys. Rev. B},
  volume = {61},
  issue = {23},
  pages = {15895--15904},
  numpages = {0},
  year = {2000},
  month = {Jun},
  publisher = {American Physical Society},
  doi = {10.1103/PhysRevB.61.15895},
  url = {https://link.aps.org/doi/10.1103/PhysRevB.61.15895}
}

@article{yann11,
  title = {Unified microscopic approach to the interplay of {pinned-Wigner-solid} and
           liquid behavior of the lowest {Landau-level} states in the neighborhood of
	   $\nu=\frac{1}{3}$},
  author = {Yannouleas, Constantine and Landman, Uzi},
  journal = {Phys. Rev. B},
  volume = {84},
  issue = {16},
  pages = {165327},
  numpages = {17},
  year = {2011},
  month = {Oct},
  publisher = {American Physical Society},
  doi = {10.1103/PhysRevB.84.165327},
  url = {https://link.aps.org/doi/10.1103/PhysRevB.84.165327}
}

@article{yann20,
  title = {Fractional quantum {Hall} physics and higher-order momentum correlations in a
           few spinful fermionic contact-interacting ultracold atoms in rotating traps},
  author = {Yannouleas, Constantine and Landman, Uzi},
  journal = {Phys. Rev. A},
  volume = {102},
  issue = {4},
  pages = {043317},
  numpages = {23},
  year = {2020},
  month = {Oct},
  publisher = {American Physical Society},
  doi = {10.1103/PhysRevA.102.043317},
  url = {https://link.aps.org/doi/10.1103/PhysRevA.102.043317}
}

@article{wign34,
  title = {On the Interaction of Electrons in Metals},
  author = {Wigner, E.},
  journal = {Phys. Rev.},
  volume = {46},
  issue = {11},
  pages = {1002--1011},
  numpages = {0},
  year = {1934},
  month = {Dec},
  publisher = {American Physical Society},
  doi = {10.1103/PhysRev.46.1002},
  url = {https://link.aps.org/doi/10.1103/PhysRev.46.1002}
}

@article{born27,
author = {Born, M. and Oppenheimer, R.},
title = {Zur Quantentheorie der Molekeln},
journal = {Annalen der Physik},
volume = {389},
issue = {20},
pages = {457-484},
doi = {https://doi.org/10.1002/andp.19273892002},
url = {https://onlinelibrary.wiley.com/doi/abs/10.1002/andp.19273892002},
year = {1927}
}

@article{thom15,
author = {Richard C. Thompson},
title = {Ion Coulomb crystals},
journal = {Contemporary Physics},
volume = {56},
issue = {1},
pages = {63--79},
year = {2015},
publisher = {Taylor \& Francis},
doi = {10.1080/00107514.2014.989715},
URL = {https://doi.org/10.1080/00107514.2014.989715},
}

@article{wine87,
  title = {Atomic-Ion Coulomb Clusters in an Ion Trap},
  author = {Wineland, D. J. and Bergquist, J. C. and Itano, Wayne M. and Bollinger, J. J.
            and Manney, C. H.},
  journal = {Phys. Rev. Lett.},
  volume = {59},
  issue = {26},
  pages = {2935--2938},
  numpages = {0},
  year = {1987},
  month = {Dec},
  publisher = {American Physical Society},
  doi = {10.1103/PhysRevLett.59.2935},
  url = {https://link.aps.org/doi/10.1103/PhysRevLett.59.2935}
}

@article{yann23,
  title = {Quantum {Wigner} molecules in moir\'e materials},
  author = {Yannouleas, Constantine and Landman, Uzi},
  journal = {Phys. Rev. B},
  volume = {108},
  issue = {12},
  pages = {L121411},
  numpages = {7},
  year = {2023},
  month = {Sep},
  publisher = {American Physical Society},
  doi = {10.1103/PhysRevB.108.L121411},
  url = {https://link.aps.org/doi/10.1103/PhysRevB.108.L121411}
}

@article{li25,
    author = {Li, Yunzhi and Li, Chen},
    title = {Exact constraint of density functional approximations at the semiclassical limit},
    journal = {The Journal of Chemical Physics},
    volume = {162},
    issue = {17},
    pages = {174112},
    year = {2025},
    month = {05},
    doi = {10.1063/5.0256369},
    url = {https://doi.org/10.1063/5.0256369}
}

@Article{perd21,
author={Perdew, John P.
and Ruzsinszky, Adrienn
and Sun, Jianwei
and Nepal, Niraj K.
and Kaplan, Aaron D.},
title={Interpretations of ground-state symmetry breaking and strong correlation
       in wavefunction and density functional theories},
journal={Proceedings of the National Academy of Sciences of the United States of America.},
year={2021},
publisher={National Academy of Sciences,},
address={Washington, D.C. :},
volume={118},
issue={4},
pages={e2017850118},
doi={10.1073/pnas.2017850118},
url={https://doi.org/10.1073/pnas.2017850118}
}

@Article{zung22,
author={Zunger, Alex},
title={Bridging the gap between density functional theory and quantum materials},
journal={Nature Computational Science},
year={2022},
month={Sep},
day={01},
volume={2},
issue={9},
pages={529-532},
doi={10.1038/s43588-022-00323-z},
url={https://doi.org/10.1038/s43588-022-00323-z}
}

@article{gori23,
author = {Vuckovic, Stefan and Gerolin, Augusto and Daas, Kimberly J. and Bahmann, Hilke
and Friesecke, Gero and Gori-Giorgi, Paola},
title = {Density functionals based on the mathematical structure of the strong-interaction
limit of DFT},
journal = {WIREs Computational Molecular Science},
volume = {13},
issue = {2},
pages = {e1634},
doi = {https://doi.org/10.1002/wcms.1634},
url = {https://wires.onlinelibrary.wiley.com/doi/abs/10.1002/wcms.1634},
year = {2023}
}

@article{kotl06,
  title = {Electronic structure calculations with dynamical mean-field theory},
  author = {Kotliar, G. and Savrasov, S. Y. and Haule, K. and Oudovenko, V. S. and Parcollet, O. and Marianetti, C. A.},
  journal = {Rev. Mod. Phys.},
  volume = {78},
  issue = {3},
  pages = {865--951},
  numpages = {0},
  year = {2006},
  month = {Aug},
  publisher = {American Physical Society},
  doi = {10.1103/RevModPhys.78.865},
  url = {https://link.aps.org/doi/10.1103/RevModPhys.78.865}
}

@misc{mori25,
      title={Ion Coulomb crystals: an exotic form of condensed matter}, 
      author={Giovanna Morigi and John Bollinger and Michael Drewsen and Daniel Podolsky and
      Efrat Shimshoni},
      year={2025},
      eprint={2508.07374},
      archivePrefix={arXiv},
      primaryClass={physics.atom-ph},
      doi={10.48550/arXiv.2508.07374},
      url={https://arxiv.org/abs/2508.07374}, 
}

@article{gira60,
    author = {Girardeau, M.},
    title = {Relationship between Systems of Impenetrable Bosons and Fermions in One Dimension},
    journal = {Journal of Mathematical Physics},
    volume = {1},
    issue = {6},
    pages = {516-523},
    year = {1960},
    month = {11},
    doi = {10.1063/1.1703687},
    url = {https://doi.org/10.1063/1.1703687},
}

@article{weis04,
author = {Toshiya Kinoshita  and Trevor Wenger  and David S. Weiss },
title = {Observation of a One-Dimensional Tonks-Girardeau Gas},
journal = {Science},
volume = {305},
issue = {5687},
pages = {1125-1128},
year = {2004},
doi = {10.1126/science.1100700},
URL = {https://www.science.org/doi/abs/10.1126/science.1100700},
}

@article{joch12,
  title = {Fermionization of Two Distinguishable Fermions},
  author = {Z\"urn, G. and Serwane, F. and Lompe, T. and Wenz, A. N. and Ries, M. G. and Bohn, J. E. and Jochim, S.},
  journal = {Phys. Rev. Lett.},
  volume = {108},
  issue = {7},
  pages = {075303},
  numpages = {5},
  year = {2012},
  month = {Feb},
  publisher = {American Physical Society},
  doi = {10.1103/PhysRevLett.108.075303},
  url = {https://link.aps.org/doi/10.1103/PhysRevLett.108.075303}
}

@article{ande72,
author = {P. W. Anderson },
title = {More Is Different: Broken symmetry and the nature of
the hierarchical structure of science},
journal = {Science},
volume = {177},
number = {4047},
pages = {393-396},
year = {1972},
doi = {10.1126/science.177.4047.393},
URL = {https://www.science.org/doi/abs/10.1126/science.177.4047.393},
}

@book{pwa_book,
  title     = "{Basic Notions of Condensed Matter Physics}",
  author    = "P. W. Anderson",
  year      = "1984",
  publisher = "The Benjamin/Cummings Publishing Co.",
  address   = "Menlo Park, CA, USA",
  URL = {https://www.amazon.com/Notions-Condensed-Physics-Advanced-Classics/dp/0201328305}
}

@book{tasaki_book,
  title     = "{Physics and Mathematics of Quantum Many-Body Systems}",
  author    = "Hal Tasaki",
  year      = "2020",
  publisher = "Springer",
  address   = "Cham, Switzerland",
  URL = {https://link.springer.com/book/10.1007/978-3-030-41265-4}
}

@misc{mcke25,
      title={Emergence: from physics to biology, sociology, and computer science}, 
      author={Ross H. McKenzie},
      year={2025},
      eprint={2508.08548},
      archivePrefix={arXiv},
      primaryClass={physics.hist-ph},
      url={https://arxiv.org/abs/2508.08548}, 
}

@article{schr26,
  title = {An Undulatory Theory of the Mechanics of Atoms and Molecules},
  author = {Schr\"odinger, E.},
  journal = {Phys. Rev.},
  volume = {28},
  issue = {6},
  pages = {1049--1070},
  numpages = {0},
  year = {1926},
  month = {Dec},
  publisher = {American Physical Society},
  doi = {10.1103/PhysRev.28.1049},
  url = {https://link.aps.org/doi/10.1103/PhysRev.28.1049}
}

@article{lowd62,
  title = {The Normal Constants of Motion in Quantum Mechanics Treated by Projection Technique},
  author = {L\"owdin, Per-Olov},
  journal = {Rev. Mod. Phys.},
  volume = {34},
  issue = {3},
  pages = {520--530},
  numpages = {0},
  year = {1962},
  month = {Jul},
  publisher = {American Physical Society},
  doi = {10.1103/RevModPhys.34.520},
  url = {https://link.aps.org/doi/10.1103/RevModPhys.34.520}
}

@article{kats23,
  title = {Emergence of Classical Magnetic Order from Anderson Towers:
           Quantum Darwinism in Action},
  author = {Sotnikov, O. M. and Stepanov, E. A. and Katsnelson, M. I. and Mila, F.
            and Mazurenko, V. V.},
  journal = {Phys. Rev. X},
  volume = {13},
  issue = {4},
  pages = {041027},
  numpages = {17},
  year = {2023},
  month = {Nov},
  publisher = {American Physical Society},
  doi = {10.1103/PhysRevX.13.041027},
  url = {https://link.aps.org/doi/10.1103/PhysRevX.13.041027}
}

@book{rs_book,
  title={The Nuclear Many-body Problem},
  author={Ring, P. and Schuck, P.},
  isbn={9783540098201},
  lccn={79025447},
  series={Texts and monographs in physics},
  address = {New York},
  url={https://books.google.com/books?id=v79PAQAAIAAJ},
  year={1980},
  publisher={Springer}
}

@article{peie57,
doi = {10.1088/0370-1298/70/5/309},
url = {https://dx.doi.org/10.1088/0370-1298/70/5/309},
year = {1957},
month = {may},
publisher = {},
volume = {70},
issue = {5},
pages = {381},
author = {R E Peierls and J Yoccoz},
title = {The Collective Model of Nuclear Motion},
journal = {Proceedings of the Physical Society. Section A},
}

@article{lowd55,
  title = {Quantum Theory of Many-Particle Systems. III. Extension of the Hartree-Fock Scheme
           to Include Degenerate Systems and Correlation Effects},
  author = {L\"owdin, Per-Olov},
  journal = {Phys. Rev.},
  volume = {97},
  issue = {6},
  pages = {1509--1520},
  numpages = {0},
  year = {1955},
  month = {Mar},
  publisher = {American Physical Society},
  doi = {10.1103/PhysRev.97.1509},
  url = {https://link.aps.org/doi/10.1103/PhysRev.97.1509}
}

@article{lowd64,
  title = {Angular Momentum Wavefunctions Constructed by Projector Operators},
  author = {L\"owdin, Per-Olov},
  journal = {Rev. Mod. Phys.},
  volume = {36},
  issue = {4},
  pages = {966--976},
  numpages = {0},
  year = {1964},
  month = {Oct},
  publisher = {American Physical Society},
  doi = {10.1103/RevModPhys.36.966},
  url = {https://link.aps.org/doi/10.1103/RevModPhys.36.966}
}

@article{fuku81,
author = {Fukutome, Hideo},
title = {Unrestricted Hartree–Fock theory and its applications to molecules and chemical
         reactions},
journal = {International Journal of Quantum Chemistry},
volume = {20},
issue = {5},
pages = {955-1065},
doi = {https://doi.org/10.1002/qua.560200502},
url = {https://onlinelibrary.wiley.com/doi/abs/10.1002/qua.560200502},
year = {1981}
}

@incollection{maye80,
title = {The Spin-Projected Extended {Hartree-Fock} Method},
editor = {Per-Olov Löwdin},
series = {Advances in Quantum Chemistry},
publisher = {Academic Press},
volume = {12},
pages = {189-262},
year = {1980},
issn = {0065-3276},
doi = {https://doi.org/10.1016/S0065-3276(08)60317-2},
url = {https://www.sciencedirect.com/science/article/pii/S0065327608603172},
author = {István Mayer},
}

@article{yann25,
  title = {Wave-function microscopy: Derivation and anatomy of exact algebraic spinful wave
           functions and full Wigner-molecular spectra of a few highly correlated rapidly
	   rotating ultracold fermionic atoms},
  author = {Yannouleas, Constantine and Landman, Uzi},
  journal = {Phys. Rev. A},
  volume = {111},
  issue = {6},
  pages = {063309},
  numpages = {14},
  year = {2025},
  month = {Jun},
  publisher = {American Physical Society},
  doi = {10.1103/6hst-g33w},
  url = {https://link.aps.org/doi/10.1103/6hst-g33w}
}

@incollection{chal06,
author = {David Chalmers},
title = {{Varieties of Emergence}},
editor = {Philip Clayton and Paul Davies},
series = {({\it The Re-Emergence of Emergence: The Emergentist Hypothesis From Science to
           Religion})},
publisher = {Oxford University Press},
chapter = {11},
year = {2006},
isbn = {9780199544318},
url = {https://consc.net/papers/emergence.pdf},
}

@Article{elli20,
author={Ellis, George F. R.},
title={Emergence in Solid State Physics and Biology},
journal={Foundations of Physics},
year={2020},
month={Oct},
day={01},
volume={50},
issue={10},
pages={1098-1139},
issn={1572-9516},
doi={10.1007/s10701-020-00367-z},
url={https://doi.org/10.1007/s10701-020-00367-z}
}

@article{ande80,
author = {Anderson, P.W. and Stein, D.L},
title = {Broken symmetry, emergent properties, dissipative structures, life},
year = {1980},
note = {(Reprinted in Ref.\ \cite{pwa_book}, p. 263)}
}

@article{wign60,
author = {Wigner, Eugene P.},
title = {The unreasonable effectiveness of mathematics in the natural sciences},
journal = {Communications on Pure and Applied Mathematics},
volume = {13},
number = {1},
pages = {1-14},
doi = {https://doi.org/10.1002/cpa.3160130102},
url = {https://onlinelibrary.wiley.com/doi/abs/10.1002/cpa.3160130102},
year = {1960}
}

@Article{Wein87,
author={Weinberg, Steven},
title={Newtonianism, reductionism and the art of congressional testimony},
journal={Nature},
year={1987},
month={Dec},
day={01},
volume={330},
number={6147},
pages={433-437},
doi={10.1038/330433a0},
url={https://doi.org/10.1038/330433a0}
}

@article{land13,
title = {Spontaneous symmetry breaking in quantum systems: Emergence or reduction?},
journal = {Studies in History and Philosophy of Science Part B: Studies in History and
Philosophy of Modern Physics},
volume = {44},
number = {4},
pages = {379-394},
year = {2013},
issn = {1355-2198},
doi = {https://doi.org/10.1016/j.shpsb.2013.07.003},
url = {https://www.sciencedirect.com/science/article/pii/S135521981300052X},
author = {Landsman, N. P.},
}

@article{laug83,
  title = {{Anomalous Quantum Hall Effect: An Incompressible Quantum Fluid with
           Fractionally Charged Excitations}},
  author = {Laughlin, R. B.},
  journal = {Phys. Rev. Lett.},
  volume = {50},
  issue = {18},
  pages = {1395--1398},
  numpages = {0},
  year = {1983},
  month = {May},
  publisher = {American Physical Society},
  doi = {10.1103/PhysRevLett.50.1395},
  url = {https://link.aps.org/doi/10.1103/PhysRevLett.50.1395}
}

@article{laug83.2,
  title = {Quantized motion of three two-dimensional electrons in a strong magnetic field},
  author = {Laughlin, R. B.},
  journal = {Phys. Rev. B},
  volume = {27},
  issue = {6},
  pages = {3383--3389},
  numpages = {0},
  year = {1983},
  month = {Mar},
  publisher = {American Physical Society},
  doi = {10.1103/PhysRevB.27.3383},
  url = {https://link.aps.org/doi/10.1103/PhysRevB.27.3383}
}

@article{laug99,
  title = {Nobel Lecture: Fractional quantization},
  author = {Laughlin, R. B.},
  journal = {Rev. Mod. Phys.},
  volume = {71},
  issue = {4},
  pages = {863--874},
  numpages = {0},
  year = {1999},
  month = {Jul},
  publisher = {American Physical Society},
  doi = {10.1103/RevModPhys.71.863},
  url = {https://link.aps.org/doi/10.1103/RevModPhys.71.863}
}

@article{laug20,
author = {R. B. Laughlin  and David Pines },
title = {The Theory of Everything},
journal = {Proceedings of the National Academy of Sciences},
volume = {97},
issue = {1},
pages = {28-31},
year = {2000},
doi = {10.1073/pnas.97.1.28},
URL = {https://www.pnas.org/doi/abs/10.1073/pnas.97.1.28},
}

@article{yann04.2,
  title = {Unified description of floppy and rigid rotating Wigner molecules formed in
           quantum dots},
  author = {Yannouleas, Constantine and Landman, Uzi},
  journal = {Phys. Rev. B},
  volume = {69},
  issue = {11},
  pages = {113306},
  numpages = {4},
  year = {2004},
  month = {Mar},
  publisher = {American Physical Society},
  doi = {10.1103/PhysRevB.69.113306},
  url = {https://link.aps.org/doi/10.1103/PhysRevB.69.113306}
}

@book{thoubook,
  title     = "The quantum mechanics of many-body systems",
  author    = "David J. Thouless",
  year      = "1961",
  publisher = "Academic",
  address   = "New York"
}

@article{pald85,
author = {Paldus, Josef and {\v C}í{\v z}ek, J.},
title = {{Hartree–Fock} stability and symmetry breaking: oxygen doubly negative ion},
journal = {Canadian Journal of Chemistry},
volume = {63},
issue = {7},
pages = {1803-1811},
year = {1985},
doi = {10.1139/v85-301},
URL = {https://doi.org/10.1139/v85-301}
}

@article{pald67,
    author = {{Čížek}, J. and Paldus, J.},
    title = {Stability Conditions for the Solutions of the {Hartree—Fock} Equations for Atomic
    and Molecular Systems. Application to the {Pi}‐Electron Model of Cyclic Polyenes},
    journal = {The Journal of Chemical Physics},
    volume = {47},
    issue = {10},
    pages = {3976-3985},
    year = {1967},
    month = {11},
    issn = {0021-9606},
    doi = {10.1063/1.1701562},
    url = {https://doi.org/10.1063/1.1701562},
}

@article{lowd63,
  title = {Discussion on The Hartree-Fock Approximation},
  author = {P. Lykos and G. W. Pratt},
  journal = {Rev. Mod. Phys.},
  volume = {35},
  pages = {496--501},
  year = {1963},
  publisher = {American Physical Society},
  doi = {10.1103/RevModPhys.35.496},
  url = {https://link.aps.org/doi/10.1103/RevModPhys.35.496},
  note = {{(See the contribution by P.O. L\"owdin on p. 496})}
}

@incollection{jone25,
title = {Density Functional Theory for the Sceptical},
editor = {E. Pavarini and E. Koch and A. Lichtenstein and D. Vollhardt},
series = {Understanding Correlated Materials with DMFT,
          Modeling and Simulation},
publisher = {Forschungzentrum J\"ulich},
volume = {15},
chapter = {2},
year = {2025},
url = {https://www.cond-mat.de/events/correl25},
author = {Robert O. Jones},
}

@article{thou60,
title = {Stability conditions and nuclear rotations in the {Hartree-Fock} theory},
journal = {Nuclear Physics},
volume = {21},
pages = {225--232},
year = {1960},
doi = {https://doi.org/10.1016/0029-5582(60)90048-1},
url = {https://www.sciencedirect.com/science/article/pii/0029558260900481},
author = {D.J. Thouless},
}

@article{lowd55.2,
  title = {Quantum Theory of Many-Particle Systems. I. Physical Interpretations by Means of
           Density Matrices, Natural Spin-Orbitals, and Convergence Problems in the Method of
	   Configurational Interaction},
  author = {L\"owdin, Per-Olov},
  journal = {Phys. Rev.},
  volume = {97},
  issue = {6},
  pages = {1474--1489},
  year = {1955},
  month = {Mar},
  publisher = {American Physical Society},
  doi = {10.1103/PhysRev.97.1474},
  url = {https://link.aps.org/doi/10.1103/PhysRev.97.1474}
}

@inbook{lowd58,
author = {L\"owdin, Per-Olov},
publisher = {John Wiley \& Sons, Ltd},
title = {Correlation Problem in Many-Electron Quantum Mechanics I.
         Review of Different Approaches and Discussion of Some Current Ideas},
booktitle = {Advances in Chemical Physics},
volume = {2},
chapter = {7},
pages = {207-322},
doi = {https://doi.org/10.1002/9780470143483.ch7},
url = {https://onlinelibrary.wiley.com/doi/abs/10.1002/9780470143483.ch7},
year = {1958}
}

@article{burk99,
  title = {Coupled quantum dots as quantum gates},
  author = {Burkard, Guido and Loss, Daniel and DiVincenzo, David P.},
  journal = {Phys. Rev. B},
  volume = {59},
  issue = {3},
  pages = {2070--2078},
  numpages = {0},
  year = {1999},
  month = {Jan},
  publisher = {American Physical Society},
  doi = {10.1103/PhysRevB.59.2070},
  url = {https://link.aps.org/doi/10.1103/PhysRevB.59.2070}
}

@article{loss98,
  title = {Quantum computation with quantum dots},
  author = {Loss, Daniel and DiVincenzo, David P.},
  journal = {Phys. Rev. A},
  volume = {57},
  issue = {1},
  pages = {120--126},
  numpages = {0},
  year = {1998},
  month = {Jan},
  publisher = {American Physical Society},
  doi = {10.1103/PhysRevA.57.120},
  url = {https://link.aps.org/doi/10.1103/PhysRevA.57.120}
}

@article{yann02,
author = {Yannouleas, Constantine and Landman, Uzi},
title = {Magnetic-field manipulation of chemical bonding in artificial molecules},
journal = {International Journal of Quantum Chemistry},
volume = {90},
issue = {2},
pages = {699-708},
doi = {https://doi.org/10.1002/qua.980},
url = {https://onlinelibrary.wiley.com/doi/abs/10.1002/qua.980},
year = {2002}
}

@Article{heit27,
author={Heitler, W. and London, F.},
title={Wechselwirkung neutraler Atome und hom{\"o}opolare Bindung nach der Quantenmechanik},
journal={Zeitschrift f{\"u}r Physik},
year={1927},
month={Jun},
day={01},
volume={44},
issue={6},
pages={455-472},
doi={10.1007/BF01397394},
url={https://doi.org/10.1007/BF01397394}
}

@book{pauncz,
  title     = "{The Construction of Spin Eigenfunctions: An Exercise Book}",
  author    = "Ruben Pauncz",
  year      = "2000",
  publisher = "Springer",
  address   = "New York",
  url       = "{https://doi.org/10.1007/978-1-4615-4291-9}"
}

@article{kouw98,
doi = {10.1088/2058-7058/11/6/26},
url = {https://doi.org/10.1088/2058-7058/11/6/26},
year = {1998},
month = {jun},
volume = {11},
issue = {6},
pages = {35},
author = {Kouwenhoven, Leo and Marcus, Charles},
title = {Quantum dots},
journal = {Physics World},
}

@article{szaf04.2,
 title = {Accuracy of the Hartree-Fock method for Wigner molecules at high magnetic fields},
 author = {Szafran, B. and Bednarek, S. and Adamowski, J. and Tavernier, M. B. and
           Anisimovas, E. and Peeters, F. M.},
 journal = {The European Physical Journal D: Atomic, Molecular, Optical and Plasma Physics},
 year = {2004},
 month = {Mar},
 volume = {28},
 issue = {3},
 pages = {373-380},
 doi = {10.1140/epjd/e2003-00320-5},
 url = {https://doi.org/10.1140/epjd/e2003-00320-5},
}

@article{garc98,
  title = {Correlation energies for two interacting electrons in a harmonic quantum dot},
  author = {Garc\'{\i}a-Castel\'an, R. M. G. and Choe, W. S. and Lee, Y. C.},
  journal = {Phys. Rev. B},
  volume = {57},
  issue = {16},
  pages = {9792--9806},
  numpages = {0},
  year = {1998},
  month = {Apr},
  publisher = {American Physical Society},
  doi = {10.1103/PhysRevB.57.9792},
  url = {https://link.aps.org/doi/10.1103/PhysRevB.57.9792}
}

@article{taut94,
doi = {10.1088/0305-4470/27/3/040},
url = {https://doi.org/10.1088/0305-4470/27/3/040},
year = {1994},
month = {feb},
publisher = {},
volume = {27},
issue = {3},
pages = {1045},
author = {M Taut},
title = {Two electrons in a homogeneous magnetic field: particular analytical solutions},
journal = {Journal of Physics A: Mathematical and General},
}

@article{yann03.2,
  title = {Group theoretical analysis of symmetry breaking in two-dimensional quantum dots},
  author = {Yannouleas, Constantine and Landman, Uzi},
  journal = {Phys. Rev. B},
  volume = {68},
  issue = {3},
  pages = {035325},
  numpages = {16},
  year = {2003},
  month = {Jul},
  publisher = {American Physical Society},
  doi = {10.1103/PhysRevB.68.035325},
  url = {https://link.aps.org/doi/10.1103/PhysRevB.68.035325}
}

@book{cotton,
  title     = "{Chemical Applications of Group Theory}",
  author    = "F. A. Cotton",
  year      = "1990",
  publisher = "Wiley",
  address   = "New York",
}

@book{hamermesh,
  title     = "{Group Theory and its Application to Physical Problems}",
  author    = "Morton Hamermesh",
  year      = "1962",
  publisher = "Addison-Wesley",
  address   = "Reading, Massachusetts",
}

@article{alon04,
doi = {10.1209/epl/i2004-10047-3},
url = {https://doi.org/10.1209/epl/i2004-10047-3},
year = {2004},
month = {jul},
publisher = {},
volume = {67},
number = {1},
pages = {8},
author = {O. E. Alon and A. I. Streltsov and K. Sakmann and L. S. Cederbaum},
title = {Continuous configuration-interaction  for condensates in a ring},
journal = {Europhysics Letters},
}

@article{hill53,
  title = {Nuclear Constitution and the Interpretation of Fission Phenomena},
  author = {Hill, David Lawrence and Wheeler, John Archibald},
  journal = {Phys. Rev.},
  volume = {89},
  issue = {5},
  pages = {1102--1145},
  numpages = {0},
  year = {1953},
  month = {Mar},
  publisher = {American Physical Society},
  doi = {10.1103/PhysRev.89.1102},
  url = {https://link.aps.org/doi/10.1103/PhysRev.89.1102}
}

@article{grif57,
  title = {Collective Motions in Nuclei by the Method of Generator Coordinates},
  author = {Griffin, James J. and Wheeler, John A.},
  journal = {Phys. Rev.},
  volume = {108},
  issue = {2},
  pages = {311--327},
  numpages = {0},
  year = {1957},
  month = {Oct},
  publisher = {American Physical Society},
  doi = {10.1103/PhysRev.108.311},
  url = {https://link.aps.org/doi/10.1103/PhysRev.108.311}
}

@article{fuku88,
    author = {Fukutome, Hideo},
    title = {Theory of Resonating Quantum Fluctuations in a Fermion System:
             Resonating Hartree-Fock Approximation},
    journal = {Progress of Theoretical Physics},
    volume = {80},
    issue = {3},
    pages = {417-432},
    year = {1988},
    month = {09},
    doi = {10.1143/PTP.80.417},
    url = {https://doi.org/10.1143/PTP.80.417},
}

@article{Okun09,
doi = {10.1143/JJAP.48.125002},
url = {https://doi.org/10.1143/JJAP.48.125002},
year = {2009},
month = {dec},
publisher = {},
volume = {48},
issue = {12R},
pages = {125002},
author = {Okunishi, Takuma and Negishi, Yuki and Muraguchi, Masakazu and Takeda, Kyozaburo},
title = {Resonating Hartree–Fock Approach for Electrons Confined in Two Dimentional Square                Quantum Dots},
journal = {Japanese Journal of Applied Physics},
}

@article{kast93,
doi = {10.1063/1.881393},
url = {https://doi.org/10.1063/1.881393},
year = {1993},
volume = {46},
issue = {1},
pages = {24-31},
author = {Kastner, Marc A. },
title = {Artificial Atoms},
journal = {Physics Today},
}

@article{kohn00,
  title = {Symmetry of the atomic electron density in Hartree, Hartree-Fock,
           and density-functional theories},
  author = {Fertig, H. A. and Kohn, W.},
  journal = {Phys. Rev. A},
  volume = {62},
  issue = {5},
  pages = {052511},
  numpages = {10},
  year = {2000},
  month = {Oct},
  publisher = {American Physical Society},
  doi = {10.1103/PhysRevA.62.052511},
  url = {https://link.aps.org/doi/10.1103/PhysRevA.62.052511}
}

@article{Nait21,
doi = {10.1088/1361-6455/ac170c},
url = {https://doi.org/10.1088/1361-6455/ac170c},
year = {2021},
month = {sep},
publisher = {IOP Publishing},
volume = {54},
issue = {16},
pages = {165201},
author = {Naito, Tomoya and Endo, Shimpei and Hagino, Kouichi and Tanimura, Yusuke},
title = {On deformability of atoms—comparative study between atoms and atomic nuclei},
journal = {Journal of Physics B: Atomic, Molecular and Optical Physics},
}

@article{kell80,
  title = {Ro-vibrational collective interpretation of supermultiplet classifications of
           intrashell levels of two-electron atoms},
  author = {Kellman, Michael E. and Herrick, David R.},
  journal = {Phys. Rev. A},
  volume = {22},
  issue = {4},
  pages = {1536--1551},
  numpages = {0},
  year = {1980},
  month = {Oct},
  publisher = {American Physical Society},
  doi = {10.1103/PhysRevA.22.1536},
  url = {https://link.aps.org/doi/10.1103/PhysRevA.22.1536}
}

@article{eich25,
  title = {Series of Molecularlike Doubly Excited States of a Quasi-Three-Body Coulomb System},
  author = {G\'en\'evriez, M. and Jungers, M. and Rosen, C. and Eichmann, U.},
  journal = {Phys. Rev. Lett.},
  volume = {135},
  issue = {15},
  pages = {153002},
  numpages = {8},
  year = {2025},
  month = {Oct},
  publisher = {American Physical Society},
  doi = {10.1103/svsd-9mj3},
  url = {https://link.aps.org/doi/10.1103/svsd-9mj3}
}

@article{Coul49,
author = {C.A. Coulson and I. Fischer},
title = {XXXIV. Notes on the molecular orbital treatment of the hydrogen molecule},
journal = {The London, Edinburgh, and Dublin Philosophical Magazine and Journal of Science},
volume = {40},
number = {303},
pages = {386--393},
year = {1949},
publisher = {Taylor \& Francis},
doi = {10.1080/14786444908521726},
URL = {https://doi.org/10.1080/14786444908521726},
}

@article{Berr89,
author = {R. Stephen Berry},
title = {How good is Niels Bohr's atomic model?},
journal = {Contemporary Physics},
volume = {30},
issue = {1},
pages = {1--19},
year = {1989},
publisher = {Taylor \& Francis},
doi = {10.1080/00107518908222587},
URL = {https://doi.org/10.1080/00107518908222587},
}

@book{bm,
  title={Nuclear Structure},
  author={\AA. Bohr and B. R. Mottelson},
  address = {Reading, Massachusetts},
  year={1975},
  publisher={Benjamin},
  volume = {II p 41},
}

@article{maks90,
  title = {Quantum dots in a magnetic field: Role of electron-electron interactions},
  author = {Maksym, P. A. and Chakraborty, Tapash},
  journal = {Phys. Rev. Lett.},
  volume = {65},
  issue = {1},
  pages = {108--111},
  numpages = {0},
  year = {1990},
  month = {Jul},
  publisher = {American Physical Society},
  doi = {10.1103/PhysRevLett.65.108},
  url = {https://link.aps.org/doi/10.1103/PhysRevLett.65.108}
}

@article{maks96,
  title = {Eckardt frame theory of interacting electrons in quantum dots},
  author = {Maksym, P. A.},
  journal = {Phys. Rev. B},
  volume = {53},
  issue = {16},
  pages = {10871--10886},
  year = {1996},
  month = {Apr},
  publisher = {American Physical Society},
  doi = {10.1103/PhysRevB.53.10871},
  url = {https://link.aps.org/doi/10.1103/PhysRevB.53.10871}
}

@article{ruan95,
  title = {Origin of magic angular momenta in few-electron quantum dots},
  author = {Ruan, W. Y. and Liu, Y. Y. and Bao, C. G. and Zhang, Z. Q.},
  journal = {Phys. Rev. B},
  volume = {51},
  issue = {12},
  pages = {7942--7945},
  numpages = {0},
  year = {1995},
  month = {Mar},
  publisher = {American Physical Society},
  doi = {10.1103/PhysRevB.51.7942},
  url = {https://link.aps.org/doi/10.1103/PhysRevB.51.7942}
}

@article{Kell78,
doi = {10.1088/0022-3700/11/24/002},
url = {https://doi.org/10.1088/0022-3700/11/24/002},
year = {1978},
month = {dec},
publisher = {},
volume = {11},
number = {24},
pages = {L755},
author = {M E Kellman and D R Herrick},
title = {Rotor-like spectra for some doubly excited two-electron states},
journal = {Journal of Physics B: Atomic and Molecular Physics},
}

@article{lin07,
  title = {Attosecond Light Pulses for Probing Two-Electron Dynamics of Helium in the
           Time Domain},
  author = {Morishita, Toru and Watanabe, Shinichi and Lin, C. D.},
  journal = {Phys. Rev. Lett.},
  volume = {98},
  issue = {8},
  pages = {083003},
  numpages = {4},
  year = {2007},
  month = {Feb},
  publisher = {American Physical Society},
  doi = {10.1103/PhysRevLett.98.083003},
  url = {https://link.aps.org/doi/10.1103/PhysRevLett.98.083003}
}

@article{feag88,
  title = {Molecular-orbital description of the states of two-electron systems},
  author = {Feagin, James M. and Briggs, John S.},
  journal = {Phys. Rev. A},
  volume = {37},
  issue = {12},
  pages = {4599--4613},
  numpages = {0},
  year = {1988},
  month = {Jun},
  publisher = {American Physical Society},
  doi = {10.1103/PhysRevA.37.4599},
  url = {https://link.aps.org/doi/10.1103/PhysRevA.37.4599}
}

@article{feag86,
  title = {Molecular Description of Two-Electron Atoms},
  author = {Feagin, James M. and Briggs, John S.},
  journal = {Phys. Rev. Lett.},
  volume = {57},
  issue = {8},
  pages = {984--987},
  numpages = {0},
  year = {1986},
  month = {Aug},
  publisher = {American Physical Society},
  doi = {10.1103/PhysRevLett.57.984},
  url = {https://link.aps.org/doi/10.1103/PhysRevLett.57.984}
}

@article{gene21,
author = {M. G\'en\'evriez},
title = {Theoretical approaches for doubly-excited Rydberg states in quasi-two-electron
         systems: two-electron dynamics far away from the nucleus},
journal = {Molecular Physics},
volume = {119},
issue = {7},
pages = {e1861353},
year = {2021},
publisher = {Taylor \& Francis},
doi = {10.1080/00268976.2020.1861353},
URL = {https://doi.org/10.1080/00268976.2020.1861353},
}

@article{lin99,
  title = {Comprehensive analysis of electron correlations in three-electron atoms},
  author = {Morishita, Toru and Lin, C. D.},
  journal = {Phys. Rev. A},
  volume = {59},
  issue = {3},
  pages = {1835--1843},
  numpages = {0},
  year = {1999},
  month = {Mar},
  publisher = {American Physical Society},
  doi = {10.1103/PhysRevA.59.1835},
  url = {https://link.aps.org/doi/10.1103/PhysRevA.59.1835}
}

@article{mads05,
  title = {Classification of atomic states by geometrical and quantum-mechanical symmetries},
  author = {Poulsen, M. D. and Madsen, L. B.},
  journal = {Phys. Rev. A},
  volume = {72},
  issue = {4},
  pages = {042501},
  numpages = {7},
  year = {2005},
  month = {Oct},
  publisher = {American Physical Society},
  doi = {10.1103/PhysRevA.72.042501},
  url = {https://link.aps.org/doi/10.1103/PhysRevA.72.042501}
}

@article{Mads03,
doi = {10.1088/0953-4075/36/20/R01},
url = {https://doi.org/10.1088/0953-4075/36/20/R01},
year = {2003},
month = {oct},
publisher = {},
volume = {36},
issue = {20},
pages = {R223},
author = {Lars Bojer Madsen},
title = {Triply excited states: electron–electron correlations in lithium},
journal = {Journal of Physics B: Atomic, Molecular and Optical Physics},
}

@article{BOLT93,
title = {Classical model of a {Wigner} crystal in a quantum dot},
journal = {Superlattices and Microstructures},
volume = {13},
issue = {2},
pages = {139},
year = {1993},
issn = {0749-6036},
doi = {https://doi.org/10.1006/spmi.1993.1026},
url = {https://www.sciencedirect.com/science/article/pii/S0749603683710268},
author = {F. Bolton and U. R\"ossler},
}

@article{kong02,
  title = {Transition between ground state and metastable states in classical
           two-dimensional atoms},
  author = {Kong, Minghui and Partoens, B. and Peeters, F. M.},
  journal = {Phys. Rev. E},
  volume = {65},
  issue = {4},
  pages = {046602},
  numpages = {13},
  year = {2002},
  month = {Mar},
  publisher = {American Physical Society},
  doi = {10.1103/PhysRevE.65.046602},
  url = {https://link.aps.org/doi/10.1103/PhysRevE.65.046602}
}

@Article{wign38,
author ="Wigner, E.",
title  ="Effects of the electron interaction on the energy levels of electrons in metals",
journal  ="Trans. Faraday Soc. ",
year  ="1938",
volume  ="34",
issue  ="0",
pages  ="678-685",
publisher  ="The Royal Society of Chemistry",
doi  ="10.1039/TF9383400678",
url  ="http://dx.doi.org/10.1039/TF9383400678",
}

@article{ezra83,
  title = {Collective and independent-particle motion in doubly excited
           two-electron atoms},
  author = {Ezra, Gregory S. and Berry, R. Stephen},
  journal = {Phys. Rev. A},
  volume = {28},
  issue = {4},
  pages = {1974--1988},
  numpages = {0},
  year = {1983},
  month = {Oct},
  publisher = {American Physical Society},
  doi = {10.1103/PhysRevA.28.1974},
  url = {https://link.aps.org/doi/10.1103/PhysRevA.28.1974}
}

@article{sala17,
  title = {Potential energy surfaces in atomic structure: The role of {Coulomb}
           correlation in the ground state of helium},
  author = {Salas, L. D. and Arce, J. C.},
  journal = {Phys. Rev. A},
  volume = {95},
  issue = {2},
  pages = {022502},
  numpages = {8},
  year = {2017},
  month = {Feb},
  publisher = {American Physical Society},
  doi = {10.1103/PhysRevA.95.022502},
  url = {https://link.aps.org/doi/10.1103/PhysRevA.95.022502}
}

@article{will02,
  title = {Phase transitions of a few-electron system in a spherical quantum dot},
  author = {Sundqvist, P. A. and Volkov, S. Yu. and Lozovik, Yu. E. and Willander, M.},
  journal = {Phys. Rev. B},
  volume = {66},
  issue = {7},
  pages = {075335},
  numpages = {10},
  year = {2002},
  month = {Aug},
  publisher = {American Physical Society},
  doi = {10.1103/PhysRevB.66.075335},
  url = {https://link.aps.org/doi/10.1103/PhysRevB.66.075335}
}

@Article{weze19,
	title={{An introduction to spontaneous symmetry breaking}},
	author={Aron J. Beekman and Louk Rademaker and Jasper van Wezel},
	journal={SciPost Phys. Lect. Notes},
	pages={11},
	year={2019},
	publisher={SciPost},
	doi={10.21468/SciPostPhysLectNotes.11},
	url={https://scipost.org/10.21468/SciPostPhysLectNotes.11},
}

@book{wein_book,
  title={ The Quantum Theory of Fields},
  author={Weinberg, S.},
  address = {Cambridge, UK},
  year={1996},
  publisher={Cambridge University Press},
  volume = {II},
}

@article{Pape15,
doi = {10.1088/0954-3899/42/10/105103},
url = {https://doi.org/10.1088/0954-3899/42/10/105103},
year = {2015},
month = {sep},
publisher = {IOP Publishing},
volume = {42},
number = {10},
pages = {105103},
author = {Papenbrock, T and Weidenmüller, H A},
title = {Effective field theory of emergent symmetry breaking in deformed atomic nuclei},
journal = {Journal of Physics G: Nuclear and Particle Physics},
}

@article{pape25,
doi = {10.1088/1361-6471/adb50b},
url = {https://doi.org/10.1088/1361-6471/adb50b},
year = {2025},
month = {mar},
publisher = {IOP Publishing},
volume = {52},
number = {3},
pages = {033001},
author = {Coello Pérez, E A and Papenbrock, T},
title = {Effective field theories for collective excitations of atomic nuclei},
journal = {Journal of Physics G: Nuclear and Particle Physics},
}

@Article{tasa94,
author={Koma, Tohru and Tasaki, Hal},
title={Symmetry breaking and finite-size effects in quantum many-body systems},
journal={Journal of Statistical Physics},
year={1994},
month={Aug},
day={01},
volume={76},
number={3},
pages={745-803},
issn={1572-9613},
doi={10.1007/BF02188685},
url={https://doi.org/10.1007/BF02188685}
}

@inbook{Land17,
author="Landsman, Klaas",
booktitle="Foundations of Quantum Theory: From Classical Concepts to Operator Algebras",
title="{Spontaneous Symmetry Breaking}",
year="2017",
publisher="Springer International Publishing",
address="Cham, Switzerland",
chapter="10",
pages="367--433",
isbn="978-3-319-51777-3",
doi="10.1007/978-3-319-51777-3_10",
url="https://doi.org/10.1007/978-3-319-51777-3_10"
}

@article{zumi69,
  title = {Structure of Phenomenological Lagrangians. I},
  author = {Coleman, S. and Wess, J. and Zumino, Bruno},
  journal = {Phys. Rev.},
  volume = {177},
  issue = {5},
  pages = {2239--2247},
  numpages = {0},
  year = {1969},
  month = {Jan},
  publisher = {American Physical Society},
  doi = {10.1103/PhysRev.177.2239},
  url = {https://link.aps.org/doi/10.1103/PhysRev.177.2239}
}

@article{zumi69.2,
  title = {Structure of Phenomenological Lagrangians. II},
  author = {Callan, Curtis G. and Coleman, Sidney and Wess, J. and Zumino, Bruno},
  journal = {Phys. Rev.},
  volume = {177},
  issue = {5},
  pages = {2247--2250},
  numpages = {0},
  year = {1969},
  month = {Jan},
  publisher = {American Physical Society},
  doi = {10.1103/PhysRev.177.2247},
  url = {https://link.aps.org/doi/10.1103/PhysRev.177.2247}
}

@Article{land20,
	title={{Quantum spin systems versus Schr\"odinger operators: A case study
	        in spontaneous symmetry breaking}},
	author={Christiaan J. F. van de Ven and Gerrit C. Groenenboom and Robin Reuvers
	and Nicolaas P. Landsman},
	journal={SciPost Phys.},
	volume={8},
	pages={022},
	year={2020},
	publisher={SciPost},
	doi={10.21468/SciPostPhys.8.2.022},
	url={https://scipost.org/10.21468/SciPostPhys.8.2.022},
}

@article{naza25,
  title = {Extraction of ground-state nuclear deformations from ultrarelativistic
           heavy-ion collisions: Nuclear structure physics context},
  author = {Dobaczewski, J. and Gade, A. and Godbey, K. and Janssens, R. V. F.
            and Nazarewicz, W.},
  journal = {Phys. Rev. Res.},
  volume = {7},
  issue = {4},
  pages = {043159},
  numpages = {9},
  year = {2025},
  month = {Nov},
  publisher = {American Physical Society},
  doi = {10.1103/kngg-1ccb},
  url = {https://link.aps.org/doi/10.1103/kngg-1ccb}
}

@book{levine_book,
  title={Quantum Chemistry},
  author={Levine, I. V.},
  address = {Boston},
  year={1970},
  publisher={Allyn and Bacon},
  volume = {II},
  }

@inbook{ande94,
author="Anderson, Philip W.",
booktitle="A Career in Theoretical Physics",
title="{Some general thoughts about broken symmetry}",
year="1994",
publisher="{World Scientific}",
address="",
chapter="",
pages="419-429",
doi="10.1142/5558",
url="https://www.worldscientific.com/worldscibooks/10.1142/5558",
}

@incollection{lauc16,
title = {{Studying Continuous Symmetry Breaking with Exact Diagonalization
          in "Quantum Materials: Experiments and Theory"}},
editor = {E. Pavarini and E. Koch and J. van den Brink and G. Sawatzky},
series = {Modeling and Simulation},
publisher = {Forschungzentrum J\"ulich},
volume = {6},
chapter = {8},
year = {2016},
url = {https://www.cond-mat.de/events/correl16},
author = {A.M. L\"auchli and M. Schuler and A. Wietek},
eprint={https://arxiv.org/abs/1704.08622},
}

@article{Misg07,
doi = {10.1088/0953-8984/19/14/145202},
url = {https://doi.org/10.1088/0953-8984/19/14/145202},
year = {2007},
month = {mar},
publisher = {},
volume = {19},
number = {14},
pages = {145202},
author = {Misguich, Gr\'egoire and Sindzingre, Philippe},
title = {Detecting spontaneous symmetry breaking in finite-size spectra of frustrated quantum
antiferromagnets},
journal = {Journal of Physics: Condensed Matter},
}

@article{bern92,
  title = {Signature of N\'eel order in exact spectra of quantum antiferromagnets
           on finite lattices},
  author = {Bernu, B. and Lhuillier, C. and Pierre, L.},
  journal = {Phys. Rev. Lett.},
  volume = {69},
  issue = {17},
  pages = {2590--2593},
  numpages = {0},
  year = {1992},
  month = {Oct},
  publisher = {American Physical Society},
  doi = {10.1103/PhysRevLett.69.2590},
  url = {https://link.aps.org/doi/10.1103/PhysRevLett.69.2590}
}

@article{yann24,
  title = {Wigner-molecule supercrystal in transition metal dichalcogenide moir\'e
           superlattices: Lessons from the bottom-up approach},
  author = {Yannouleas, Constantine and Landman, Uzi},
  journal = {Phys. Rev. B},
  volume = {109},
  issue = {12},
  pages = {L121302},
  numpages = {7},
  year = {2024},
  month = {Mar},
  publisher = {American Physical Society},
  doi = {10.1103/PhysRevB.109.L121302},
  url = {https://link.aps.org/doi/10.1103/PhysRevB.109.L121302}
}

@article{yann24.2,
  title = {{Crystal-Field Effects in the Formation of Wigner-Molecule Supercrystals in
           Moir\'e Transition Metal Dichalcogenide Superlattices}},
  author = {Yannouleas, Constantine and Landman, Uzi},
  journal = {Phys. Rev. Lett.},
  volume = {133},
  issue = {24},
  pages = {246502},
  numpages = {9},
  year = {2024},
  month = {Dec},
  publisher = {American Physical Society},
  doi = {10.1103/PhysRevLett.133.246502},
  url = {https://link.aps.org/doi/10.1103/PhysRevLett.133.246502}
}

@article{crom24,
author = {Hongyuan Li and Ziyu Xiang and Aidan P. Reddy and Trithep Devakul and
          Renee Sailus and Rounak Banerjee and Takashi Taniguchi and Kenji Watanabe
	  and Sefaattin Tongay and Alex Zettl and Liang Fu and Michael F. Crommie
	  and Feng Wang},
title = {Wigner molecular crystals from multielectron moir\'e artificial atoms},
journal = {Science},
volume = {385},
number = {6704},
pages = {86-91},
year = {2024},
doi = {10.1126/science.adk1348},
URL = {https://www.science.org/doi/abs/10.1126/science.adk1348},
eprint = {},
}

@Article{Hu25,
author={Hu, Tianyi
and Zhang, Tingfeng
and Zhang, Yongqi
and Liu, Bing
and Wang, Zhengfei},
title={Topological Wigner Molecule Crystal in Transition-Metal Dichalcogenide Moir\'e
       Superlattices},
journal={ACS Nano},
year={2025},
month={Sep},
day={16},
publisher={American Chemical Society},
volume={19},
number={36},
pages={32499-32506},
doi={10.1021/acsnano.5c09463},
url={https://doi.org/10.1021/acsnano.5c09463}
}

@misc{khal25,
      title={Spin-orbital magnetism in moir\'e Wigner molecules}, 
      author={Ahmed Khalifa and Rokas Veitas and Francisco Machado and Shubhayu Chatterjee},
      year={2025},
      eprint={2507.06307},
      archivePrefix={arXiv},
      primaryClass={cond-mat.str-el},
      url={https://arxiv.org/abs/2507.06307}, 
}

@article{Ke25,
doi = {10.1088/2633-4356/add075},
url = {https://doi.org/10.1088/2633-4356/add075},
year = {2025},
month = {may},
publisher = {IOP Publishing},
volume = {5},
number = {2},
pages = {022001},
author = {Ke, Yifan and Hu, Wei},
title = {Exploring Wigner crystals in two-dimensional and moir\'e systems:
         from spectroscopy to theoretical modeling},
journal = {Materials for Quantum Technology},
}

@Article{Liqi25,
author={Li, Xiang
and Qian, Yubing
and Ren, Weiluo
and Xu, Yang
and Chen, Ji},
title={Emergent Wigner phases in moir\'e superlattice from deep learning},
journal={Communications Physics},
year={2025},
month={Sep},
day={02},
volume={8},
number={1},
pages={364},
issn={2399-3650},
doi={10.1038/s42005-025-02282-z},
url={https://doi.org/10.1038/s42005-025-02282-z}
}

@article{redd23,
  title = {Artificial Atoms, Wigner Molecules, and an Emergent Kagome Lattice in
           Semiconductor Moir\'e Superlattices},
  author = {Reddy, Aidan P. and Devakul, Trithep and Fu, Liang},
  journal = {Phys. Rev. Lett.},
  volume = {131},
  issue = {24},
  pages = {246501},
  numpages = {6},
  year = {2023},
  month = {Dec},
  publisher = {American Physical Society},
  doi = {10.1103/PhysRevLett.131.246501},
  url = {https://link.aps.org/doi/10.1103/PhysRevLett.131.246501}
}

@Article{yazd24,
author={Tsui, Yen-Chen
and He, Minhao
and Hu, Yuwen
and Lake, Ethan
and Wang, Taige
and Watanabe, Kenji
and Taniguchi, Takashi
and Zaletel, Michael P.
and Yazdani, Ali},
title={Direct observation of a magnetic-field-induced Wigner crystal},
journal={Nature},
year={2024},
month={Apr},
day={01},
volume={628},
number={8007},
pages={287-292},
doi={10.1038/s41586-024-07212-7},
url={https://doi.org/10.1038/s41586-024-07212-7}
}

@Article{Jona81,
author={Jona-Lasinio, G.
and Martinelli, F.
and Scoppola, E.},
title={New approach to the semiclassical limit of quantum mechanics},
journal={Communications in Mathematical Physics},
year={1981},
month={Jun},
day={01},
volume={80},
number={2},
pages={223-254},
doi={10.1007/BF01213012},
url={https://doi.org/10.1007/BF01213012}
}

@article{jona84,
doi = {10.1088/0305-4470/17/15/011},
url = {https://doi.org/10.1088/0305-4470/17/15/011},
year = {1984},
month = {oct},
publisher = {},
volume = {17},
number = {15},
pages = {2935},
author = {S Graffi and V Grecchi and G Jona-Lasinio},
title = {Tunnelling instability via perturbation theory},
journal = {Journal of Physics A: Mathematical and General},
}

@article{SIMO85,
title = {Semiclassical analysis of low lying eigenvalues. IV. The flea on the elephant},
journal = {Journal of Functional Analysis},
volume = {63},
number = {1},
pages = {123-136},
year = {1985},
doi = {https://doi.org/10.1016/0022-1236(85)90101-6},
url = {https://www.sciencedirect.com/science/article/pii/0022123685901016},
author = {Barry Simon},
}

@article{zung25,
  title = {Symmetry breaking forms split-off flat bands in quantum oxides controlling
           metal versus insulator phases},
  author = {Xiong, Jia-Xin and Zhang, Xiuwen and Zunger, Alex},
  journal = {Phys. Rev. B},
  volume = {111},
  issue = {3},
  pages = {035154},
  numpages = {18},
  year = {2025},
  month = {Jan},
  publisher = {American Physical Society},
  doi = {10.1103/PhysRevB.111.035154},
  url = {https://link.aps.org/doi/10.1103/PhysRevB.111.035154}
}

@article{AKMA99,
title = {The Wigner molecule in a 2D quantum dot},
journal = {Physica E: Low-dimensional Systems and Nanostructures},
volume = {4},
number = {4},
pages = {277-285},
year = {1999},
issn = {1386-9477},
doi = {https://doi.org/10.1016/S1386-9477(99)00019-3},
url = {https://www.sciencedirect.com/science/article/pii/S1386947799000193},
author = {N. Akman and M. Tomak},
}

@article{woot00,
  title = {Distributed entanglement},
  author = {Coffman, Valerie and Kundu, Joydip and Wootters, William K.},
  journal = {Phys. Rev. A},
  volume = {61},
  issue = {5},
  pages = {052306},
  numpages = {5},
  year = {2000},
  month = {Apr},
  publisher = {American Physical Society},
  doi = {10.1103/PhysRevA.61.052306},
  url = {https://link.aps.org/doi/10.1103/PhysRevA.61.052306}
}

@article{lida06,
doi = {10.1088/0953-8984/18/21/S02},
url = {https://doi.org/10.1088/0953-8984/18/21/S02},
year = {2006},
month = {may},
publisher = {},
volume = {18},
number = {21},
pages = {S721},
author = {Woodworth, Ryan and Mizel, Ari and Lidar, Daniel A},
title = {Few-body spin couplings and their implications for universal quantum computation},
journal = {Journal of Physics: Condensed Matter},
}

@book{auerbook,
  title={Interacting Electrons and Quantum Magnetism},
  author={Auerbach, A},
  address = {New York},
  year={1998},
  publisher={Springer},
  }

@article{dago94,
  title = {Correlated electrons in high-temperature superconductors},
  author = {Dagotto, Elbio},
  journal = {Rev. Mod. Phys.},
  volume = {66},
  issue = {3},
  pages = {763--840},
  numpages = {0},
  year = {1994},
  month = {Jul},
  publisher = {American Physical Society},
  doi = {10.1103/RevModPhys.66.763},
  url = {https://link.aps.org/doi/10.1103/RevModPhys.66.763}
}

@article{jona86,
  title = {Instability of tunneling and the concept of molecular structure in quantum
           mechanics: The case of pyramidal molecules and the enantiomer problem},
  author = {Claverie, P. and Jona-Lasinio, G.},
  journal = {Phys. Rev. A},
  volume = {33},
  issue = {4},
  pages = {2245--2253},
  numpages = {0},
  year = {1986},
  month = {Apr},
  publisher = {American Physical Society},
  doi = {10.1103/PhysRevA.33.2245},
  url = {https://link.aps.org/doi/10.1103/PhysRevA.33.2245}
}

@Article{kim25,
author={Jang, Wonjin
and Kim, Jehyun
and Park, Jaemin
and Cho, Min-Kyun
and Jang, Hyeongyu
and Sim, Sangwoo
and Jung, Hwanchul
and Umansky, Vladimir
and Kim, Dohun},
title={True decoherence-free-subspace derived from a semiconductor double
       quantum dot Heisenberg spin-trimer},
journal={npj Quantum Information},
year={2025},
month={Dec},
day={18},
volume={12},
issue={1},
pages={1},
doi={10.1038/s41534-025-01151-5},
url={https://doi.org/10.1038/s41534-025-01151-5}
}

@Article{Nogu14,
author={Noguchi, Atsushi
and Shikano, Yutaka
and Toyoda, Kenji
and Urabe, Shinji},
title={Aharonov--Bohm effect in the tunnelling of a quantum rotor in a linear Paul trap},
journal={Nature Communications},
year={2014},
month={May},
day={13},
volume={5},
issue={1},
pages={3868},
doi={10.1038/ncomms4868},
url={https://doi.org/10.1038/ncomms4868},
}

@article{copp12,
  title = {Pulse-Gated Quantum-Dot Hybrid Qubit},
  author = {Koh, Teck Seng and Gamble, John King and Friesen, Mark and Eriksson, M. A.
            and Coppersmith, S. N.},
  journal = {Phys. Rev. Lett.},
  volume = {109},
  issue = {25},
  pages = {250503},
  numpages = {5},
  year = {2012},
  month = {Dec},
  publisher = {American Physical Society},
  doi = {10.1103/PhysRevLett.109.250503},
  url = {https://link.aps.org/doi/10.1103/PhysRevLett.109.250503}
}

@Article{copp14,
author={Shi, Zhan
and Simmons, C. B.
and Ward, Daniel R.
and Prance, J. R.
and Wu, Xian
and Koh, Teck Seng
and Gamble, John King
and Savage, D. E.
and Lagally, M. G.
and Friesen, Mark
and Coppersmith, S. N.
and Eriksson, M. A.},
title={Fast coherent manipulation of three-electron states in a double quantum dot},
journal={Nature Communications},
year={2014},
month={Jan},
day={06},
volume={5},
issue={1},
pages={3020},
doi={10.1038/ncomms4020},
url={https://doi.org/10.1038/ncomms4020}
}

@Article{loss24,
author={De Palma, Franco
and Oppliger, Fabian
and Jang, Wonjin
and Bosco, Stefano
and Jan{\'i}k, Mari{\'a}n
and Calcaterra, Stefano
and Katsaros, Georgios
and Isella, Giovanni
and Loss, Daniel
and Scarlino, Pasquale},
title={Strong hole-photon coupling in planar Ge for probing charge degree and
       strongly correlated states},
journal={Nature Communications},
year={2024},
month={Nov},
day={23},
volume={15},
issue={1},
pages={10177},
doi={10.1038/s41467-024-54520-7},
url={https://doi.org/10.1038/s41467-024-54520-7}
}

@article{dzur13,
  title = {Silicon quantum electronics},
  author = {Zwanenburg, Floris A. and Dzurak, Andrew S. and Morello, Andrea and Simmons,
            Michelle Y. and Hollenberg, Lloyd C. L. and Klimeck, Gerhard and Rogge, Sven
	    and Coppersmith, Susan N. and Eriksson, Mark A.},
  journal = {Rev. Mod. Phys.},
  volume = {85},
  issue = {3},
  pages = {961--1019},
  numpages = {0},
  year = {2013},
  month = {Jul},
  publisher = {American Physical Society},
  doi = {10.1103/RevModPhys.85.961},
  url = {https://link.aps.org/doi/10.1103/RevModPhys.85.961}
}

@article{cao16,
  title = {Tunable Hybrid Qubit in a GaAs Double Quantum Dot},
  author = {Cao, Gang and Li, Hai-Ou and Yu, Guo-Dong and Wang, Bao-Chuan and Chen,
            Bao-Bao and Song, Xiang-Xiang and Xiao, Ming and Guo, Guang-Can and Jiang,
	    Hong-Wen and Hu, Xuedong and Guo, Guo-Ping},
  journal = {Phys. Rev. Lett.},
  volume = {116},
  issue = {8},
  pages = {086801},
  numpages = {5},
  year = {2016},
  month = {Feb},
  publisher = {American Physical Society},
  doi = {10.1103/PhysRevLett.116.086801},
  url = {https://link.aps.org/doi/10.1103/PhysRevLett.116.086801}
}

@Article{copp17,
author={Thorgrimsson, Brandur
and Kim, Dohun
and Yang, Yuan-Chi
and Smith, L. W.
and Simmons, C. B.
and Ward, Daniel R.
and Foote, Ryan H.
and Corrigan, J.
and Savage, D. E.
and Lagally, M. G.
and Friesen, Mark
and Coppersmith, S. N.
and Eriksson, M. A.},
title={Extending the coherence of a quantum dot hybrid qubit},
journal={npj Quantum Information},
year={2017},
month={Aug},
day={21},
volume={3},
issue={1},
pages={32},
doi={10.1038/s41534-017-0034-2},
url={https://doi.org/10.1038/s41534-017-0034-2}
}

@ARTICLE{kouw01,
  author = {L.P. Kouwenhoven and D.G. Austing and S. Tarucha},
  title = {Few-electron quantum dots},
  journal = {Rep. Prog. Phys.},
  year = {2001},
  volume = {64},
  pages = {701},
  doi = {doi.org/10.1088/0034-4885/64/6/201},
  url = {https://doi.org/10.1088/0034-4885/64/6/201},
}

@ARTICLE{bers21,
  author = {Isaac B. Bersuker},
  title = {The Jahn–Teller and Pseudo-Jahn–Teller Effects: A Unique and Only Source
           of Spontaneous Symmetry Breaking in Atomic Matter},
  journal = {Symmetry},
  year = {2021},
  volume = {13},
  issue = {9},
  pages = {1577},
  doi = {doi.org/10.3390/sym13091577},
  url = {https://doi.org/10.3390/sym13091577},
}

@book{bersbook,
  title={The Jahn-Teller Effect},
  author={Bersuker, I.B.},
  address = {Cambridge, UK},
  year={2006},
  publisher={Cambridge University Press},
  }

@article{jahn37,
    author = {Jahn, H. A. and Teller, E.},
    title = {Stability of polyatomic molecules in degenerate electronic states -
             I {—-} Orbital degeneracy},
    journal = {Proceedings of the Royal Society of London.
               A. Mathematical and Physical Sciences},
    volume = {161},
    issue = {905},
    pages = {220-235},
    year = {1937},
    month = {07},
    issn = {0080-4630},
    doi = {10.1098/rspa.1937.0142},
    url = {https://doi.org/10.1098/rspa.1937.0142},
}

@article{dunn92,
doi = {10.1088/0953-8984/4/32/014},
url = {https://doi.org/10.1088/0953-8984/4/32/014},
year = {1992},
month = {aug},
publisher = {},
volume = {4},
issue = {32},
pages = {6775},
author = {L D Hallam and C A Bates and J L Dunn},
title = {Symmetry-adapted states for T(X)(e+t$_2$) Jahn-Teller systems},
journal = {Journal of Physics: Condensed Matter},
}

@article{Dunn12,
doi = {10.1088/1367-2630/14/8/083038},
url = {https://doi.org/10.1088/1367-2630/14/8/083038},
year = {2012},
month = {aug},
publisher = {IOP Publishing},
volume = {14},
issue = {8},
pages = {083038},
author = {Dunn, Janette L and Lakin, Andrew J and Hands, Ian D},
title = {Manifestation of dynamic {Jahn–Teller} distortions and surface interactions in
         scanning tunnelling microscopy images of the fullerene anion C$_{60}^-$},
journal = {New Journal of Physics},
}

@misc{zung25.2,
      title={Symmetry breaking transforms strong to normal correlation and
             false metals to true insulators}, 
      author={Alex Zunger and Jia-Xin Xiong and John P. Perdew},
      year={2025},
      eprint={2512.18236},
      archivePrefix={arXiv},
      primaryClass={cond-mat.mtrl-sci},
      url={https://arxiv.org/abs/2512.18236}, 
}

@ARTICLE{cao17,
  author = {B.-B. Chen and B.-C. Wang and G. Cao and H.-O. Li and M. Xiao and G.-C. Guo and 
            H.-W. Jiang and X. Hu and G.-P. Guo},
  title = {Spin Blockade and Coherent Dynamics of High-Spin States in a Three-Electron 
           Double Quantum Dot},
  journal = {Phys. Rev. B},
  year = {2017},
  volume = {95},
  pages = {035408},
  issue = { },
  doi = {10.1103/PhysRevB.95.035408},
  url = {https://doi.org/10.1103/PhysRevB.95.035408},
}

@article{copp12.2,
  title = {Fast Hybrid Silicon Double-Quantum-Dot Qubit},
  author = {Shi, Zhan and Simmons, C. B. and Prance, J. R. and Gamble, John King and
            Koh, Teck Seng and Shim, Yun-Pil and Hu, Xuedong and Savage, D. E. and
	    Lagally, M. G. and Eriksson, M. A. and Friesen, Mark and Coppersmith, S. N.},
  journal = {Phys. Rev. Lett.},
  volume = {108},
  issue = {14},
  pages = {140503},
  numpages = {5},
  year = {2012},
  month = {Apr},
  publisher = {American Physical Society},
  doi = {10.1103/PhysRevLett.108.140503},
  url = {https://link.aps.org/doi/10.1103/PhysRevLett.108.140503}
}

@Article{Vinc00,
author={DiVincenzo, D. P.
and Bacon, D.
and Kempe, J.
and Burkard, G.
and Whaley, K. B.},
title={Universal quantum computation with the exchange interaction},
journal={Nature},
year={2000},
month={Nov},
day={01},
volume={408},
issue={6810},
pages={339-342},
doi={10.1038/35042541},
url={https://doi.org/10.1038/35042541}
}

@article{burk17,
doi = {10.1088/1361-648X/aa761f},
url = {https://doi.org/10.1088/1361-648X/aa761f},
year = {2017},
month = {aug},
publisher = {IOP Publishing},
volume = {29},
issue = {39},
pages = {393001},
author = {Russ, Maximilian and Burkard, Guido},
title = {Three-electron spin qubits},
journal = {Journal of Physics: Condensed Matter},
}

@Article{Cao13,
author={Cao, Gang
and Li, Hai-Ou
and Tu, Tao
and Wang, Li
and Zhou, Cheng
and Xiao, Ming
and Guo, Guang-Can
and Jiang, Hong-Wen
and Guo, Guo-Ping},
title={Ultrafast universal quantum control of a quantum-dot charge qubit using
       {Landau--Zener--St{\"u}ckelberg} interference},
journal={Nature Communications},
year={2013},
month={Jan},
day={29},
volume={4},
issue={1},
pages={1401},
doi={10.1038/ncomms2412},
url={https://doi.org/10.1038/ncomms2412}
}

@article{burk13,
  title = {Interplay of charge and spin coherence in {Landau-Zener-St\"uckelberg-Majorana}
           interferometry},
  author = {Ribeiro, Hugo and Petta, J. R. and Burkard, Guido},
  journal = {Phys. Rev. B},
  volume = {87},
  issue = {23},
  pages = {235318},
  numpages = {12},
  year = {2013},
  month = {Jun},
  publisher = {American Physical Society},
  doi = {10.1103/PhysRevB.87.235318},
  url = {https://link.aps.org/doi/10.1103/PhysRevB.87.235318}
}

@article{darw31,
  title = {The Diamagnetism of the Free Electron},
  author = {Darwin, C. G.},
  journal = {Mathematical Proceedings of the Cambridge Philosophical Society},
  volume = {27},
  issue = {1},
  pages = {86--90},
  year = {1931},
  doi = {10.1017/S0305004100009373},
  url = {https://doi.org/10.1017/S0305004100009373}
}

@Article{Fock28,
author={Fock, V.},
title={Bemerkung zur Quantelung des harmonischen Oszillators im Magnetfeld},
journal={Zeitschrift {f\"ur} Physik},
year={1928},
month={May},
day={01},
volume={47},
issue={5},
pages={446-448},
doi={10.1007/BF01390750},
url={https://doi.org/10.1007/BF01390750}
}

@article{joch24,
  title = {Realization of a Laughlin State of Two Rapidly Rotating Fermions},
  author = {Lunt, Philipp and Hill, Paul and Reiter, Johannes and Preiss, Philipp M.
            and Ga\l{}ka, Maciej and Jochim, Selim},
  journal = {Phys. Rev. Lett.},
  volume = {133},
  issue = {25},
  pages = {253401},
  numpages = {7},
  year = {2024},
  month = {Dec},
  publisher = {American Physical Society},
  doi = {10.1103/PhysRevLett.133.253401},
  url = {https://link.aps.org/doi/10.1103/PhysRevLett.133.253401}
}

@article{popp04,
  title = {Adiabatic path to fractional quantum Hall states of a few bosonic atoms},
  author = {Popp, M. and Paredes, B. and Cirac, J. I.},
  journal = {Phys. Rev. A},
  volume = {70},
  issue = {5},
  pages = {053612},
  numpages = {6},
  year = {2004},
  month = {Nov},
  publisher = {American Physical Society},
  doi = {10.1103/PhysRevA.70.053612},
  url = {https://link.aps.org/doi/10.1103/PhysRevA.70.053612}
}

@article{Coop08,
author = {N. R. Cooper},
title = {Rapidly rotating atomic gases},
journal = {Advances in Physics},
volume = {57},
issue = {6},
pages = {539--616},
year = {2008},
publisher = {Taylor \& Francis},
doi = {10.1080/00018730802564122},
URL = {https://doi.org/10.1080/00018730802564122},
}

@article{hazz08,
  title = {Stirring trapped atoms into fractional quantum Hall puddles},
  author = {Baur, Stefan K. and Hazzard, Kaden R. A. and Mueller, Erich J.},
  journal = {Phys. Rev. A},
  volume = {78},
  issue = {6},
  pages = {061608},
  numpages = {4},
  year = {2008},
  month = {Dec},
  publisher = {American Physical Society},
  doi = {10.1103/PhysRevA.78.061608},
  url = {https://link.aps.org/doi/10.1103/PhysRevA.78.061608}
}

@article{Palm20,
doi = {10.1088/1367-2630/aba30e},
url = {https://doi.org/10.1088/1367-2630/aba30e},
year = {2020},
month = {aug},
publisher = {IOP Publishing},
volume = {22},
issue = {8},
pages = {083037},
author = {Palm, L and Grusdt, F and Preiss, P M},
title = {Skyrmion ground states of rapidly rotating few-fermion systems},
journal = {New Journal of Physics},
}

@article{yann10,
  title = {Quantal molecular description and universal aspects of the spectra of bosons
           and fermions in the lowest {Landau} level},
  author = {Yannouleas, Constantine and Landman, Uzi},
  journal = {Phys. Rev. A},
  volume = {81},
  issue = {2},
  pages = {023609},
  numpages = {15},
  year = {2010},
  month = {Feb},
  publisher = {American Physical Society},
  doi = {10.1103/PhysRevA.81.023609},
  url = {https://link.aps.org/doi/10.1103/PhysRevA.81.023609}
}

@Article{grus23,
author={L{\'e}onard, Julian
and Kim, Sooshin
and Kwan, Joyce
and Segura, Perrin
and Grusdt, Fabian
and Repellin, C{\'e}cile
and Goldman, Nathan
and Greiner, Markus},
title={Realization of a fractional quantum {Hall} state with ultracold atoms},
journal={Nature},
year={2023},
month={Jul},
day={01},
volume={619},
issue={7970},
pages={495-499},
doi={10.1038/s41586-023-06122-4},
url={https://doi.org/10.1038/s41586-023-06122-4}
}

@article{yann02.3,
  title = {Trial wave functions with long-range Coulomb correlations for two-dimensional
  {$N$-electron} systems in high magnetic fields},
  author = {Yannouleas, Constantine and Landman, Uzi},
  journal = {Phys. Rev. B},
  volume = {66},
  issue = {11},
  pages = {115315},
  numpages = {5},
  year = {2002},
  month = {Sep},
  publisher = {American Physical Society},
  doi = {10.1103/PhysRevB.66.115315},
  url = {https://link.aps.org/doi/10.1103/PhysRevB.66.115315}
}

@article{halp83,
  title ={Theory of the quantized {Hall} conductance},
  author ={Bertrand I. Halperin},
  journal ={Helvetica Physica Acta},
  year ={1983},
  volume ={56},
  pages ={75-102},
  url ={https://api.semanticscholar.org/CorpusID:117562779}
}

@article{grus24,
doi={10.1103/Physics.17.178},
url = {https://doi.org/10.1103/Physics.17.178 },
year = {2024},
volume = {17},
pages = {178},
author = {Fabian Grusdt},
title = {Ultracold Fermions Enter the Fractional Quantum Hall Arena},
journal = {Physics},
}

@article{girv83,
  title = {Interacting electrons in two-dimensional Landau levels: Results for
           small clusters},
  author = {Girvin, S. M. and Jach, Terrence},
  journal = {Phys. Rev. B},
  volume = {28},
  issue = {8},
  pages = {4506--4509},
  numpages = {0},
  year = {1983},
  month = {Oct},
  publisher = {American Physical Society},
  doi = {10.1103/PhysRevB.28.4506},
  url = {https://link.aps.org/doi/10.1103/PhysRevB.28.4506}
}

@article{ston92,
  title = {Edge waves in the quantum Hall effect and quantum dots},
  author = {Stone, Michael and Wyld, H. W. and Schult, R. L.},
  journal = {Phys. Rev. B},
  volume = {45},
  issue = {24},
  pages = {14156--14161},
  numpages = {0},
  year = {1992},
  month = {Jun},
  publisher = {American Physical Society},
  doi = {10.1103/PhysRevB.45.14156},
  url = {https://link.aps.org/doi/10.1103/PhysRevB.45.14156}
}

@article{mott99,
  title = {Yrast Spectra of Weakly Interacting Bose-Einstein Condensates},
  author = {Mottelson, B.},
  journal = {Phys. Rev. Lett.},
  volume = {83},
  issue = {14},
  pages = {2695--2698},
  numpages = {0},
  year = {1999},
  month = {Oct},
  publisher = {American Physical Society},
  doi = {10.1103/PhysRevLett.83.2695},
  url = {https://link.aps.org/doi/10.1103/PhysRevLett.83.2695}
}

@article{pape01,
  title = {Rotational spectra of weakly interacting Bose-Einstein condensates},
  author = {Papenbrock, Thomas and Bertsch, George F.},
  journal = {Phys. Rev. A},
  volume = {63},
  issue = {2},
  pages = {023616},
  numpages = {5},
  year = {2001},
  month = {Jan},
  publisher = {American Physical Society},
  doi = {10.1103/PhysRevA.63.023616},
  url = {https://link.aps.org/doi/10.1103/PhysRevA.63.023616}
}

@article{ueda01,
  title = {Low-lying excitations from the yrast line of weakly interacting trapped bosons},
  author = {Nakajima, Tatsuya and Ueda, Masahito},
  journal = {Phys. Rev. A},
  volume = {63},
  issue = {4},
  pages = {043610},
  numpages = {4},
  year = {2001},
  month = {Mar},
  publisher = {American Physical Society},
  doi = {10.1103/PhysRevA.63.043610},
  url = {https://link.aps.org/doi/10.1103/PhysRevA.63.043610}
}

@article{trug85,
  title = {Exact results for the fractional quantum Hall effect with general interactions},
  author = {Trugman, S. A. and Kivelson, S.},
  journal = {Phys. Rev. B},
  volume = {31},
  issue = {8},
  pages = {5280--5284},
  numpages = {0},
  year = {1985},
  month = {Apr},
  publisher = {American Physical Society},
  doi = {10.1103/PhysRevB.31.5280},
  url = {https://link.aps.org/doi/10.1103/PhysRevB.31.5280}
}

@article{enge07,
    author = {Jiang, Hong and Engel, Eberhard},
    title = {Random-phase-approximation-based correlation energy functionals:
             Benchmark results for atoms},
    journal = {The Journal of Chemical Physics},
    volume = {127},
    issue = {18},
    pages = {184108},
    year = {2007},
    month = {11},
    doi = {10.1063/1.2795707},
    url = {https://doi.org/10.1063/1.2795707},
}

@incollection{simo20,
  title = "{Wavefunctionology: The Special Structure of Certain Fractional
            Quantum Hall Wavefunctions} in",
  booktitle     = "{{Fractional Quantum Hall Effects: New Developments}}",
  author = "Steven H. Simon",
  editor    = "B. I. Halperin and J. K. Jain",
  chapter = {8},
  pages = {377-434},
  year      = "2020",
  publisher = "World Scientific",
  address   = "Hackensack",
  doi = "10.1142/9789811217494_0008",
  url = "https://doi.org/10.1142/9789811217494_0008"
}

@article{vlec36,
  title = {Nonorthogonality and Ferromagnetism},
  author = {van Vleck, J. H.},
  journal = {Phys. Rev.},
  volume = {49},
  issue = {3},
  pages = {232--240},
  numpages = {0},
  year = {1936},
  month = {Feb},
  publisher = {American Physical Society},
  doi = {10.1103/PhysRev.49.232},
  url = {https://link.aps.org/doi/10.1103/PhysRev.49.232}
}

@article{ande67,
  title = {Infrared Catastrophe in Fermi Gases with Local Scattering Potentials},
  author = {Anderson, P. W.},
  journal = {Phys. Rev. Lett.},
  volume = {18},
  issue = {24},
  pages = {1049--1051},
  numpages = {0},
  year = {1967},
  month = {Jun},
  publisher = {American Physical Society},
  doi = {10.1103/PhysRevLett.18.1049},
  url = {https://link.aps.org/doi/10.1103/PhysRevLett.18.1049}
}

@article{kohn99,
  title = {Nobel Lecture: Electronic structure of matter-wave functions and
           density functionals},
  author = {Kohn, W.},
  journal = {Rev. Mod. Phys.},
  volume = {71},
  issue = {5},
  pages = {1253--1266},
  numpages = {0},
  year = {1999},
  month = {Oct},
  publisher = {American Physical Society},
  doi = {10.1103/RevModPhys.71.1253},
  url = {https://link.aps.org/doi/10.1103/RevModPhys.71.1253}
}

@article{deml12,
  title = {Time-Dependent Impurity in Ultracold Fermions:
           Orthogonality Catastrophe and Beyond},
  author = {Knap, Michael and Shashi, Aditya and Nishida, Yusuke and Imambekov,
            Adilet and Abanin, Dmitry A. and Demler, Eugene},
  journal = {Phys. Rev. X},
  volume = {2},
  issue = {4},
  pages = {041020},
  numpages = {17},
  year = {2012},
  month = {Dec},
  publisher = {American Physical Society},
  doi = {10.1103/PhysRevX.2.041020},
  url = {https://link.aps.org/doi/10.1103/PhysRevX.2.041020}
}

@article{ares18,
  title = {Orthogonality catastrophe and fractional exclusion statistics},
  author = {Ares, Filiberto and Gupta, Kumar S. and de Queiroz, Amilcar R.},
  journal = {Phys. Rev. E},
  volume = {97},
  issue = {2},
  pages = {022133},
  numpages = {6},
  year = {2018},
  month = {Feb},
  publisher = {American Physical Society},
  doi = {10.1103/PhysRevE.97.022133},
  url = {https://link.aps.org/doi/10.1103/PhysRevE.97.022133}
}

@article{fukuy78,
  title = {Pinning and conductivity of two-dimensional charge-density waves in
           magnetic fields},
  author = {Fukuyama, Hidetoshi and Lee, Patrick A.},
  journal = {Phys. Rev. B},
  volume = {18},
  issue = {11},
  pages = {6245--6252},
  numpages = {0},
  year = {1978},
  month = {Dec},
  publisher = {American Physical Society},
  doi = {10.1103/PhysRevB.18.6245},
  url = {https://link.aps.org/doi/10.1103/PhysRevB.18.6245}
}

@article{maki83,
  title = {Static and dynamic properties of a two-dimensional Wigner crystal in a
           strong magnetic field},
  author = {Maki, Kazumi and Zotos, Xenophon},
  journal = {Phys. Rev. B},
  volume = {28},
  issue = {8},
  pages = {4349--4356},
  numpages = {0},
  year = {1983},
  month = {Oct},
  publisher = {American Physical Society},
  doi = {10.1103/PhysRevB.28.4349},
  url = {https://link.aps.org/doi/10.1103/PhysRevB.28.4349}
}

@article{girv84,
  title = {Liquid-solid transition and the fractional quantum-Hall effect},
  author = {Lam, Pui K. and Girvin, S. M.},
  journal = {Phys. Rev. B},
  volume = {30},
  issue = {1},
  pages = {473--475},
  numpages = {0},
  year = {1984},
  month = {Jul},
  publisher = {American Physical Society},
  doi = {10.1103/PhysRevB.30.473},
  url = {https://link.aps.org/doi/10.1103/PhysRevB.30.473}
}

@article{tsui10,
  title = {Observation of a Pinning Mode in a Wigner Solid with $\ensuremath{\nu}=1/3$
           Fractional Quantum Hall Excitations},
  author = {Zhu, Han and Chen, Yong P. and Jiang, P. and Engel, L. W. and Tsui, D. C.
            and Pfeiffer, L. N. and West, K. W.},
  journal = {Phys. Rev. Lett.},
  volume = {105},
  issue = {12},
  pages = {126803},
  numpages = {4},
  year = {2010},
  month = {Sep},
  publisher = {American Physical Society},
  doi = {10.1103/PhysRevLett.105.126803},
  url = {https://link.aps.org/doi/10.1103/PhysRevLett.105.126803}
}

@article{liu26,
author ={Liu, Yufeng and Gu, Yu and Bao, Ting and Mao, Ning and Jiang, Shudan and
         Liu, Liang and Guan, Dandan and Li, Yaoyi and Zheng, Hao and Liu, Canhua,
	 {\it et al}},
 title = {Imaging moiré flat bands and Wigner molecular crystals in twisted bilayer MoTe2},
 journal = {National Science Review},
 volume = {13},
 issue = {4},
 pages = {nwag014},
 year = {2026},
 month = {01},
    doi = {10.1093/nsr/nwag014},
    url = {https://doi.org/10.1093/nsr/nwag014},
}

@article{schu93,
  title = {Wigner crystal in one dimension},
  author = {Schulz, H. J.},
  journal = {Phys. Rev. Lett.},
  volume = {71},
  issue = {12},
  pages = {1864--1867},
  numpages = {0},
  year = {1993},
  month = {Sep},
  publisher = {American Physical Society},
  doi = {10.1103/PhysRevLett.71.1864},
  url = {https://link.aps.org/doi/10.1103/PhysRevLett.71.1864}
}

@article{Jaur93,
doi = {10.1209/0295-5075/24/7/013},
url = {https://doi.org/10.1209/0295-5075/24/7/013},
year = {1993},
month = {dec},
publisher = {},
volume = {24},
issue = {7},
pages = {581},
author = {K. Jauregui and W. Häusler and B. Kramer},
title = {Wigner Molecules in Nanostructures},
journal = {Europhysics Letters},
}

@article{szaf04.3,
  title = {Spatial ordering of charge and spin in quasi-one-dimensional Wigner molecules},
  author = {Szafran, B. and Peeters, F. M. and Bednarek, S. and Chwiej, T.
            and Adamowski, J.},
  journal = {Phys. Rev. B},
  volume = {70},
  issue = {3},
  pages = {035401},
  numpages = {9},
  year = {2004},
  month = {Jul},
  publisher = {American Physical Society},
  doi = {10.1103/PhysRevB.70.035401},
  url = {https://link.aps.org/doi/10.1103/PhysRevB.70.035401}
}

@Article{esco19,
	title={{A Wigner molecule at extremely low densities: a numerically exact study}},
	author={Miguel Escobar Azor and Léa Brooke and Stefano Evangelisti and
	        Thierry Leininger and Pierre-François Loos and Nicolas Suaud and
		J. A. Berger},
	journal={SciPost Phys. Core},
	volume={1},
	pages={001},
	year={2019},
	publisher={SciPost},
	doi={10.21468/SciPostPhysCore.1.1.001},
	url={https://scipost.org/10.21468/SciPostPhysCore.1.1.001},
}

@article{vu20,
  title = {One-dimensional few-electron effective Wigner crystal in quantum and
           classical regimes},
  author = {Vu, DinhDuy and Das Sarma, S.},
  journal = {Phys. Rev. B},
  volume = {101},
  issue = {12},
  pages = {125113},
  numpages = {14},
  year = {2020},
  month = {Mar},
  publisher = {American Physical Society},
  doi = {10.1103/PhysRevB.101.125113},
  url = {https://link.aps.org/doi/10.1103/PhysRevB.101.125113}
}

@article{haus93,
  title = {Interacting electrons in a one-dimensional quantum dot},
  author = {H\"ausler, Wolfgang and Kramer, Bernhard},
  journal = {Phys. Rev. B},
  volume = {47},
  issue = {24},
  pages = {16353--16357},
  numpages = {0},
  year = {1993},
  month = {Jun},
  publisher = {American Physical Society},
  doi = {10.1103/PhysRevB.47.16353},
  url = {https://link.aps.org/doi/10.1103/PhysRevB.47.16353}
}

@Article{Wend96,
author={Wendler, L.
and Fomin, V. M.
and Chaplik, A. V.
and Govorov, A. O.},
title={Energy spectra of two interacting electrons in a quantum ring:
       rotating Wigner molecule},
journal={Zeitschrift f{\"u}r Physik B: Condensed Matter},
year={1996},
month={Mar},
day={01},
volume={100},
issue={2},
pages={211-221},
doi={10.1007/s002570050115},
url={https://doi.org/10.1007/s002570050115}
}

@article{bao06,
  title = {Few-electron quantum rings in a magnetic field: Ground-state properties},
  author = {Liu, Y. M. and Bao, C. G. and Shi, T. Y.},
  journal = {Phys. Rev. B},
  volume = {73},
  issue = {11},
  pages = {113313},
  numpages = {4},
  year = {2006},
  month = {Mar},
  publisher = {American Physical Society},
  doi = {10.1103/PhysRevB.73.113313},
  url = {https://link.aps.org/doi/10.1103/PhysRevB.73.113313},
}

@article{szaf05,
  title = {Few-electron eigenstates of concentric double quantum rings},
  author = {Szafran, B. and Peeters, F. M.},
  journal = {Phys. Rev. B},
  volume = {72},
  issue = {15},
  pages = {155316},
  numpages = {9},
  year = {2005},
  month = {Oct},
  publisher = {American Physical Society},
  doi = {10.1103/PhysRevB.72.155316},
  url = {https://link.aps.org/doi/10.1103/PhysRevB.72.155316}
}

@article{yann17,
  title = {Trial wave functions for ring-trapped ions and neutral atoms: Microscopic
           description of the quantum space-time crystal},
  author = {Yannouleas, Constantine and Landman, Uzi},
  journal = {Phys. Rev. A},
  volume = {96},
  issue = {4},
  pages = {043610},
  numpages = {12},
  year = {2017},
  month = {Oct},
  publisher = {American Physical Society},
  doi = {10.1103/PhysRevA.96.043610},
  url = {https://link.aps.org/doi/10.1103/PhysRevA.96.043610}
}

@article{piac04,
  title = {Generic properties of a quasi-one-dimensional classical Wigner crystal},
  author = {Piacente, G. and Schweigert, I. V. and Betouras, J. J. and Peeters, F. M.},
  journal = {Phys. Rev. B},
  volume = {69},
  issue = {4},
  pages = {045324},
  numpages = {17},
  year = {2004},
  month = {Jan},
  publisher = {American Physical Society},
  doi = {10.1103/PhysRevB.69.045324},
  url = {https://link.aps.org/doi/10.1103/PhysRevB.69.045324}
}

@article{crem11,
  title = {Signatures of Wigner localization in epitaxially grown nanowires},
  author = {Kristinsd\'ottir, L. H. and Cremon, J. C. and Nilsson, H. A. and
            Xu, H. Q. and Samuelson, L. and Linke, H. and Wacker, A. and Reimann, S. M.},
  collaboration = {Nanometer Structure Consortium, nmC@LU},
  journal = {Phys. Rev. B},
  volume = {83},
  issue = {4},
  pages = {041101},
  numpages = {4},
  year = {2011},
  month = {Jan},
  publisher = {American Physical Society},
  doi = {10.1103/PhysRevB.83.041101},
  url = {https://link.aps.org/doi/10.1103/PhysRevB.83.041101}
}

@article{wuns08,
  title = {Electron-electron interactions and charging effects in graphene quantum dots},
  author = {Wunsch, B. and Stauber, T. and Guinea, F.},
  journal = {Phys. Rev. B},
  volume = {77},
  issue = {3},
  pages = {035316},
  numpages = {9},
  year = {2008},
  month = {Jan},
  publisher = {American Physical Society},
  doi = {10.1103/PhysRevB.77.035316},
  url = {https://link.aps.org/doi/10.1103/PhysRevB.77.035316}
}

@article{shap19,
author = {I. Shapir and A. Hamo and S. Pecker and C. P. Moca and {\"O}. Legeza and G. Zarand
          and S. Ilani},
title = {Imaging the electronic {Wigner crystal} in one dimension},
journal = {Science},
volume = {364},
issue = {6443},
pages = {870--875},
year = {2019},
doi = {10.1126/science.aat0905},
URL = {https://www.science.org/doi/abs/10.1126/science.aat0905},
}

@article{roy12,
  title = {Effective mass theory of interacting electron states in semiconducting
           carbon nanotube quantum dots},
  author = {Roy, Mervyn and Maksym, P. A.},
  journal = {Phys. Rev. B},
  volume = {85},
  issue = {20},
  pages = {205432},
  numpages = {12},
  year = {2012},
  month = {May},
  publisher = {American Physical Society},
  doi = {10.1103/PhysRevB.85.205432},
  url = {https://link.aps.org/doi/10.1103/PhysRevB.85.205432}
}

@article{Paga18,
doi = {10.1088/2058-9565/aae0fe},
url = {https://doi.org/10.1088/2058-9565/aae0fe},
year = {2018},
month = {oct},
publisher = {IOP Publishing},
volume = {4},
number = {1},
pages = {014004},
author = {Pagano, G and Hess, P W and Kaplan, H B and Tan, W L and Richerme, P and
          Becker, P and Kyprianidis, A and Zhang, J and Birckelbaw, E and
	  Hernandez, M R and Wu, Y and Monroe, C},
title = {Cryogenic trapped-ion system for large scale quantum simulation},
journal = {Quantum Science and Technology},
}

@article{haef17,
  title = {Realization of Translational Symmetry in Trapped Cold Ion Rings},
  author = {Li, Hao-Kun and Urban, Erik and Noel, Crystal and Chuang, Alexander and
            Xia, Yang and Ransford, Anthony and Hemmerling, Boerge and Wang, Yuan and
	    Li, Tongcang and H\"affner, Hartmut and Zhang, Xiang},
  journal = {Phys. Rev. Lett.},
  volume = {118},
  issue = {5},
  pages = {053001},
  numpages = {5},
  year = {2017},
  month = {Jan},
  publisher = {American Physical Society},
  doi = {10.1103/PhysRevLett.118.053001},
  url = {https://link.aps.org/doi/10.1103/PhysRevLett.118.053001}
}

@Article{ejte23,
author={Zhang, J.
and Chow, B. T.
and Ejtemaee, S.
and Haljan, P. C.},
title={Spectroscopic characterization of the quantum linear-zigzag transition in trapped ions},
journal={npj Quantum Information},
year={2023},
month={Jul},
day={13},
volume={9},
issue={1},
pages={68},
doi={10.1038/s41534-023-00741-5},
url={https://doi.org/10.1038/s41534-023-00741-5}
}

@article{yann24.3,
  title = {Electronic {Wigner-molecule} polymeric chains in elongated silicon quantum dots
           and finite-length quantum wires},
  author = {Goldberg, Arnon and Yannouleas, Constantine and Landman, Uzi},
  journal = {Phys. Rev. Appl.},
  volume = {21},
  issue = {6},
  pages = {064063},
  numpages = {10},
  year = {2024},
  month = {Jun},
  publisher = {American Physical Society},
  doi = {10.1103/PhysRevApplied.21.064063},
  url = {https://link.aps.org/doi/10.1103/PhysRevApplied.21.064063}
}

@misc{gull26.2,
      title={Interacting electrons in silicon quantum interconnects}, 
      author={Anantha S. Rao and Christopher David White and Sean R. Muleady and
              Anthony Sigillito and Michael J. Gullans},
      year={2026},
      eprint={2601.05306},
      archivePrefix={arXiv},
      primaryClass={cond-mat.mes-hall},
      url={https://arxiv.org/abs/2601.05306}, 
}

@Article{crom24.2,
author={Li, Hongyuan
and Xiang, Ziyu
and Wang, Tianle
and Naik, Mit H.
and Kim, Woochang
and Nie, Jiahui
and Li, Shiyu
and Ge, Zhehao
and He, Zehao
and Ou, Yunbo
and Banerjee, Rounak
and Taniguchi, Takashi
and Watanabe, Kenji
and Tongay, Sefaattin
and Zettl, Alex
and Louie, Steven G.
and Zaletel, Michael P.
and Crommie, Michael F.
and Wang, Feng},
title={Imaging tunable {Luttinger} liquid systems in van der {Waals} heterostructures},
journal={Nature},
year={2024},
month={Jul},
day={01},
volume={631},
issue={8022},
pages={765-770},
doi={10.1038/s41586-024-07596-6},
url={https://doi.org/10.1038/s41586-024-07596-6}
}

@article{pepp18,
  title = {Imaging the Zigzag Wigner Crystal in Confinement-Tunable Quantum Wires},
  author = {Ho, Sheng-Chin and Chang, Heng-Jian and Chang, Chia-Hua and Lo, Shun-Tsung
        and Creeth, Graham and Kumar, Sanjeev and Farrer, Ian and Ritchie, David
	and Griffiths, Jonathan and Jones, Geraint and Pepper, Michael and Chen, Tse-Ming},
  journal = {Phys. Rev. Lett.},
  volume = {121},
  issue = {10},
  pages = {106801},
  numpages = {5},
  year = {2018},
  month = {Sep},
  publisher = {American Physical Society},
  doi = {10.1103/PhysRevLett.121.106801},
  url = {https://link.aps.org/doi/10.1103/PhysRevLett.121.106801}
}

@article{warr22,
  title = {Silicon qubits move a step closer to achieving error correction},
  author = {Ada Warren and Sophia E. Economou},
  journal = {Nature},
  volume = {601},
  pages = {320--322},
  year = {2022},
  doi = {10.1038/d41586-022-00047-0},
  url = {https://doi.org/10.1038/d41586-022-00047-0}
}

@article{gull26,
  title = {Electrical interconnects for silicon spin qubits},
  author = {White, Christopher David and Sigillito, Anthony and Gullans, Michael J.},
  journal = {Phys. Rev. B},
  volume = {113},
  issue = {8},
  pages = {085301},
  numpages = {25},
  year = {2026},
  month = {Feb},
  publisher = {American Physical Society},
  doi = {10.1103/tmpl-pjvw},
  url = {https://link.aps.org/doi/10.1103/tmpl-pjvw}
}

@article{awsc21,
  title = {Development of Quantum Interconnects (QuICs) for Next-Generation Information
           Technologies},
  author = {Awschalom, David and Berggren, Karl K. and Bernien, Hannes and Bhave, Sunil and
  Carr, Lincoln D. and Davids, Paul and Economou, Sophia E. and Englund, Dirk and
  Faraon, Andrei {\it et al\/}},
  journal = {PRX Quantum},
  volume = {2},
  issue = {1},
  pages = {017002},
  numpages = {21},
  year = {2021},
  month = {Feb},
  publisher = {American Physical Society},
  doi = {10.1103/PRXQuantum.2.017002},
  url = {https://link.aps.org/doi/10.1103/PRXQuantum.2.017002}
}

@article{dzur23,
author = {Wang, Zeheng and Feng, MengKe and Serrano, Santiago and Gilbert, William and
          Leon, Ross C. C. and Tanttu, Tuomo and Mai, Philip and Liang, Dylan and
	  Huang, Jonathan Y. and Su, Yue and Lim, Wee Han and Hudson, Fay E. and
	  Escott, Christopher C. and Morello, Andrea and Yang, Chih Hwan and
	  Dzurak, Andrew S. and Saraiva, Andre and Laucht, Arne},
title = {Jellybean Quantum Dots in Silicon for Qubit Coupling and On-Chip Quantum Chemistry},
journal = {Advanced Materials},
volume = {35},
issue = {19},
pages = {2208557},
doi = {https://doi.org/10.1002/adma.202208557},
url = {https://onlinelibrary.wiley.com/doi/abs/10.1002/adma.202208557},
year = {2023}
}

@article{kuem24,
  title = {Elongated quantum dot as a distributed charge sensor},
  author = {Patom\"aki, S. M. and Williams, J. and Berritta, F. and Lain\'e, C. and
            Fogarty, M. A. and Leon, R. C. C. and Jussot, J. and Kubicek, S. and
	    Chatterjee, A. and Govoreanu, B. and Kuemmeth, F. and Morton, J. J. L. and
	    Gonzalez-Zalba, M. F.},
  journal = {Phys. Rev. Appl.},
  volume = {21},
  issue = {5},
  pages = {054042},
  numpages = {16},
  year = {2024},
  month = {May},
  publisher = {American Physical Society},
  doi = {10.1103/PhysRevApplied.21.054042},
  url = {https://link.aps.org/doi/10.1103/PhysRevApplied.21.054042}
}

@article{schi93,
  title = {Phase transitions in anisotropically confined ionic crystals},
  author = {Schiffer, J. P.},
  journal = {Phys. Rev. Lett.},
  volume = {70},
  issue = {6},
  pages = {818--821},
  numpages = {0},
  year = {1993},
  month = {Feb},
  publisher = {American Physical Society},
  doi = {10.1103/PhysRevLett.70.818},
  url = {https://link.aps.org/doi/10.1103/PhysRevLett.70.818}
}

@phdthesis{ejte15,
    title        = {{Dynamics of Trapped Ions Near the Linear–Zigzag Structural
                     Phase Transition}},
    author       = {Sara Ejtemaee},
    year         = 2015,
    month        = {},
    address      = {},
    note         = {Available at \url{https://summit.sfu.ca/item/16248}},
    school       = {Simon Fraser University},
    type         = {PhD thesis}
}

@Article{yan16,
author={Yan, L. L.
and Wan, W.
and Chen, L.
and Zhou, F.
and Gong, S. J.
and Tong, X.
and Feng, M.},
title={Exploring structural phase transitions of ion crystals},
journal={Scientific Reports},
year={2016},
month={Feb},
day={11},
volume={6},
issue={1},
pages={21547},
issn={2045-2322},
doi={10.1038/srep21547},
url={https://doi.org/10.1038/srep21547}
}

@article{paca70,
  title = {Slater Determinants, Parity Projection, and {Hartree-Fock} Calculations},
  author = {Pacati, F. D. and Boffi, S.},
  journal = {Phys. Rev. C},
  volume = {2},
  issue = {4},
  pages = {1205--1210},
  numpages = {0},
  year = {1970},
  month = {Oct},
  publisher = {American Physical Society},
  doi = {10.1103/PhysRevC.2.1205},
  url = {https://link.aps.org/doi/10.1103/PhysRevC.2.1205}
}

@article{Meye08,
doi = {10.1088/0953-8984/21/2/023203},
url = {https://doi.org/10.1088/0953-8984/21/2/023203},
year = {2008},
month = {dec},
publisher = {},
volume = {21},
issue = {2},
pages = {023203},
author = {Meyer, Julia S and Matveev, K A},
title = {Wigner crystal physics in quantum wires},
journal = {Journal of Physics: Condensed Matter},
}

@article{nish06,
  title = {Intermediate low spin states in a few-electron quantum dot in the
  $\nu \leq 1$ regime},
  author = {Nishi, Y. and Maksym, P. A. and Austing, D. G. and Hatano, T. and
            Kouwenhoven, L. P. and Aoki, H. and Tarucha, S.},
  journal = {Phys. Rev. B},
  volume = {74},
  issue = {3},
  pages = {033306},
  numpages = {4},
  year = {2006},
  month = {Jul},
  publisher = {American Physical Society},
  doi = {10.1103/PhysRevB.74.033306},
  url = {https://link.aps.org/doi/10.1103/PhysRevB.74.033306}
}

@article{retz08,
  title = {Double Well Potentials and Quantum Phase Transitions in Ion Traps},
  author = {Retzker, A. and Thompson, R. C. and Segal, D. M. and Plenio, M. B.},
  journal = {Phys. Rev. Lett.},
  volume = {101},
  issue = {26},
  pages = {260504},
  numpages = {4},
  year = {2008},
  month = {Dec},
  publisher = {American Physical Society},
  doi = {10.1103/PhysRevLett.101.260504},
  url = {https://link.aps.org/doi/10.1103/PhysRevLett.101.260504}
}

@article{fras16,
 URL = {https://www.jstor.org/stable/26551758},
 author = {James D. Fraser},
 journal = {Philosophy of Science},
 number = {4},
 pages = {585--605},
 publisher = {[The University of Chicago Press, Philosophy of Science Association]},
 title = {Spontaneous Symmetry Breaking in Finite Systems},
 urldate = {2026-03-03},
 volume = {83},
 year = {2016}
}

@Article{Seif20,
author={Seifert, Vanessa A.},
title={The strong emergence of molecular structure},
journal={European Journal for Philosophy of Science},
year={2020},
month={Oct},
day={01},
volume={10},
issue={3},
pages={45},
doi={10.1007/s13194-020-00308-7},
url={https://doi.org/10.1007/s13194-020-00308-7}
}

@incollection{hend19,
  title = "{ Prospects for strong emergence in chemistry} in",
  booktitle     = "{{Philosophical and scientific perspectives on downward causation}}",
  author = "Hendry, R. F.",
  editor    = "M. P. Paoletti and F. Orilia",
  pages = {146-163},
  year      = "2019",
  publisher = "Routledge",
  address   = "New York",
  url = "https://www.routledge.com/Philosophical-and-Scientific-Perspectives-on-Downward-Causation/PaoliniPaoletti-Orilia/p/book/9780367372309"
}

@Article{Lomb25,
author={Lombardi, Olimpia
and Villani, Giovanni},
title={About the Concept of Molecular Structure},
journal={Foundations of Science},
year={2025},
month={Dec},
day={01},
volume={30},
issue={4},
pages={1003-1020},
doi={10.1007/s10699-024-09963-y},
url={https://doi.org/10.1007/s10699-024-09963-y}
}

@Article{Wool25,
author={Woolley, R. Guy},
title={Comment on The Born-Oppenheimer approximation and its role in the reduction
       of chemistry},
journal={Foundations of Chemistry},
year={2025},
month={Dec},
day={01},
volume={27},
issue={3},
pages={373-380},
doi={10.1007/s10698-025-09552-2},
url={https://doi.org/10.1007/s10698-025-09552-2}
}

@Article{Scer25,
author={Scerri, Eric},
title={The Born-Oppenheimer Approximation and its role in the reduction of chemistry},
journal={Foundations of Chemistry},
year={2025},
month={Aug},
day={01},
volume={27},
issue={2},
pages={183-197},
doi={10.1007/s10698-025-09543-3},
url={https://doi.org/10.1007/s10698-025-09543-3}
}

@article{maty11,
  title = {On the emergence of molecular structure},
  author = {M\'atyus, Edit and Hutter, J\"urg and M\"uller-Herold, Ulrich and Reiher, Markus},
  journal = {Phys. Rev. A},
  volume = {83},
  issue = {5},
  pages = {052512},
  numpages = {5},
  year = {2011},
  month = {May},
  publisher = {American Physical Society},
  doi = {10.1103/PhysRevA.83.052512},
  url = {https://link.aps.org/doi/10.1103/PhysRevA.83.052512}
}

@article{maty11.2,
    author = {M\'atyus, Edit and Hutter, Jürg and M\"uller-Herold, Ulrich and
              Reiher, Markus},
    title = {Extracting elements of molecular structure from the all-particle wave function},
    journal = {The Journal of Chemical Physics},
    volume = {135},
    issue = {20},
    pages = {204302},
    year = {2011},
    month = {11},
    doi = {10.1063/1.3662487},
    url = {https://doi.org/10.1063/1.3662487},
}

@article{Maty19,
author = {Edit M\'atyus},
title = {Pre-Born–Oppenheimer molecular structure theory},
journal = {Molecular Physics},
volume = {117},
issue = {5},
pages = {590--609},
year = {2019},
publisher = {Taylor \& Francis},
doi = {10.1080/00268976.2018.1530461},
URL = {https://doi.org/10.1080/00268976.2018.1530461},
}

@article{naka11,
    author = {Hoshino, Minoru and Nishizawa, Hiroaki and Nakai, Hiromi},
    title = {Rigorous non-Born-Oppenheimer theory: Combination of explicitly correlated
             Gaussian method and nuclear orbital plus molecular orbital theory},
    journal = {The Journal of Chemical Physics},
    volume = {135},
    issue = {2},
    pages = {024111},
    year = {2011},
    month = {07},
    doi = {10.1063/1.3609806},
    url = {https://doi.org/10.1063/1.3609806},
}

@article{lang24,
author = {Lang, Lucas and Cezar, Henrique M. and Adamowicz, Ludwik and Pedersen, Thomas B.},
title = {Quantum Definition of Molecular Structure},
journal = {Journal of the American Chemical Society},
volume = {146},
issue = {3},
pages = {1760-1764},
year = {2024},
doi = {10.1021/jacs.3c11467},
URL = {https://doi.org/10.1021/jacs.3c11467},
}

@article{agos22,
    author = {Agostini, Federica and Curchod, Basile F. E.},
    title = {Chemistry without the Born–Oppenheimer approximation},
    journal = {Philosophical Transactions of the Royal Society A: Mathematical,
               Physical and Engineering Sciences},
    volume = {380},
    pages = {20200375},
    year = {2022},
    doi = {10.1098/rsta.2020.0375},
    url = {https://doi.org/10.1098/rsta.2020.0375},
}

@article{reih19,
    author = {Muolo, Andrea and M\'atyus, Edit and Reiher, Markus},
    title = {H$_3^+$ as a five-body problem described with explicitly correlated
             Gaussian basis sets},
    journal = {The Journal of Chemical Physics},
    volume = {151},
    issue = {15},
    pages = {154110},
    year = {2019},
    month = {10},
    doi = {10.1063/1.5121318},
    url = {https://doi.org/10.1063/1.5121318},
}

@article{reih23,
author = {Feldmann, Robin and Baiardi, Alberto and Reiher, Markus},
title = {Symmetry-Projected Nuclear-Electronic Hartree–Fock: Eliminating Rotational
         Energy Contamination},
journal = {The Journal of Physical Chemistry A},
volume = {127},
issue = {42},
pages = {8943-8954},
year = {2023},
doi = {10.1021/acs.jpca.3c04822},
URL = {https://doi.org/10.1021/acs.jpca.3c04822},
}

@article{reye25,
    author = {Moncada, Félix and Reyes, Andr\'es and Pettersson, Lars G. M.},
    title = {Restoring rotational symmetry of multicomponent wavefunctions with
             nuclear orbitals},
    journal = {The Journal of Chemical Physics},
    volume = {162},
    issue = {2},
    pages = {024110},
    year = {2025},
    month = {01},
    doi = {10.1063/5.0244318},
    url = {https://doi.org/10.1063/5.0244318},
}

@article{okuy11,
  title = {Optical Aharonov-Bohm effect on Wigner molecules in type-II semiconductor
           quantum dots},
  author = {Okuyama, Rin and Eto, Mikio and Hyuga, Hiroyuki},
  journal = {Phys. Rev. B},
  volume = {83},
  issue = {19},
  pages = {195311},
  numpages = {9},
  year = {2011},
  month = {May},
  publisher = {American Physical Society},
  doi = {10.1103/PhysRevB.83.195311},
  url = {https://link.aps.org/doi/10.1103/PhysRevB.83.195311}
}

@article{okuy16,
author = {Kim, Hee Dae and Okuyama, Rin and Kyhm, Kwangseuk and Eto, Mikio and
          Taylor, Robert A. and Nicolet, Aurelien L. and Potemski, Marek and
	  Nogues, Gilles and Dang, Le Si and Je, Ku-Chul and Kim, Jongsu and
	  Kyhm, Ji-Hoon and Yoen, Kyu Hyoek and Lee, Eun Hye and Kim, Jun Young and
	  Han, Il Ki and Choi, Wonjun and Song, Jindong},
title = {Observation of a Biexciton Wigner Molecule by Fractional Optical Aharonov-Bohm
         Oscillations in a Single Quantum Ring},
journal = {Nano Letters},
volume = {16},
issue = {1},
pages = {27-33},
year = {2016},
doi = {10.1021/acs.nanolett.5b02419},
URL = {https://doi.org/10.1021/acs.nanolett.5b02419},
}

@article{okuy18,
author = {Kim, Heedae and Park, Seongho and Okuyama, Rin and Kyhm, Kwangseuk and
          Eto, Mikio and Taylor, Robert A. and Nogues, Gilles and Dang, Le Si and
	  Potemski, Marek and Je, Koochul and Kim, Jongsu and Kyhm, Jihoon and Song, Jindong},
title = {Light Controlled Optical Aharonov-Bohm Oscillations in a Single Quantum Ring},
journal = {Nano Letters},
volume = {18},
issue = {10},
pages = {6188-6194},
year = {2018},
doi = {10.1021/acs.nanolett.8b02131},
URL = {https://doi.org/10.1021/acs.nanolett.8b02131},
}

@article{yann07.4,
author = {Ihn, Thomas and Ellenberger, Christoph and Ensslin, Klaus and Yannouleas, Constantine
          and Landman, Uzi and Driscoll, Dan C. and Gossard, Art C.},
title = {Quantum dots based on parabolic quantum wells: importance of electronic correlations},
journal = {International Journal of Modern Physics B},
volume = {21},
issue = {08n09},
pages = {1316-1325},
year = {2007},
doi = {10.1142/S0217979207042781},
URL = {https://doi.org/10.1142/S0217979207042781},
}

@article{kouw97.2,
author = {L. P. Kouwenhoven  and T. H. Oosterkamp  and M. W. S. Danoesastro and M. Eto and
          D. G. Austing  and T. Honda  and S. Tarucha },
title = {Excitation Spectra of Circular, Few-Electron Quantum Dots},
journal = {Science},
volume = {278},
issue = {5344},
pages = {1788-1792},
year = {1997},
doi = {10.1126/science.278.5344.1788},
URL = {https://www.science.org/doi/abs/10.1126/science.278.5344.1788},
}

@article{chap92,
  title = {Spin-singlet--spin-triplet oscillations in quantum dots},
  author = {Wagner, M. and Merkt, U. and Chaplik, A. V.},
  journal = {Phys. Rev. B},
  volume = {45},
  issue = {4},
  pages = {1951--1954},
  numpages = {0},
  year = {1992},
  month = {Jan},
  publisher = {American Physical Society},
  doi = {10.1103/PhysRevB.45.1951},
  url = {https://link.aps.org/doi/10.1103/PhysRevB.45.1951}
}

@article{shi07,
  title = {Composite fermion solid and liquid states in two component quantum dots},
  author = {Shi, Chuntai and Jeon, Gun Sang and Jain, Jainendra K.},
  journal = {Phys. Rev. B},
  volume = {75},
  issue = {16},
  pages = {165302},
  numpages = {10},
  year = {2007},
  month = {Apr},
  publisher = {American Physical Society},
  doi = {10.1103/PhysRevB.75.165302},
  url = {https://link.aps.org/doi/10.1103/PhysRevB.75.165302}
}

@article{yosh83,
  title = {Ground-state energy of a two-dimensional charge-density-wave state in a
           strong magnetic field},
  author = {Yoshioka, D. and Lee, P. A.},
  journal = {Phys. Rev. B},
  volume = {27},
  issue = {8},
  pages = {4986--4996},
  numpages = {0},
  year = {1983},
  month = {Apr},
  publisher = {American Physical Society},
  doi = {10.1103/PhysRevB.27.4986},
  url = {https://link.aps.org/doi/10.1103/PhysRevB.27.4986}
}

@article{hawr93,
  title = {Magnetoluminescence from correlated electrons in quantum dots},
  author = {Hawrylak, Pawel and Pfannkuche, Daniela},
  journal = {Phys. Rev. Lett.},
  volume = {70},
  issue = {4},
  pages = {485--488},
  numpages = {0},
  year = {1993},
  month = {Jan},
  publisher = {American Physical Society},
  doi = {10.1103/PhysRevLett.70.485},
  url = {https://link.aps.org/doi/10.1103/PhysRevLett.70.485}
}

@article{Ruan99,
doi = {10.1088/0953-8984/11/2/010},
url = {https://doi.org/10.1088/0953-8984/11/2/010},
year = {1999},
month = {jan},
publisher = {},
volume = {11},
issue = {2},
pages = {435},
author = {W Y Ruan and Ho-Fai Cheung},
title = {Correlations and ground-state transitions of few-electron quantum dots in a
         strong magnetic field},
journal = {Journal of Physics: Condensed Matter},
}

@article{Maks00,
doi = {10.1088/0953-8984/12/22/201},
url = {https://doi.org/10.1088/0953-8984/12/22/201},
year = {2000},
month = {jun},
publisher = {},
volume = {12},
issue = {22},
pages = {R299},
author = {P A Maksym and H Imamura and G P Mallon and H Aoki},
title = {Molecular aspects of electron correlation in quantum dots},
journal = {Journal of Physics: Condensed Matter},
}

@article{yang93,
  title = {Addition spectra of quantum dots in strong magnetic fields},
  author = {Yang, S.-R. Eric and MacDonald, A. H. and Johnson, M. D.},
  journal = {Phys. Rev. Lett.},
  volume = {71},
  issue = {19},
  pages = {3194--3197},
  numpages = {0},
  year = {1993},
  month = {Nov},
  publisher = {American Physical Society},
  doi = {10.1103/PhysRevLett.71.3194},
  url = {https://link.aps.org/doi/10.1103/PhysRevLett.71.3194}
}

@article{jeon04,
  title = {Composite fermion theory of correlated electrons in semiconductor quantum dots
           in high magnetic fields},
  author = {Jeon, Gun Sang and Chang, Chia-Chen and Jain, Jainendra K.},
  journal = {Phys. Rev. B},
  volume = {69},
  issue = {24},
  pages = {241304},
  numpages = {4},
  year = {2004},
  month = {Jun},
  publisher = {American Physical Society},
  doi = {10.1103/PhysRevB.69.241304},
  url = {https://link.aps.org/doi/10.1103/PhysRevB.69.241304}
}

@Article{lozo99,
author={Astrakharchik, G. E.
and Belousov, A. I.
and Lozovik, Yu. E.},
title={Two-dimensional mesoscopic dusty plasma clusters: Structure and phase transitions},
journal={Journal of Experimental and Theoretical Physics},
year={1999},
month={Oct},
day={01},
volume={89},
issue={4},
pages={696-703},
doi={10.1134/1.559030},
url={https://doi.org/10.1134/1.559030}
}

@article{beda94,
  title = {Ordering and phase transitions of charged particles in a classical finite
           two-dimensional system},
  author = {Bedanov, Vladimir M. and Peeters, F. M.},
  journal = {Phys. Rev. B},
  volume = {49},
  issue = {4},
  pages = {2667--2676},
  numpages = {0},
  year = {1994},
  month = {Jan},
  publisher = {American Physical Society},
  doi = {10.1103/PhysRevB.49.2667},
  url = {https://link.aps.org/doi/10.1103/PhysRevB.49.2667}
}

@article{LOZO90,
title = {Coulomb clusters in a trap},
journal = {Physics Letters A},
volume = {145},
issue = {5},
pages = {269-271},
year = {1990},
issn = {0375-9601},
doi = {https://doi.org/10.1016/0375-9601(90)90362-R},
url = {https://www.sciencedirect.com/science/article/pii/037596019090362R},
author = {Yu.E. Lozovik and V.A. Mandelshtam},
}

@MISC{hydr25,
    TITLE = {Hydrogen wave function for electron orbitals},
    AUTHOR = {Claire},
    year = {2025},
    HOWPUBLISHED = {{\it Physics Stack Exchange\/}},
    URL = {https://physics.stackexchange.com/q/740549}
}

@misc{gali26,
  author = {{David D. Nolte}},
  title = {Galileo unbound -- 100 Years of Quantum Physics: Schr\"odinger’s
           Wave Mechanics (1926)},
  url = {https://wp.me/pa5Pv3-2s6},
  year = {2026}
}

@misc{cen26,
  author = {{Ananya Palivela}},
  title = {{C\&EN news -- The Schr\"odinger equation turns 100}},
  url = {https://cen.acs.org/physical-chemistry/computational-chemistry/Schrdinger-equation-turns-100/104/web/2026/02},
  year = {2026}
}

@conference{esi26,
  organization = {Organizers: Christoph Dellago, Wolfgang Reiter, Norbert Schuch and
               Jakob Yngvason, ESI, Univerity of Vienna},
  title = {{The World in One Line – Schr\"odinger’s Equation Turns 100}},
  url = {https://www.esi.ac.at/events/e596/},
  year = {2026}
}

\end{document}